\documentclass[12pt, a4paper]{article}
\usepackage{graphicx}
\usepackage[labelfont=bf]{caption}
\usepackage{subcaption}
\graphicspath{{Figures/}}
\usepackage[inkscapelatex=false]{svg}
\usepackage[percent]{overpic}
\newcommand{\InFigTextSize}{\fontsize{8}{9.6}\selectfont}
\usepackage{psfrag}
\usepackage[utf8]{inputenc}
\usepackage{makecell}
\usepackage{ragged2e}
\usepackage{setspace}
\usepackage{multirow}
\usepackage{amsmath}
\usepackage{color}
\usepackage{epstopdf}
\usepackage{siunitx}
\usepackage{booktabs}
\usepackage{authblk}
\usepackage{ragged2e}
\usepackage{chngcntr}
\usepackage[margin=1in]{geometry}
\newcolumntype{L}{>{$}l<{$}}

\usepackage{float}
\usepackage{hyperref}
\usepackage[backend=biber,style=numeric-comp,sorting=none]{biblatex}
\defbibheading{bibliography}[]{}

\hypersetup{
    linktoc=all,        
    colorlinks=true,   
    linkcolor=black,    
    citecolor=blue,   
    urlcolor=blue,  
    filecolor=blue     
}

\title{\textbf{Transient Vibroacoustic Control of a Shock-Loaded Inter-Connected Cylindrical Double Shell}}
\author[1]{Rahim Vesal}
\author[2]{Seyyed M. Hasheminejad}
\author[1]{Hervé Lissek}

\affil[1]{Laboratory of Wave Engineering, Ecole Polytechnique Fédérale de Lausanne, 1015 Lausanne, Switzerland}
\affil[2]{Acoustics Research Laboratory, Iran University of Science and Technology, Narmak, Tehran 16846-13114, Iran}

\date{\today}

\begin{document}
\maketitle
\noindent This manuscript is a preprint and has been submitted for publication.
\renewcommand\thesubfigure{\alph{subfigure}}
\makeatletter
\renewcommand\p@subfigure{\thefigure-}
\makeatother
\section*{Abstract}
\pagenumbering{arabic}
Double-wall cylindrical shells are widely used in applications where resistance to acoustic shock loading is critical. While the transient vibroacoustic behavior of single-walled shells has been extensively investigated, extending these analyses to double-wall cylindrical configurations introduces increased complexity due to multiple inter-shell acoustic reflections and strong coupling between acoustic fields and structural vibrations. These structures often feature mechanical interconnections between the shells to ensure structural integrity, load sharing, alignment, and enhanced resilience against static and dynamic loads. These links introduce additional pathways for vibration transmission and significantly influence the overall behavior of the system, thus making the analytical description of the coupled vibroacoustic response even more challenging. This study investigates the transient vibroacoustics of an inter-connected double-wall cylindrical shell subjected to an acoustic shock, considering fully coupled fluid–structure interactions. A comprehensive two-dimensional acoustoelastic model is developed in polar coordinates, incorporating the surrounding medium, the fluid occupying the inter-shell gap, and the fluid inside the inner shell. A semi-analytical solution method is employed to capture the time-domain evolution of acoustic fields and shell vibrations. The model's accuracy is verified by benchmarking against available data reported in the literature. Leveraging the passive dynamics of the system, we present a hybrid mechanism that integrates optimized nonlinear vibration absorbers with piezoelectric actuators to control the vibroacoustic behavior. The results demonstrate the effectiveness of the proposed hybrid mechanism in mitigating shock-induced acoustic pressure waves and enhancing the structural resilience of the double-wall cylindrical shell.\\
\vspace{-10pt}\\
\textbf{Keywords:} Vibroacoustics, Active Structural Acoustic Control, Double-wall Structure, Hybrid Control, Nonlinear Vibration Absorber, Cylindrical Shell

\section*{Introduction}
Double-wall cylindrical shells find a wide range of applications across various industries. Aircraft fuselages, submarines, heat exchangers, and double-wall storage tanks are prominent examples of these structures, which can be modeled as coaxial cylindrical shells. One of the most crucial design considerations for these structures is ensuring their resilience against shock waves. Although substantial research has explored the response of single-walled shells to acoustic shocks \cite{mindlin1953response,haywood1958response,gregson2006coupled}, a review of the literature reveals a notable gap in studies specifically addressing double-wall configurations. Huang's 1979 analytical study is one of the most well-known studies on the response of double-wall shells to acoustic shocks \cite{huang1979transient}. Addressing several fundamental questions about the interaction between double-wall shell systems and acoustic pulses, this research has served as a key reference for subsequent investigations in the field \cite{mair1999benchmarks}. While emphasizing structural dynamics, this research overlooks the critical influence of fluid dynamics, which advances show to be the primary driver of interactions between structural vibrations and acoustic waves. Key phenomena such as the propagation of acoustic disturbances, the reflection of high-pressure acoustic fronts from solid surfaces, and the concentration of reflected waves at focal points contribute significantly to the intricate vibroacoustic behavior of these systems. Despite some experimental efforts to explore the response of double-wall shells, capturing the intricate role of fluid dynamics in structural-acoustic shock interactions remains a formidable challenge in the advancement of research in this field. In a system of two shells with additional weighted rings, Stultz et al. \cite{stultz1996single} study the influence of the outer shell dynamics on the inner shell without investigating the hydrodynamic fields arising from interactions with structural vibrations. Wardlaw and Luton \cite{wardlaw2000fluid} examine the vibroacoustic response of a double-wall shell subjected to an internal acoustic impulse, particularly to determine the potential for cavitation. Sigrist and Leblond \cite{sigrist2008methode} propose a semi-analytical model to study the interactions between an underwater impulse and a double-wall shell. Taking an application-driven approach, this research assesses the feasibility of implementing the developed model in practice and provides valuable insights into the emerging hydrodynamic fields and the corresponding structural responses. Expanding beyond the double shells, some other studies explore the response of cylindrical structures to acoustic shock loads, offering valuable insights into their interaction with structural vibrations. For example, Oakley et al. \cite{oakley1999investigation,oakley2001shock} explore shock waves' interactions with multiple cylinders enclosed within a larger cylindrical chamber and delve into the complex emerging reflection patterns. In a similar study, Sigrist et al. \cite{sigrist2007fluid} present a numerical model to examine vibroacoustics in an analogous configuration. Jialing and Hongli \cite{jialing1997study} also investigate shock wave reflections from cylindrical surfaces in a channel, providing a deeper understanding of wave scattering patterns.

While studies in the field of double-wall structures vibroacoustics are scarce, they mostly focus on frequency-domain analysis, such as those by Liu and He \cite{liu2016analytical,liu2016diffuse} and Danshjo et al. \cite{daneshjou2018investigation}, whereas transient behavior remains even more neglected. Although frequency domain studies yield valuable information, for example in sound transmission analysis, it is essential to inspect unsteady phenomena in the time domain. Rapid dynamics of the system, combined with the complexity of the vibroacoustic response shortly after an acoustic shock impacts the shell surface, are key features of shell-shock interactions that underscore the importance of studying this phenomenon in the time domain. Among the limited number of studies in this field, Iakovlev’s works stand out for their valuable contributions. At the core of these studies lies a solution method based on response functions to compute transient acoustic pressure with significantly improved computational efficiency. His main research on the transient response of single-walled cylindrical shells is briefly summarized as follows: development of a response function-based solution method \cite{iakovlev2002singular,iakovlev2007inverse}; shock wave scattering from the surface of a rigid cylinder \cite{iakovlev2011modeling}; vibroacoustic analysis of a submerged hollow cylindrical shell \cite{iakovlev2008interaction_1,iakovlev2008interaction_2}; transient response of a submerged solid cylindrical shell \cite{iakovlev2009interaction,iakovlev2006external,iakovlev2014transient,iakovlev2014resonance}. Throughout their work, Iakovlev et al. represent the validity of the developed solution method against experimental data and provide a detailed exploration of the complex vibroacoustic phenomena and the resulting hydrodynamic field, considering the effect of the fluid properties. Their research also ventures into the transient response of double-wall cylindrical structures. For example, they examine the vibroacoustics of a cylindrical shell with a rigid core at its center in a series of studies and explore how variations in the core dimensions influence the system’s response to acoustic shock excitation \cite{iakovlev2011modeling,iakovlev2004influence,iakovlev2010hydrodynamic}. Notably, Iakovlev et al. delve into the transient vibroacoustic behavior of a double-wall system featuring a hollow shell nested within another under acoustic shock loading \cite{iakovlev2015shock}. With a detailed examination of the intricate interplay between structural vibrations of both shells and acoustic media, their study draws a particular relevance to the present research. Yet, incorporating the inner-shell interior fluid into the model and its remarkable impact on the system’s vibroacoustics remain as unaddressed research gaps that the current study aims to bridge through its core contributions.

Upon establishing a clear understanding of double-wall structures dynamics, the path forward clearly points toward controlling their vibroacoustic response. Accordingly, several studies suggest equipping double-beam or double-panel structures with piezoelectric layers or inter-coupling them through tuned vibration absorbers to enhance their resistance against impact loading and control their transient vibroacoustic response \cite{hasheminejad2021numerical,hasheminejad2024energy}. However, this critical pursuit has also seen surprisingly little attention for double-wall shells, with only a handful of notable efforts emerging in the field. For example, Hasheminjad et al. implement sliding mode control method to mitigate sound transmission through electrorheological cylindrical double shells \cite{hasheminejad2020active}, and Hasheminjad and Jamalpour combine a set of active and semi-active controllers to develop a hybrid system for controlling the sound passing through a double-wall cylindrical shell \cite{hasheminejad2022control}. As evidenced by the literature, even the few available studies tend to focus on controlling the steady-state behavior, rather than addressing the more demanding challenges of the transient response under shock loading. 

In the present study, we explore the transient behavior of a double-wall cylindrical shell subjected to an acoustic shock, and introduce an efficient hybrid mechanism that integrates optimized nonlinear vibration absorbers (NVAs) with piezoelectric actuators to control the response of the system. For this purpose, a comprehensive two-dimensional acoustoelastic model is developed, where the fluid-solid-interactions are described through the governing fully coupled mathematical formulation, and a semi-analytical solution method is applied to obtain the numerical results. The accuracy of the developed model as well as the solution methodology is verified in comparison to the available data in the literature. By addressing the areas overlooked in prior studies, this work represents a refined model for double cylindrical shell vibroacoustics. The results capture the system’s vulnerability to deformation or potential failure, and evaluate the effect of hybrid control parameters to devise a robust strategy for enhancing structural resilience against acoustic shocks.

\section*{Development of vibroacoustic model for a double circular shell}
\begin{figure}[H]
\centering
    \includegraphics[width=0.99\columnwidth]{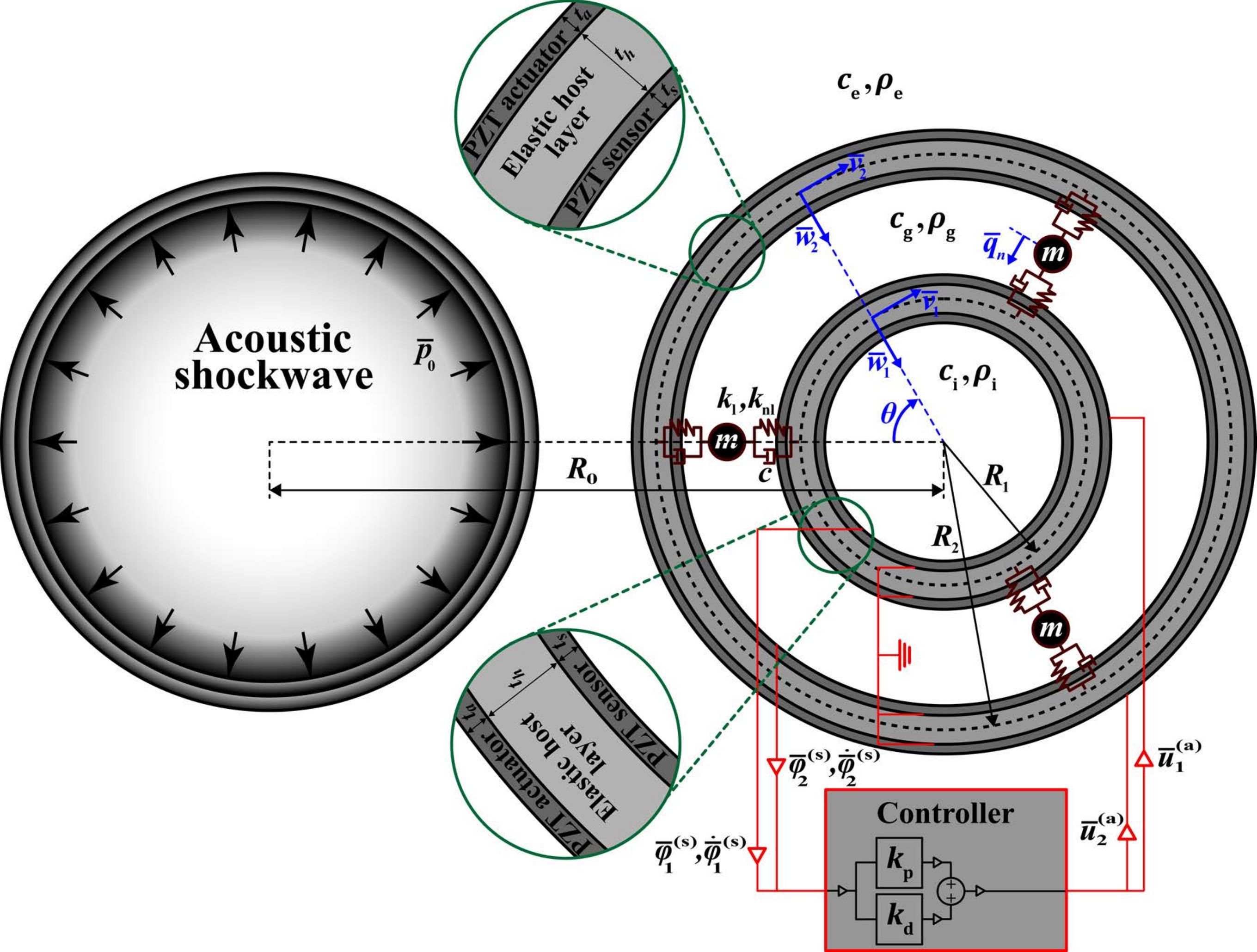}
    \caption{Sketch (cut view) of the double-shell structure}
    \label{fig:1}
\end{figure}

Fig. \ref{fig:1} illustrates a two-dimensional acoustoelastic model, where two coaxial cylindrical shells with radii of $R_1$ and $R_2$ (referred to the median surface of the shell) divide the acoustic medium into three compartments - namely interior, gap and exterior fluids. Each shell integrates a three-layered architecture; an elastic core sandwiched between an inner layer of piezoelectric sensor and an outer layer of piezoelectric actuator.The thicknesses of the elastic core, sensor, and actuator layers are denoted by $t_{\text{h}}$, $t_{\text{s}}$, and $t_{\text{a}}$, respectively. The piezoelectric layers are coated with thin electrodes, the mechanical effect of which is neglected due to their minimal thickness. The shells are mechanically interconnected via an array of $N$ NVAs arranged at equal angular intervals. Each absorber is characterized by a mass $m$, damping coefficient $c$, and linear and nonlinear stiffness coefficients $k_\text{l}$ and $k_{\text{nl}}$, respectively. Acoustic domains are assumed to consist of linear, compressible, inviscid, irrotational fluids, with $\rho_n$ and $c_n$ ($n={\text{i}},{\text{g}},{\text{e}}$) denoting the mass density and speed of sound at interior, gap and exterior media, respectively. The system is subjected to an incident acoustic disturbance, which is originated at a radial distance of $R_0$ from the axis of the double-shell structure.

Here, the governing equations underlying the dynamics of the system are presented. First, the wave propagation is described in polar coordinates as given by Eq.(\ref{Eq:1}), where $\bar{p}_j(\bar{r}, \theta, \tau)= -\rho_j \frac{\partial }{\partial \tau}\bar{\psi}_j(\bar{r}, \theta, \tau)$ denotes the acoustic pressure and $\bar{\psi}$ represents the corresponding velocity potential, as functions of the radial distance $\bar{r}$, angular position $\theta$, and time $\tau$ \cite{fahy2007sound}.
\begin{equation}
\left.
\begin{aligned}
\frac{\partial^2 \bar{p}_j}{\partial \bar{r}^2}
+ \frac{1}{\bar{r}} \frac{\partial \bar{p}_j}{\partial \bar{r}}
+ \frac{1}{\bar{r}^2} \frac{\partial^2 \bar{p}_j}{\partial \theta^2}
= \frac{1}{c_j^2} \frac{\partial^2 \bar{p}_j}{\partial \tau^2}\\
\frac{\partial^2 \bar{\psi}_j}{\partial \bar{r}^2}
+ \frac{1}{\bar{r}} \frac{\partial \bar{\psi}_j}{\partial \bar{r}}
+ \frac{1}{\bar{r}^2} \frac{\partial^2 \bar{\psi}_j}{\partial \theta^2}
= \frac{1}{c_j^2} \frac{\partial^2 \bar{\psi}_j}{\partial \tau^2}
\end{aligned}
\right\}
, \quad j \in \{\text{i}, \text{g}, \text{e}\}.
\label{Eq:1}
\end{equation}
The acoustic pressure in each subdomain can be expressed as a scalar sum of its contributing components. For the interior medium, it arises exclusively from the radiations from the inner shell vibrations, i.e., $\bar{p}_\text{i} = \bar{p}_\text{i}^{(r)}$. Likewise, in the inter-shell region, the total acoustic pressure results from superposition of wave fields radiated by both the inner and outer shells, i.e., $\bar{p}_\text{g} = \bar{p}_\text{g}^{(r)}$. The total pressure in the exterior acoustic field comprises the incident wave ($\bar{p}_0$) and its diffraction by the shell ($\bar{p}_\text{d}$) in addition to the radiated pressure from structural vibrations ($\bar{p}_\text{e}^{(r)}$) , i.e., $\bar{p}_\text{e} = \bar{p}_0 + \bar{p}_\text{d} + \bar{p}_\text{e}^{(r)}$. The incident shockwave can be mathematically expressed as a circular acoustic pulse in the form of Eq.(\ref{Eq:2}) \cite{iakovlev2009interaction,iakovlev2006external}:
\begin{equation}
\begin{gathered}
\bar{p}_0(\bar{r}, \theta, \tau) = \frac{\bar{p}_\alpha \bar{S}}{\bar{R}} \text{e}^{-\lambda^{-1}\left[\tau - {c_\text{e}}^{-1}(\bar{R} - \bar{S})\right]}  \text{H}\left[\tau - {c_\text{e}}^{-1}(\bar{R} - \bar{S})\right], \\
\bar{\psi}_0(\bar{r}, \theta, \tau) = -\frac{\lambda \bar{p}_\alpha \bar{S}}{\rho_\text{e} \bar{R}} \text{e}^{-\lambda^{-1}\left[\tau - {c_\text{e}}^{-1}(\bar{R} - \bar{S})\right]}  \text{H}\left[\tau - {c_\text{e}}^{-1}(\bar{R} - \bar{S})\right],
\end{gathered}
\label{Eq:2}
\end{equation}
where, $\bar{p}_\alpha$ is the peak incident pressure at the wavefront, $\lambda$ is the exponential decay constant, $\bar{R} = \sqrt{R_0^2 + \bar{r}^2 - 2 R_0 \bar{r} \cos \theta}$ denotes the radial distance from the source, $\bar{S} = R_0 - R_2$ is the minimum distance between the disturbance source and the outer shell. In addition, $\text{H}(\cdot)$ is the Heaviside function that constrains the wavefront to travel at the speed of sound. Examining the impact of spherical acoustic disturbances near the shell surface is both intellectually stimulating and practically valuable for engineering applications. Nevertheless, framing the incident wave through the simplified formulation of Eq.(\ref{Eq:2}) yields notable analytical advantages. First, the stress induced by a spherical acoustic excitation on cylindrical shells reaches its maximum value at the shell’s mid-plane. Considering the incident wave in the proposed form, the resulting mid-plane stress on the shell wall shows strong agreement with that obtained from a spherical excitation \cite{iakovlev2004influence}. Second, replacing the sophisticated spherical wave model with the simplified 2D approximation of Eq.(\ref{Eq:2}) substantially reduces computational costs while maintaining sufficient accuracy. As detailed in \hyperref[sec:appendixA]{Appendix A}, by replacing the effect of each absorber with an equivalent concentrated force at the attachment point (see \cite{forbes2008resonance}) and applying Hamilton’s principle, along with Kirchhoff-Love theory for thin shells and Maxwell’s electrodynamic equations, the system's electromechanical behavior can be described in four primary governing equations of (\ref{Eq:3}) through (\ref{Eq:6}) as \cite{hasheminejad2022control,ke2014thermo}:
\begin{equation}
\begin{gathered}
I_i \frac{d^2}{d\tau^2} \bar{v}_i(\theta, \tau) = \left( \frac{d_{22}}{R_i^4} + \frac{a_{22}}{R_i^2} \right) \frac{\partial^2}{\partial \theta^2} \bar{v}_i(\theta, \tau) + \left( \frac{d_{22}}{R_i^4} \frac{\partial^3}{\partial \theta^3} - \frac{a_{22}}{R_i^2} \frac{\partial}{\partial \theta} \right) \bar{w}_i(\theta, \tau) \\ 
+ \frac{E_{32}^{(\text{s})}}{R_i^2} \frac{\partial}{\partial \theta} \bar{\varphi}_i^{(\text{s})}(\theta, \tau) + \frac{E_{32}^{(\text{a})}}{R_i^2} \frac{\partial}{\partial \theta} \bar{\varphi}_i^{(\text{a})}(\theta, \tau), \quad i \in \{\text{1}, \text{2}\},
\end{gathered}
\label{Eq:3}
\end{equation}
\begin{equation}
\begin{gathered}
I_i \frac{d^2}{d\tau^2}\bar{w}_i(\theta, \tau) = 
\left( \frac{a_{22}}{R_i^2} \frac{\partial}{\partial \theta} - \frac{d_{22}}{R_i^4} \frac{\partial^3}{\partial \theta^3} \right) \bar{v}_i(\theta, \tau) \\
- \left( \frac{d_{22}}{R_i^4} \frac{\partial^4}{\partial \theta^4} + \frac{a_{22}}{R_i^2} \right)\bar{w}_i(\theta, \tau)
-\frac{E_{32}^{(\text{s})}}{R_i^2} \frac{\partial^2}{\partial \theta^2} \bar{\varphi}_i^{(\text{s})}(\theta, \tau) 
- \frac{E_{32}^{(\text{a})}}{R_i^2} \frac{\partial^2}{\partial \theta^2} \bar{\varphi}_i^{(\text{a})}(\theta, \tau) \\
- \frac{e_{32}^{(\text{a})}}{R_i} \bar{u}_i^{(\text{a})}(\theta, \tau) + \bar{p}_i^{\text{net}} + \frac{1}{R_i} \sum_{n=1}^N \delta(\theta - \theta_n) \bar{f}_n^{(i)}(\tau), \quad i \in \{\text{1}, \text{2}\},
\end{gathered}
\label{Eq:4}
\end{equation}
\begin{equation}
\begin{gathered}
\left( \frac{\zeta_{i,22}^{(j)}}{R_i^2} \frac{\partial^2}{\partial \theta^2} - \zeta_{i,33}^{(j)} \right) \bar{\varphi}_i^{(j)}(\theta, \tau) 
+ \frac{E_{32}^{(j)}}{R_i^2} \frac{\partial}{\partial \theta} \bar{v}_i(\theta, \tau) 
+ \frac{E_{32}^{(j)}}{R_i^2} \frac{\partial^2}{\partial \theta^2} \bar{w}_i(\theta, \tau) = 0, \\
\quad i \in \{1, 2\},\quad j \in \{\text{a}, \text{s}\},
\end{gathered}
\label{Eq:5}
\end{equation}
\begin{equation}
\begin{gathered}
m \frac{d^2}{d \tau^2} \bar{q}_n(\tau) + \bar{f}_n^{(1)}(\tau) + \bar{f}_n^{(2)}(\tau) = 0, \quad n \in \{1, 2, \dots, N\},
\end{gathered}
\label{Eq:6}
\end{equation}
where, \( \bar{v}_{1,2} \) and \( \bar{w}_{1,2} \) denote the tangential and normal displacements of each shell at the neutral axis, respectively. \( \bar{q}_n \) represents the transverse displacement of the \( n \)-th dynamic vibration absorber, and \( \theta_n \) refers to the angular position of the attachment point. The electric potential distribution in piezoelectric sensor and actuator layers are indicated by \( \bar{\varphi}_{1,2}^{(\text{a},\text{s})} \). \( \bar{u}_{1,2}^{(\text{a})} \) denotes the corresponding control voltage applied to the actuator layer on each shell, and \( \delta(.) \) is the Dirac delta function. \( e \) indicates the piezoelectric coefficient, and \( I \), \( \zeta \), \( a \), \( d \), and \( E \) are all system parameters described in \hyperref[sec:appendixA]{Appendix A}. \( \bar{f}_n^{(1,2)} \) shows the equivalent forces exerted by the \( n \)-th absorber on the corresponding shell. Similarly, \( \bar{p}_{1,2}^{\text{net}} \) refer to the net acoustic pressure acting on each shell. \( \bar{f}^{(1,2)} \) and \( \bar{p}^{\text{net}} \) can be mathematically expressed as Eqs. (\ref{Eq:7}) and (\ref{Eq:8}), respectively.
\begin{equation}
\begin{gathered}
\bar{f}_n^{(i)}(\tau) = c \frac{d}{d\tau} \bar{\Delta}_{i,n}(\tau) + k_\text{l} \bar{\Delta}_{i,n}(\tau) + k_{\text{nl}} \bar{\Delta}_{i,n}^3(\tau), \\
\bar{\Delta}_{i,n}(\tau) = \bar{q}_n(\tau) - \bar{w}_i(\theta_n, \tau), \quad i \in \{1,2\},\quad n \in \{1,2,\dots,N\},
\end{gathered}
\label{Eq:7}
\end{equation}
\begin{equation}
\begin{gathered}
\bar{p}_1^{\text{net}} = \left[ \bar{p}_\text{g} - \bar{p}_\text{i} \right] \big|_{\bar{r} = R_1}, \quad \bar{p}_2^{\text{net}} = \left[ \bar{p}_\text{e} - \bar{p}_\text{g} \right] \big|_{\bar{r} = R_2}.
\end{gathered}
\label{Eq:8}
\end{equation}
Given the physical essence of the system, the acoustic pressure must retain a finite amplitude at the center while attenuating with distance. Moreover, the cyclic geometry of the structure naturally introduces periodicity in the acoustic pressure, displacement field, and electric potential. Based on the fluid continuity at the structure interface, the boundary conditions governing the acoustic wave equations can be described by Eq. (\ref{Eq:9}).
\begin{equation}
\begin{gathered}
\left. \frac{1}{\rho_\text{e}} \frac{\partial}{\partial \bar{r}} \bar{p}_\text{e}^{(\text{r})} \right|_{\bar{r}=R_2} = \left. \frac{1}{\rho_\text{g}} \frac{\partial}{\partial \bar{r}} \bar{p}_\text{g}^{(\text{r})} \right|_{\bar{r}=R_2} = \frac{\partial^2}{\partial \tau^2} \bar{w}_2, \\
\left. \frac{1}{\rho_\text{g}} \frac{\partial}{\partial \bar{r}} \bar{p}_\text{g}^{(\text{r})} \right|_{\bar{r}=R_1} = \left. \frac{1}{\rho_\text{i}} \frac{\partial}{\partial \bar{r}} \bar{p}_\text{i}^{(\text{r})} \right|_{\bar{r}=R_1} = \frac{\partial^2}{\partial \tau^2} \bar{w}_1, \\
\left. \frac{\partial}{\partial \bar{r}} \bar{p}_0 \right|_{\bar{r}=R_2} = -\left. \frac{\partial }{\partial \bar{r}}\bar{p}_\text{d} \right|_{\bar{r}=R_2}.
\end{gathered}
\label{Eq:9}
\end{equation}
Prior to outlining the solution methodology, it is essential to first reformulate the governing equations in a non-dimensional form. An appropriate choice of the reference time scale allows us to precisely capture rapid and short-term acoustic and vibrational phenomena without the computational burden of impractically small time steps, and thereby significantly reduces computational cost and error. Additionally, presenting the results in a non-dimensional form facilitates comparison with the literature data and also provides a more comprehensible interpretation for the reader. Accordingly, the non-dimensional variables and functions used for this purpose are defined as given in Eq. (\ref{Eq:10}).
\begin{equation}
\begin{gathered}
\begin{bmatrix}
t \\
r \\
w(\theta,t) \\
v(\theta,t) \\
q(t)
\end{bmatrix}
= \frac{1}{R_2}
\begin{bmatrix}
c_\text{e} \tau \\
\bar{r} \\
\bar{w}(\theta,\tau) \\
\bar{v}(\theta,\tau) \\
\bar{q}(\tau)
\end{bmatrix},
\quad
\begin{bmatrix}
\phi^{(\text{s})}(\theta,t) \\
\phi^{(\text{a})}(\theta,t) \\
u^{(\text{a})}(\theta,t)
\end{bmatrix}
= \frac{1}{U_0}
\begin{bmatrix}
\bar{\phi}^{(\text{s})}(\theta,\tau) \\
\bar{\phi}^{(\text{a})}(\theta,\tau) \\
\bar{u}^{(\text{a})}(\theta,\tau)
\end{bmatrix},
\\
p(r,\theta,t) = \frac{\bar{p}(\bar{r},\theta,\tau)}{\rho_\text{e} c_\text{e}^2},
\quad
\psi(r,\theta,t) = \frac{\bar{\psi}(\bar{r},\theta,\tau)}{c_\text{e} R_2},
\end{gathered}
\label{Eq:10}
\end{equation}
where, \( U_{0} \) represents an arbitrary chosen reference electric potential. According to Eq. (\ref{Eq:10}) a dimensionless time unit corresponds to the time required for an acoustic wave to travel a distance equivalent to the outer shell's radius in the surrounding medium. Taking the boundary conditions into account and substituting Eq. (\ref{Eq:10}) into (\ref{Eq:1}) and (\ref{Eq:3}) through (\ref{Eq:6}), the governing equations can be reformulated in a non-dimensional structure, as shown in Eqs. (\ref{Eq:11}) to (\ref{Eq:15}).
\begin{equation}
\begin{gathered}
\frac{\partial^2 p_j}{\partial r^2} + \frac{1}{r} \frac{\partial p_j}{\partial r} + \frac{1}{r^2} \frac{\partial^2 p_j}{\partial \theta^2} = \alpha_j^2 \frac{\partial^2 p_j}{\partial t^2}, \\
\frac{\partial^2 \psi_j}{\partial r^2} + \frac{1}{r} \frac{\partial \psi_j}{\partial r} + \frac{1}{r^2} \frac{\partial^2 \psi_j}{\partial \theta^2} = \alpha_j^2 \frac{\partial^2 \psi_j}{\partial t^2}, \quad j \in \{\text{i},\text{g},\text{e}\}
\end{gathered}
\label{Eq:11}
\end{equation}
\begin{equation}
\begin{gathered}
\ddot{v}_i(\theta,t) = \\
\frac{R_2^2}{I_i c_{\text{e}}^2 R_i^2} \left[ \left( \frac{d_{22}}{R_i^2} + a_{22} \right) \frac{\partial^2}{\partial \theta^2}v_i(\theta,t) + \left( \frac{d_{22}}{R_i^2} \frac{\partial^3}{\partial \theta^3} - a_{22} \frac{\partial}{\partial \theta} \right) w_i(\theta,t) \right. \\
\left. + \frac{U_0}{R_2} \left( E_{32}^{(\text{s})} \frac{\partial}{\partial \theta} \varphi_i^{(\text{s})}(\theta,t) + E_{32}^{(\text{a})} \frac{\partial}{\partial \theta}\varphi_i^{(\text{a})}(\theta,t) \right) \right], \quad i \in \{1,2\},
\end{gathered}
\label{Eq:12}
\end{equation}
\begin{equation}
\begin{gathered}
\ddot{w}_i(\theta,t) = \\
\frac{R_2^2}{I_i c_{\text{e}}^2 R_i^2} \left[ \left( a_{22} \frac{\partial}{\partial \theta} - \frac{d_{22}}{R_i^2} \frac{\partial^3}{\partial \theta^3} \right)v_i(\theta,t) - \left( \frac{d_{22}}{R_i^2} \frac{\partial^4}{\partial \theta^4} + a_{22} \right) w_i(\theta,t) \right. \\
- \frac{U_0}{R_2} \left( E_{32}^{(\text{s})} \frac{\partial^2}{\partial \theta^2} \varphi_i^{(\text{s})}(\theta,t) + E_{32}^{(\text{a})} \frac{\partial^2}{\partial \theta^2} \varphi_i^{(\text{a})}(\theta,t) + R_i e_{32}^{(\text{a})} u_i^{(\text{a})}(\theta,t) \right) \\
+ R_i \sum_{n=1}^{N} \delta(\theta-\theta_n) \left( \frac{c c_\text{e}}{R_2} \dot{\Delta}_{i,n}(t) + k_\text{l} \Delta_{i,n}(t) + k_{\text{nl}} R_2^2 \Delta_{i,n}^3(t) \right) \\
+ \left. \frac{\rho_\text{e} c_{\text{e}}^2 R_i^2}{R_2} p_i^\text{net} \right], \quad i \in \{1,2\},
\end{gathered}
\label{Eq:13}
\end{equation}
\begin{equation}
\begin{gathered}
\frac{U_0}{E_{32}^{(j)} R_2} \left( {\zeta_{22}^{(j)}} \frac{\partial^2}{\partial \theta^2} - \zeta_{33}^{(j)}R_i^2 \right) \varphi_i^{(j)}(\theta,t) + \frac{\partial v_i(\theta,t)}{\partial \theta}  + \frac{\partial^2 w_i(\theta,t)}{\partial \theta^2} = 0, \\
p_1^\text{net} = \left. (p_\text{g} - p_\text{i}) \right|_{r=r_1}, \quad p_2^\text{net} = \left. (p_\text{e} - p_\text{g}) \right|_{r=r_2}, \\
\Delta_{i,n}(t) = q_n(t) - w_i(\theta_n,t), \quad i \in \{1,2\}, \quad j \in \{\text{a},\text{s}\},
\end{gathered}
\label{Eq:14}
\end{equation}
\begin{equation}
\begin{gathered}
\ddot{q}_n(t) + \frac{R_2}{m c_\text{e}^2} \sum_{i=1}^{2} \left[ c c_\text{e} \dot{\Delta}_{i,n}(t) + k_\text{l} R_2 \Delta_{i,n}(t) + k_\text{nl} R_2^3 \Delta_{i,n}^3(t) \right] = 0, \\
n \in \{1,2,\dots,N\},
\end{gathered}
\label{Eq:15}
\end{equation}
where \( \dot{x}(t) = \frac{\text{d} x(t)}{\text{d} t} \) and \( \ddot{x}(t) = \frac{\text{d}^2 x(t)}{\text{d} t^2} \) represent the first and second derivatives of a given function of \( x(t) \) with respect to the dimensionless time variable \( t \), respectively. Additionally, \( r_1 = \frac{R_1}{R_2} \) and \( r_2 = 1 \) denote the dimensionless radii of the inner and outer shells, and \( \alpha_j = \frac{c_{\text{e}}}{c_j} \) shows the sound speed ratio at each acoustic subdomain. The solution of the dimensionless wave equation (\ref{Eq:11}) is carried out using a response function-based approach, which is thoroughly discussed in Yakovlev's works \cite{iakovlev2009interaction,iakovlev2006external,iakovlev2014transient}. While these studies detail the evolution of the method across different acoustic subdomains, a unified framework for double-shell elastoacoustic models incorporating the interior, gap, and exterior media is yet to be established. Therefore, we present the principal steps of this approach in \hyperref[sec:appendixB]{Appendix B} to provide clarity and consolidate the methodology. Following the solution methodology, and considering the system’s angular periodicity and its boundary conditions, each component of the dimensionless displacement, electric potential, and acoustic pressure can be expressed as a modal expansion in the form of the equation set (\ref{Eq:16}).
\begin{equation}
\begin{gathered}
p_0(r,\theta,t) = \sum_{m=0}^{M} P_m^{(0)}(r,t) \cos m\theta, \quad p_\text{d}(r,\theta,t) = \sum_{m=0}^{M} P_m^{(\text{d})}(r,t) \cos m\theta, \\
p_\text{e}^{(\text{r})}(r,\theta,t) = \sum_{m=0}^{M} P_m^{(\text{r},\text{e})}(r,t) \cos m\theta, \quad p_\text{g}^{(\text{r})}(r,\theta,t) = \sum_{m=0}^{M} P_m^{(\text{r},\text{g})}(r,t) \cos m\theta, \\
p_\text{i}^{(\text{r})}(r,\theta,t) = \sum_{m=0}^{M} P_m^{(\text{r},\text{i})}(r,t) \cos m\theta, \\
w_i(\theta,t) = \sum_{m=0}^{M} W_m^{(i)}(t) \cos m\theta, \quad v_i(\theta,t) = \sum_{m=0}^{M} V_m^{(i)}(t) \sin m\theta, \\
\varphi_i^{(j)}(\theta,t) = \sum_{m=0}^{M} \varPhi_m^{(i,j)}(t) \cos m\theta, \quad i \in \{1,2\}, \quad j \in \{\text{a}, \text{s}\}.
\end{gathered}
\label{Eq:16}
\end{equation}
where \(M\) represents the truncation order of the series. To suppress the vibrations of piezoelectric shells and the resulting acoustic radiation, a PD control scheme is implemented that combines proportional and derivative feedback from the voltage output of the sensor layers to supply the actuator layers with control voltage, as described in Eq. (\ref{Eq:17}) \cite{bodaghi2012non,hasheminejad2022control}:
\begin{equation}
\begin{gathered}
u_i^{(\text{a})} = k_\text{p} \varphi_i^{(\text{s})} + k_\text{d} \dot{\varphi}_i^{(\text{s})}, \quad i \in \{1,2\},
\label{Eq:17}
\end{gathered}
\end{equation}
where \(k_\text{p}\) and \(k_\text{d}\) denote the proportional and derivative feedback gain constants, respectively. Incorporating the control law of Eq. (\ref{Eq:17}) as well as the modal series expansion of Eq. (\ref{Eq:16}) into the dimensionless motion equations (\ref{Eq:12}–\ref{Eq:15}) and applying modal orthogonality, the system reduces to a set of ordinary differential equations as shown in Eqs. (\ref{Eq:18})–(\ref{Eq:21}):
\begin{equation}
\begin{gathered}
\ddot{V}_m^{(i)}(t) = \frac{R_2^2}{I_i c_{\text{e}}^2 R_i^2} \left[ - \frac{m U_0}{R_2} \left( E_{32}^{(\text{s})} \varPhi_m^{(i,\text{s})}(t) + E_{32}^{(\text{a})} \varPhi_m^{(i,\text{a})}(t) \right) \right. \\
\left. -m^2 \left( \frac{d_{22}}{R_i^2} + a_{22} \right) V_m^{(i)}(t) + \left( \frac{m^3 d_{22}}{R_i^2} + m a_{22} \right) W_m^{(i)}(t) \right],
\end{gathered}
\label{Eq:18}
\end{equation}
\begin{equation}
\begin{gathered}
\ddot{W}_m^{(i)}(t) = \frac{R_2^2}{I_i c_{\text{e}}^2 R_i^2} \left\{ U_0 \frac{m^2}{R_2} \left(  E_{32}^{(\text{s})} \varPhi_m^{(i,\text{s})}(t) + E_{32}^{(\text{a})} \varPhi_m^{(i,\text{a})}(t) \right) \right. \\
+ e_{32}^{(\text{a})} U_0 \frac{R_i}{R_2} \left( k_\text{p} \varPhi_m^{(i,\text{s})}(t) + k_\text{d} \dot{\varPhi}_m^{(i,\text{s})}(t) \right) + \frac{R_i^2 c_{\text{e}}^2 \rho_\text{e}}{R_2 } P_m^{(i,\text{net})}\\
+ R_i \cos(m \theta_n) \sum_{n=1}^{N} \left[ k_\text{l} \left( q_n(t) - \sum_{k=0}^{M} W_k^{(i)}(t) \cos(k \theta_n) \right)  \right. \\
 + k_{\text{nl}} R_2^2 \left( q_n(t) - \sum_{k=0}^{M} W_k^{(i)}(t) \cos(k \theta_n) \right)^3
 \left. + \frac{c c_\text{e}}{R_2} \left( \dot{q}_n(t) - \sum_{k=0}^{M} \dot{W}_k^{(i)}(t) \cos(k \theta_n) \right) \right]  \\
+ \left. \left( m a_{22} + \frac{m^3 d_{22}}{R_i^2} \right) V_m^{(i)}(t)  - \left( \frac{m^4 d_{22}}{R_i^2} + a_{22} \right) W_m^{(i)}(t) \right\}, \\
P_m^{(1,\text{net})} = P_m^{(r,\text{g})}(r_1, t) - P_m^{(r,\text{i})}(r_1, t), \\
P_m^{(2,\text{net})} = P_m^{(0)}(r_2, t) + P_m^{(\text{d})}(r_2, t) + P_m^{(r,\text{e})}(r_2, t) - P_m^{(r,\text{g})}(r_2, t),
\end{gathered}
\label{Eq:19}
\end{equation}
\begin{equation}
\begin{gathered}
\varPhi_m^{(i,j)}(t) = \frac{E_{32}^{(j)} R_2}{U_0} \left( \frac{1} {m^2 \zeta_{22}^{(j)} + \zeta_{33}^{(j)} R_i^2} \right) \left[m V_m^{(i)}(t) - m^2 W_m^{(i)}(t)\right],
\end{gathered}
\label{Eq:20}
\end{equation}
\begin{equation}
\begin{gathered}
\ddot{q}_n(t) = - \frac{R_2}{m c_\text{e}^2} \sum_{i=1}^{2} \left[ k_\text{l} R_2 \left( q_n(t) - \sum_{k=0}^{M} W_k^{(i)}(t) \cos(k \theta_n) \right) \right. \\
+ k_\text{nl} R_2^3 \left( q_n(t) - \sum_{k=0}^{M} W_k^{(i)}(t) \cos(k \theta_n) \right)^3 
\left. + c c_\text{e} \left( \dot{q}_n(t) - \sum_{k=0}^{M} \dot{W}_k^{(i)}(t) \cos(k \theta_n) \right) \right],
\end{gathered}
\label{Eq:21}
\end{equation}
where $i \in \{1,2\}$, $j \in \{\text{a},\text{s}\}$, $n \in \{1,2,\dots,N\}$, and $m \in \{0,1,\dots,M\}$, and the modal components $P^{(0)}$, $P^{(\text{d})}$, $P^{(\text{r},\text{e})}$, $P^{(\text{r},\text{g})}$, and $P^{(\text{r},\text{i})}$ are defined in \hyperref[sec:appendixB]{Appendix B}.

\section*{Numerical simulation}
The solution methodology proceeds in two main stages. At each time step, first, the integrals in Eqs. (\ref{Eq:B.13}) and (\ref{Eq:B.14}) are numerically evaluated using the trapezoidal rule to obtain the modal acoustic pressure components. These components are then substituted into Eqs.(\ref{Eq:18})–(\ref{Eq:21}), where finite difference method is employed to discretize the differential operators and determine the unknowns at the next time step. Conveniently, the displacements of the vibration absorbers ($q_n$, $n \in {1,2,\dots,N}$), are directly evaluated at each time step as part of the solution procedure. Finally, once the modal components $W^{(1)}$, $V^{(1)}$, $W^{(2)}$, $V^{(2)}$, $P^{(0)}$, $P^{(\text{d})}$, $P^{(\text{r},\text{e})}$, $P^{(\text{r},\text{g})}$, and $P^{(\text{r},\text{i})}$ are determined, the displacements and electric potentials in the shells, as well as the acoustic pressure fields across all subdomains, can be readily computed using Eq. (\ref{Eq:16}). A Matlab code \cite{MATLAB:R2023b_u3} is developed to numerically solve the governing coupled system of ordinary differential equations and implement the control algorithm. Due to the large number of parameters and limitations in our computational resources, the numerical results are restricted to a limited set of parameters under specified conditions. Accordingly, piezoelectric layers are all composed of commercially available PZT-4, whereas the elastic layers in both shells are made of aluminum \cite{hasheminejad2021sound,farsangi2012levy,khorshidi2020theoretical}. The fluid in all three acoustic subdomains is considered to be water. Throughout the numerical simulations, the reference time ($t = 0$) is set to the instant when the incident wave initially reaches the outer shell. The performance of the finite difference method is closely tied to appropriate time-step selection that ensures the convergence and precision of the solution as well as computational efficiency. Given that acoustic fields generally exhibit more intricate behavior than structural displacements, the wave equation imposes the primary constraint on time-step selection. Following the discussion on the convergence of numerical solution in \cite{iakovlev2010hydrodynamic,iakovlev2008interaction_2,iakovlev2009interaction}, a dimensionless time step of $\delta t = 0.001$ is adopted throughout this chapter. Furthermore, the series in Eq. (\ref{Eq:16}) are truncated to $M = 150$ modes to optimize computational performance, while ensuring reliable representation of both the shell dynamics and the surrounding acoustic fields. Unless otherwise specified, all default numerical values corresponding to the quantities involved in the numerical simulations are summarized in Table \ref{table:1}.
\begin{table}[H]
\caption{Default values for double shell model parameters.}
\centering
\renewcommand{\arraystretch}{1.4}
\setlength{\tabcolsep}{12pt}
\begin{tabular}{@{}l@{\hspace{0.1cm}}l@{\hspace{0.1cm}}l@{}}
\toprule
\begin{tabular}[t]{l}
\textbf{Vibration absorbers} \\
\cmidrule(lr){1-1} \\ [-5ex]
$m = \SI{10}{\kilogram}$ \\ [-1ex]
$c = \SI{1e6}{\newton\!\cdot\!\second\per\meter}$ \\ [-1ex]
$k_\text{l} = \SI{1e9}{\newton/\meter}$ \\ [-1ex]
$k_\text{nl} = \SI{1e20}{\newton/\meter^3}$ \\ [0.5em]
\textbf{Incident Shockwave}\\
\cmidrule(lr){1-1} \\ [-5ex]
$R_0 = \SI{5}{\meter}$ \\ [-1ex]
$\bar{p}_\alpha = \SI{2.5e4}{\pascal}$ \\ [-1ex]
$\lambda = \SI{1.314e-4}{\second}$  \\ [0.5em]
\textbf{Acoustic fields} \\
\cmidrule(lr){1-1} \\ [-5ex]
$\rho_j = \SI{1000}{\kilo\gram\per\meter^3}$ \\ [-1ex]
$c_j = \SI{1400}{\meter/\second}$ \\ [0.1em]
\end{tabular}
\begin{tabular}[t]{l}
\textbf{PZT layers} \\
\cmidrule(lr){1-1} \\ [-5ex]
$t_i = \SI{3}{\milli\meter}$ \\ [-1ex]
$\rho_i = \SI{7500}{\kilo\gram\per\meter^3}$ \\ [-0.5ex]
$V_0 = \SI{1}{\volt}$ \\ [0.09em]
$e_{31}^{(i)} = e_{32}^{(i)} = \SI{-4.1}{\coulomb\per\meter^2}$ \\ [0.09em]
$e_{33}^{(i)} = \SI{14.1}{\coulomb\per\meter^2}$ \\ [0.09em]
$c_{11}^{(i)} = c_{22}^{(i)} = \SI{132}{\giga\pascal}$ \\ [0.09em]
$c_{12}^{(i)} = \SI{71}{\giga\pascal}$ \\ [0.09em]
$c_{13}^{(i)} = c_{23}^{(i)} = \SI{73}{\giga\pascal}$ \\ [0.09em]
$c_{33}^{(i)} = \SI{115}{\giga\pascal}$ \\ [0.09em]
$c_{66}^{(i)} = \SI{30.5}{\giga\pascal}$ \\
\end{tabular}
\begin{tabular}[t]{l}
\textbf{ } \\ [0.2em]
$\mu_{11}^{(i)} = \mu_{22}^{(i)} = \SI{7.124}{\nano\farad/\meter}$ \\ [0.09em]
$\mu_{33}^{(i)} = \SI{5.841}{\nano\farad/\meter}$ \\ [0.5em]
\textbf{Elastic layers} \\
\cmidrule(lr){1-1} \\ [-5ex]
$t_h = \SI{6}{\milli\meter}$ \\ [-1ex]
$\rho_h = \SI{2700}{\kilo\gram\per\meter^3}$ \\ [-1ex]
$E_h = \SI{70}{\giga\pascal}$ \\ [-1ex]
$\nu = 0.33$ \\ [-1ex]
$R_1 = \SI{0.5}{\meter}$ \\ [-1ex]
$R_2 = \SI{1}{\meter}$ \\ 
\cmidrule(lr){1-1} \\ [-5ex]
$i \in \{\text{a}, \text{s}\}$, \\ [-1ex]
$j \in \{\text{i}, \text{g}, \text{e}\}$.
\end{tabular}
\\
\bottomrule
\end{tabular}
\label{table:1}
\end{table}

\subsection*{Model verification}
\begin{figure}[H]
\centering
\begin{minipage}[t]{0.5\textwidth}
        \centering
        \begin{overpic}[height=0.2\textheight,trim= 0cm 0cm 0cm 0.0cm,clip]{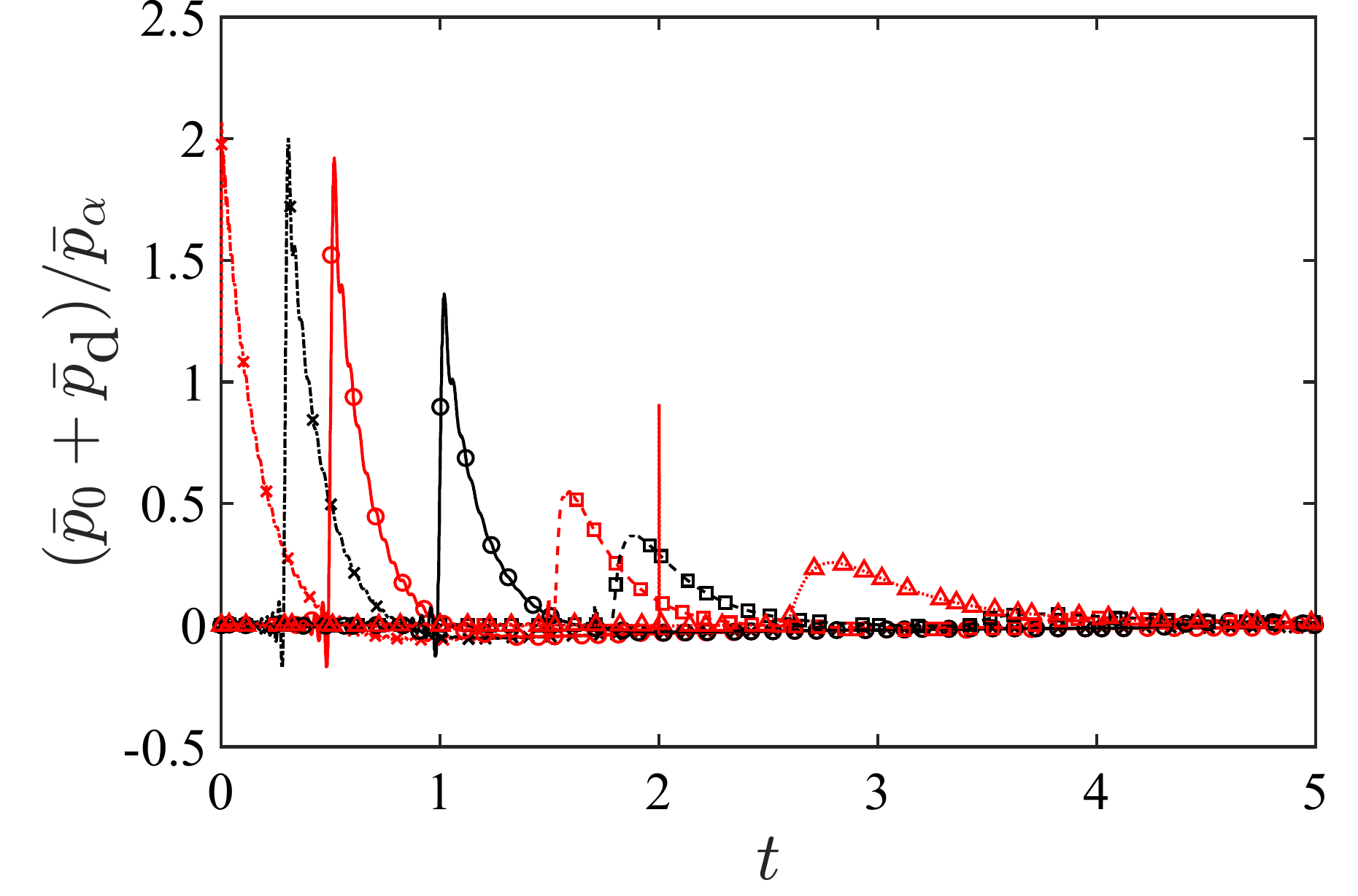}
        \put(60,27){\includegraphics[width=0.4\textwidth]{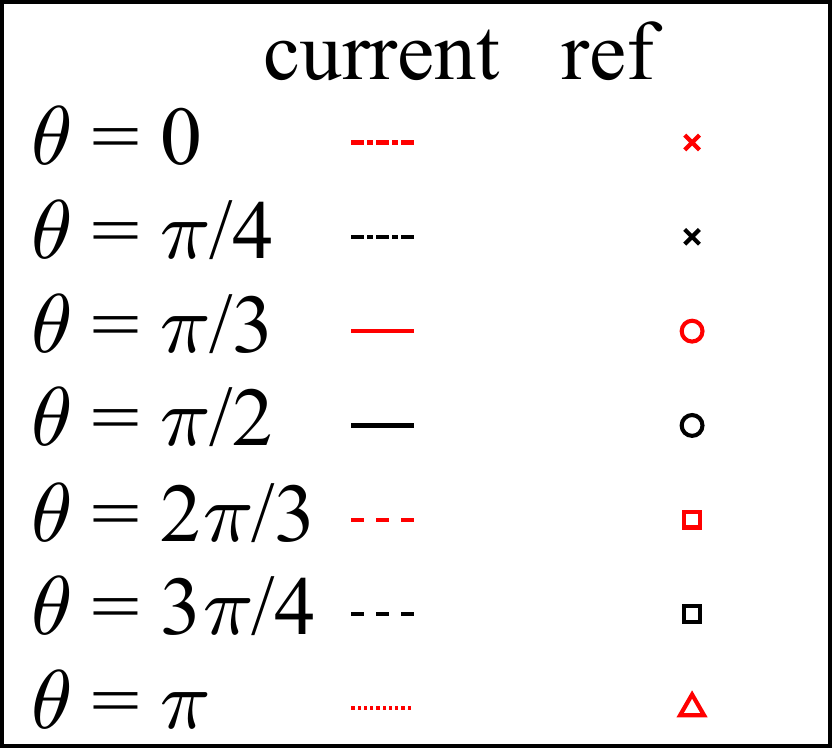}}
        \put(95.0,61.3){\color{black}{\InFigTextSize \cite{iakovlev2008interaction_2}}}
        \end{overpic}
        \subcaption{Normalized incident and diffracted pressure}
        \label{fig:2-a}
    \end{minipage}%
    \hfill
    \begin{minipage}[t]{0.5\textwidth}
        \centering
        \includegraphics[height=0.2\textheight]{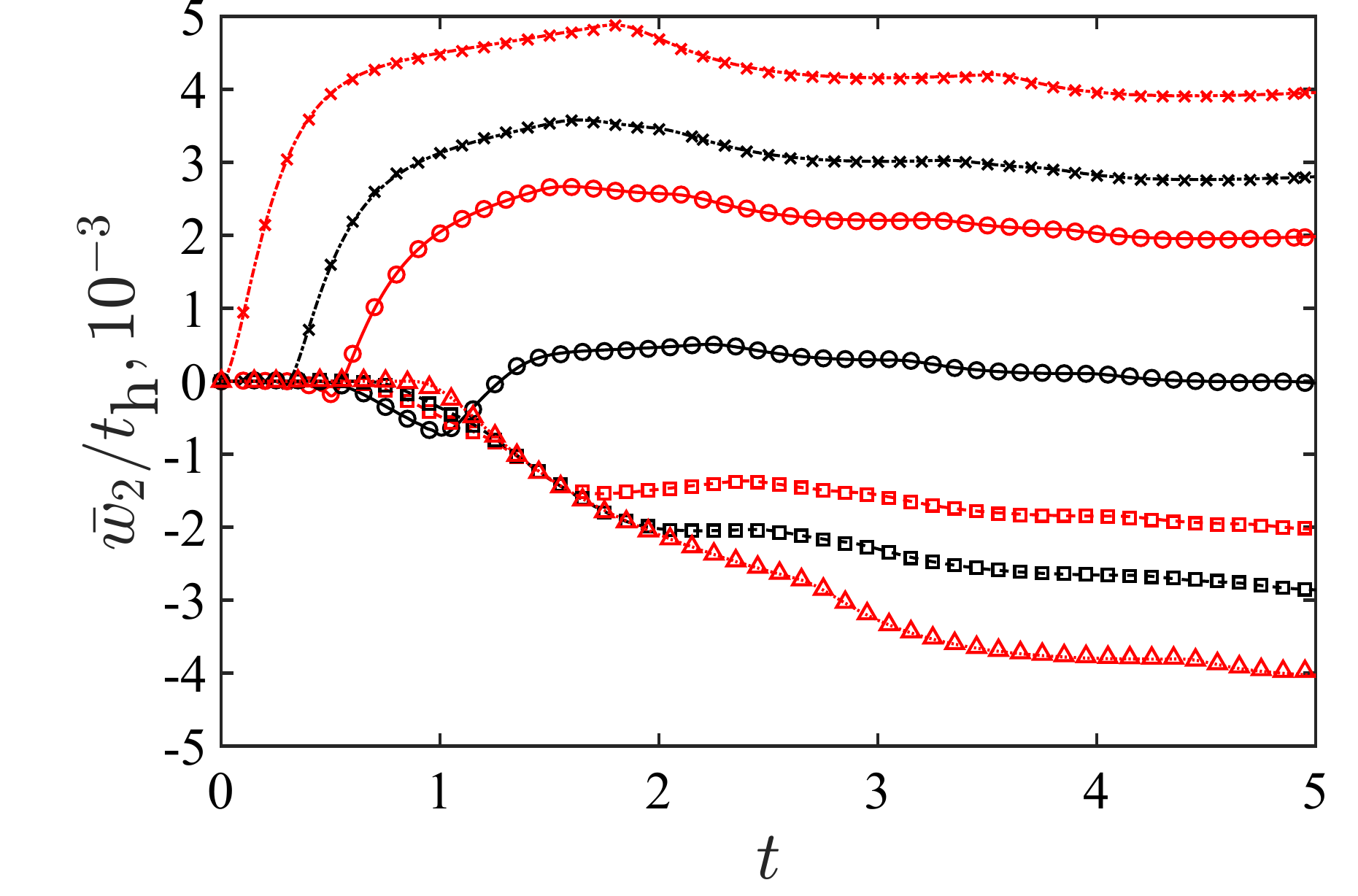}
        \subcaption{Normalized shell displacement}
        \label{fig:2-b}
    \end{minipage}%
    \vspace{0.5cm}
    \begin{minipage}[t]{0.5\textwidth}
        \centering
        \begin{overpic}[height=0.2\textheight,trim= 0cm 0cm 0cm 0.0cm,clip]{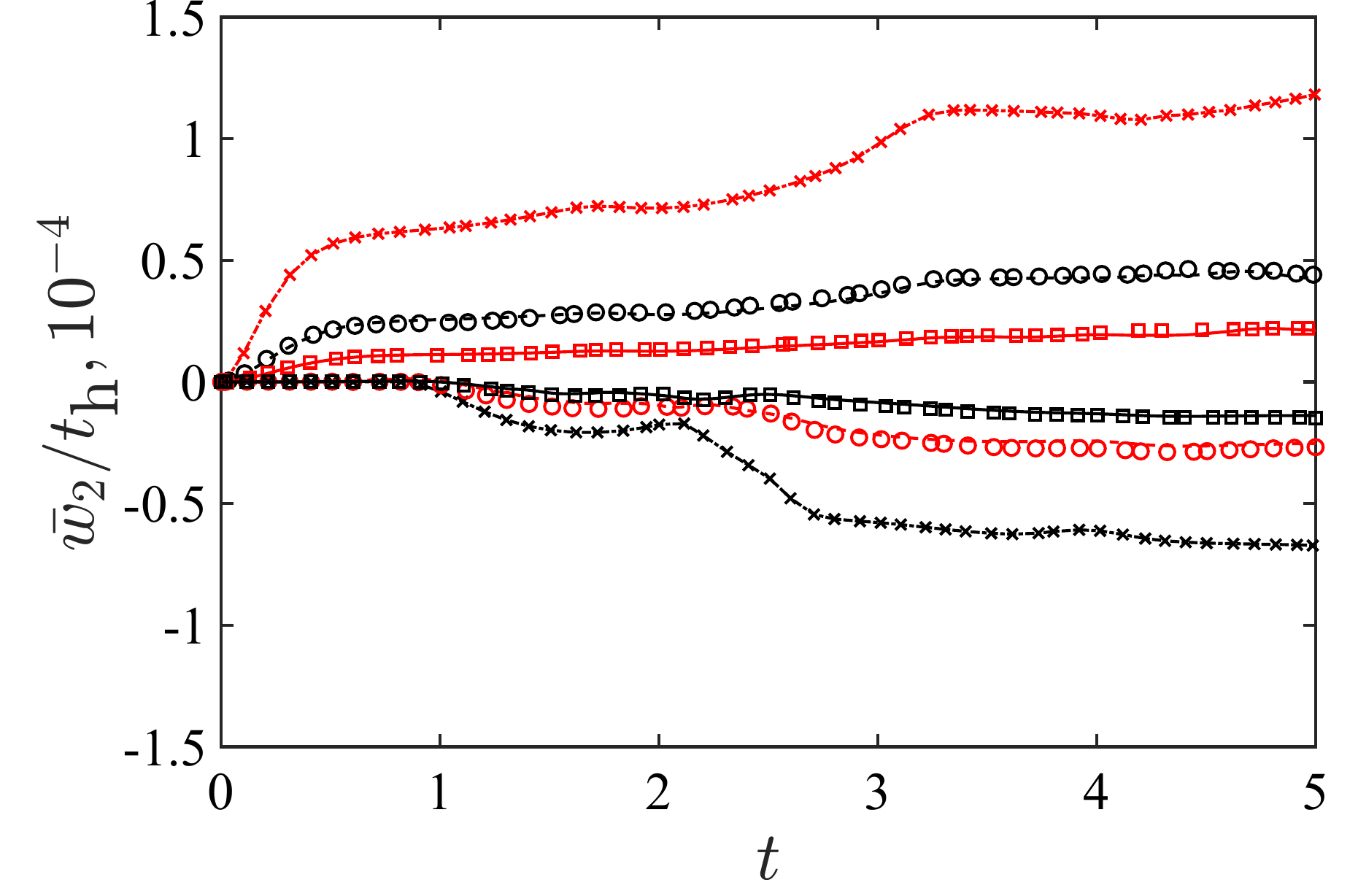}
        \put(102,23){\includegraphics[width=0.56\textwidth]{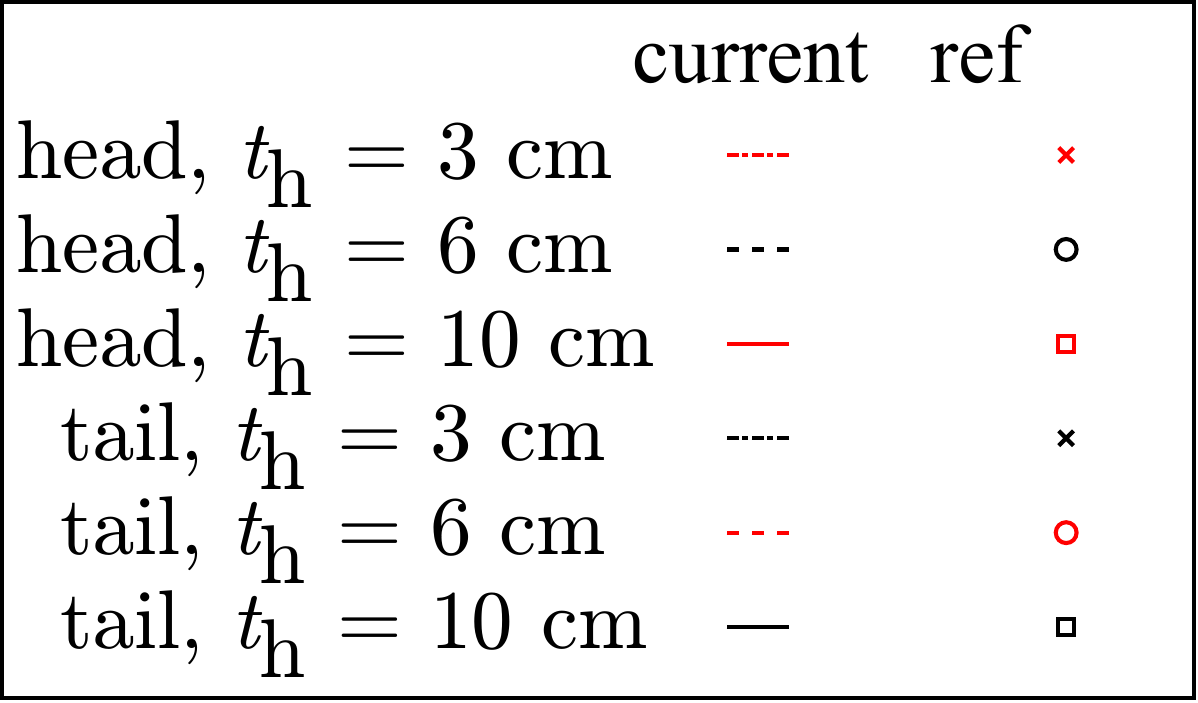}}
        \put(154.7,53.7){\color{black}{\InFigTextSize \cite{iakovlev2014transient}}}
        \end{overpic}
        \subcaption{Normalized shell displacement}
        \label{fig:2-c}
    \end{minipage}%
    \begin{minipage}[t]{0.2\textwidth}
        \centering
        \rule{0pt}{0.2\textheight}
    \end{minipage}%
    \vspace{0.5cm}
    \hspace{0.5cm}
    \begin{minipage}[t]{0.5\textwidth}
        \centering
        \begin{overpic}[height=0.2\textheight,trim= 0cm 0cm 0cm 0.0cm,clip]{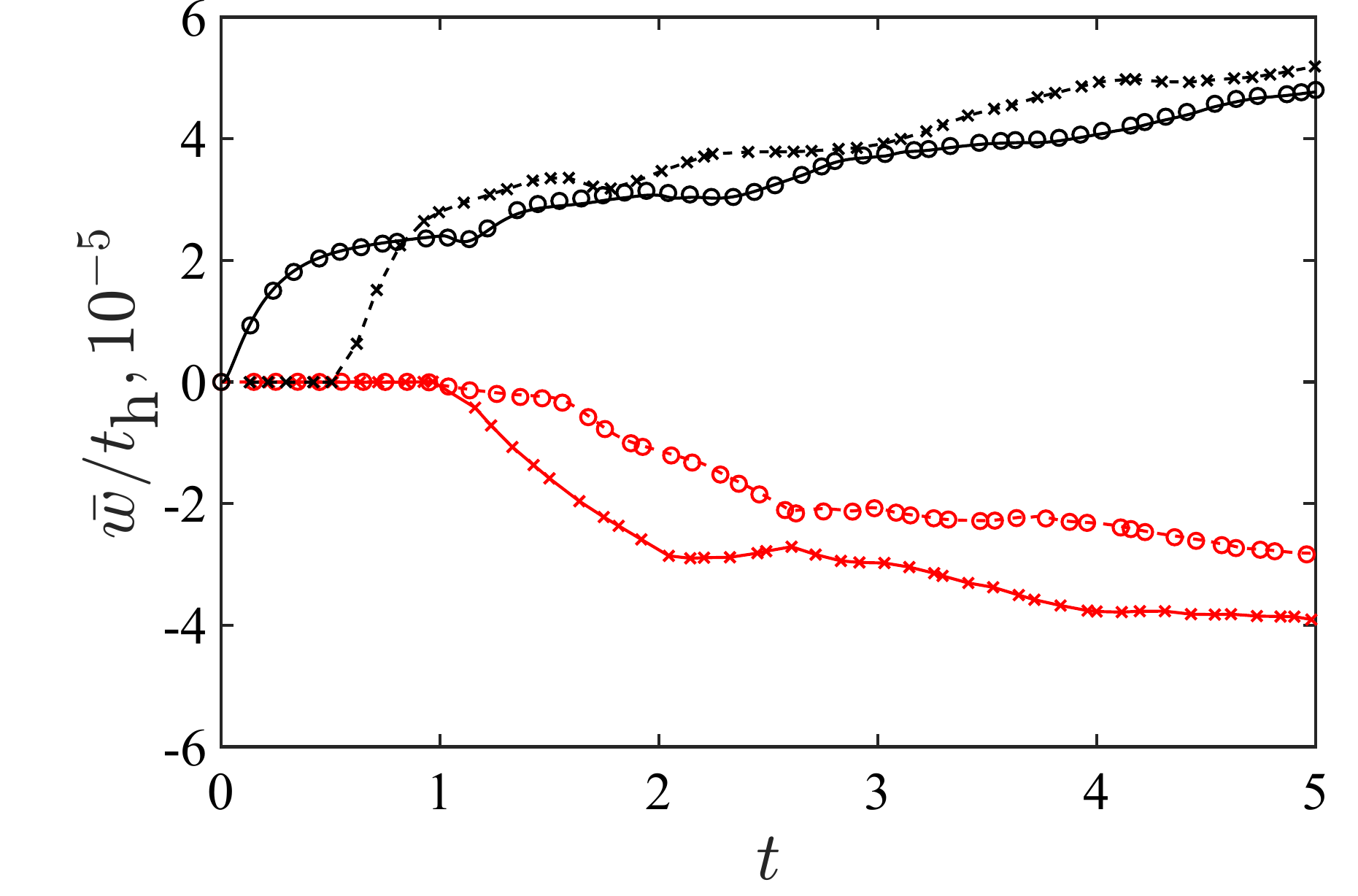}
        \end{overpic}
        \subcaption{Normalized shell displacement}
        \label{fig:2-d}
    \end{minipage}%
    \hfill
    \begin{minipage}[t]{0.5\textwidth}
        \centering
        \begin{overpic}[height=0.2\textheight,trim= 0cm 0cm 0cm 0.0cm,clip]{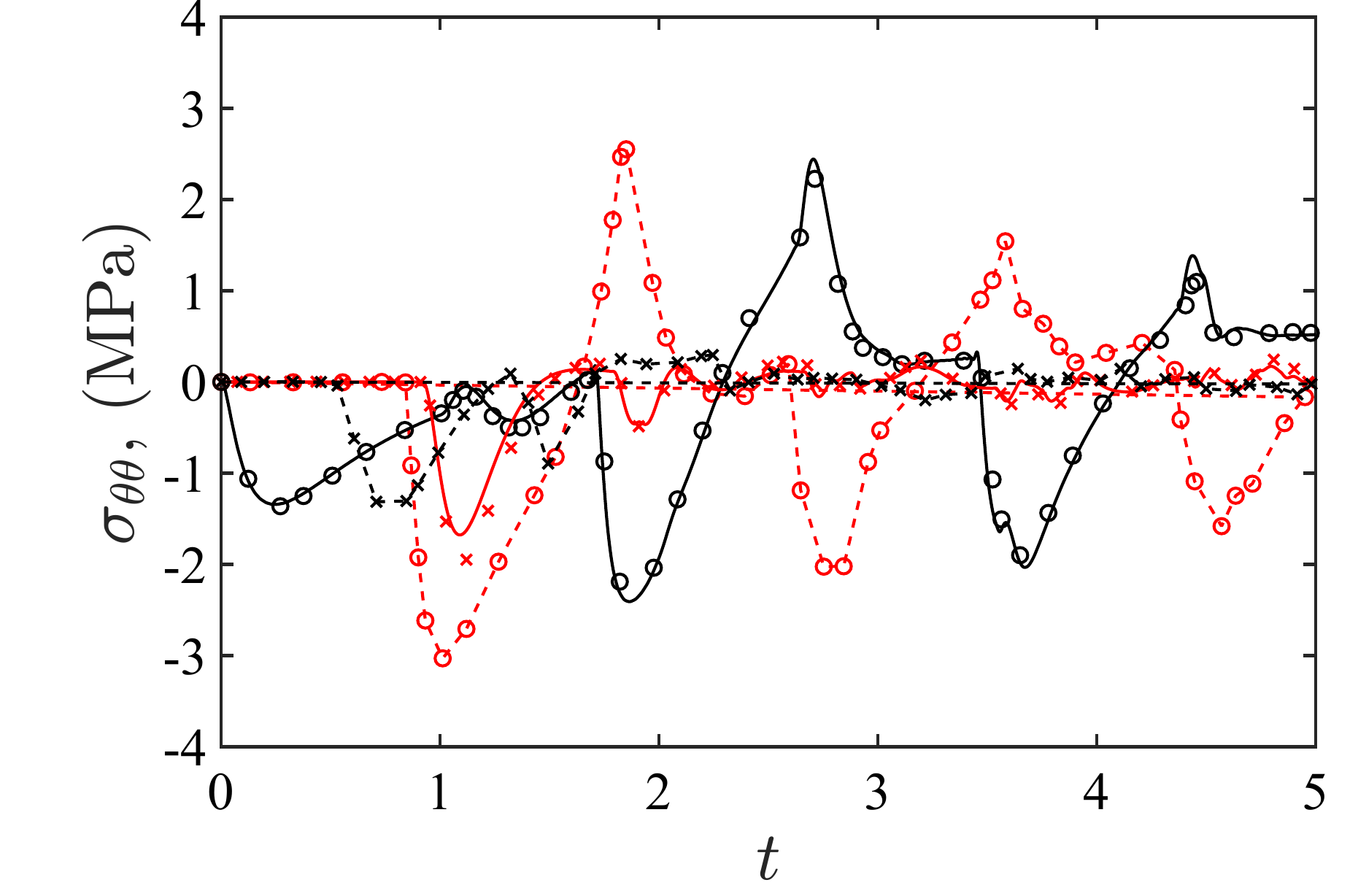}
        \put(18,57){\includegraphics[width=0.545\textwidth]{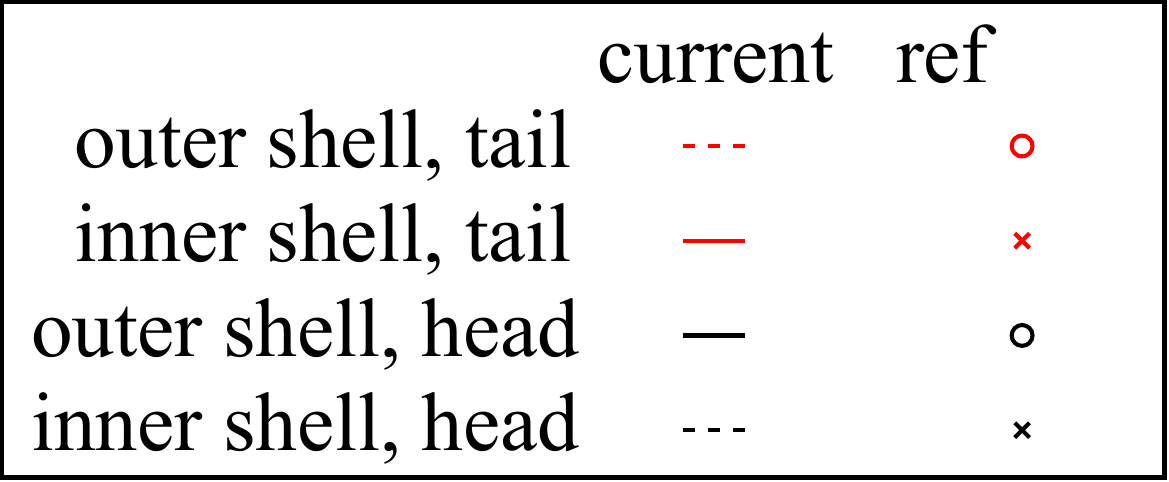}}
        \put(69.0,76.8){\color{black}{\InFigTextSize \cite{iakovlev2015shock}}}
        \end{overpic}
        \subcaption{Induced mechanical stress}
        \label{fig:2-e}
    \end{minipage}%
\caption{Verification of the developed model for double-shell vibroacoustics against available data in the literature.}
    \label{fig:2}
\end{figure}

Prior to discussing the main results, the validity of the mathematical formulation as well as the MATLAB implementation is evaluated under limiting scenarios in comparison to the published literature. First, the accuracy of the proposed solution method in analyzing acoustic wave scattering is examined against benchmark results in \cite{iakovlev2008interaction_2} through a simplified model of a single shell without interior fluid. In order to emulate this scenario and isolate the dynamics of the outer shell from the inner components, the fluid density and speed of sound in the gap $(\rho_\text{g}, c_\text{g})$ are set to negligibly small values. Vibration absorbers are excluded from the model by setting $N = 0$, and the piezoelectric layers are considered extremely thin to be ineffective. The shell is assumed to be composed of steel with $\nu = 0.3$, $\rho_\text{h} = 7800~\mathrm{kg/m^3}$, $E_\text{h} = 177.45~\mathrm{GPa}$, and thickness $t_\text{h} = 1~\mathrm{cm}$, and the incident wave's peak acoustic pressure is set to $\bar{p}_\alpha = 250~\mathrm{kPa}$. In addition, following \cite{iakovlev2008interaction_2}, a large source-to-shell distance $(R_0 \gg R_2)$ is assumed to approximate a plane incident wave. Other parameters conform to Table \ref{table:1}. The summation of normalized incident and diffracted acoustic pressures on the shell surface, $(\bar{p}_0 + \bar{p}_d)/\bar{p}_\alpha$, is illustrated in Fig. \ref{fig:2-a} and compared with Fig. 1 of \cite{iakovlev2008interaction_2}, while Fig. \ref{fig:2-b} compares the normalized displacement $w_2 / t_\text{h}$ with Fig. 3 of \cite{iakovlev2008interaction_2}. The agreement between the results confirm the model's accuracy in predicting acoustic wave scattering from the shell surface.

The next limiting case involves a single shell filled with interior fluid, and the results are compared with those reported in \cite{iakovlev2014transient}. To replicate this configuration and suppress the influence of inner shell dynamics on the acoustic field within the inter-shell gap, its radius is reduced to a negligible value in our model. With this adjustment, and according to the mathematical definition of the response functions in Eqs. (\ref{Eq:B.7}) to (\ref{Eq:B.10}) of \hyperref[sec:appendixB]{Appendix B}, $\varXi^{(\text{e})}$ remains constant, while $\varXi^{(1)}$ diminishes, and $\varXi^{(2)}$ approaches $\varXi^{(\text{i})}$. To avoid numerical divergence caused by abrupt acoustic pressure build-up within the inner shell, the internal fluid's density and sound speed ($\rho_\text{i}$, $c_\text{i}$) are chosen to be very small. Vibration absorbers are omitted from the model by setting $N = 0$, and the piezoelectric layers are assumed to be negligibly thin. The shell host layer is assumed to be composed of steel with $\nu = 0.3$, $\rho_\text{h} = 7900~\mathrm{kg/m^3}$, and $E_\text{h} = 241.0~\mathrm{GPa}$. Furthermore, the sound speed in both the exterior and gap fluids is set to $c_\text{e} = c_\text{g} = 1470~\text{m/s}$, and the exponential decay constant of the incident wave is assumed to be $\lambda = \SI{3.76e-3}{\second}$. Other model parameters follow the default values listed in Table \ref{table:1}. Fig. \ref{fig:2-c} depicts the normalized displacement of the shell ($w_2 / t_\text{h}$) evaluated at the head and tail points (i.e., the nearest and farthest locations relative to the acoustic excitation source) for various shell thicknesses. The results are compared with the reference data provided in Figs. 12 and 13 of \cite{iakovlev2014transient}. The observed consistency between datasets supports the reliability of the proposed model in capturing the shell response in the presence of internal acoustic medium.

To complement the model verification and confirm its fidelity in capturing the effect of gap medium as well as both shells' dynamics, a final benchmark is conducted considering a limiting scenario of a double-shell with a hollow inner shell. For this purpose, the obtained results from this configuration are compared with those reported in \cite{iakovlev2015shock}. In this case, the interior medium is modeled with an extremely low density and sound speed ($\rho_\text{i}$, $c_\text{i}$) to ensure its presence has a minimal impact on the acoustic response. Both shells are assumed to be made of steel with $\nu = 0.3$, $\rho_\text{h} = 7800~\mathrm{kg/m^3}$, $E_\text{h} = 177.45~\mathrm{GPa}$, and thickness of $t_\text{h} = 1~\mathrm{cm}$. The incident wave first impinges on the outer shell with a peak pressure of $\bar{p}_\alpha = 250~\mathrm{kPa}$. Numerical values for all other parameters are consistent with Table \ref{table:1}. As shown in in Fig. \ref{fig:2-d}, the normalized displacements of the inner and outer shells ($\bar{w}_1 / t_\text{h}$ and $\bar{w}_2 / t_\text{h}$) are evaluated at the head and tail points and compared with the corresponding results reported in Figs. 4 and 5 of \cite{iakovlev2015shock}. Furthermore, mechanical stress at head and tail points of each shell is calculated using (\ref{Eq:22}), and compared with Figs. 7 and 8 of \cite{iakovlev2015shock} in Fig. \ref{fig:2-e}. The close match observed between the results in Figs. \ref{fig:2-d} and \ref{fig:2-e} reveals the reliability of the numerical framework for further analysis.
\begin{equation}
\sigma_{\theta\theta}^{(i)}(\theta,\tau) = \frac{E_\text{h}}{R_i(1 - \nu^2)} \left( \frac{\partial \bar{v}_i(\theta,\tau)}{\partial \theta} - \bar{w}_i(\theta,\tau) \right), \quad i \in \{1, 2\}.
\label{Eq:22}
\end{equation}

\subsection*{Main results}
Here, we analyze the behavior of the double-shell structure model, shown in Fig. \ref{fig:1}, under circular acoustic excitation, and investigate how vibration absorbers and piezoelectric actuation influence its vibroacoustic characteristics.
\addcontentsline{toc}{subsubsection}{Bare system}

\begin{figure}[H]
\centering
\begin{minipage}[t]{0.5\textwidth}
        \centering
        \includegraphics[height=0.2\textheight]{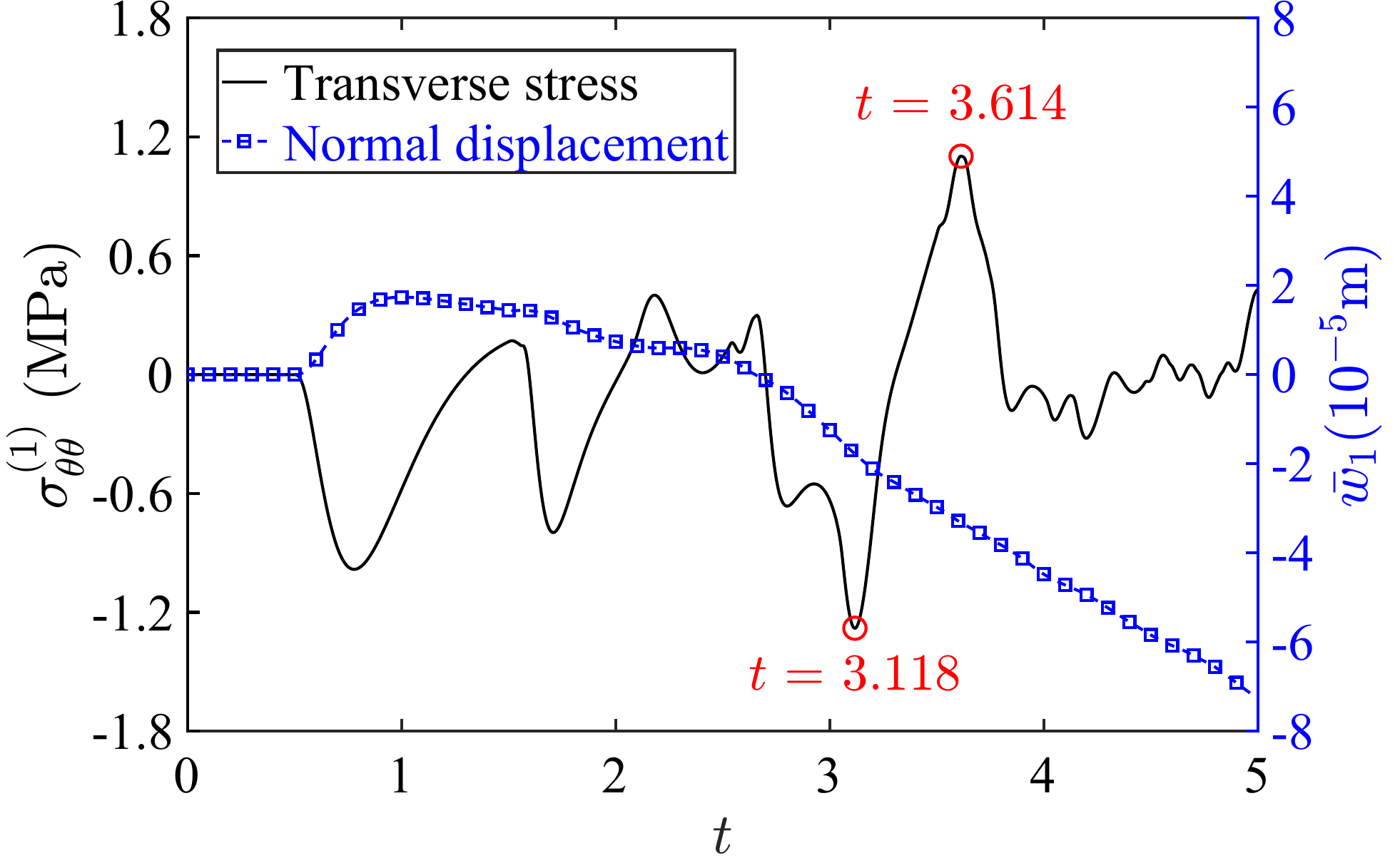}
        \subcaption{Inner shell head point}
        \label{fig:3-a}
    \end{minipage}%
    \hfill
    \begin{minipage}[t]{0.5\textwidth}
        \centering
        \includegraphics[height=0.2\textheight]{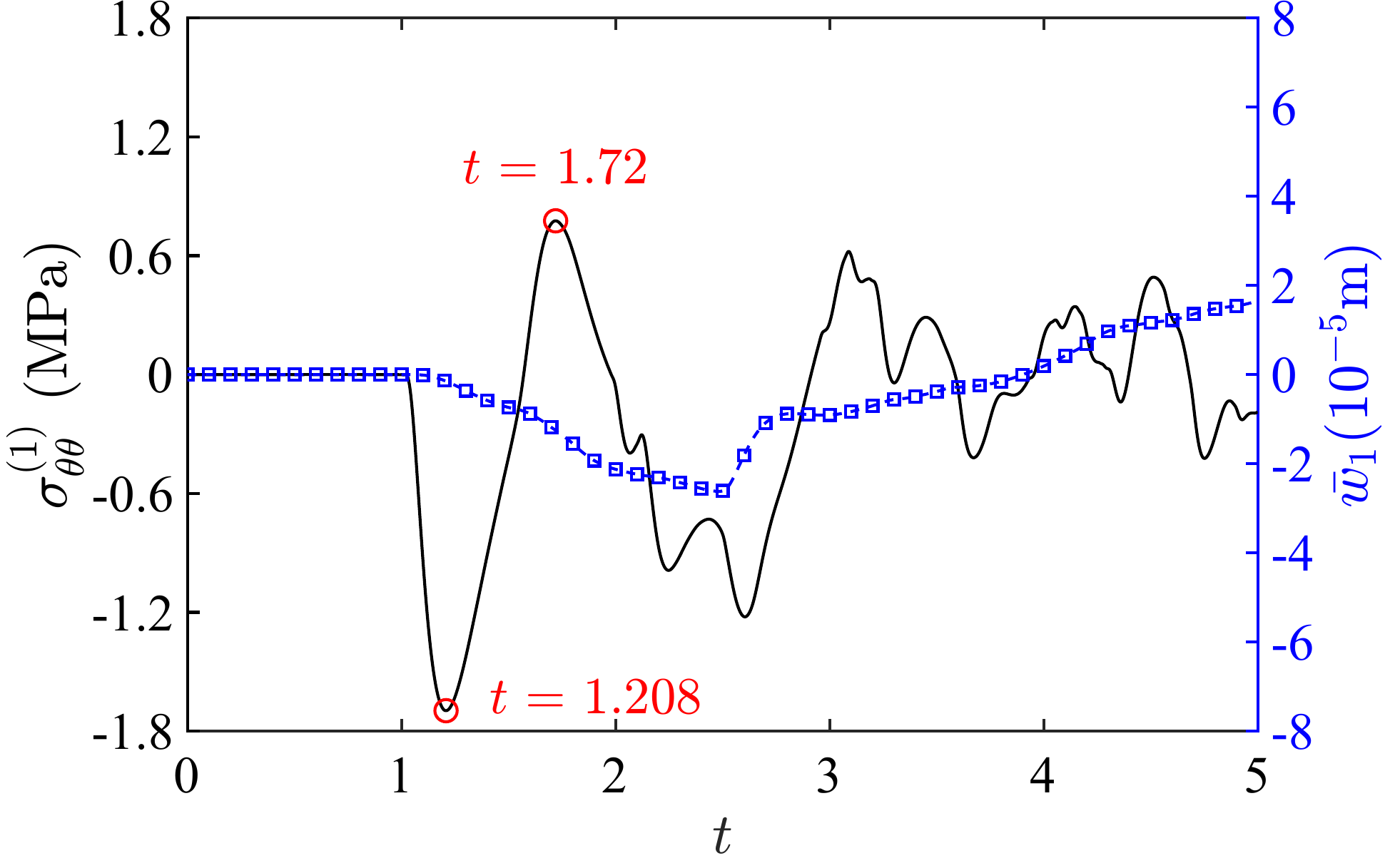}
        \subcaption{Inner shell tail point}
        \label{fig:3-b}
    \end{minipage}%
    \vspace{0.5cm}
    \begin{minipage}[t]{0.5\textwidth}
        \centering
        \includegraphics[height=0.2\textheight]{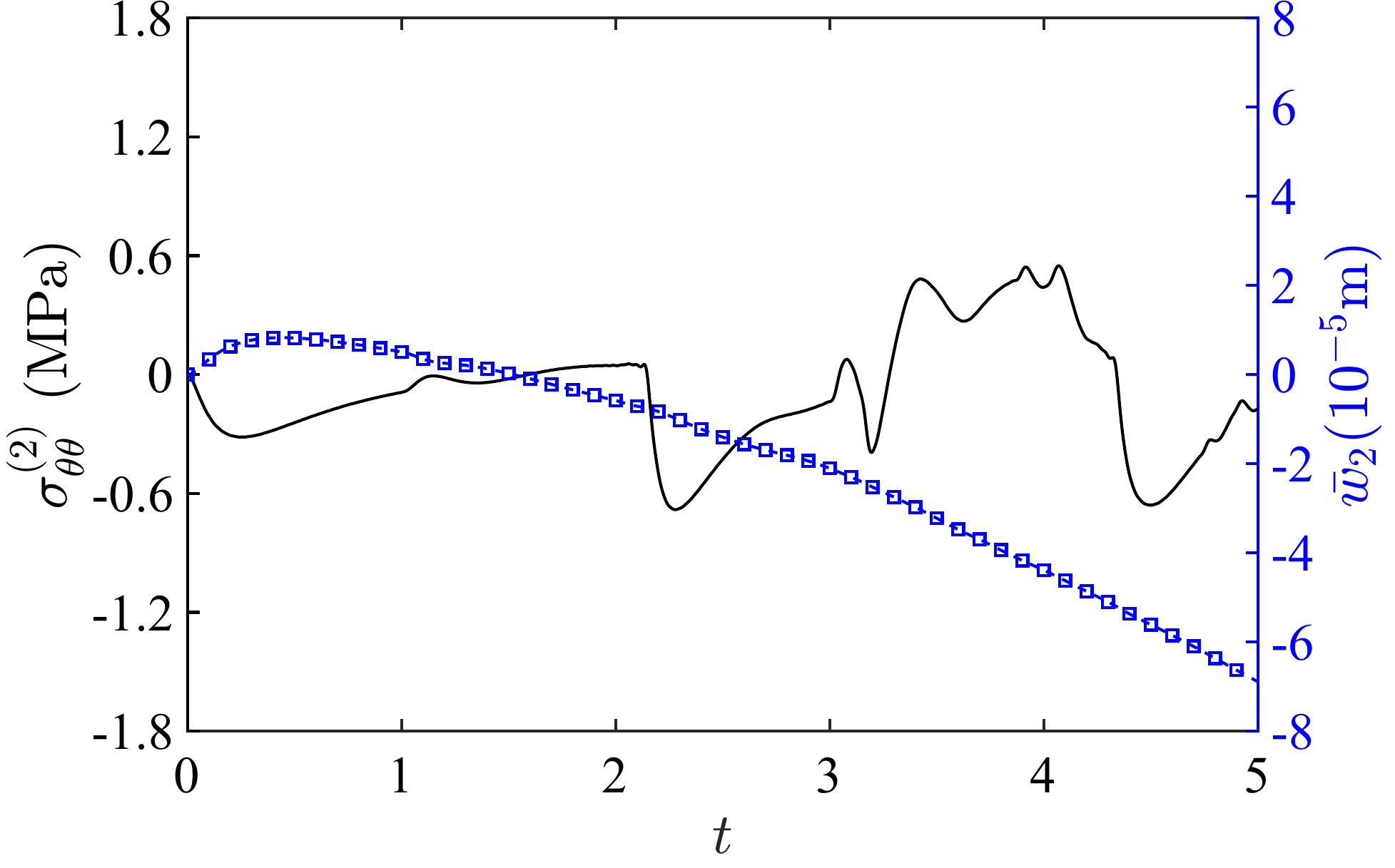}
        \subcaption{Outer shell head point}
        \label{fig:3-c}
    \end{minipage}%
    \hfill
    \begin{minipage}[t]{0.5\textwidth}
        \centering
        \includegraphics[height=0.2\textheight]{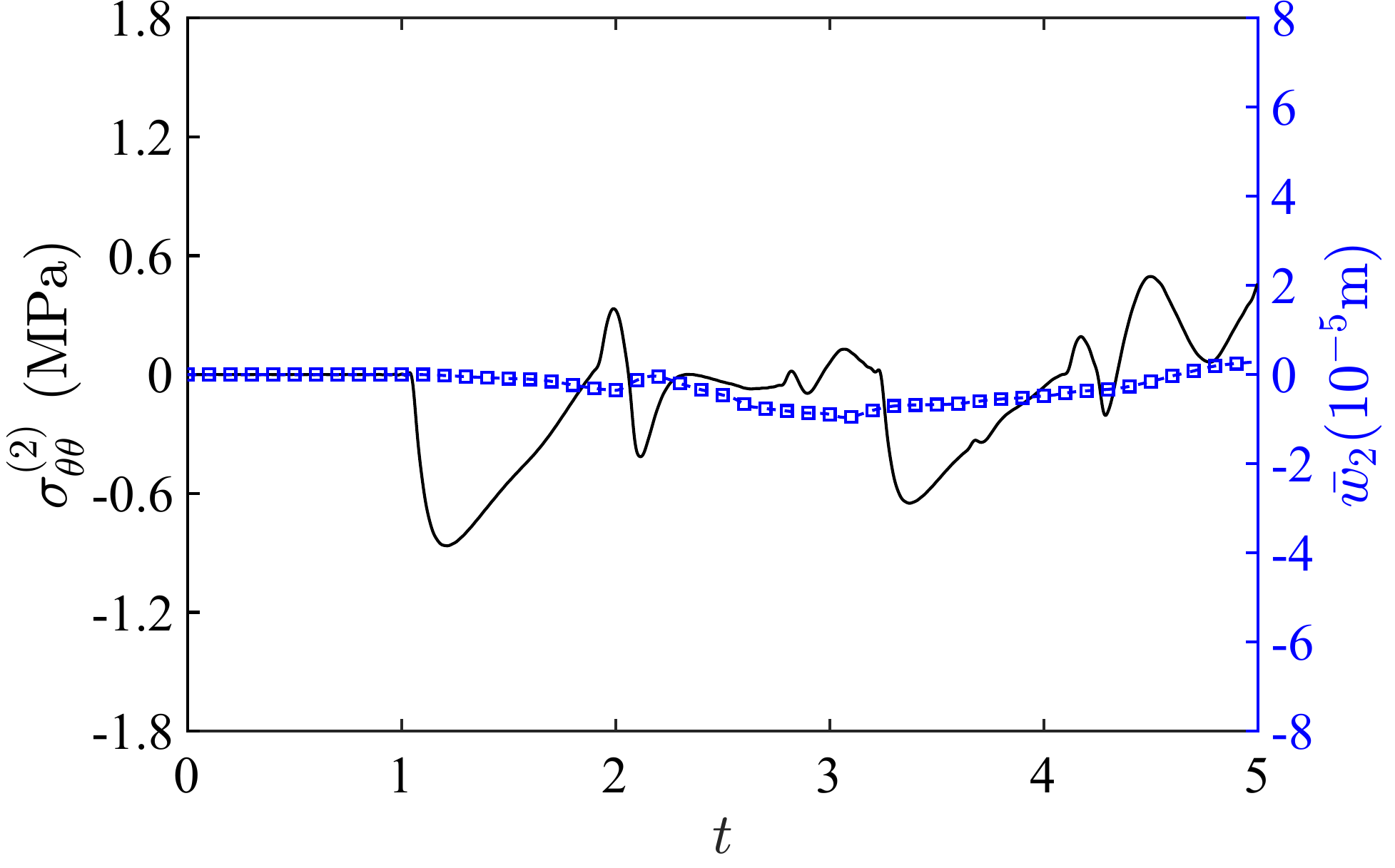}
        \subcaption{Outer shell tail point}
        \label{fig:3-d}
    \end{minipage}%
    \caption{Induced stress and displacement in the bare system at head and tail points of each shell.}
    \label{fig:3}
\end{figure}

\noindent \textbf{Bare system:} as a foundation for subsequent evaluations, we begin by analyzing the system's dynamics in the absence of any control mechanism or vibration absorbers --- hereafter referred to as the “bare system”. Findings in \cite{iakovlev2004influence, iakovlev2010hydrodynamic}, identify the head and tail points on each shell as the most affected regions by fluid-structure interaction, making them highly susceptible to deformation and damage due to stress concentration. Fig. \ref{fig:3} shows the transverse displacement and stress at these locations, where Figs. \ref{fig:3-a} and \ref{fig:3-b} correspond to the head and tail points of the inner shell, respectively, while Figs. \ref{fig:3-c} and \ref{fig:3-d} representing those on the outer shell. Identical scale ranges are utilized in all four plots to facilitate comparison. Perhaps, the most notable observation from these plots is the larger fluctuations in the stress curves. Besides that, mechanical stress accounts for both tangential and normal displacement according to Eq. (\ref{Eq:22}), which suggests it as a more suitable design metric than relying on normal displacement alone. Fig. \ref{fig:3} also highlights that the inner shell undergoes more pronounced stress and displacement. The intensity of the incident acoustic shock wave barely attenuates by traversing the thin outer shell. Once subjected to approximately the same level of excitation, the lighter inner shell undergoes greater displacement, which in turn leads to higher stress. In addition, given that Eq. (\ref{Eq:22}) links stress intensity inversely to shell radius, the inner shell naturally bears more stress. Therefore, one may conclude that in order to improve the double-shell structure's resilience to acoustic shocks, design efforts must prioritize protecting the inner shell as the system’s Achilles’ heel! Consequently, the primary design objective of the passive vibration absorbers or piezoelectric actuators, is set to minimize peak stress levels especially on the inner shell.

The stress curves also show that the peak stress is more likely to occur at the tail point of each shell, indicating that these regions are more vulnerable to shock excitations. Figs. \ref{fig:3-a} and \ref{fig:3-c} show that vibrations at the head points of both shells initiate at $t=0$ and $t=0.5$, which correspond to the time when the incident wave hits these points. In contrast, the tail points begin vibrating before the wavefront reaches them (see Figs. \ref{fig:3-b} and \ref{fig:3-d}). This is due to the faster wave propagation through the solid domain. As noted earlier, a major contribution of this chapter of the thesis is the integration of the interior fluid's influence into the system's governing equations. Although often overlooked, it may play a critical role in the vibroacoustic response of double-shell structures. The mismatch in sound speeds between the fluid and solid domains causes a phase shift between the interior fluid acoustic pressure and the shell vibrations, leading to low-amplitude high-frequency oscillations in the inner shell. For instance, the high-frequency stress fluctuations is evidenced by Figs. \ref{fig:3-a} and \ref{fig:3-b}, whereas this effect is absent in Fig. \ref{fig:2-e} when the interior fluid is excluded.

\begin{figure}[H]
\centering
    \begin{minipage}[t]{0.25\textwidth}
        \centering
        \includegraphics[height=0.18\textheight]{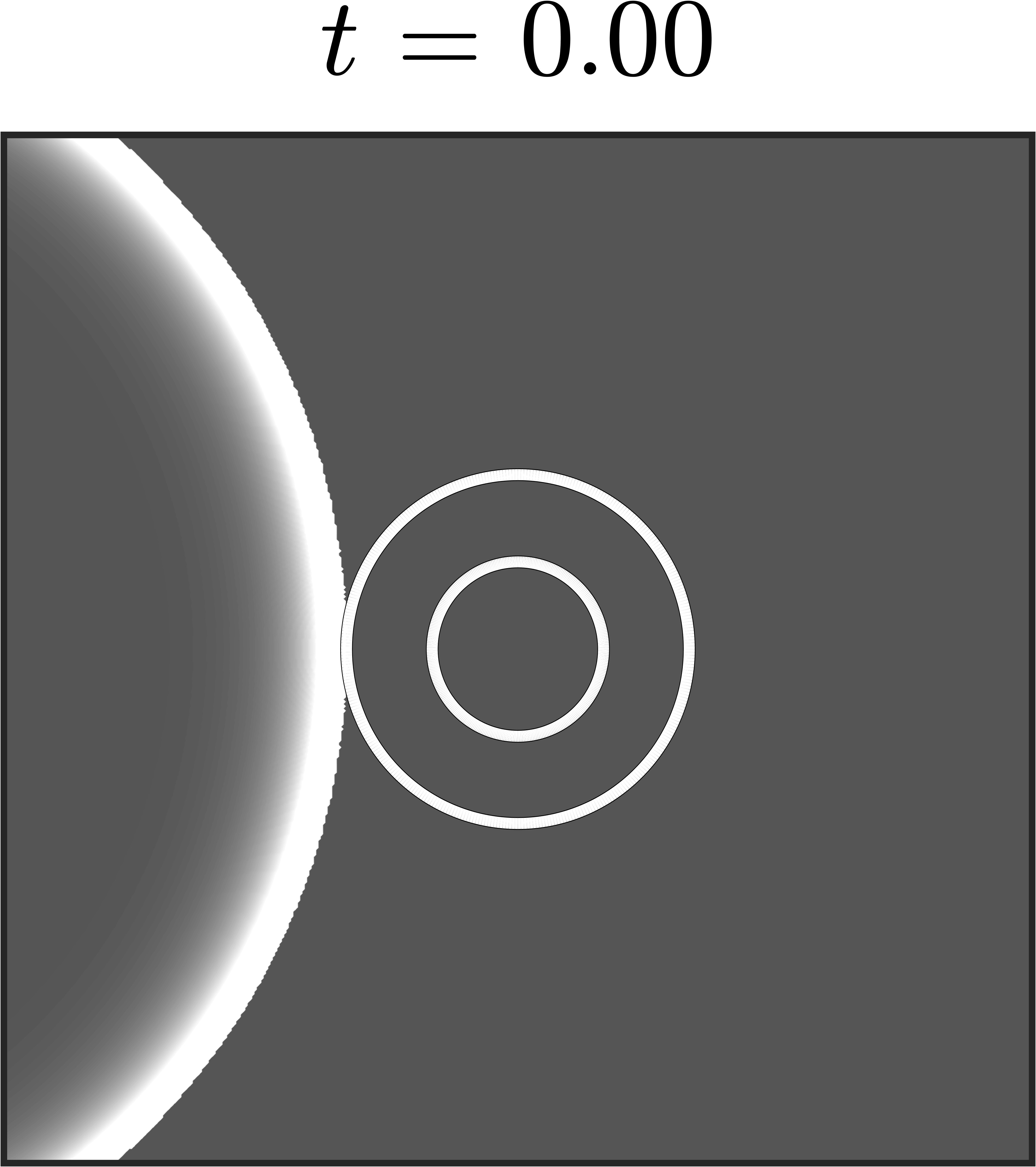}
    \end{minipage}%
    \begin{minipage}[t]{0.25\textwidth}
        \centering
        \includegraphics[height=0.18\textheight]{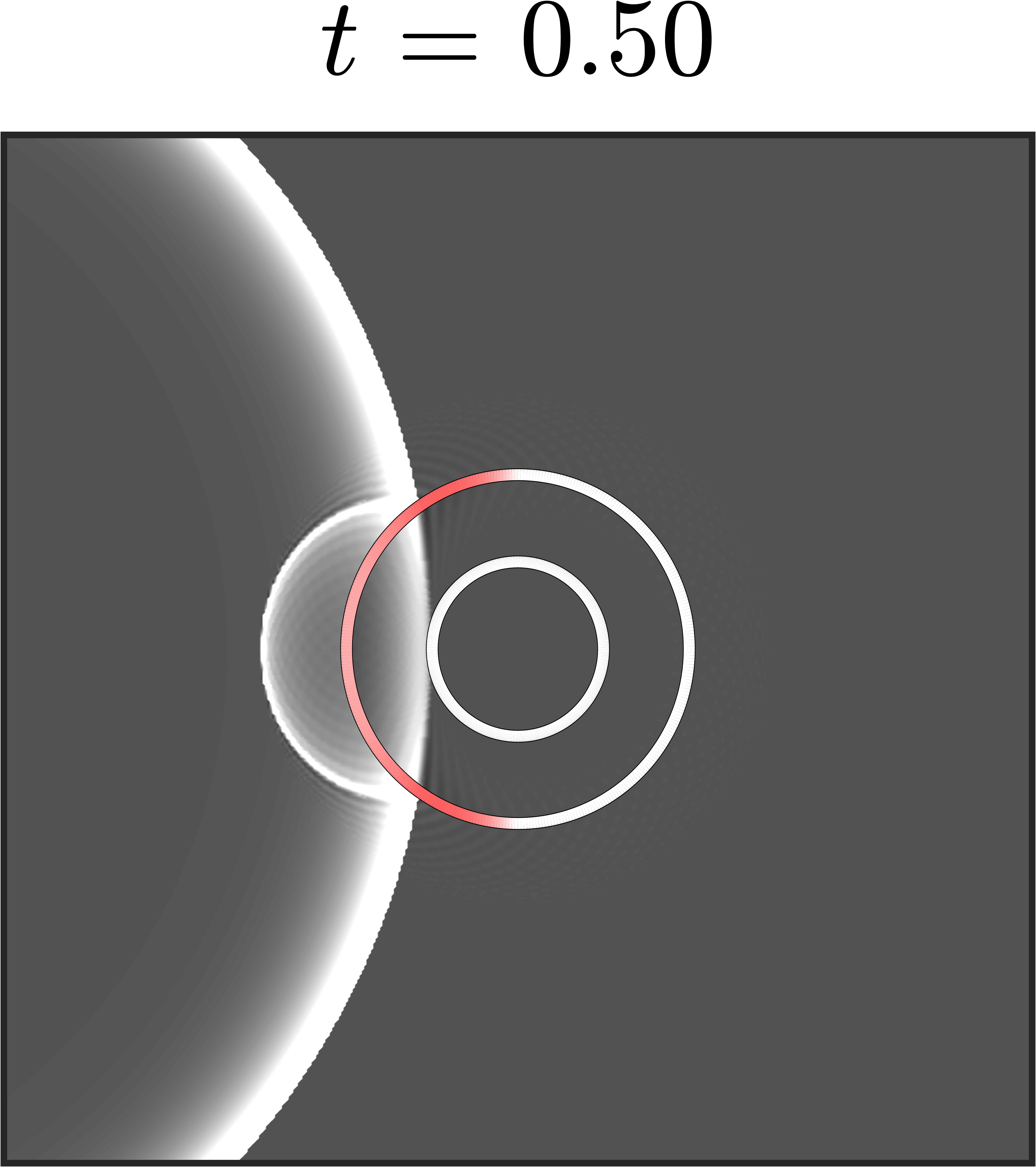}
    \end{minipage}%
    \begin{minipage}[t]{0.25\textwidth}
        \centering
        \includegraphics[height=0.18\textheight]{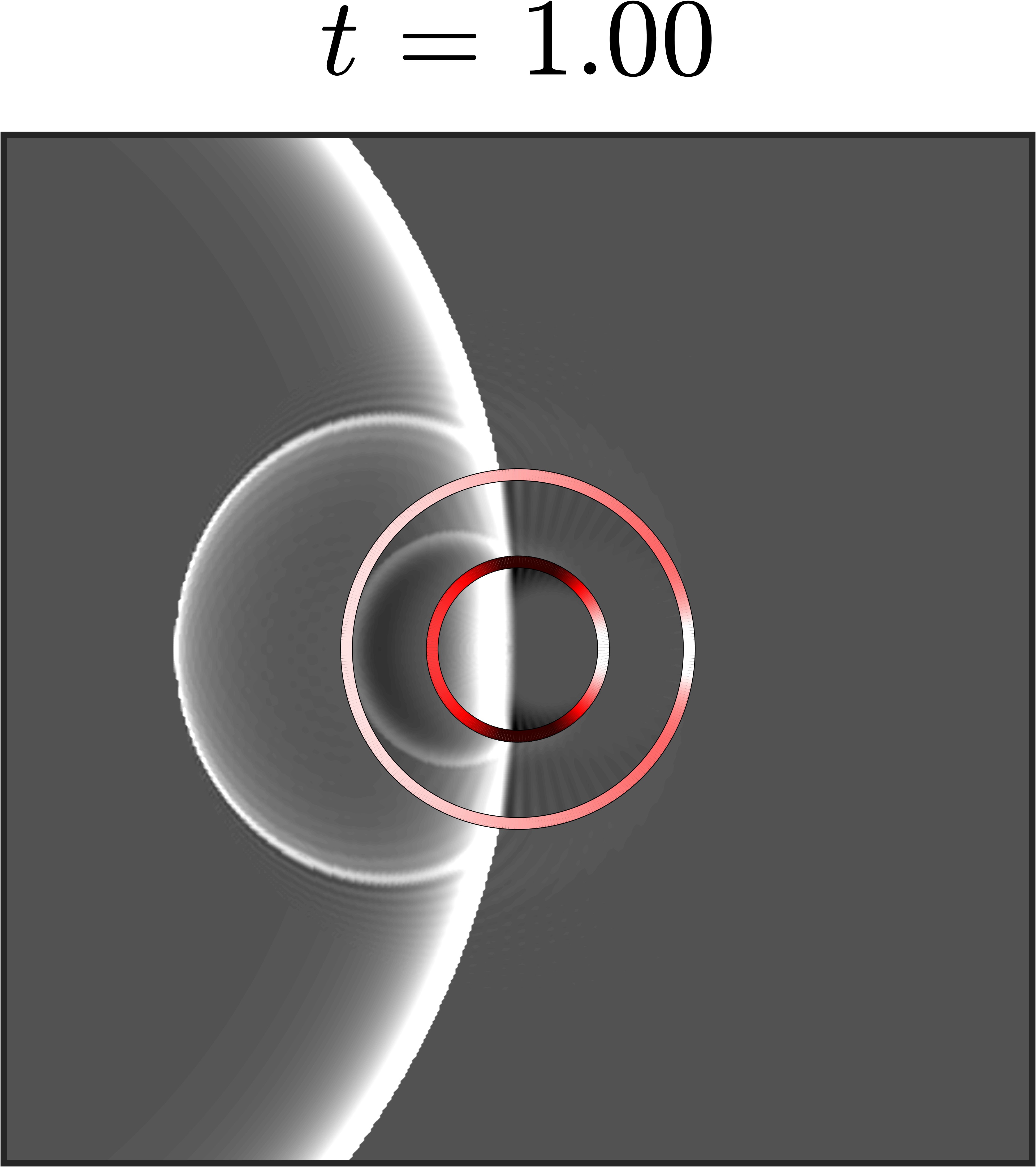}
    \end{minipage}%
    \begin{minipage}[t]{0.25\textwidth}
        \centering
        \includegraphics[height=0.18\textheight]{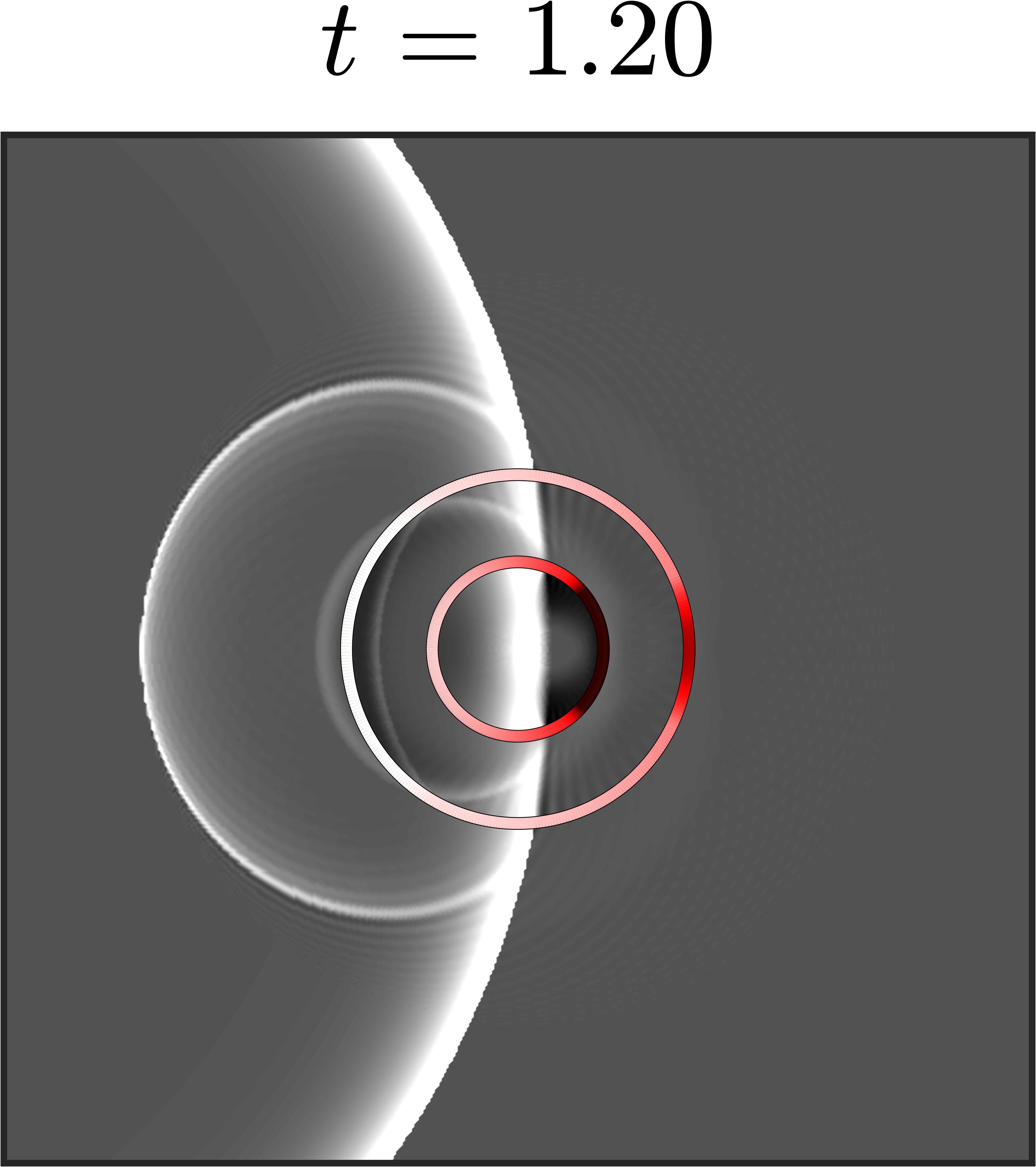}
    \end{minipage}%
    \vspace{0.3cm}
    \begin{minipage}[t]{0.25\textwidth}
        \centering
        \includegraphics[height=0.18\textheight]{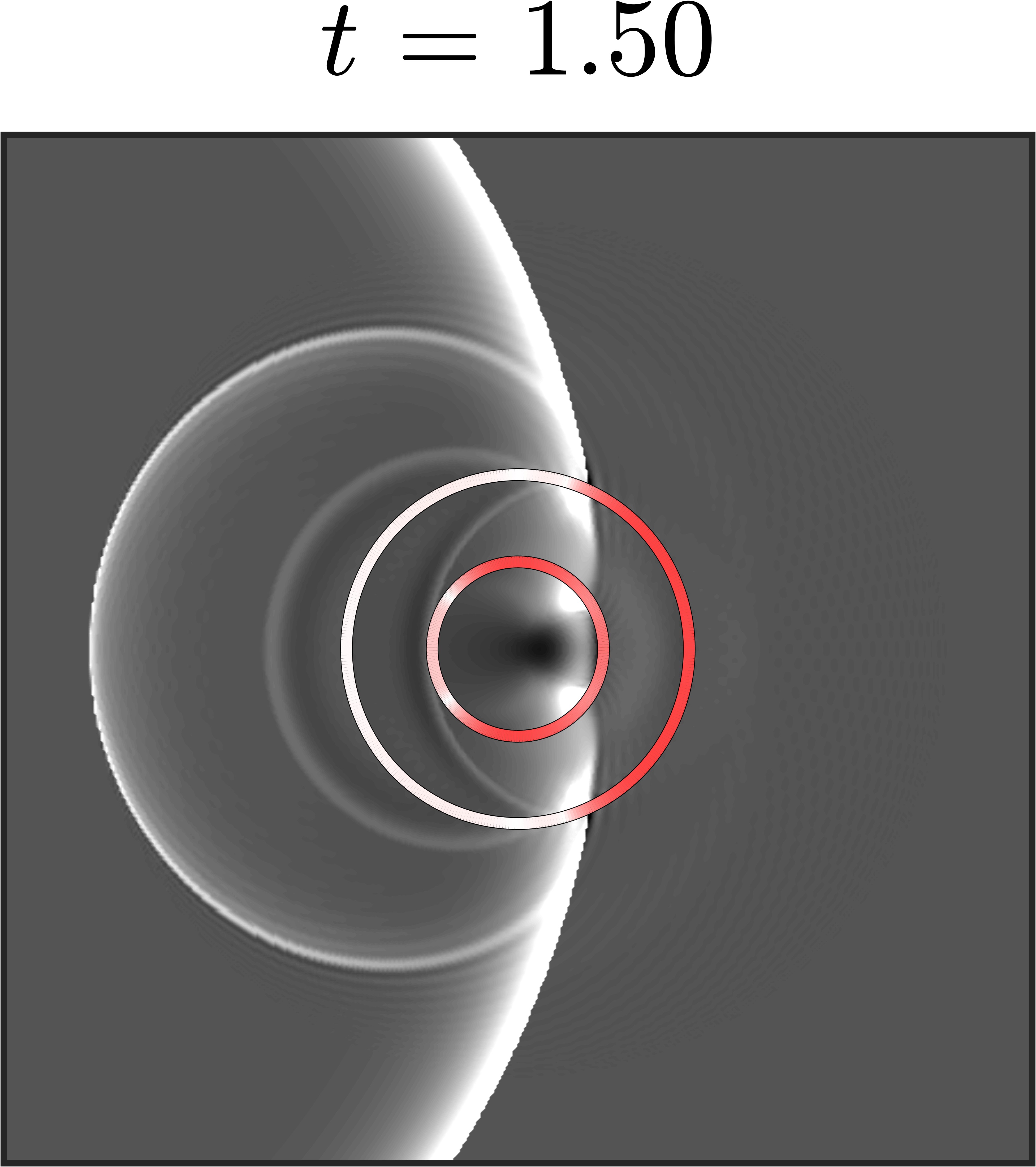}
    \end{minipage}%
    \begin{minipage}[t]{0.25\textwidth}
        \centering
        \includegraphics[height=0.18\textheight]{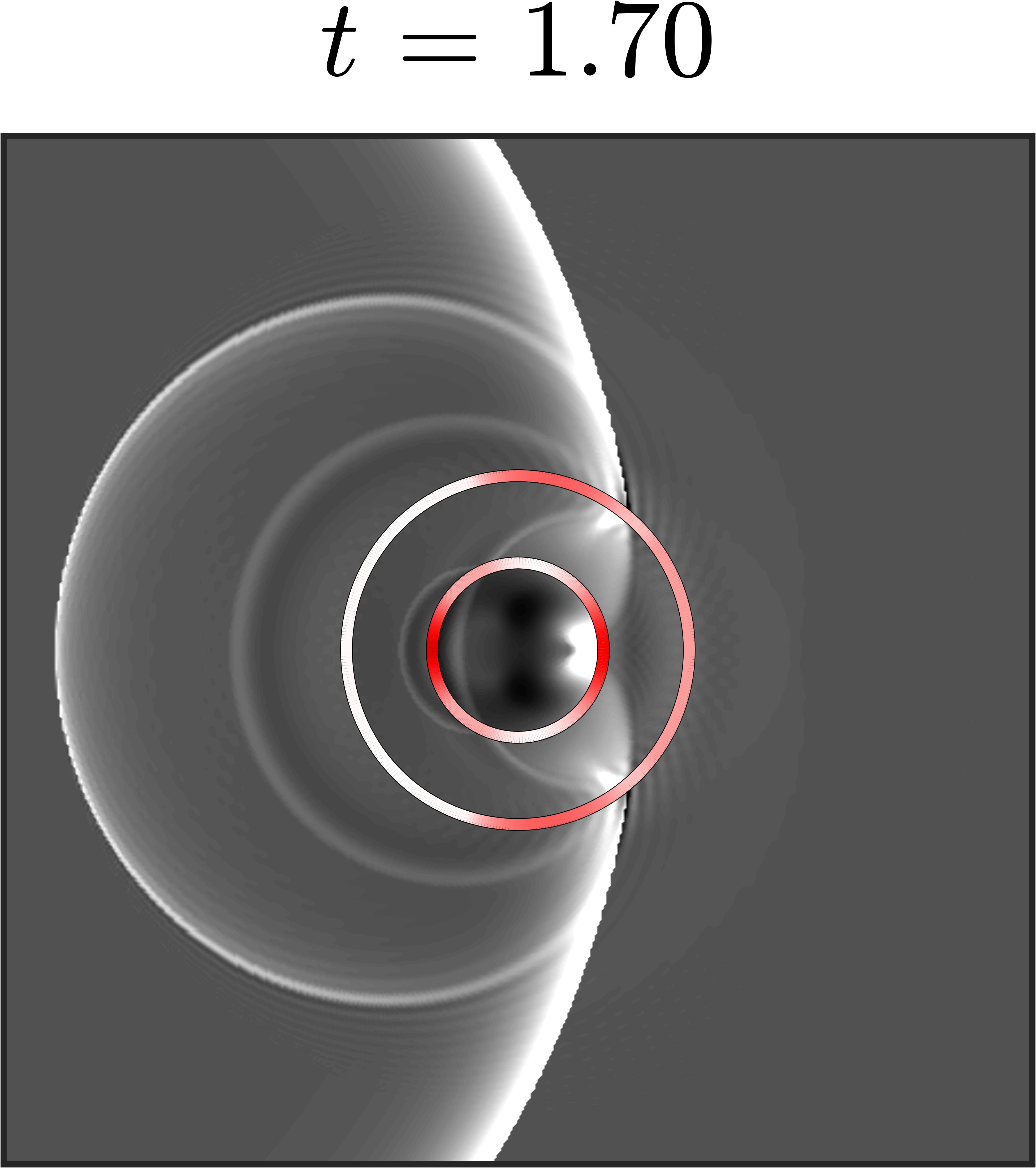}
    \end{minipage}%
    \begin{minipage}[t]{0.25\textwidth}
        \centering
        \includegraphics[height=0.18\textheight]{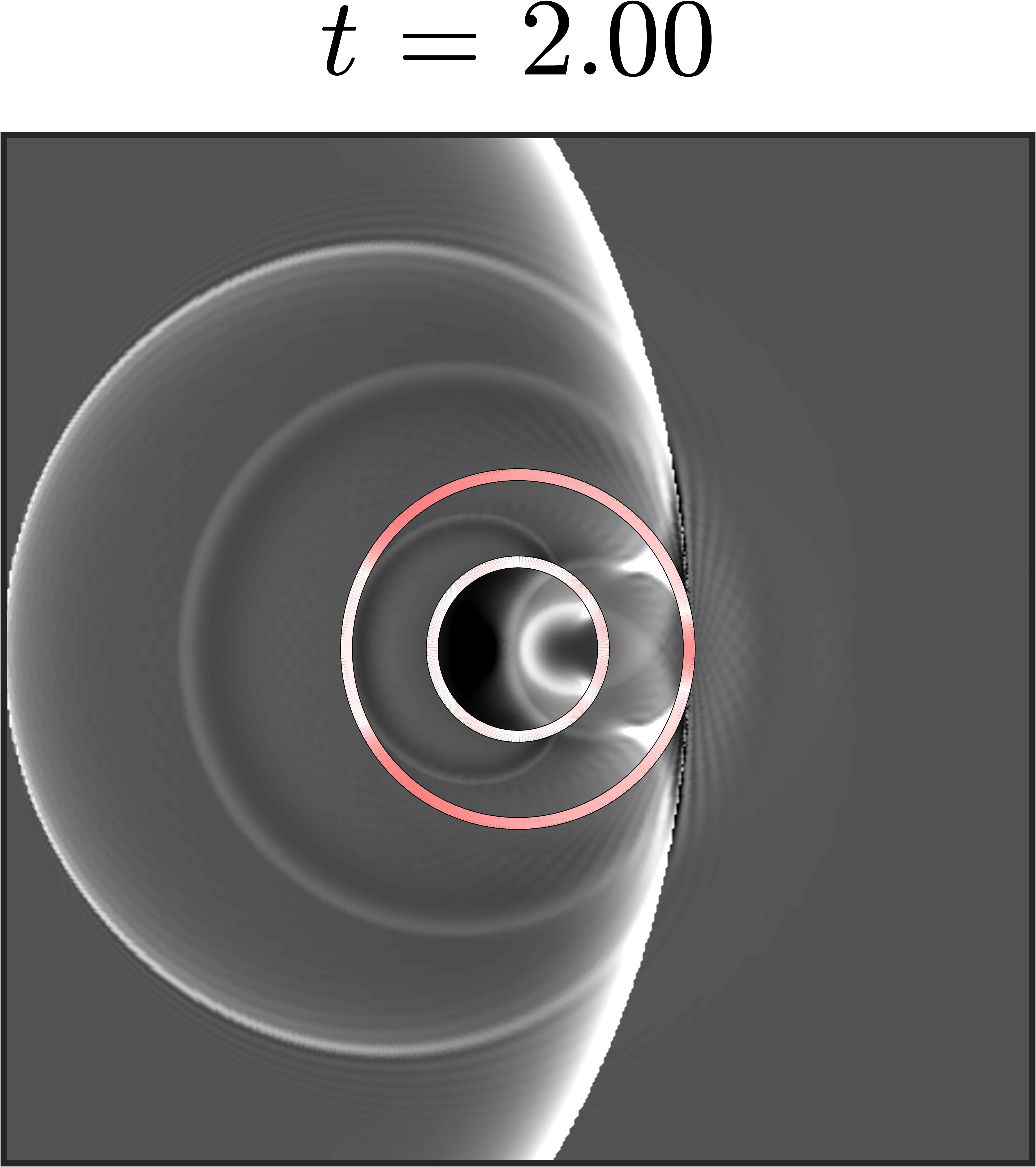}
    \end{minipage}%
    \begin{minipage}[t]{0.25\textwidth}
        \centering
        \includegraphics[height=0.18\textheight]{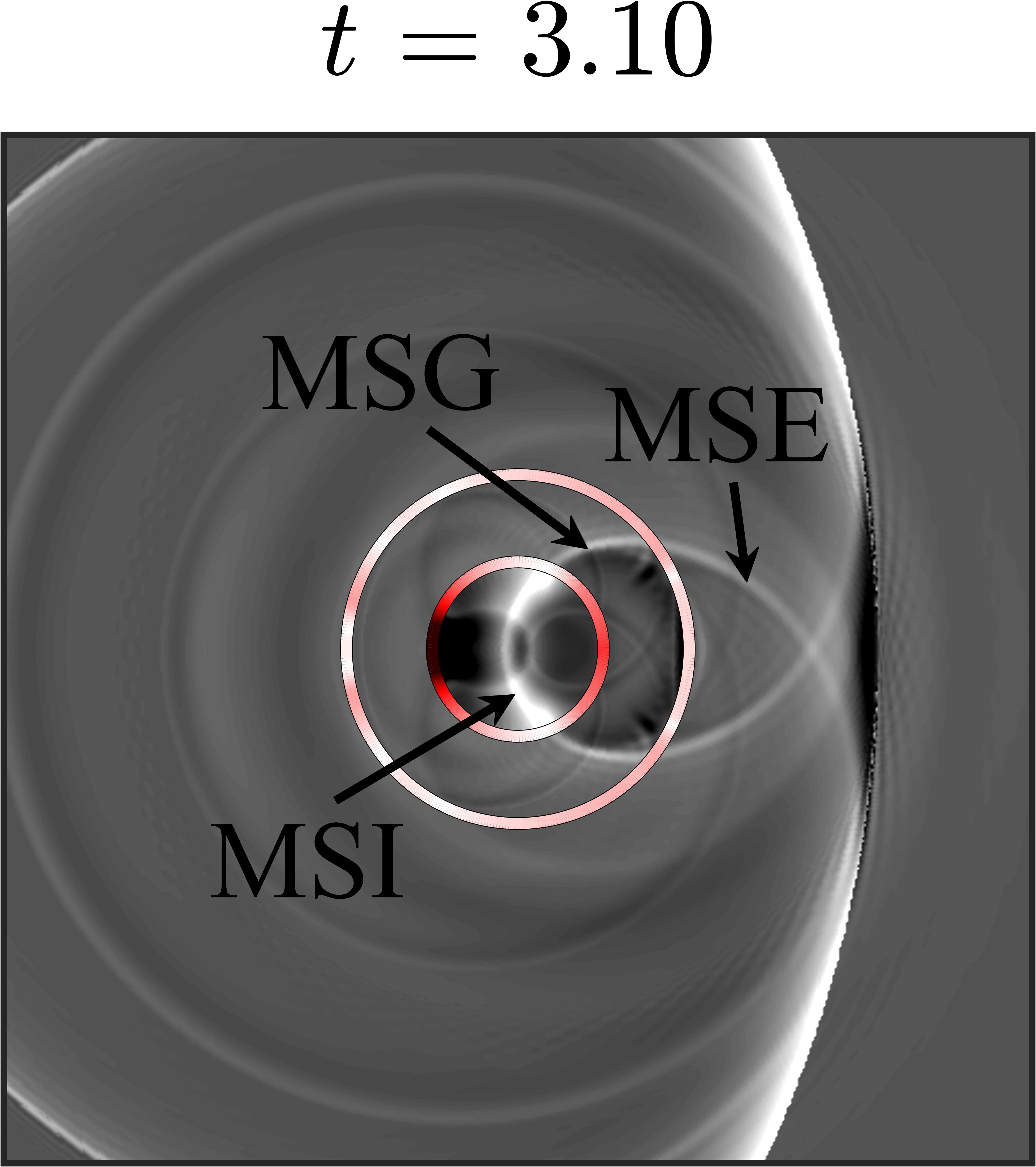}
    \end{minipage}%
    \vspace{0.3cm}
    \begin{minipage}[t]{0.25\textwidth}
        \centering
        \includegraphics[height=0.18\textheight]{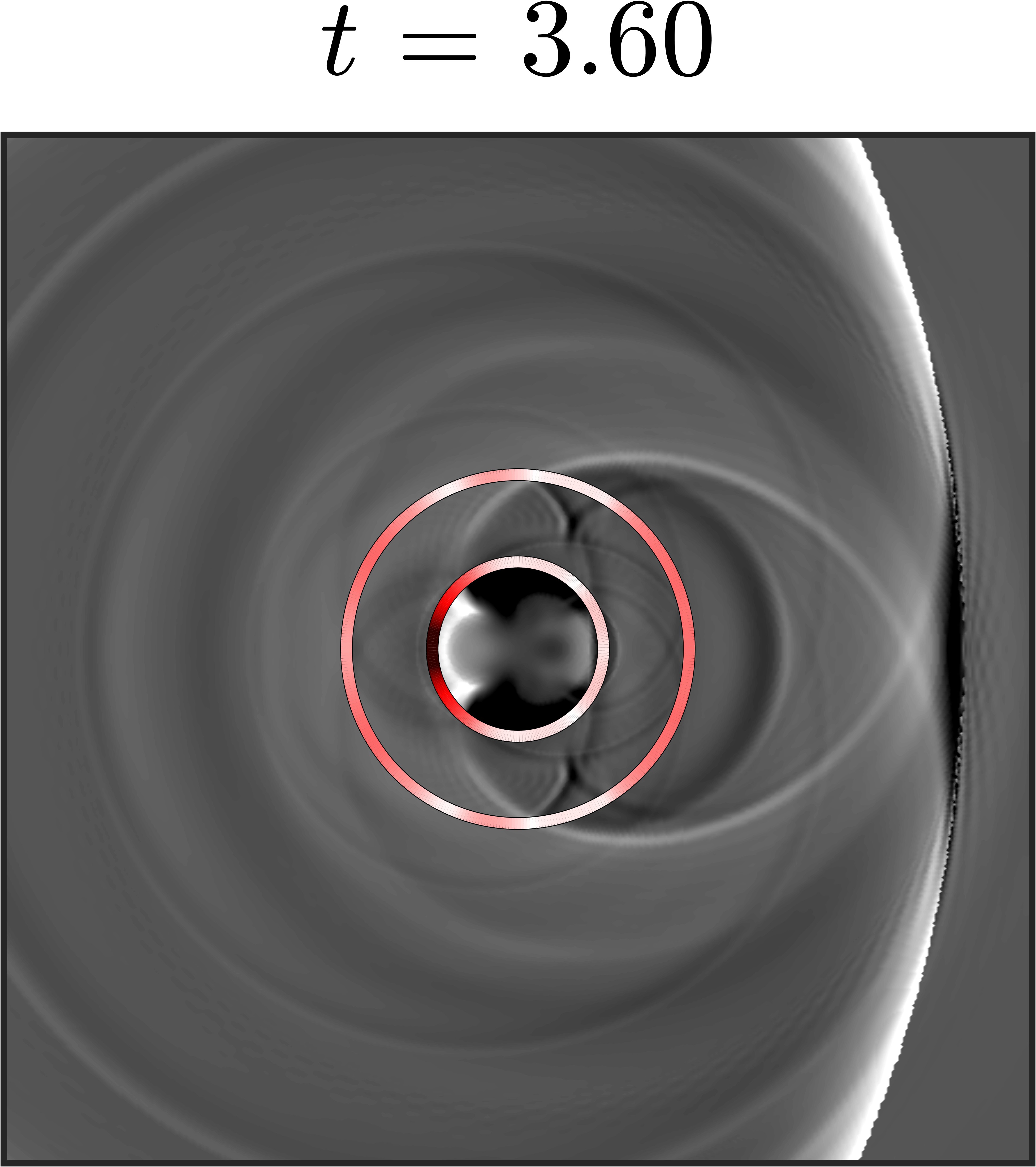}
    \end{minipage}%
    \begin{minipage}[t]{0.25\textwidth}
        \centering
        \includegraphics[height=0.18\textheight]{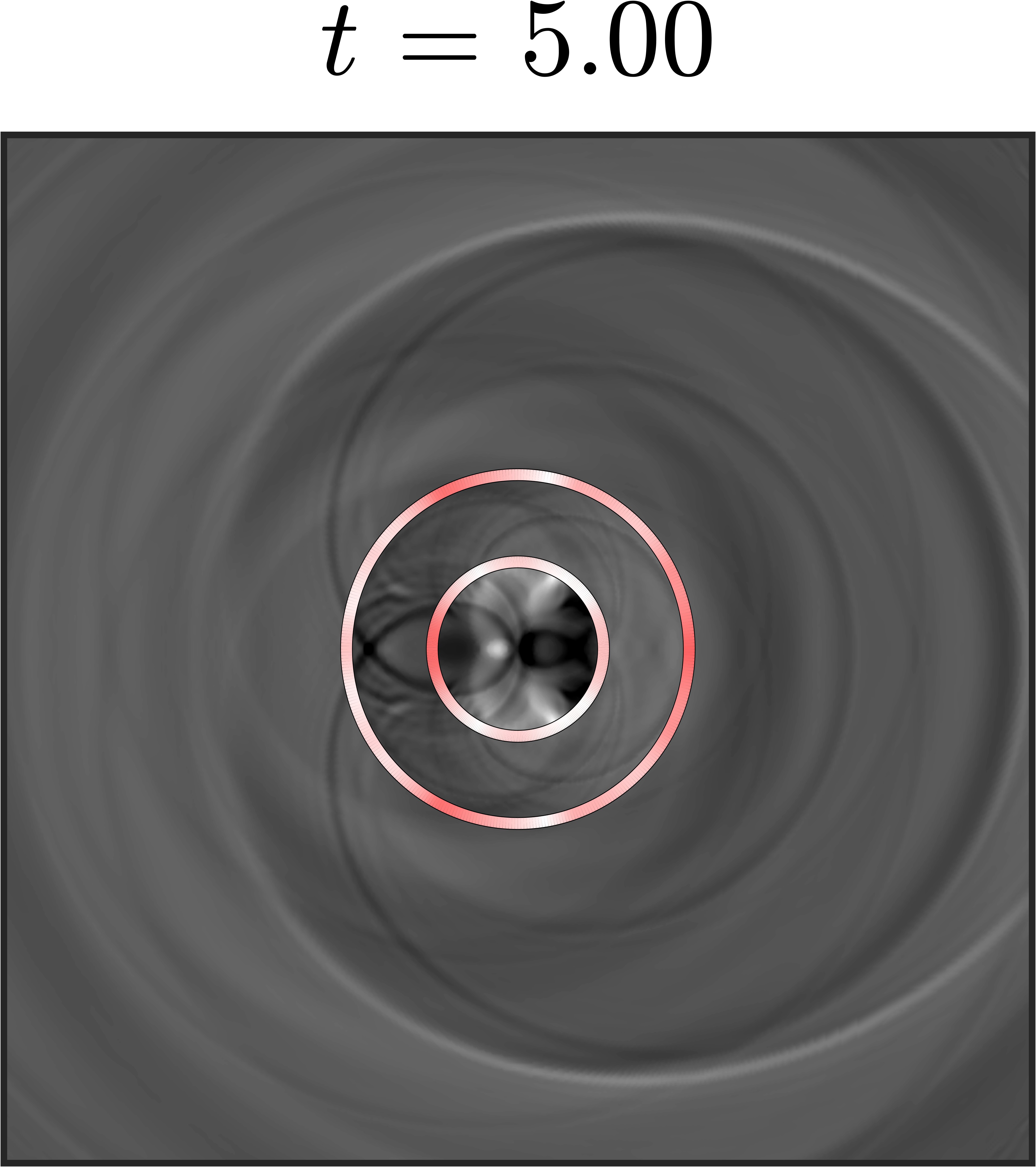}
    \end{minipage}%
    \begin{minipage}[t]{0.5\textwidth}
        \centering
        \raisebox{2.5ex}{\includegraphics[width=0.25\textheight]{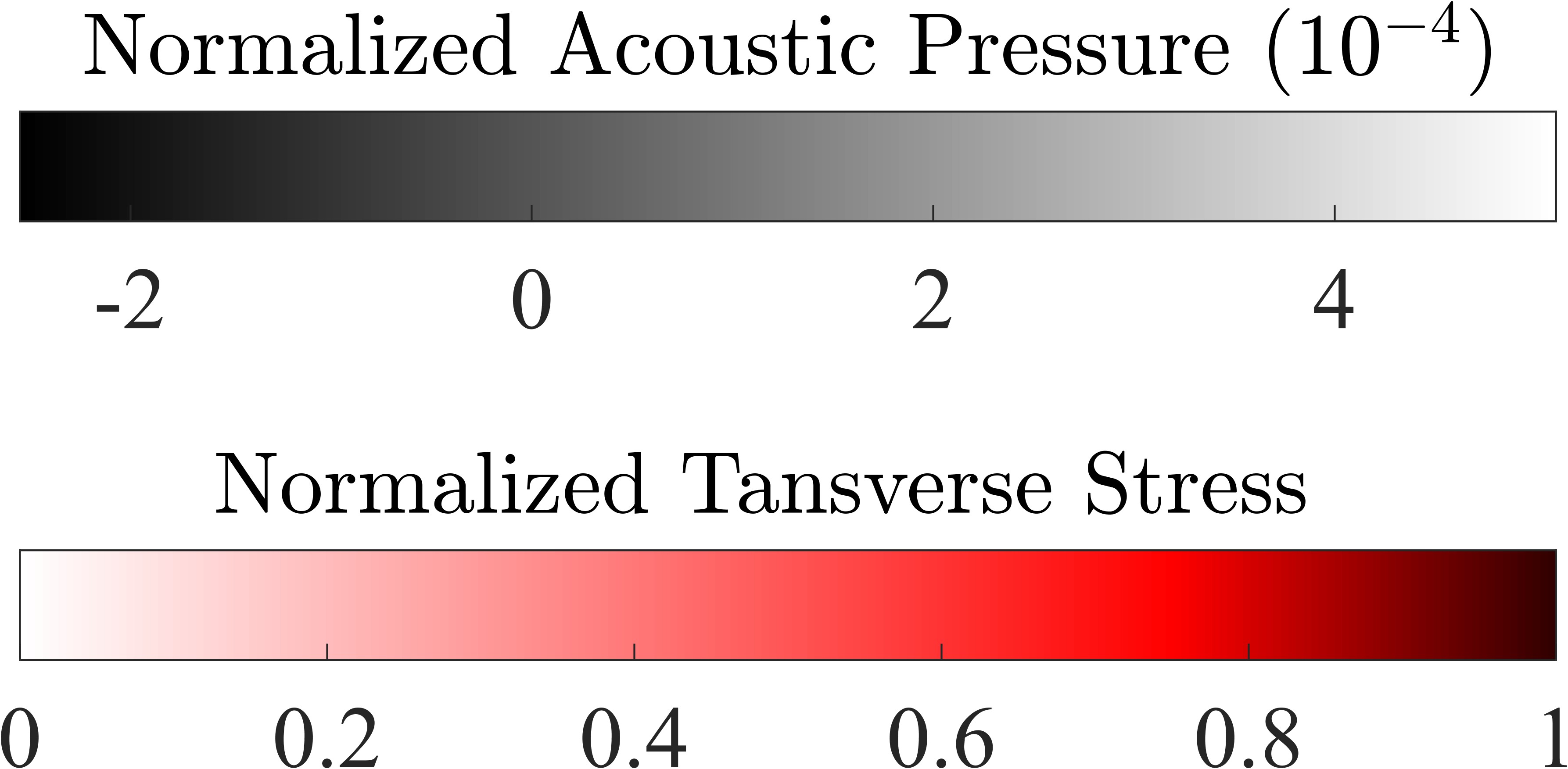}}
    \end{minipage}%
    \caption{Snapshots of transient acoustic field and induced stress in the bare system.}
    \label{fig:4}
\end{figure}

Fig. \ref{fig:4} provides an intuitive visual narrative of vibroacoustic dynamics in the double-shell system. Through a series of snapshots, it depicts dimensionless sound pressure field and normalized transverse stress distribution (with respect to reference mechanical stress of $\sigma_{\text{ref}} = 1~\text{MPa}$). For image processing in Fig. \ref{fig:4}, the acoustic pressure is mapped onto a square slice with a dimensionless size of $6\times6$ and $450\times450 = 202{,}500$ pixels, while the shear stress is color-coded along each shell at 1° angular intervals. Despite the results accuracy, slight visual artifacts in the acoustic fields indicate residual error. These errors are tied to the finite number of acoustic modes used for numerical solution and diminish by increasing the mode count. Due to our computational constraints, these small errors are neglected. Beyond offering a global picture of the system's vibroacoustic behavior, these snapshots help better interpret the results of Fig. \ref{fig:3}. For instance, faster wave propagation in solid media, stress concentration at the head and tail points, and the inner shell’s vulnerability are clearly observable. Moreover, the snapshots at $t = 1.2$, $1.7$, $3.1$, and $3.6$ correspond to the time instants at which the stress at the head and tail of the inner shell reaches its peak—highlighted by red circles in Figs. \ref{fig:3-a} and \ref{fig:3-b} to better illustrate what is going on at those instants. Moreover, the selected snapshots at $t = 1.2$, $1.7$, $3.1$, and $3.6$ correspond to the peak stress moments at the head and tail points of the inner shell to clarify the system’s dynamic at those moments. 

Snapshots in Fig. \ref{fig:4} also provide valuable insights from an acoustic standpoint into the system's dynamics. The acoustic field inside a double-shell structure is characterized by a cascade of reflections within the inner cavity or echos between the shells in the gap fluid. Each time a wave hits a surface and reflects back, it exchanges a portion of its energy with the shell vibrations. From a broader perspective, one may consider the double-shell assembly along with the interior and gap media as a control volume interacting only with the exterior fluid surrounding the outer shell. Due to the infinite extent of the surrounding acoustic medium, any energy radiated from the control volume is lost irreversibly. Therefore, the system gradually dissipates energy and approaches a state of rest. Following the discussion on Fig. \ref{fig:3}, it is important to note that the effect of the interior fluid on the inner shell dynamics is not limited to high-frequency and low-amplitude stress fluctuations. Multiple reflections within the confined space of the inner shell not only complicate the acoustic field but also enable constructive interference between transmitted, detracted, and radiated waves from the shell surface. As a result, high-pressure wavefronts, commonly referred to as Mach stems, begin to emerge within the internal fluid. The intensity of these Mach stems are often comparable to that of the primary incident wave, allowing them to penetrate through the inner shell and exert a pronounced influence the entire system's vibroacoustic. Snapshot at $t = 2.0$ and $3.1$ in Fig. \ref{fig:4} are examples, where the transmission of intense Mach stems into the gap medium is clearly identifiable. Another notable observation in Fig. \ref{fig:4} is the preservation of acoustic wave intensity by passing through the shells. The acoustic pressure field on either side of each shell appears to be a geometric continuation of that on the opposite side, and wavefronts on both sides intersect the shell surface at the same points. For example, the snapshot at $t = 3.1$ clearly demonstrates this geometric continuity, where Mach stems within the interior medium (MSI), inter-shell gap fluid (MSG), and the exterior medium (MSE) align along a coherent trajectory. Often termed "shell transparency to shock wave", this phenomenon occurs when the same fluid surrounds both sides of a shell \cite{iakovlev2007submerged}.

\noindent \textbf{Vibration absorber attachment:} with the primary objective of minimizing the mechanical stress especially in the inner shell, here, we investigate the performance of vibration absorbers in controlling the vibroacoustics of the double-shell system. For this purpose, the effect of key parameters, including the mass, linear and nonlinear stiffness, and damping as well as the attachment location and the number of absorbers used, is evaluated. It should be noted that by analyzing the role of each parameter individually, we are constrained to comparing only a limited number of reasonable configurations, while holding the remaining  parameters constant. Consequently, this procedure does not guarantee an optimal set of absorber parameters. For achieving an optimized configuration, one may consider employing standard optimization protocols. For instance, in our previous studies \cite{hasheminejad2024energy, hasheminejad2021numerical}, we utilized Genetic Algorithm to determine optimal absorber parameters for better suppressing acoustic radiation from double-beam and double-plate structures. Such tools, although effective in finding high-performing solutions, often function as opaque solvers without providing transparent links between the obtained results and the underlying physics. For this reason, we rather follow a step-by-step optimization protocol in this chapter. While acknowledging the suboptimal performance of the adopted approach, we believe it aligns well with our focus on interpreting the physical effects of vibration absorber parameters.

\begin{figure}[H]
\centering
\begin{minipage}[t]{0.5\textwidth}
        \centering
        \begin{overpic}[height=0.2\textheight,trim= 0cm 0cm 0cm 0.0cm,clip]{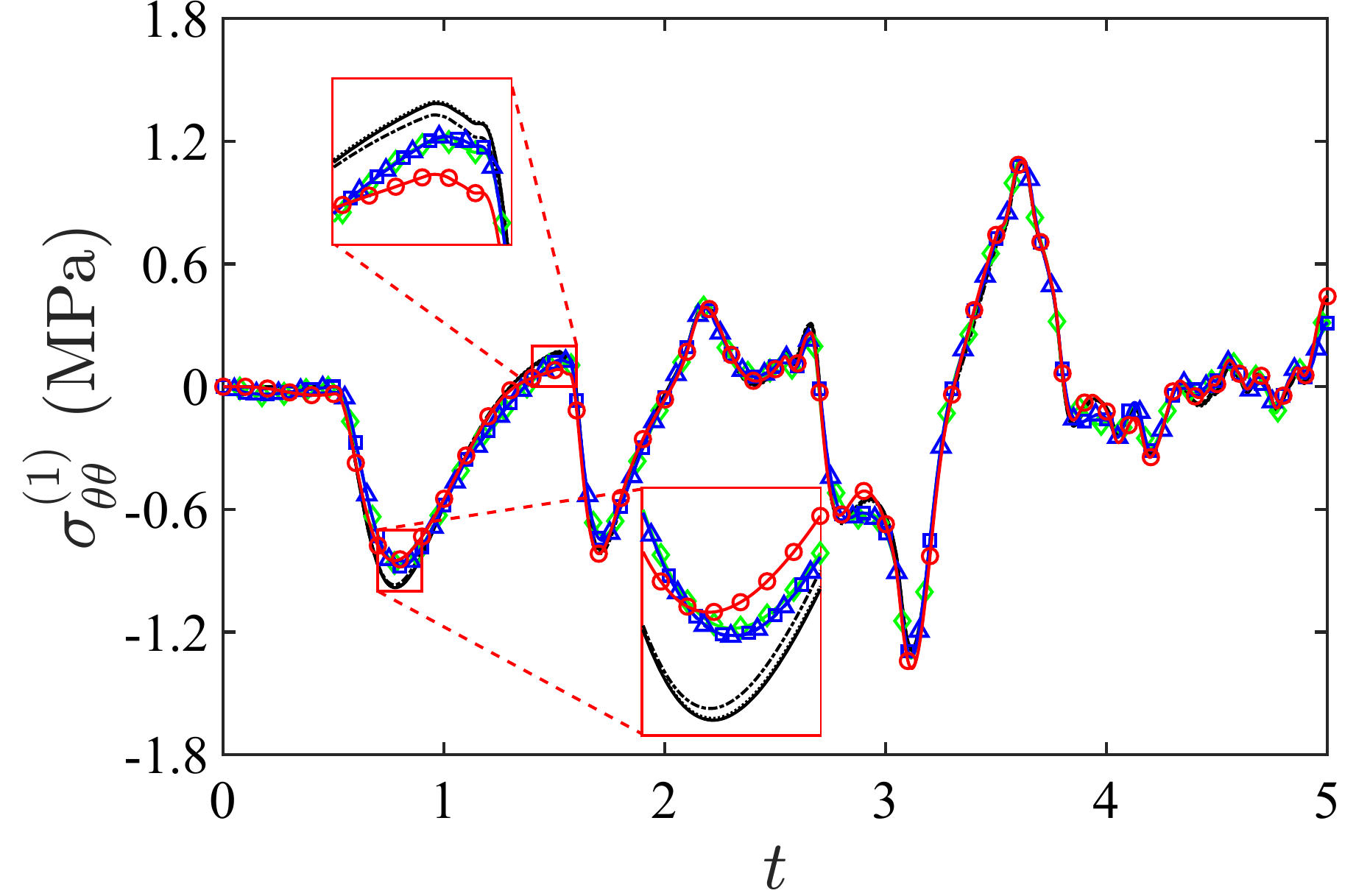}
        \end{overpic}
        \subcaption{Inner shell head point}
        \label{fig:5-a}
    \end{minipage}%
    \begin{minipage}[t]{0.5\textwidth}
        \centering
        \begin{overpic}[height=0.2\textheight,trim= 0cm 0cm 0cm 0.0cm,clip]{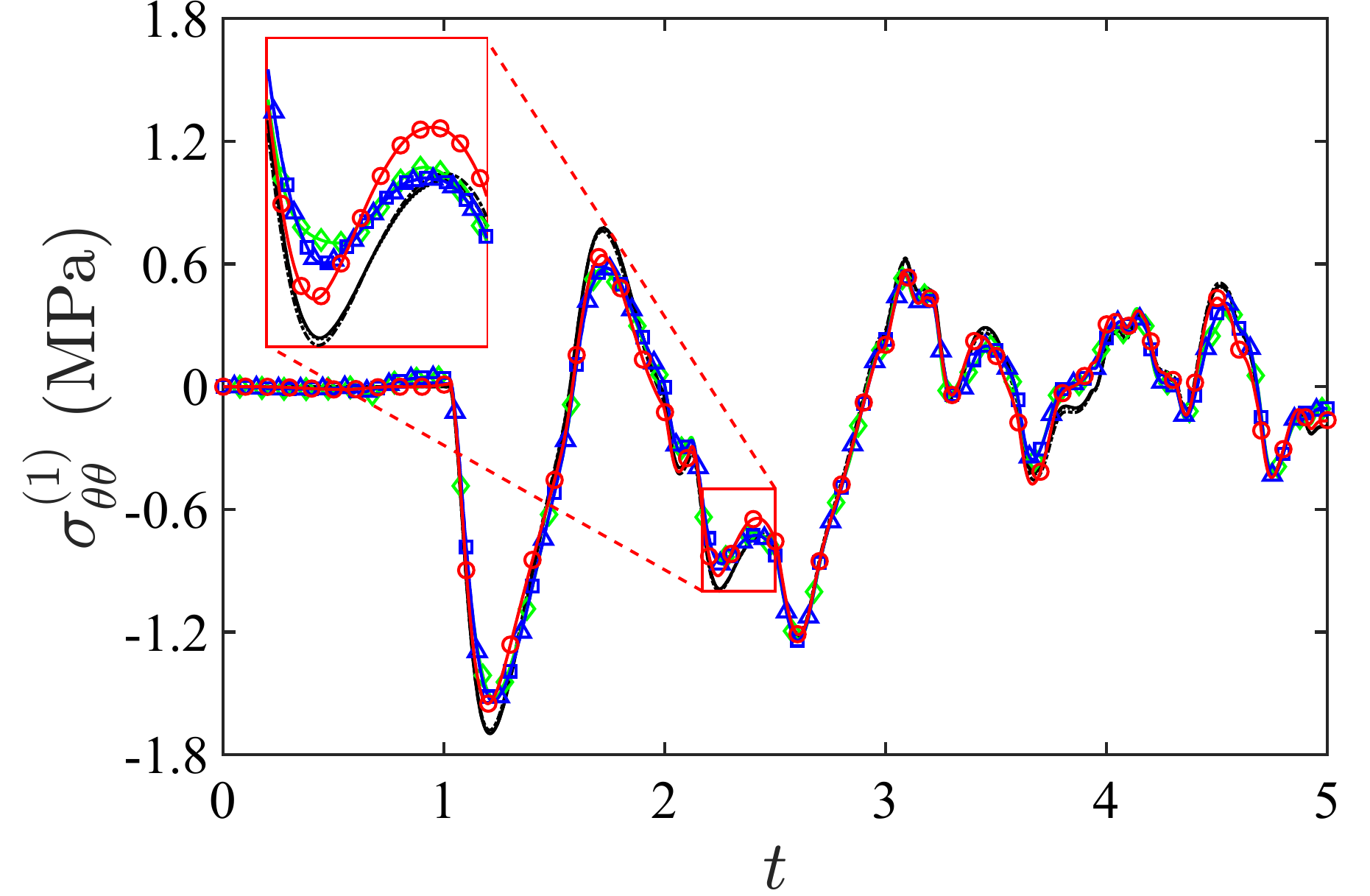}
        \put(46,47){\includegraphics[width=0.5\textwidth]{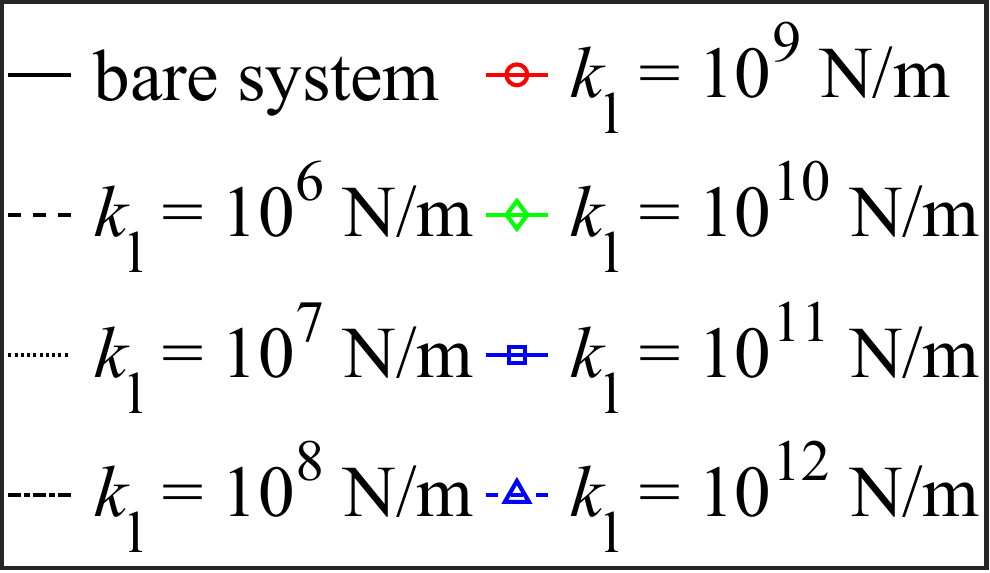}}
        \end{overpic}
        \subcaption{Inner shell tail point}
        \label{fig:5-b}
    \end{minipage}%
    \caption{Effect of absorber linear stiffness on stress induced at inner shell's head and tail.}
    \label{fig:5}
\end{figure}

\noindent $\bullet$ Linear stiffness: the effect of the vibration absorber linear stiffness is studied in Fig. \ref{fig:5}. To do so, a single undamped linear absorber of mass $m = 10.0\ \text{kg}$ is assumed to be linking the shells at their head points. Accordingly, different values are considered for the linear stiffness $k_\text{l}$, and for each case, the stress induced on the inner shell at the head and tail points is compared with that of the bare system in Figs. \ref{fig:5-a} and \ref{fig:5-b}, respectively. A prominent feature observed here is the earlier initiation of vibrations in the inner shell by adding a dynamic absorber to the model. Unlike the bare system, where oscillations in the inner shell emerges only after the incident wave reaches it at $t = 0.5$, the coupled configuration responds immediately at $t = 0$. This behavior is attributed to the absorber acting as a mechanical link and enabling direct force transmission between the shells. As illustrated in Fig. \ref{fig:5} and further examined in the subsequent results, the presence of a single absorber may introduce marginal changes to the early vibroacoustic response with only slight perturbations near the transient extrema. Achieving effective structural control, on the other hand, necessitates the incorporation of multiple absorbers in the model configuration. Therefore, our objective here in not to quantitatively assess the absorber’s efficiency in vibration suppression, but rather to identify any deviations, even though slight and only at extrema, from the bare system's response as indicators of its influence. According to \ref{fig:5}, for linear stiffness values $k_\text{l} \leq 10^8\ \text{N/m}$, the presence of the absorber has practically no effect on the system dynamics. Increasing the stiffness to $k_\text{l} = 10^9\ \text{N/m}$ results in a noticeable mitigation of stress fluctuations in the inner shell. By further increasing the stiffness values by another order of magnitude or more, i.e., $k_\text{l} \geq 10^{10}\ \text{N/m}$, the system behavior approaches a limiting condition, in which the absorber behaves as a rigid link connecting the shells at the attachment location. To preserve the absorber's effectiveness and avoid excessive stiffening, the linear stiffness is henceforth fixed at $k_\text{l} = 10^9\ \text{N/m}$ for the remainder of the numerical analysis. Interestingly, this value is physically justifiable choice, as it approximates the tensile stiffness of a steel rod with a circular cross-section of radius $2\ \text{cm}$ and length $25\ \text{cm}$.

\begin{figure}[H]
\centering
\begin{minipage}[t]{0.5\textwidth}
        \centering
        \includegraphics[height=0.2\textheight]{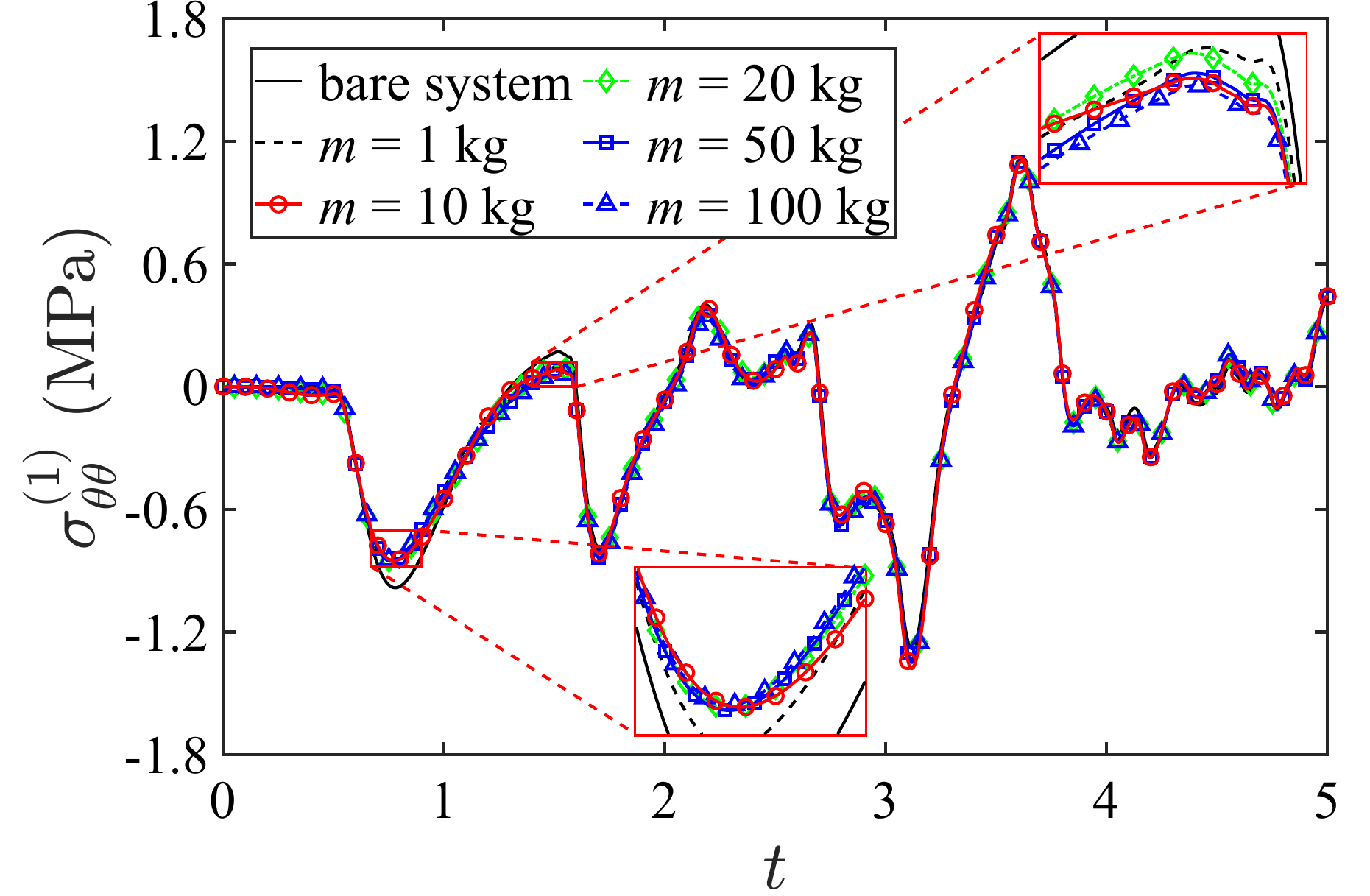}
        \subcaption{Inner shell head point}
        \label{fig:6-a}
    \end{minipage}%
    \begin{minipage}[t]{0.5\textwidth}
        \centering
        \includegraphics[height=0.2\textheight]{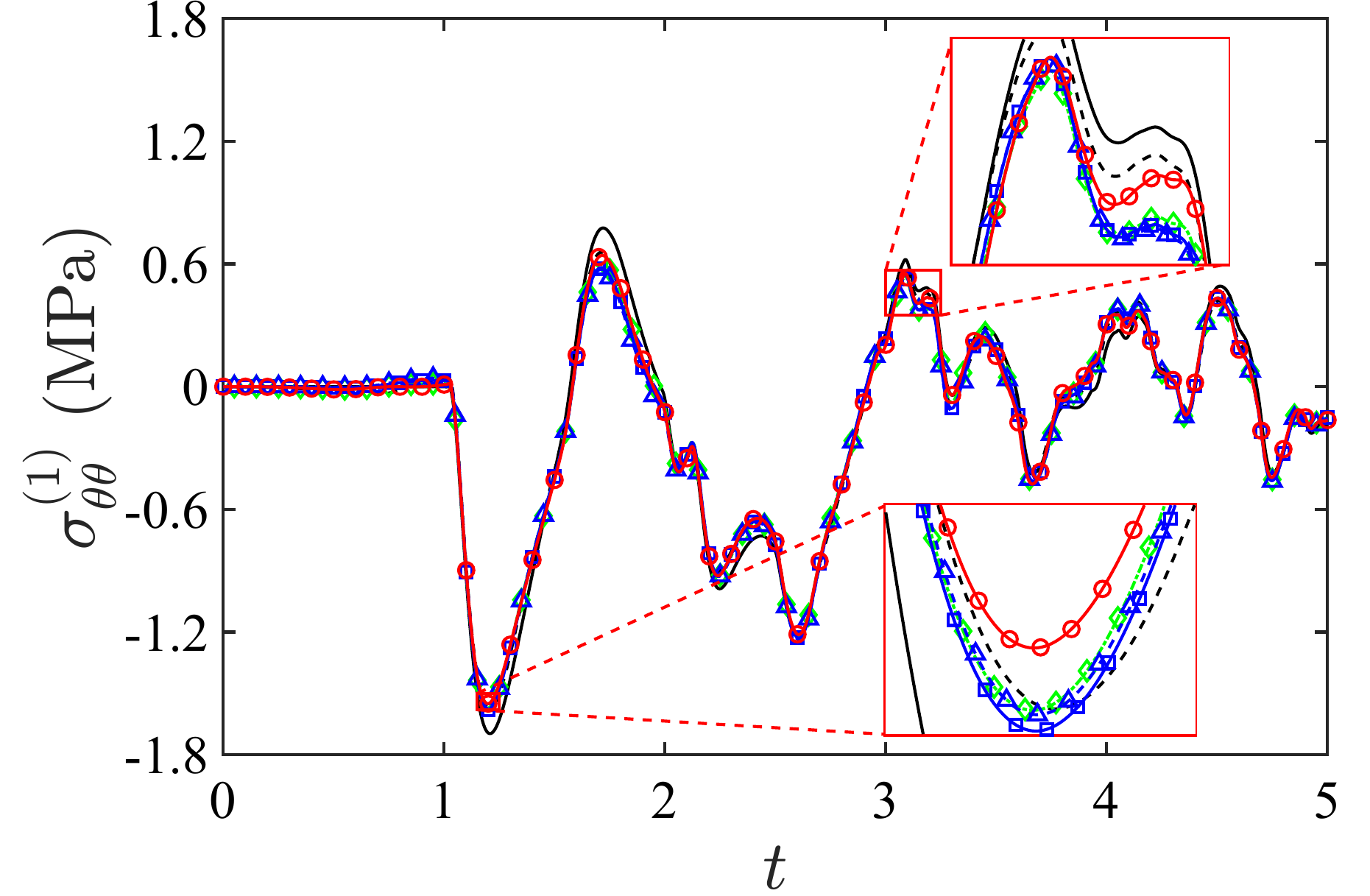}
        \subcaption{Inner shell tail point}
        \label{fig:6-b}
    \end{minipage}%
    \caption{Effect of absorber mass on  stress induced at inner shell's head and tail points.}
    \label{fig:6}
\end{figure}

\noindent $\bullet$ Mass: Fig. \ref{fig:6} examines the effect of mass on the performance of a linear undamped absorber, interconnecting the shells at head points with a linear spring stiffness of $k_\text{l} = 10^9\ \text{N/m}$. The analysis considers five different values for the absorber mass $m$, and compares the induced stress on the inner shell at the head and tail points with that of the bare system in Figs. \ref{fig:6-a} and \ref{fig:6-b}, respectively. Similar to the findings on linear stiffness in Fig. \ref{fig:5}, the results for different mass values exhibit trends closely matching that of the bare system, with only slight deviations at the extrema. However, the system’s behavior at limiting conditions for the absorber mass is fundamentally different. Unlike in the stiffness case, reducing the absorber mass doesn’t recover the bare system response. Instead, with diminishing inertia, the absorber behaves as a pair of springs connected in series, yielding an equivalent stiffness of $0.5k_\text{l}$ between the shells at the attachment point. On the other hand, as the mass becomes significantly large ($m \geq 50\ \text{kg}$), the  dominant inertia suppresses the dynamics of the absorber to a quasi-static behavior over short time scales. Consequently, the absorber mass acts as a rigid anchor, with each shell tethered to it independently via a spring of stiffness $k_\text{l}$. According to the magnified views in Figs. \ref{fig:6-a} and \ref{fig:6-b}, hereafter, a mass of $m = 10\ \text{kg}$ is adopted to achieve an acceptable performance in the subsequent analyses. It is worth noting that the modeled double-shell system has a total mass of approximately $577\ \text{kg}$ per unit length, and therefore, adding a $10\ \text{kg}$ absorber will only result in a modest mass increase not more than 1.7$\%$. 

\begin{figure}[H]
\centering
\begin{minipage}[t]{0.5\textwidth}
        \centering
        \begin{overpic}[height=0.2\textheight,trim= 0cm 0cm 0cm 0.0cm,clip]{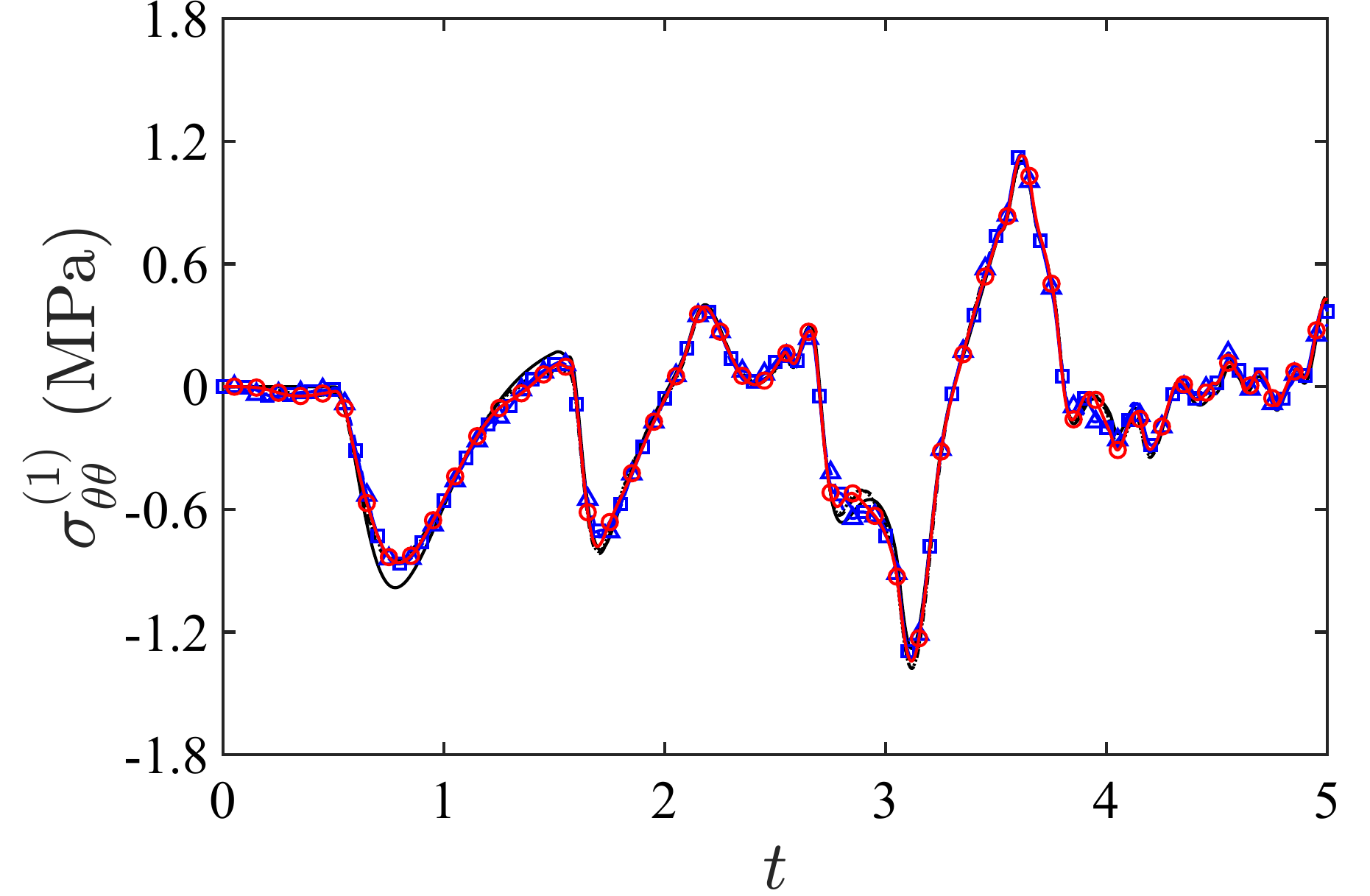}
        \put(9,43.5){\includegraphics[width=0.57\textwidth]{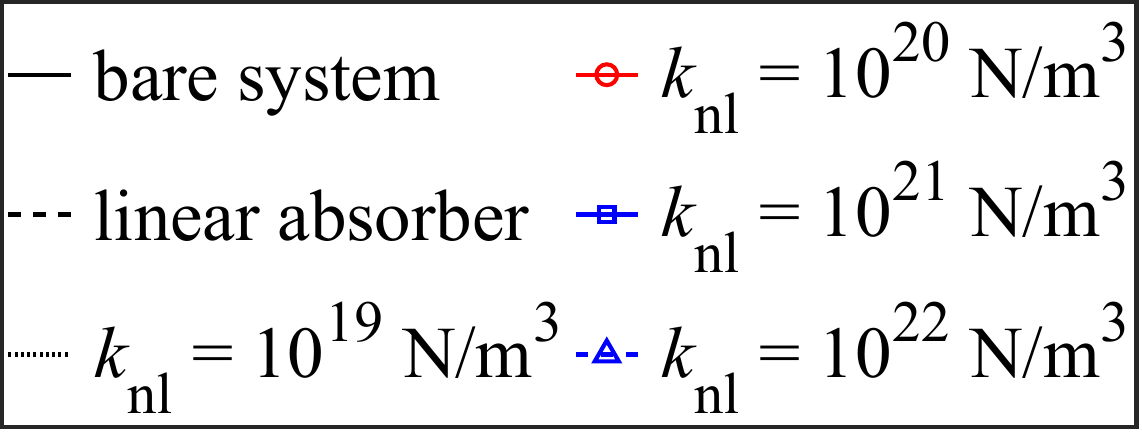}}
        \end{overpic}
        \subcaption{Inner shell head point}
        \label{fig:7-a}
    \end{minipage}%
    \begin{minipage}[t]{0.5\textwidth}
        \centering
        \begin{overpic}[height=0.2\textheight,trim= 0cm 0cm 0cm 0.0cm,clip]{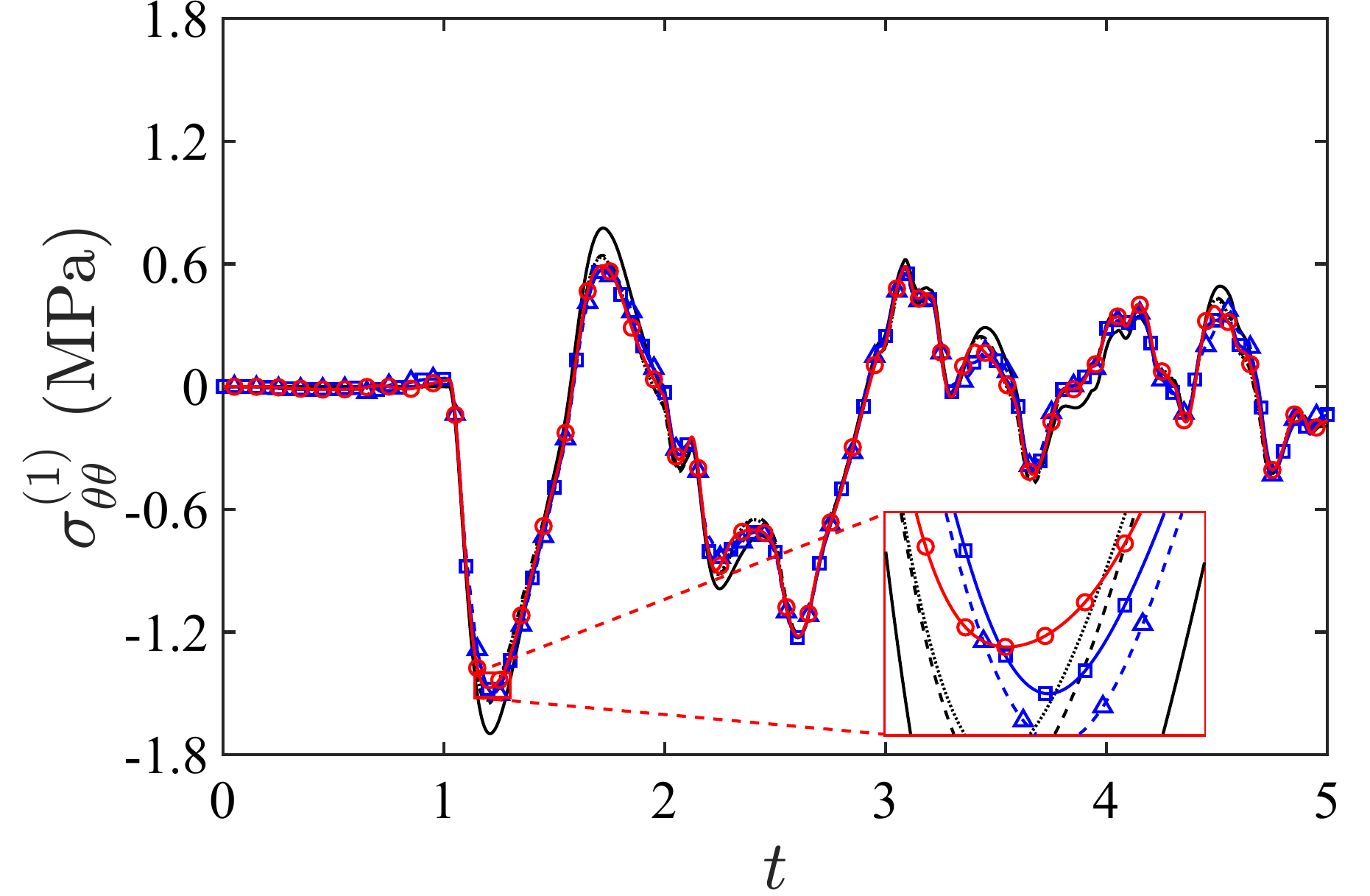}
        \end{overpic}
        \subcaption{Inner shell tail point}
        \label{fig:7-b}
    \end{minipage}%
    \caption{Effect of absorber nonlinear stiffness on stress at inner shell's head and tail.}
    \label{fig:7}
\end{figure}

\noindent $\bullet$ Nonlinear stiffness: Fig. \ref{fig:7} investigates the effect of nonlinear stiffness of a absorber linking the shells at head points with a linear spring stiffness of $k_\text{l} = 10^9\ \text{N/m}$, and mass of $m = 10\ \text{kg}$. Four different values are considered for the nonlinear stiffness component of the absorber spring. For each case, the induced stress on the inner shell is evaluated at the head and tail points. As shown in Figs. \ref{fig:7-a} and \ref{fig:7-b}, the results are then compared with those obtained from the bare system as well as the system equipped with a linear absorber, respectively. For small values of nonlinear stiffness ($k_\text{nl} \leq 10^{19}\ \text{N/m}^3$), the system’s response remains nearly identical to that of a linear absorber. Once the nonlinear stiffness increases to $k_\text{nl} = 10^{20}\ \text{N/m}^3$, the absorber begins to perform more effectively. Considering the displacement amplitudes of the shells, this choice of $k_\text{nl}$ results in a comparable force contributions from both the linear and nonlinear components of the spring. Further increasing the nonlinear stiffness beyond this level, i.e., $k_\text{nl} \geq 10^{21}\ \text{N/m}^3$, introduces strong nonlinearity into the double-shell system. Similar to the linear stiffness case, the absorber’s behavior converges toward that of a rigid link in the limit, which impairs its stress mitigation capability.

\begin{figure}[H]
\centering
\begin{minipage}[t]{0.5\textwidth}
        \centering
        \includegraphics[height=0.2\textheight]{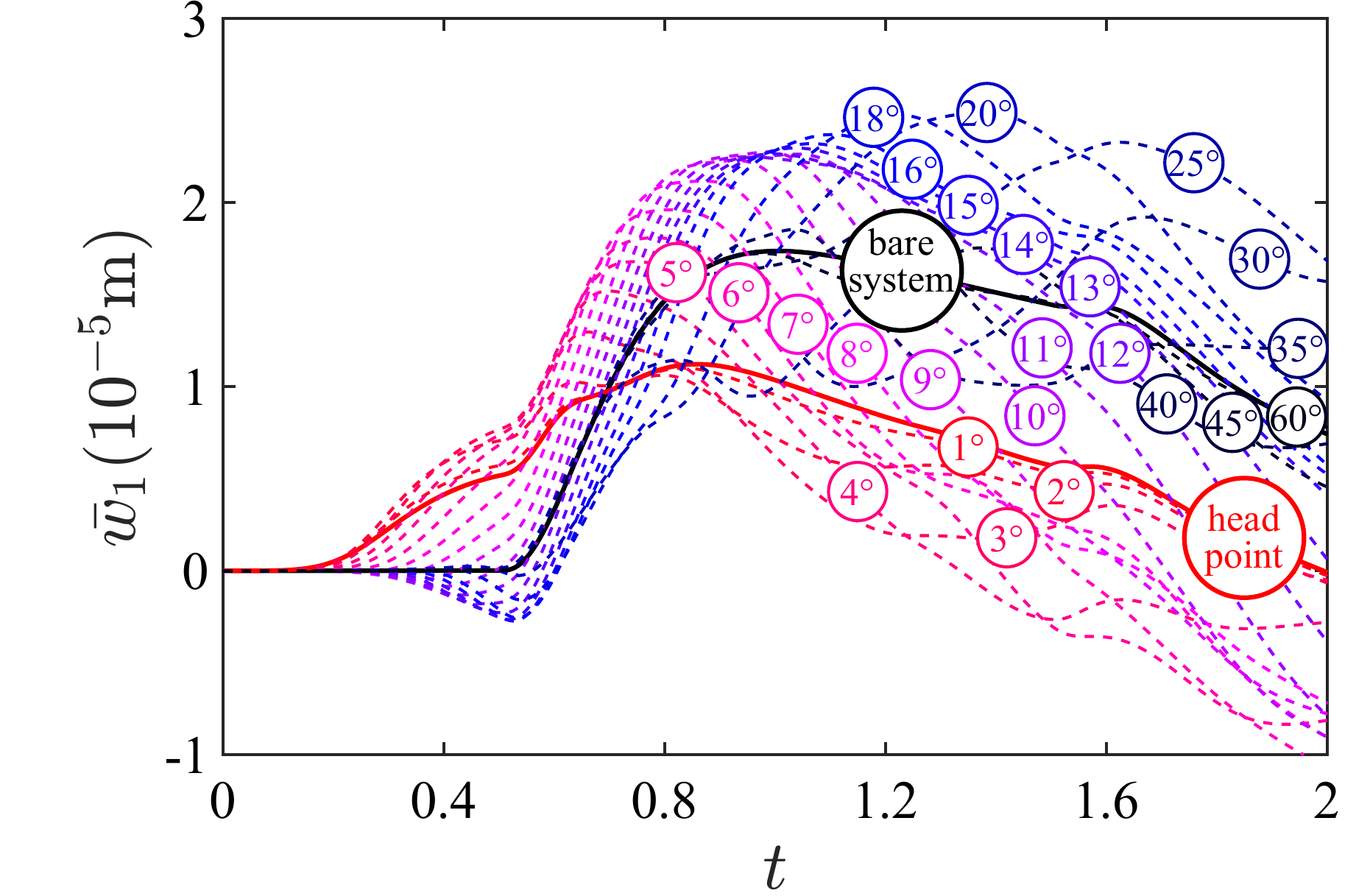}
        \subcaption{Inner shell head point}
        \label{fig:8-a}
    \end{minipage}%
    \begin{minipage}[t]{0.5\textwidth}
        \centering
        \includegraphics[height=0.2\textheight]{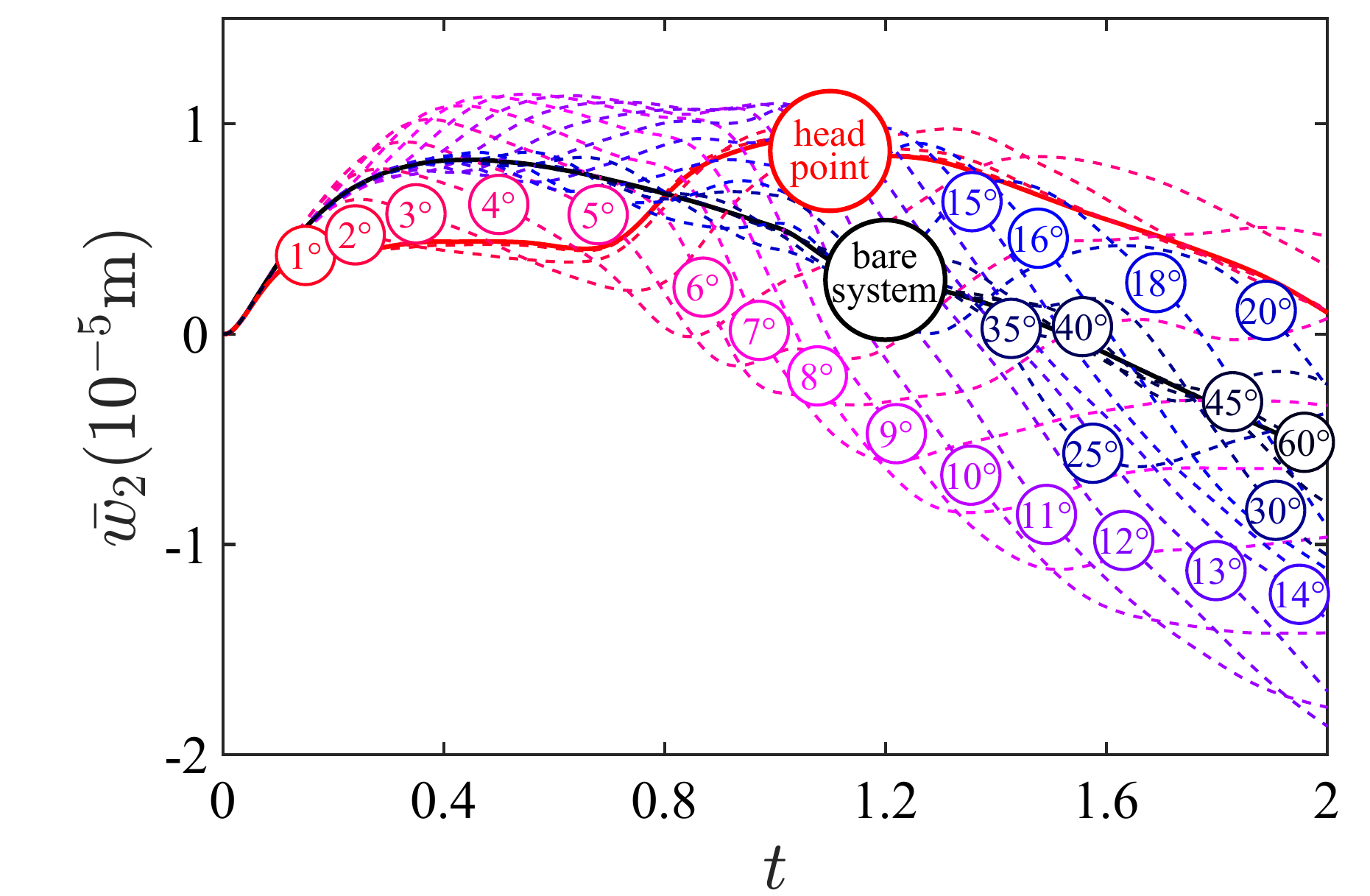}
        \subcaption{Outer shell head point}
        \label{fig:8-b}
    \end{minipage}%
    \caption{Effect of absorber attachment point on displacement of inner and outer shells at head points.}
    \label{fig:8}
\end{figure}

\noindent $\bullet$ Attachment position: as noted earlier, deploying multiple vibration absorbers in the system configuration enables more efficient structural control. It is evident that each absorber exerts a dominant local influence on the shell dynamics near its attachment point, while this effect gradually fades by increasing distance. It is essential to explore the vicinity of effectiveness for a single absorber prior to analyzing configurations with multiple absorbers. To this end, an angular offset is introduced to the attachment location of the undamped absorber with $k_\text{l} = 10^9\ \text{N/m}$, $k_\text{nl} = 10^{20}\ \text{N/m}^3$ and $m = 10\ \text{kg}$. Figs. \ref{fig:8-a} and \ref{fig:8-b} depict the transverse displacement at the head points of the inner and outer shells for various attachment positions over the time interval $0 \leq t \leq 2$, respectively. In both figures, solid black lines show the bare system's response, whereas solid red lines reflect the model’s behavior with an absorber fixed at the head point, where its influence is maximized. The remaining dashed curves represent the system's response for absorber placements at various angular offsets from the main axis, with color gradients proportionally scaled to the magnitude of the offset. According to Fig. \ref{fig:8}, even slight deviations in the absorber’s attachment point can significantly affect its performance during the early stages following the shockwave's impact. For instance, an angular offset as small as $3^\circ$ leads to clearly noticeable changes in the system response. The discrepancy widens with greater angular offsets. Beyond $30^\circ$, the absorber's effect becomes marginal, and at $60^\circ$, the transient behavior becomes almost indistinguishable from that of the bare system.

\begin{figure}[H]
\centering
\begin{minipage}[t]{0.5\textwidth}
        \centering
        \includegraphics[height=0.2\textheight]{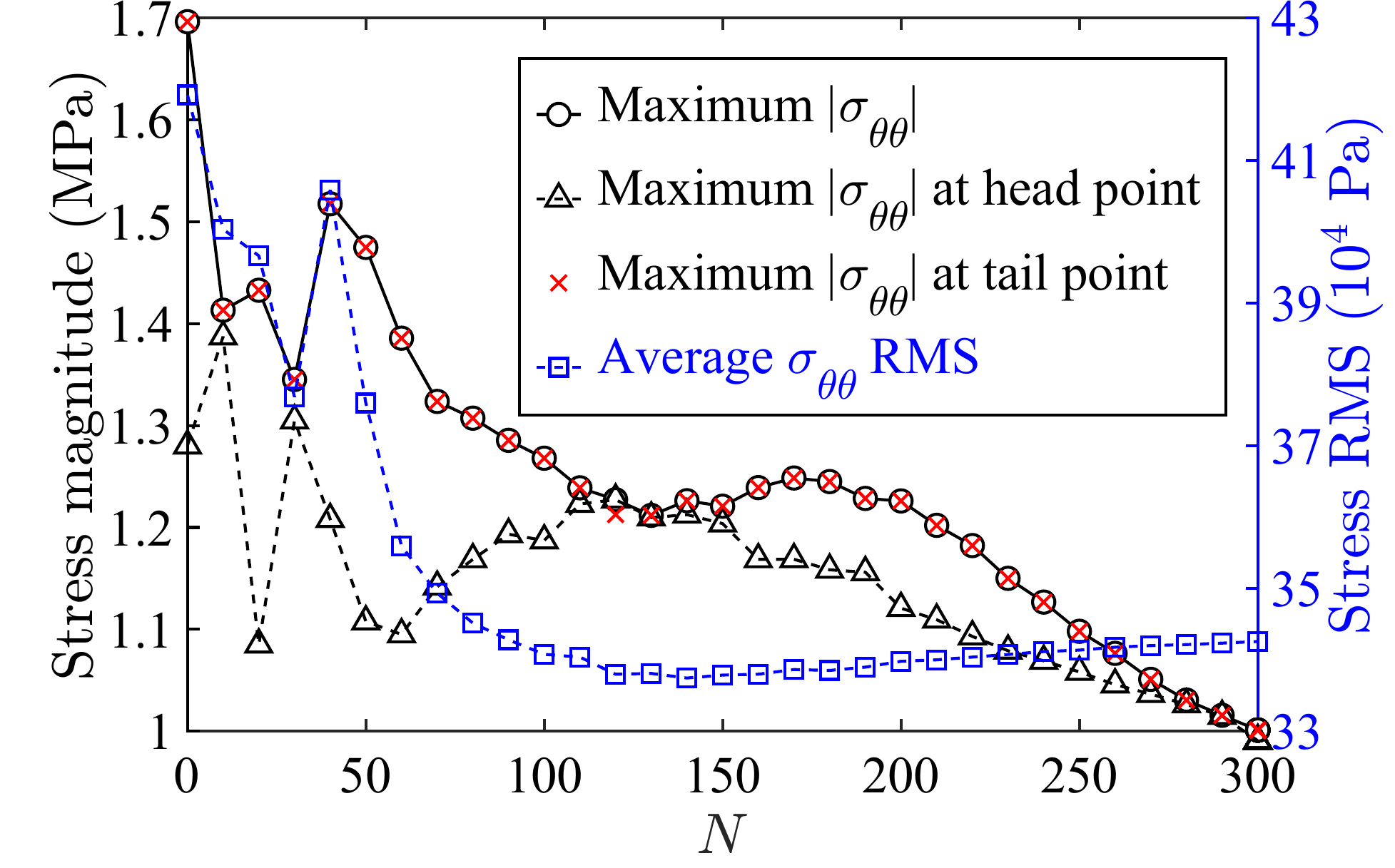}
        \subcaption{Inner shell}
        \label{fig:9-a}
    \end{minipage}%
    \hfill
    \begin{minipage}[t]{0.5\textwidth}
        \centering
        \includegraphics[height=0.2\textheight]{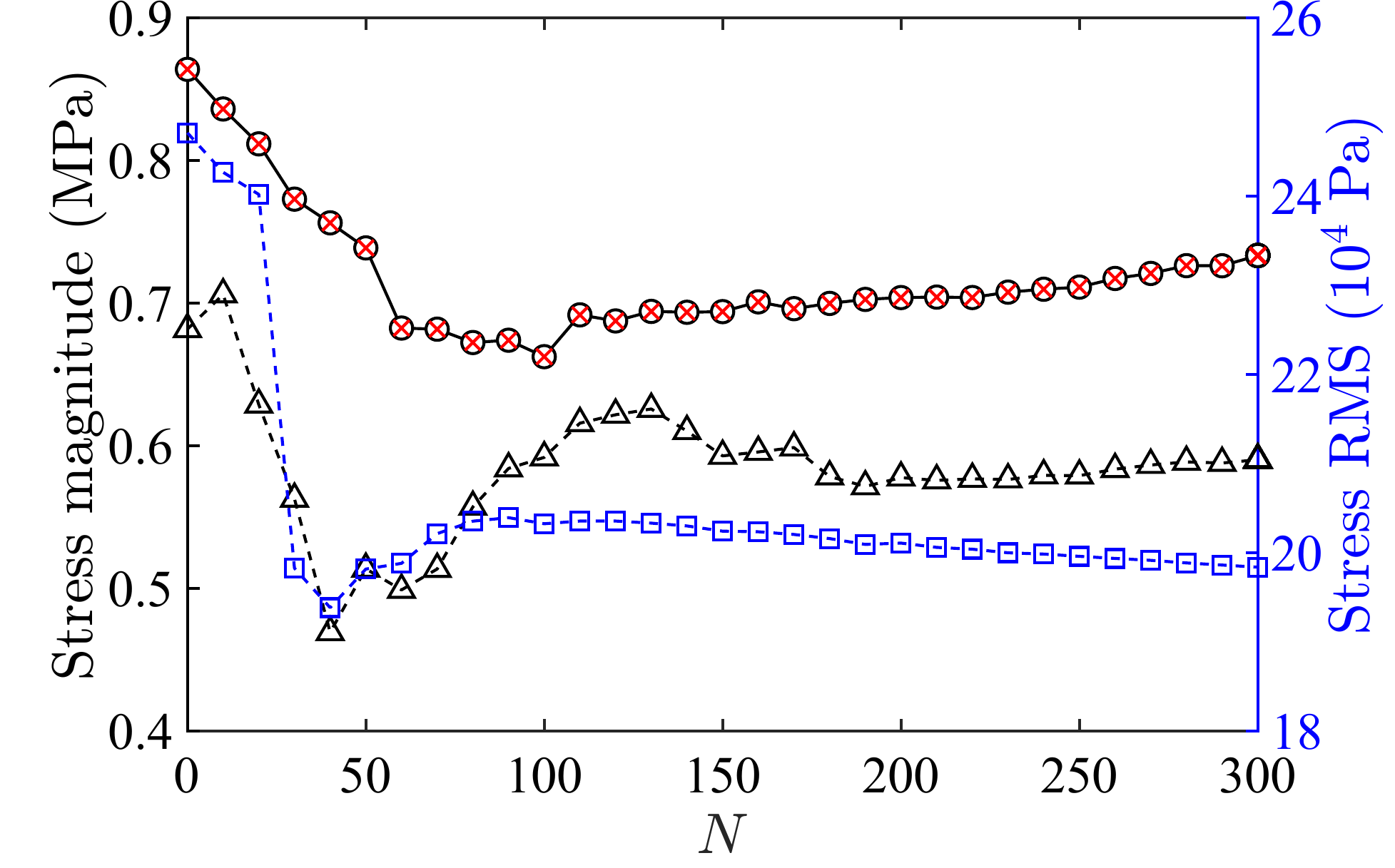}
        \subcaption{Outer shell}
        \label{fig:9-b}
    \end{minipage}%
    \vspace{0.5cm}
    \begin{minipage}[t]{0.5\textwidth}
        \centering
        \includegraphics[height=0.2\textheight]{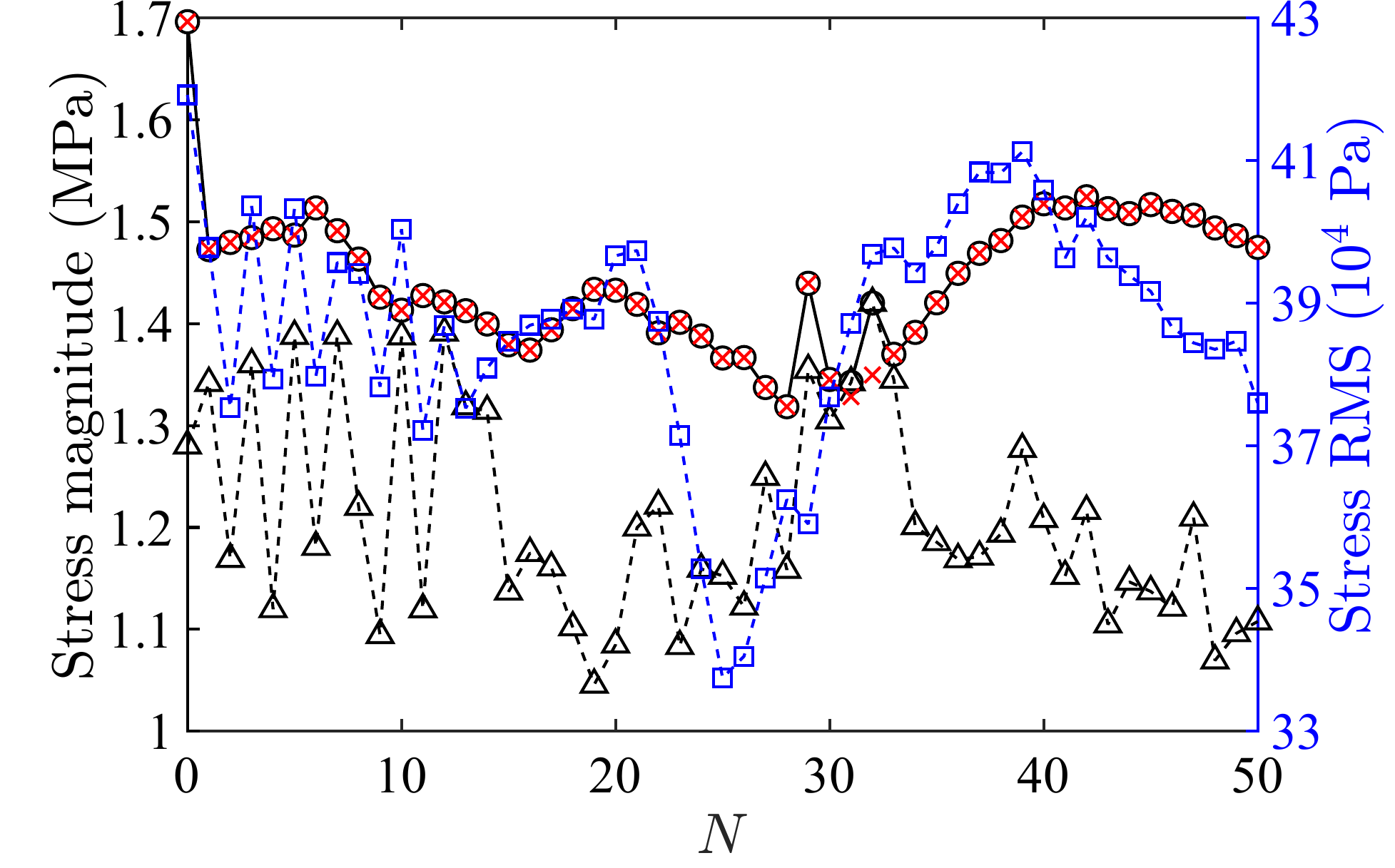}
        \subcaption{Inner shell}
        \label{fig:9-c}
    \end{minipage}%
    \hfill
    \begin{minipage}[t]{0.5\textwidth}
        \centering
        \includegraphics[height=0.2\textheight]{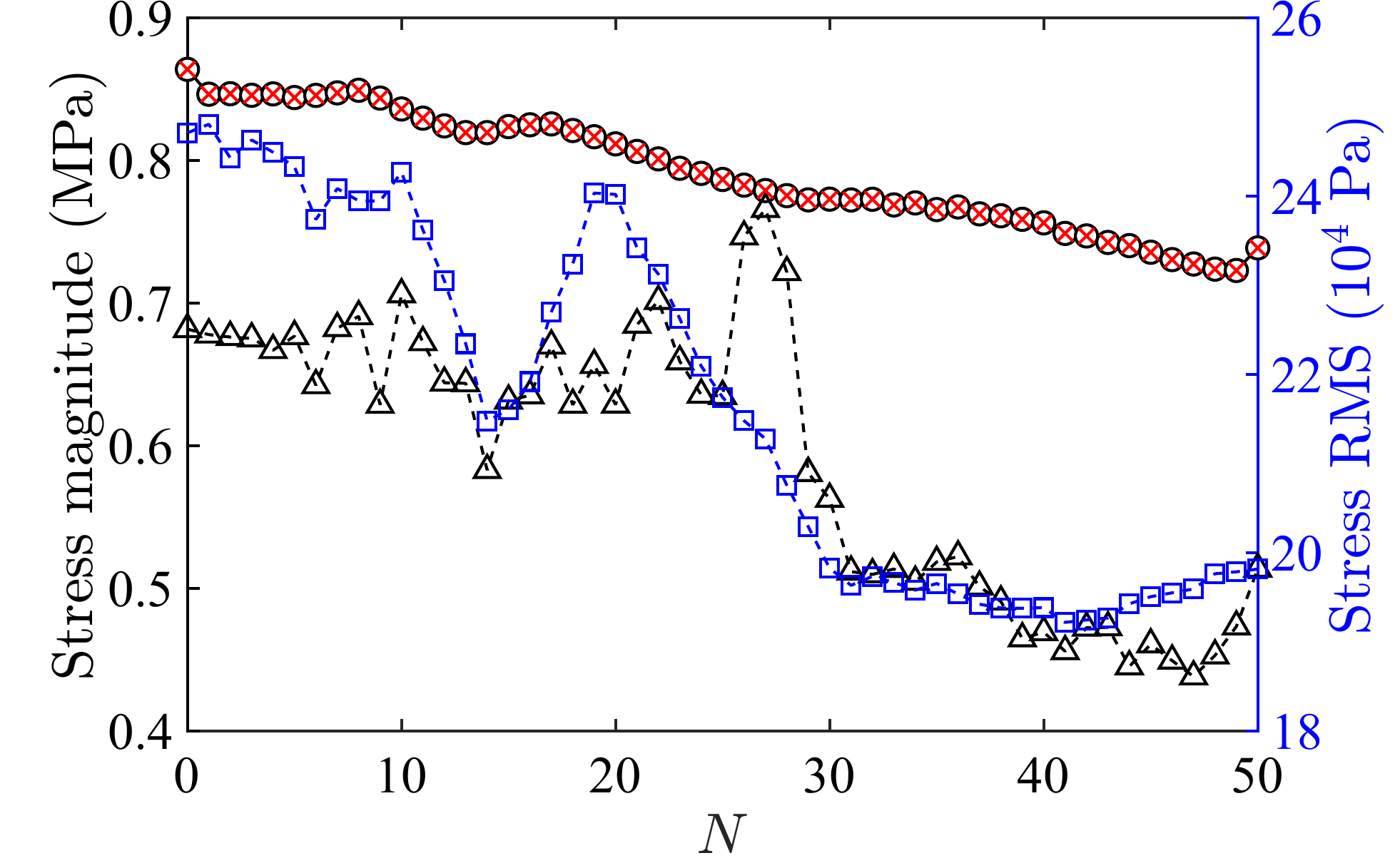}
        \subcaption{Outer shell}
        \label{fig:9-d}
    \end{minipage}%
    \caption{Effect of number of employed absorbers on the stress induced at both shells.}
    \label{fig:9}
\end{figure}

\noindent $\bullet$ Multiple absorbers: with a qualitative understanding of how absorber position shapes its performance, we now examine the effect of incorporating multiple absorbers within the system configuration. To do so, we summarize two sets of stress metrics in Fig. \ref{fig:9}: first, the maximum transverse stress magnitude over time interval $0 \leq t \leq 5$ induced on the corresponding shell surface as well as head and tail points (left y-axis); and second, the RMS stress averaged over the shell surface during the same time period (right y-axis). All absorbers are assumed to have identical properties of $k_\text{l} = 10^9~\text{N/m}$, $m = 10~\text{kg}$, and $k_\text{nl} = 10^{20}~\text{N/m}^3$. The first absorber in every configuration connects the head points of the shells, with all subsequent absorbers positioned at regular spacing (see Fig. \ref{fig:1}). Figs. \ref{fig:9-a} and \ref{fig:9-b} depict the variation of transverse stress in the inner and outer shells as a function of absorber count, sampled at intervals of 10 from $N = 0$ to $300$. The results suggest that increasing the number of absorbers initially leads to a noticeable reduction in RMS stress across both shells. Beyond a certain threshold, however, additional absorbers only contribute to structural mass without enhancing stress mitigation. Interestingly, Fig. \ref{fig:9} reinforces an earlier hypothesis from Fig. \ref{fig:3}. The close match between the global peak stress values and local maxima at the tail points suggests that, stress accumulation is most severe at the tail of each shell, regardless of number of absorbers. This pattern persists across most configurations, except for rare deviations wherein the head point overtakes the tail in stress magnitude with a marginal difference.
Figs. \ref{fig:9-a} and \ref{fig:9-b} reveal a pronounced local minimum in both shells' RMS stress profiles for $N \leq 50$. To facilitate a more detailed evaluation, all four stress indicators are reassessed for every integer value of $N$ within this range, and presented in Figs. \ref{fig:9-c} and \ref{fig:9-d}. Sharper fluctuations observed in the inner shell results underscore its higher sensitivity to absorbers' arrangement. Another striking feature in Fig. \ref{fig:9-c} is the zigzagging up-and-down pattern in the stress curves for low values of $N$. Although drawing a definitive general conclusion would be challenging, this pattern appears to be strongly influenced by the parity of the absorber count. In configurations with an even number of absorbers, one is positioned directly at the tail, a structurally critical point as discussed earlier. For odd values of $N$, however, no absorber is located  in this region. By increasing the number of attached absorbers, the spacing between consecutive units decreases, and thereby diminishes the system’s sensitivity to the parity of $N$. Fig. \ref{fig:9-c} shows a sharp reduction in RMS value of stress at $N=25$ for the inner shell. Considering stress mitigation in the inner shell as the performance benchmark for the control mechanisms, this configuration is selected for the remaining analyses in this study. With this choice of $N$, both the RMS and peak absolute transverse stress values in the inner shell reduce by approximately $20\%$ compared to the bare system.

\begin{figure}[H]
\centering
\begin{minipage}[t]{0.5\textwidth}
        \centering
        \includegraphics[height=0.2\textheight]{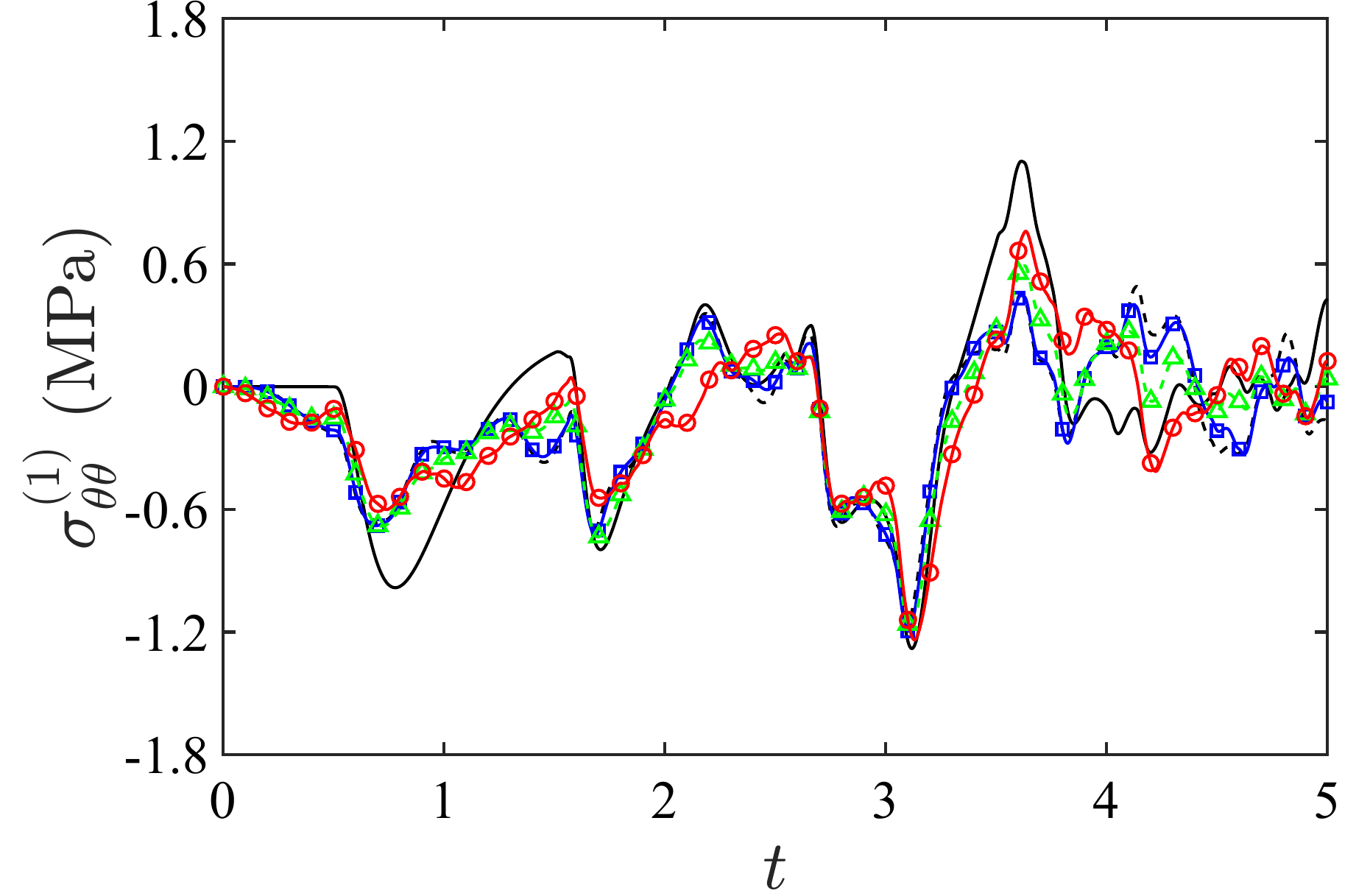}
        \subcaption{Inner shell head point}
        \label{fig:10-a}
    \end{minipage}%
    \hfill
    \begin{minipage}[t]{0.5\textwidth}
        \centering
        \begin{overpic}[height=0.2\textheight,trim= 0cm 0cm 0cm 0.0cm,clip]{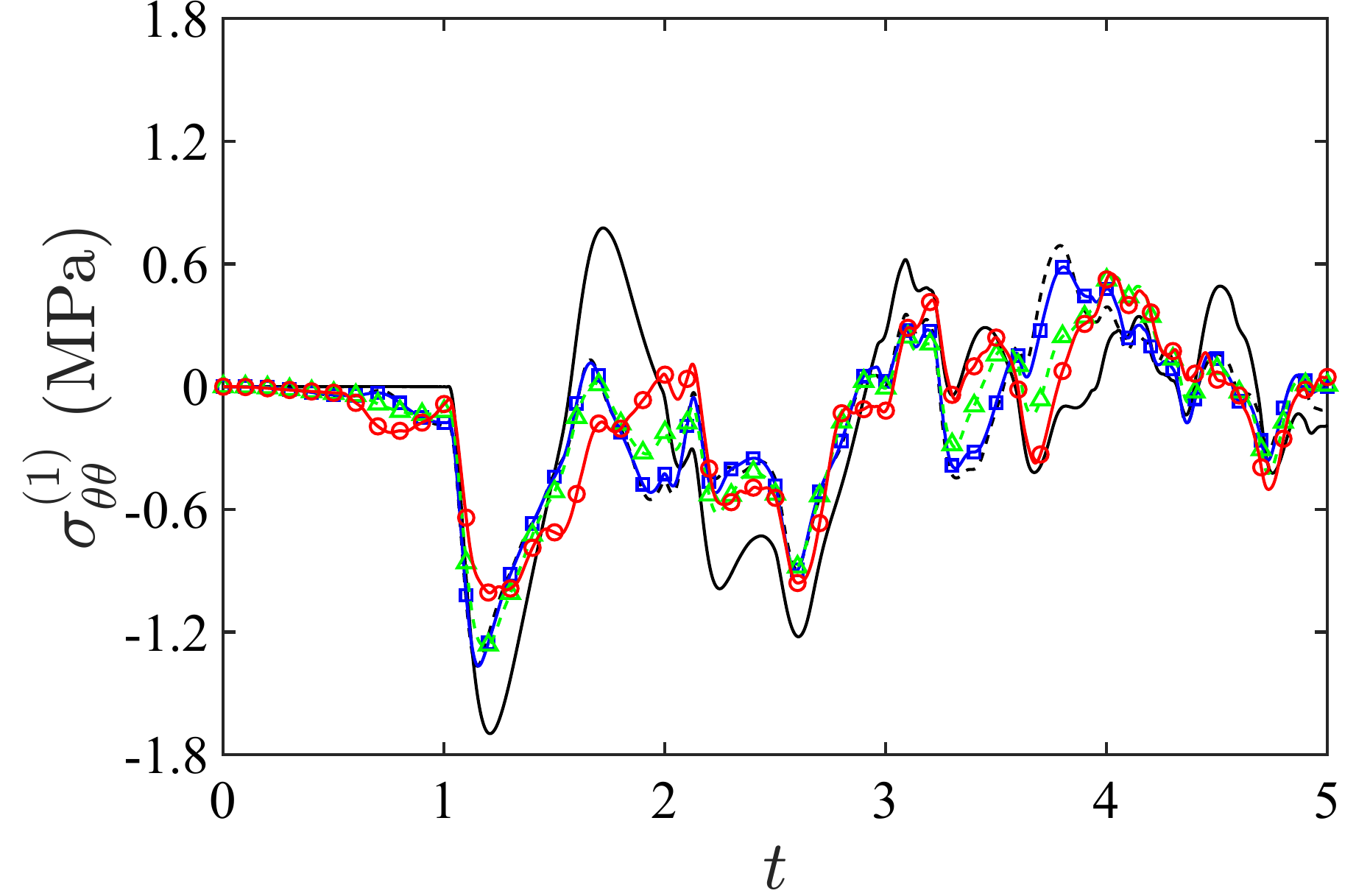}
        \put(-28,48.5){\includegraphics[width=0.59\textwidth]{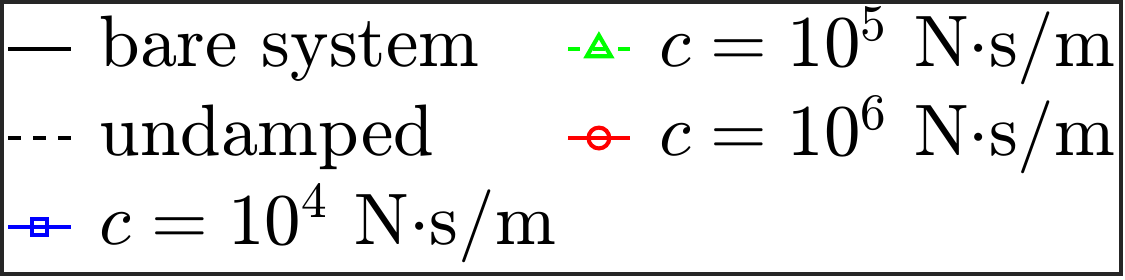}}
        \end{overpic}
        \subcaption{Inner shell tail point}
        \label{fig:10-b}
    \end{minipage}%
    \caption{Effect of damping coefficient of 25 identical absorbers, employed in the structure of double-shell, on the stress induced at inner shell's head and tail points.}
    \label{fig:10}
\end{figure}

\noindent $\bullet$ Damping: here, we investigate how damping influences the efficiency in a system of 25 identical absorbers interlinking the double-shell structure, each characterized by $k_\text{l} = 10^9~\text{N/m}$, $m = 10~\text{kg}$, and $k_\text{nl} = 10^{20}~\text{N/m}^3$.
Figs. \ref{fig:10-a} and \ref{fig:10-b} present a comparison of stress magnitudes at the head and tail points of the inner shell for three different damping values $c$, against undamped absorbers as well as the bare system. As illustrated, low damping values ($c \leq 10^5~\text{N}\!\cdot\!\text{s}/\text{m}$), exert a negligible influence on the stress induced in the structure. Nevertheless, increasing the damping by one order of magnitude to $c = 10^6\ {\text{N}\!\cdot\!\text{s}/\text{m}}$ yields a marked suppression of stress, particularly at the inner shell's tail point. Thus, the damping coefficient is fixed at $c = 10^6\ {\text{N}\!\cdot\!\text{s}/\text{m}}$ for the subsequent analyses in this study.

\begin{figure}[H]
\centering
\begin{minipage}[t]{0.5\textwidth}
        \centering
        \includegraphics[height=0.2\textheight]{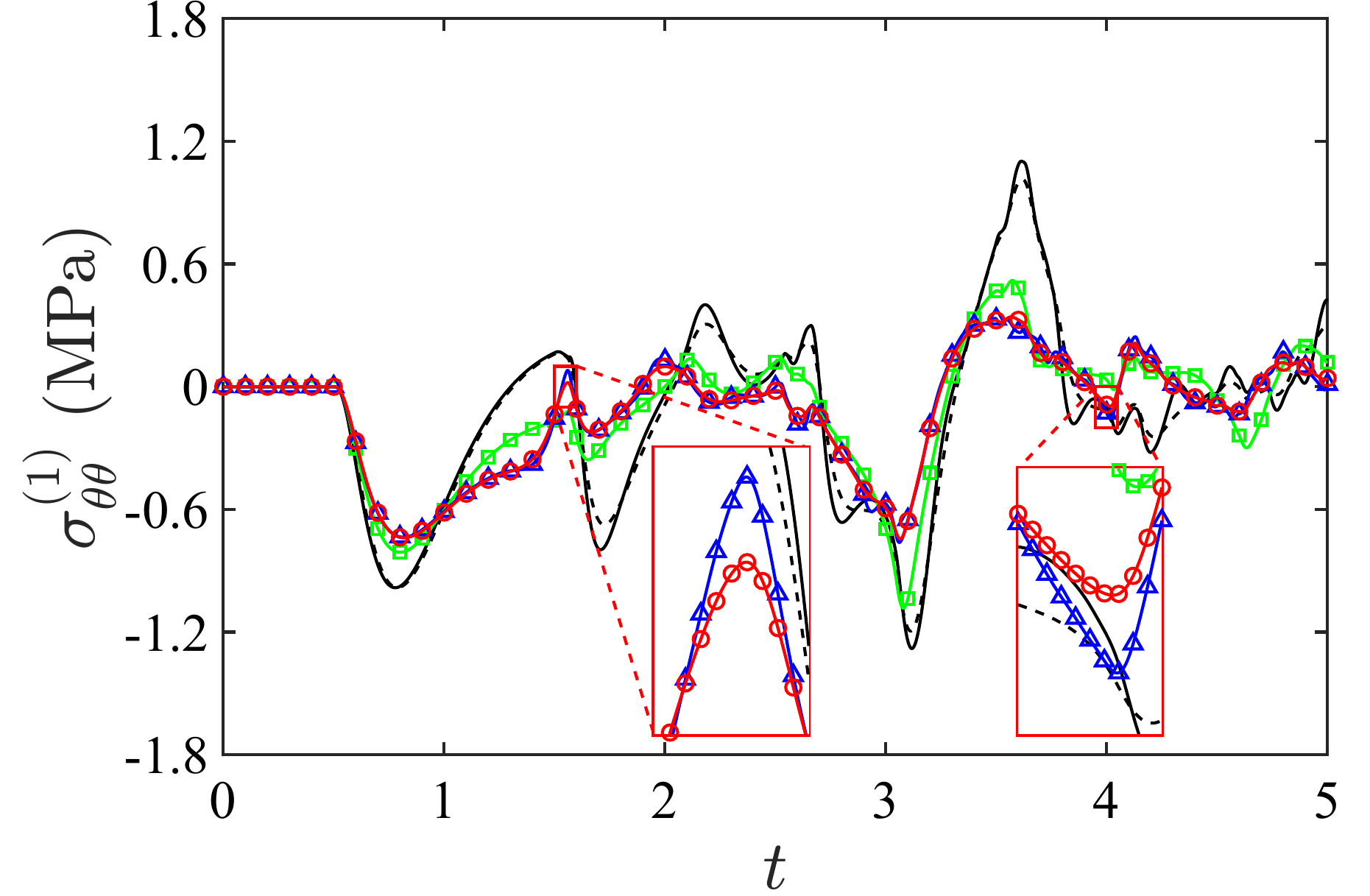}
        \subcaption{Inner shell head point}
        \label{fig:11-a}
    \end{minipage}%
    \hfill
    \begin{minipage}[t]{0.5\textwidth}
        \centering
        \includegraphics[height=0.2\textheight]{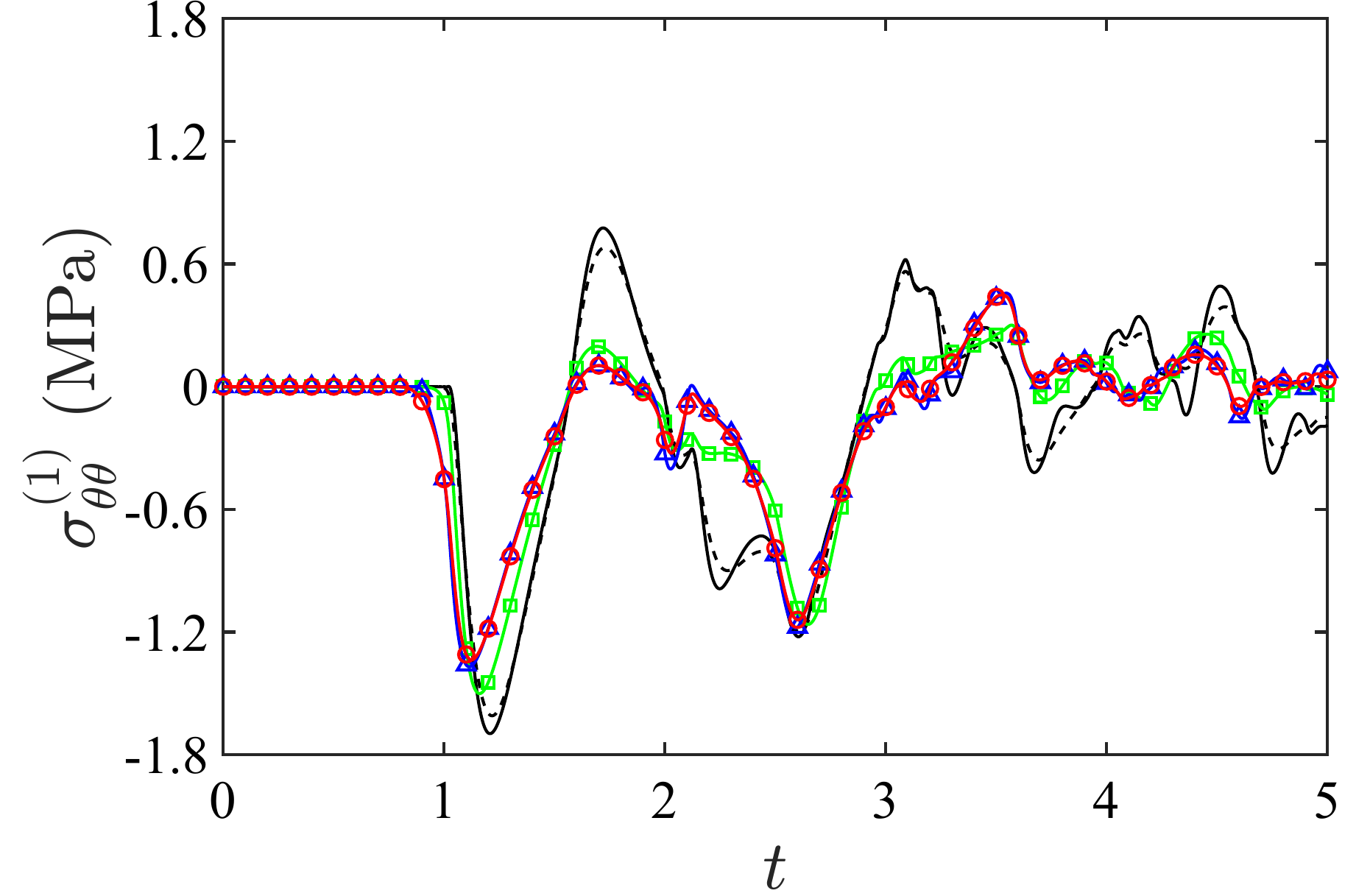}
        \subcaption{Inner shell tail point}
        \label{fig:11-b}
    \end{minipage}%
    \vspace{0.5cm}
    \begin{minipage}[t]{0.5\textwidth}
        \centering
        \includegraphics[height=0.2\textheight]{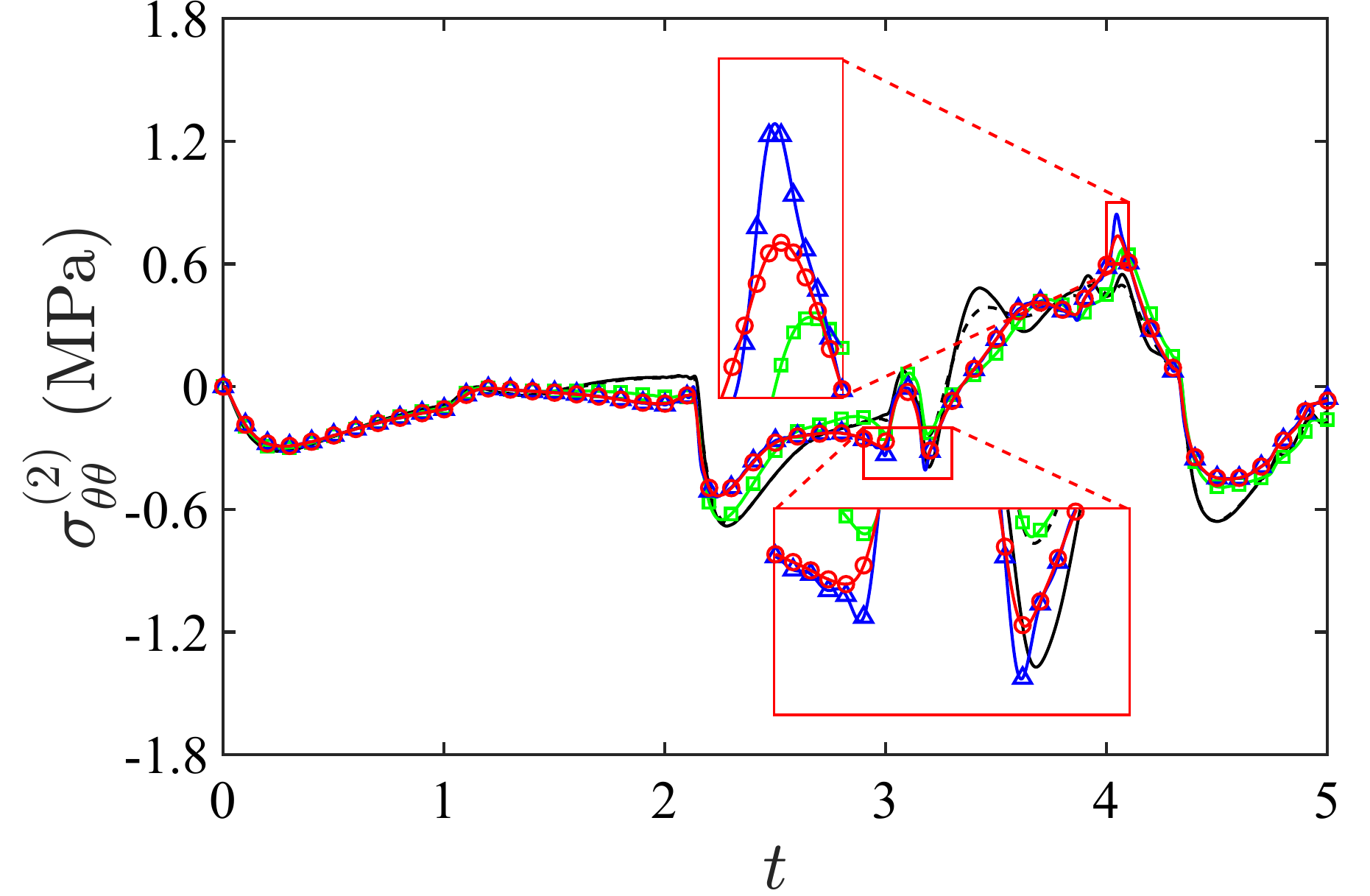}
        \subcaption{Outer shell head point}
        \label{fig:11-c}
    \end{minipage}%
    \hfill
    \begin{minipage}[t]{0.5\textwidth}
        \centering
        \begin{overpic}[height=0.2\textheight,trim= 0cm 0cm 0cm 0.0cm,clip]{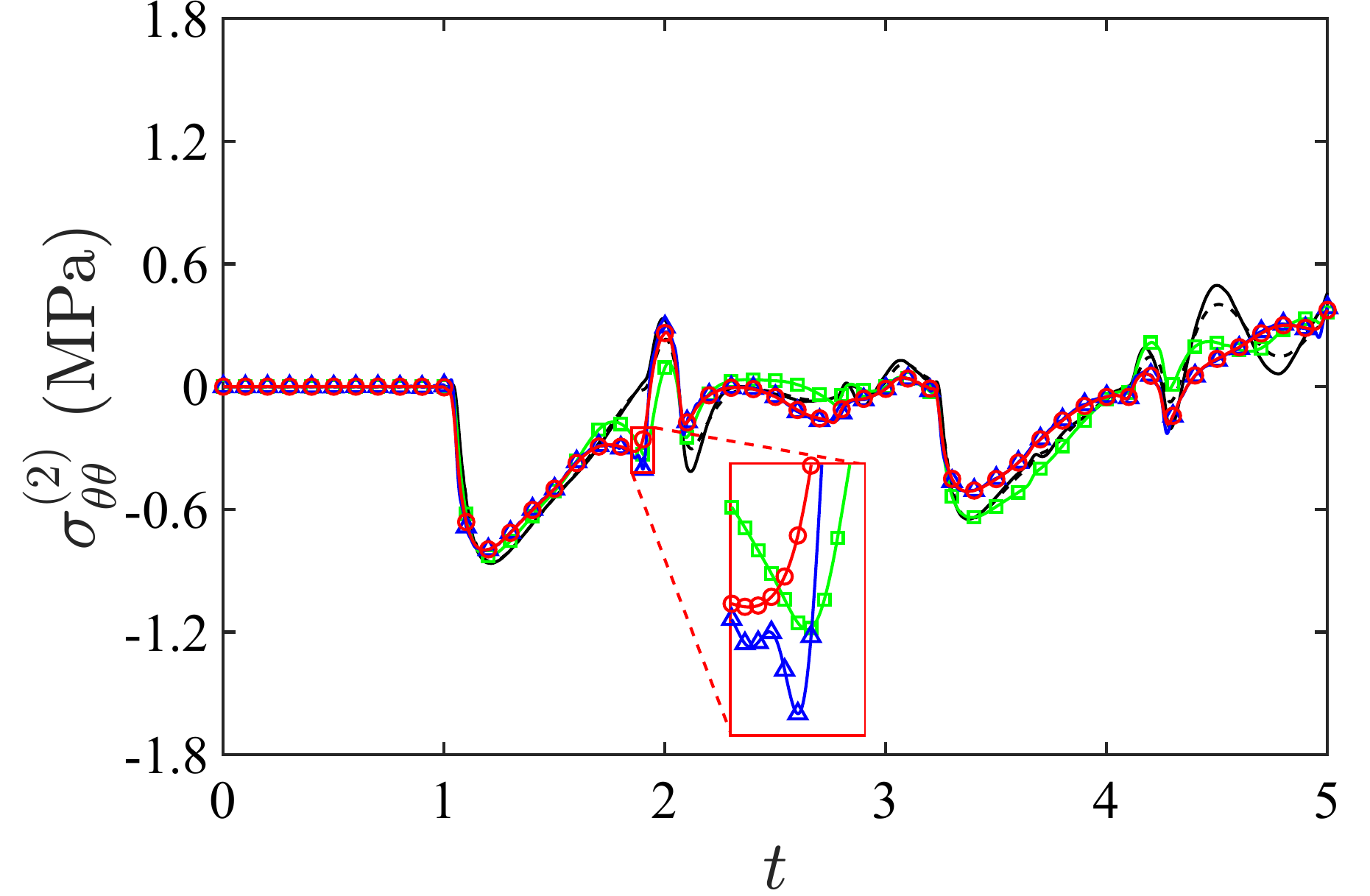}
        \put(-20,47.3){\includegraphics[width=0.79\textwidth]{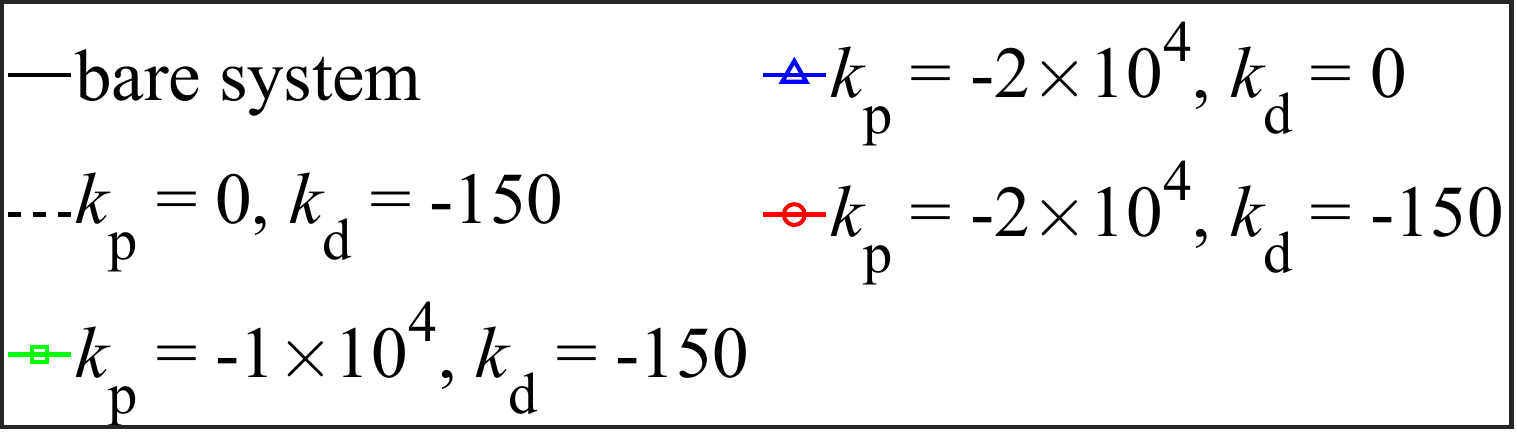}}
        \end{overpic}
        \subcaption{Outer shell tail point}
        \label{fig:11-d}
    \end{minipage}%
    \caption{Effect of active control parameters on induced stress at head and tail points of each shell.}
    \label{fig:11}
\end{figure}

\noindent \textbf{Active control:} here, we delve into the active control of piezoelectric layers and evaluate the influence of control parameters on the system's capability to suppress the stress amplitudes. Fig. \ref{fig:11} depicts how variations in proportional ($k_\text{p}$) and derivative ($k_\text{d}$) feedback gains affect the performance of the active control mechanism. Based on the voltage limit reported in \cite{bodaghi2012non}, we note that the actuator input must be constrained by $\lvert u_i^{(\text{a})}/t_i \rvert \leq 590~\text{V}/\text{mm}$, $i \in \{1, 2\}$
to prevent potential damage to the piezoelectric layers. As shown in Fig. \ref{fig:11}, the proportional control gain $k_\text{p}$ plays the most decisive role in reducing stress fluctuations. Sole by setting $k_\text{p}$ to $-2 \times 10^4$, the control scheme yields a noticeable decline in stress amplitude in the inner shell compared to the passive bare system. The derivative gain $k_\text{d}$, on the other hand, mainly contributes to the suppression of rapid oscillations in the induced mechanical stress. In the absence of a proportional feedback ($k_\text{p} = 0$), $k_\text{d}$ alone yields only a marginal reduction in stress amplitudes, as illustrated by the slight decrease in peak stress in Figs. \ref{fig:11-a} and \ref{fig:11-b}. When applied jointly with $k_\text{p}$, however, the combined action leads to markedly improved control performance. According to the obtained results, the values $k_\text{p} = -2 \times 10^4$ and $k_\text{d} = -150$ are adopted for the subsequent analyses in this study.

\begin{figure}[H]
\centering
\begin{minipage}[t]{0.5\textwidth}
        \centering
        \includegraphics[height=0.2\textheight]{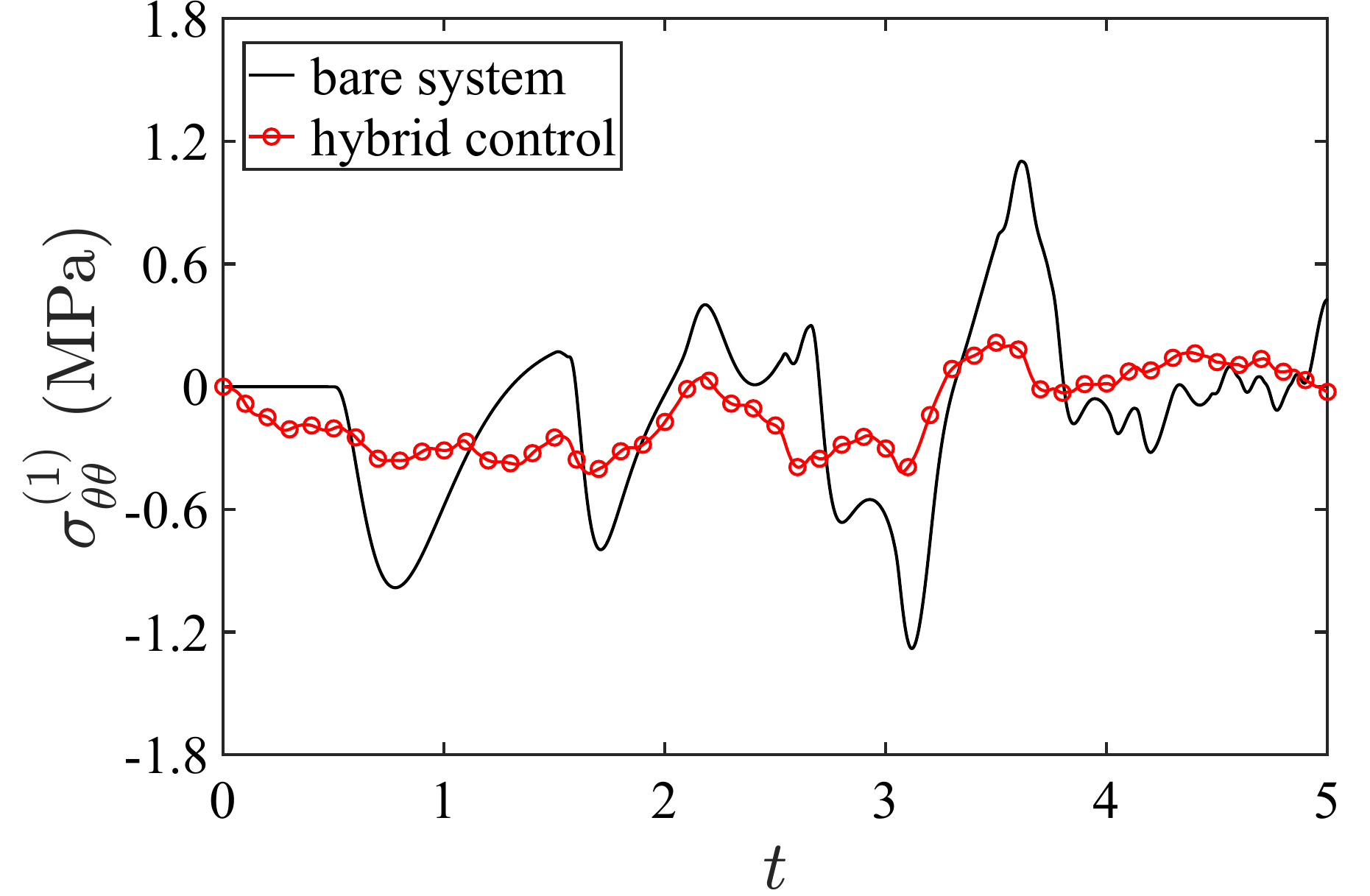}
        \subcaption{Inner shell head point}
        \label{fig:12-a}
    \end{minipage}%
    \hfill
    \begin{minipage}[t]{0.5\textwidth}
        \centering
        \includegraphics[height=0.2\textheight]{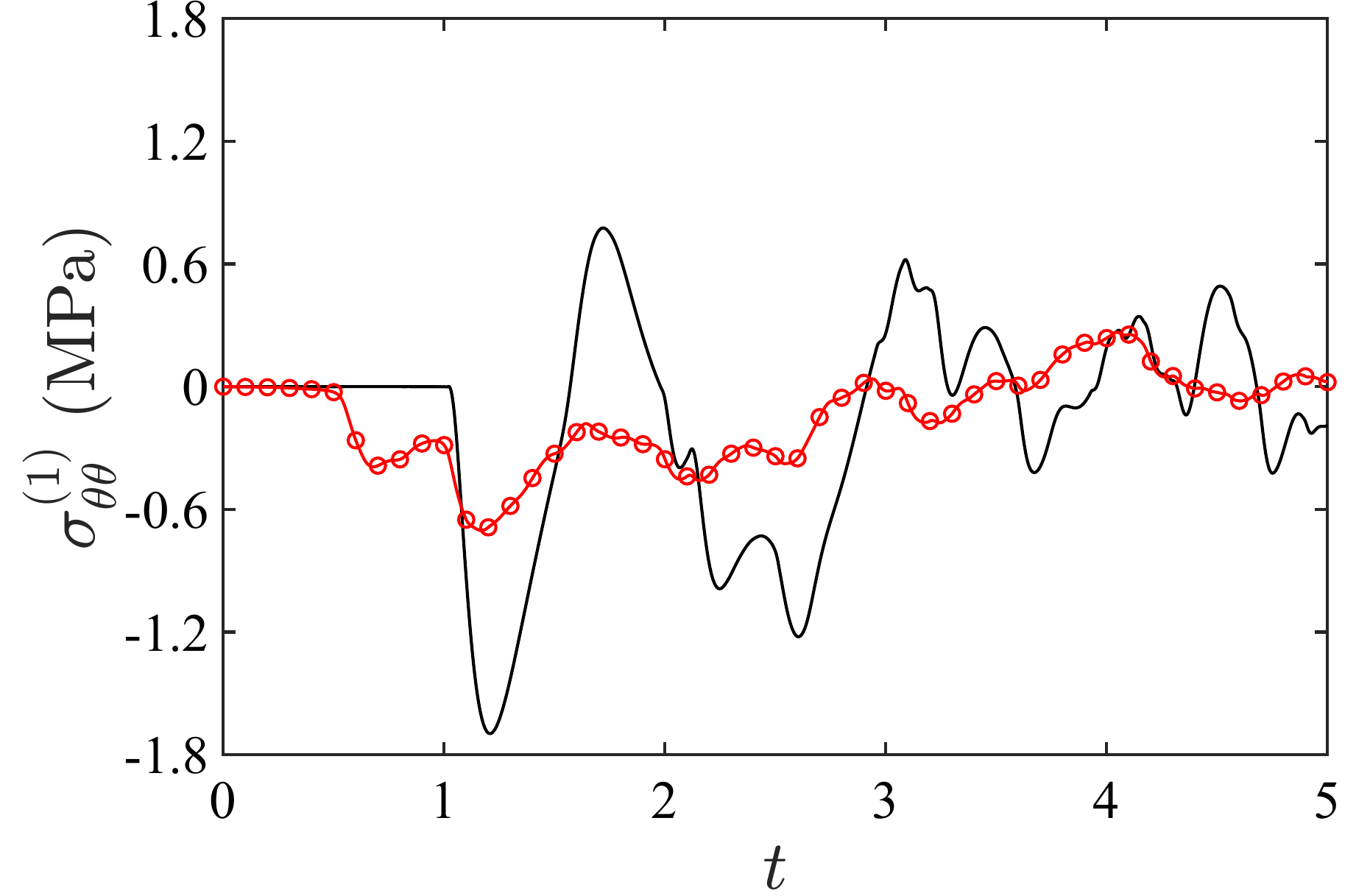}
        \subcaption{Inner shell tail point}
        \label{fig:12-b}
    \end{minipage}%
    \vspace{0.5cm}
    \begin{minipage}[t]{0.5\textwidth}
        \centering
        \includegraphics[height=0.2\textheight]{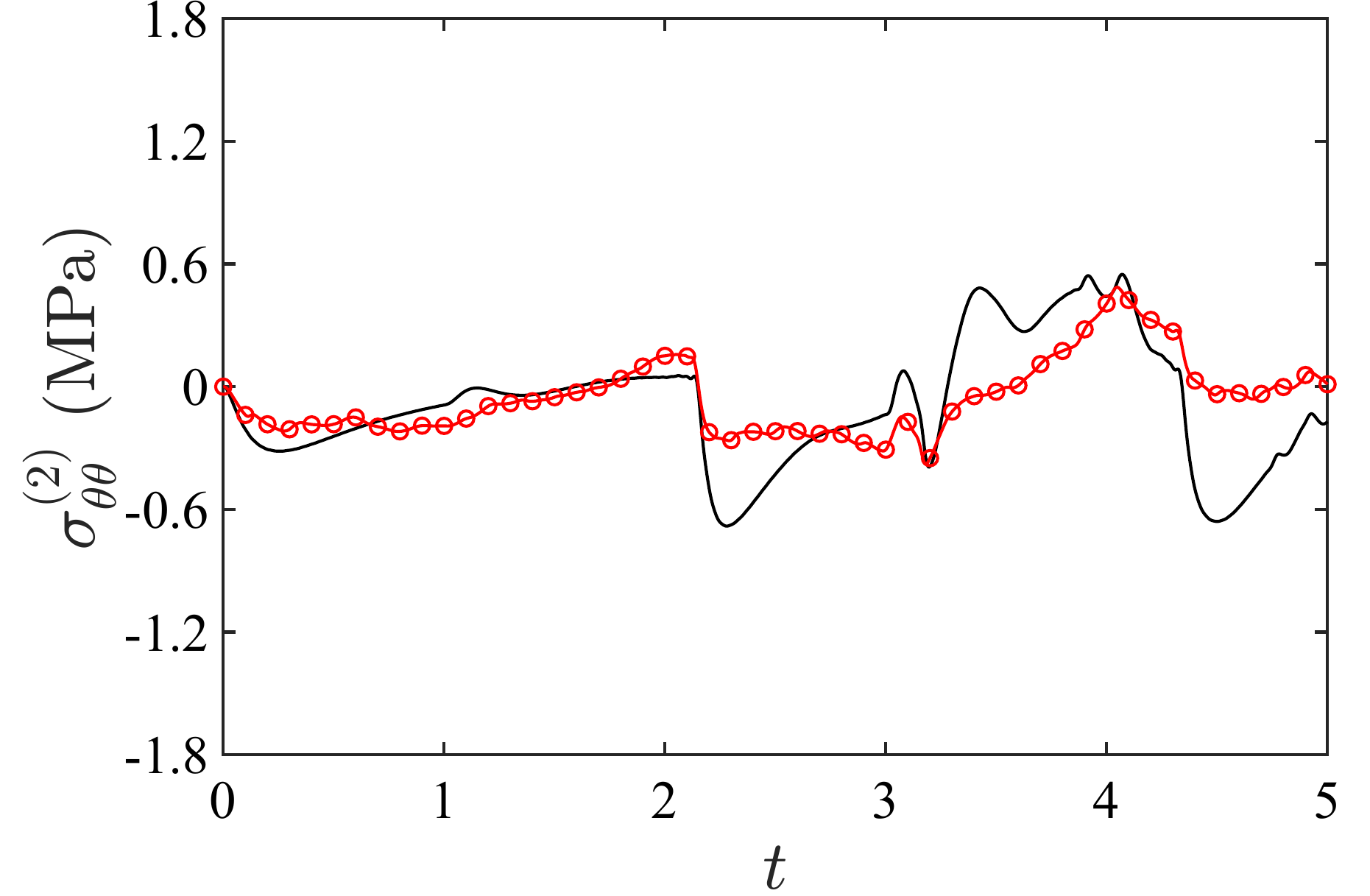}
        \subcaption{Outer shell head point}
        \label{fig:12-c}
    \end{minipage}%
    \hfill
    \begin{minipage}[t]{0.5\textwidth}
        \centering
        \includegraphics[height=0.2\textheight]{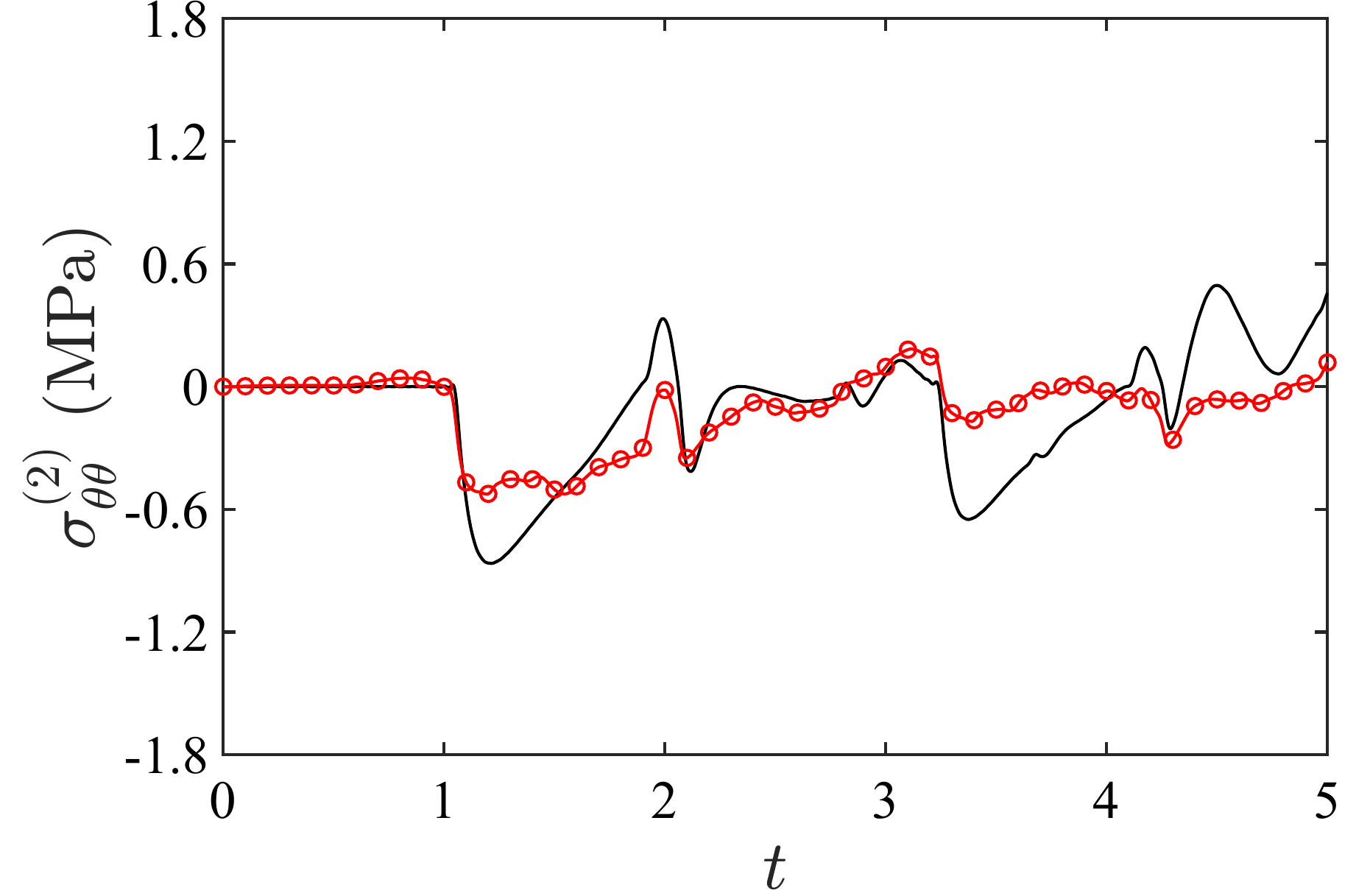}
        \subcaption{Outer shell tail point}
        \label{fig:12-d}
    \end{minipage}%
    \caption{Performance of the hybrid control system in stress mitigation at head and tail points of each shell.}
    \label{fig:12}
\end{figure}

\noindent \textbf{Hybrid control:} following the individual assessments of passive dynamic absorbers and active piezoelectric control, we now adopt a hybrid control scheme that unifies both mechanisms to enhance vibroacoustic suppression in the double-shell model. Accordingly, the dynamic absorber parameters are adopted from previous sections as
$k_\text{l} = 10^9~\text{N/m}$,
$k_\text{nl} = 10^{20}~\text{N/m}^3$,
$m = 10~\text{kg}$, $c = 10^6~\text{N}\!\cdot\!\text{s}/\text{m}$, and
$N = 25$, while the active control gains are taken as
$k_\text{p} = -2 \times 10^4$ and $k_\text{d} = -150$. Figs. \ref{fig:12-a} through \ref{fig:12-d} demonstrate the effective performance of the developed hybrid control system in attenuation of the stress at the head and tail points of the inner and outer shells.

Fig. \ref{fig:13} captures the hybrid controller’s performance through a time-resolved comparison of dimensionless acoustic fields and normalized transverse stress intensities (scaled by $\sigma_\text{ref} = 1~\text{MPa}$), against those of the bare system. To better monitor of the system dynamics, results are displayed in sequential snapshots, taken with time steps of $\Delta t = 0.1$ during early times and $\Delta t = 0.5$ at later stages. As discussed earlier, the use of vibration absorbers in the double-shell assembly causes an instantaneous transfer of force to the inner shell,  preceding the arrival of the incident wave. This effect can be observed from as early as $t \geq 0.1$, when the application of stress to the inner shell and the subsequent emission of acoustic waves from its surface are detectable shortly after the incident wave impinges on the outer shell. Fig. \ref{fig:13} shows more radiation irregularities from the hybrid-controlled structure. The forces exerted by the absorbers and piezoelectric actuators contribute to altering the acceleration of adjacent shell segments. This acceleration heterogeneity leads to disturbances in the emitted acoustic waves, which is evident, for instance at $t = 0.5$. A comparison of results (e.g., at $t = 1.0$) reveal that despite similar acoustic patterns in the exterior medium, the fields' shape and intensity within the gap medium and particularly the interior fluid, are significantly different. The disturbed radiation pattern and multiple reflections before the incident wave reaches the inner shell, reshape the acoustic field inside and reduce its intensity. The attenuation of acoustic waves results in a notable drop in the induced stress amplitudes. This decrease in acoustic intensity and consequently in stress is evident in the snapshots captured at $t = 1.0$ and $3.0$. The hybrid control system's impact becomes more apparent over time, effectively reducing the stress amplitude in both shells and the acoustic fields intensity, as shown for $t \geq 4.0$ in Fig. \ref{fig:13}. To better demonstrate the effect of the hybrid control mechanism on the vibroacoustic response of the double-shell, a short animation of the sound wave propagation and stress distribution has been created and is accessible at \cite{hybridControlVideo2025}.
\newpage
\begin{figure}[H]
\centering
\makebox[0.25\linewidth][c]{\text{Bare system}}%
\hfill
\makebox[0.25\linewidth][c]{\text{Hybrid control}}%
\hfill
\makebox[0.25\linewidth][c]{\text{Bare system}}%
\hfill
\makebox[0.25\linewidth][c]{\text{Hybrid control}}

\vspace{0.5cm} 

    \begin{minipage}[t]{0.25\textwidth}
        \centering
        \includegraphics[height=0.18\textheight]{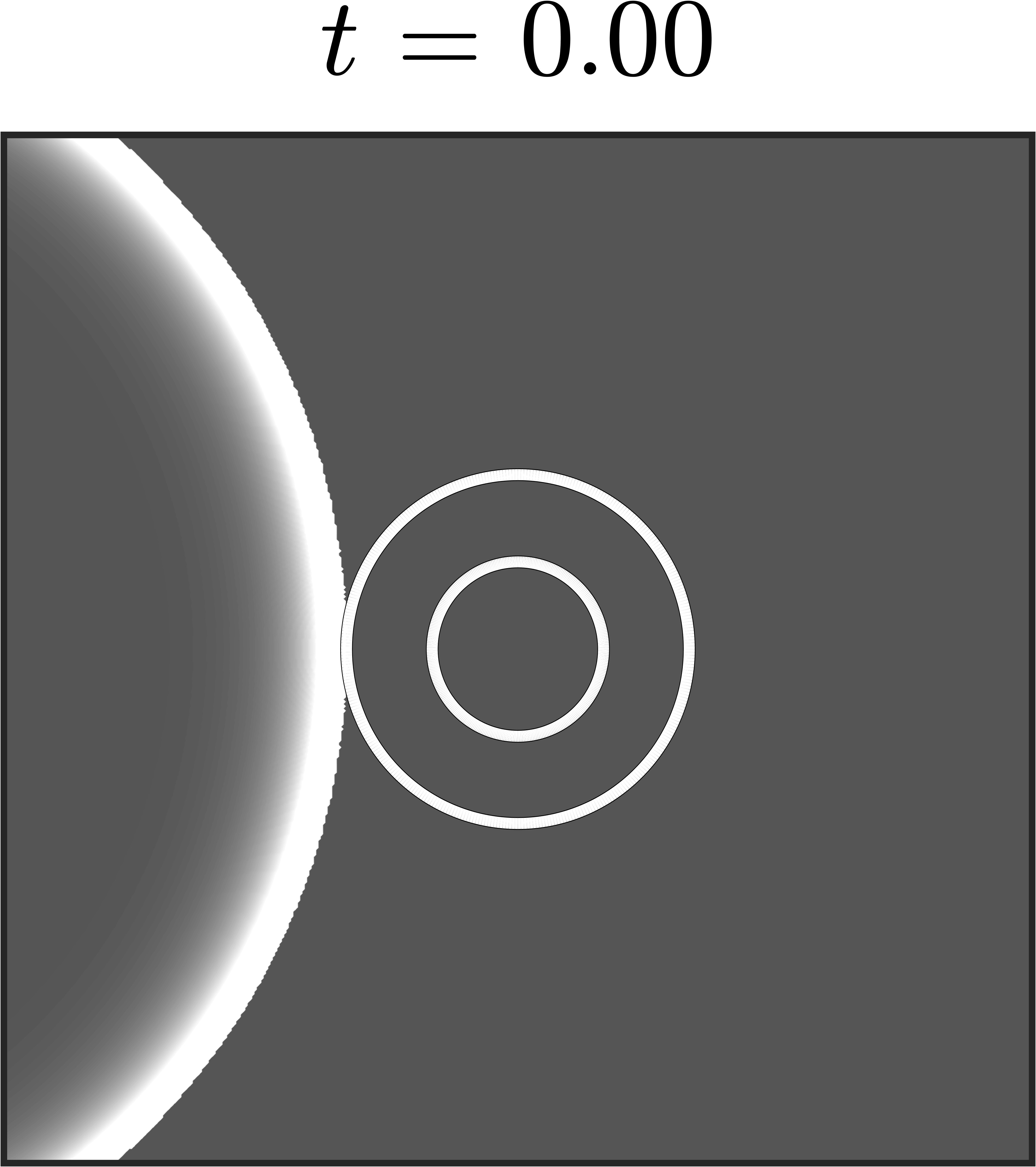}
    \end{minipage}%
    \begin{minipage}[t]{0.25\textwidth}
        \centering
        \includegraphics[height=0.18\textheight]{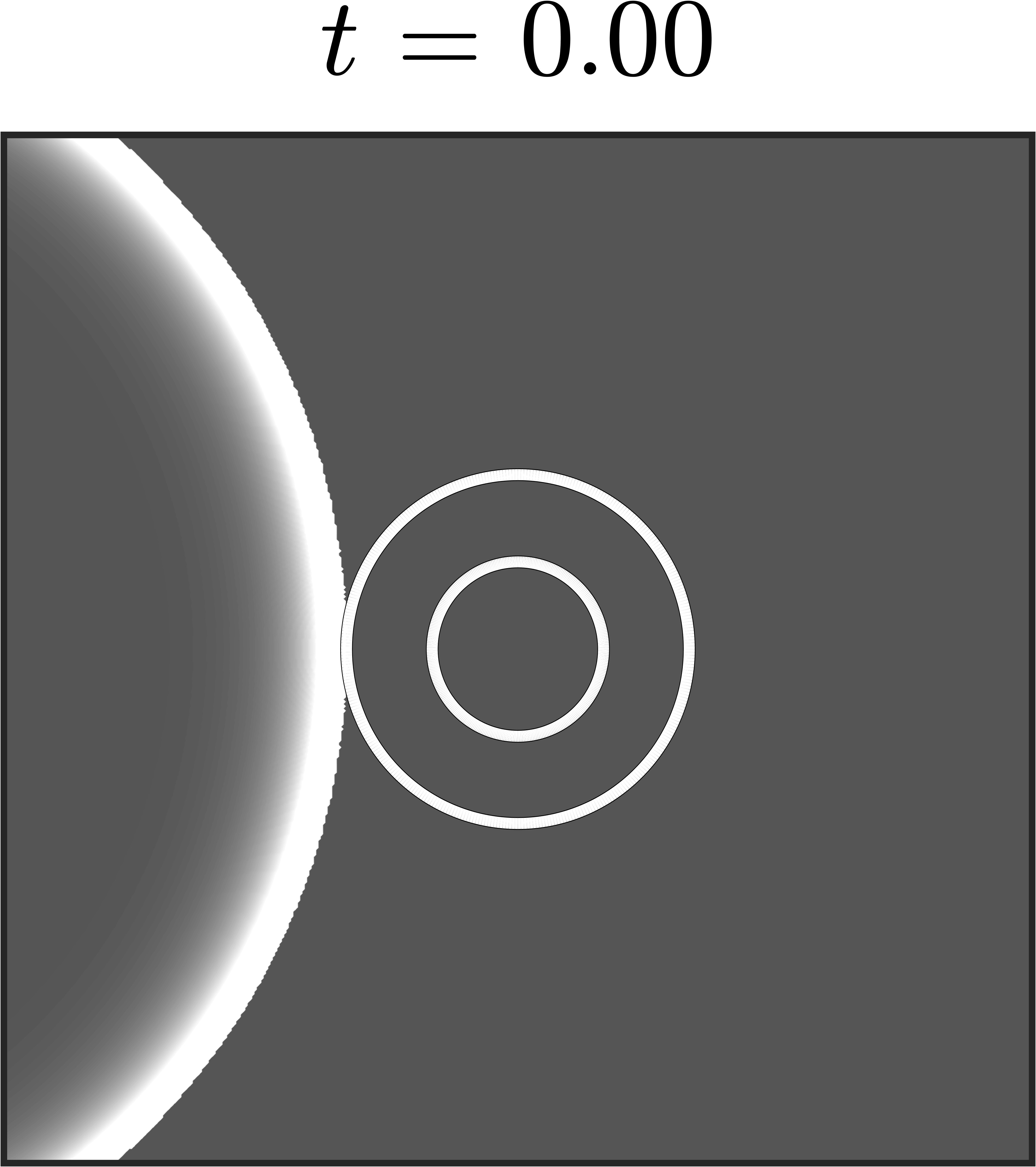}
    \end{minipage}%
    \begin{minipage}[t]{0.25\textwidth}
        \centering
        \includegraphics[height=0.18\textheight]{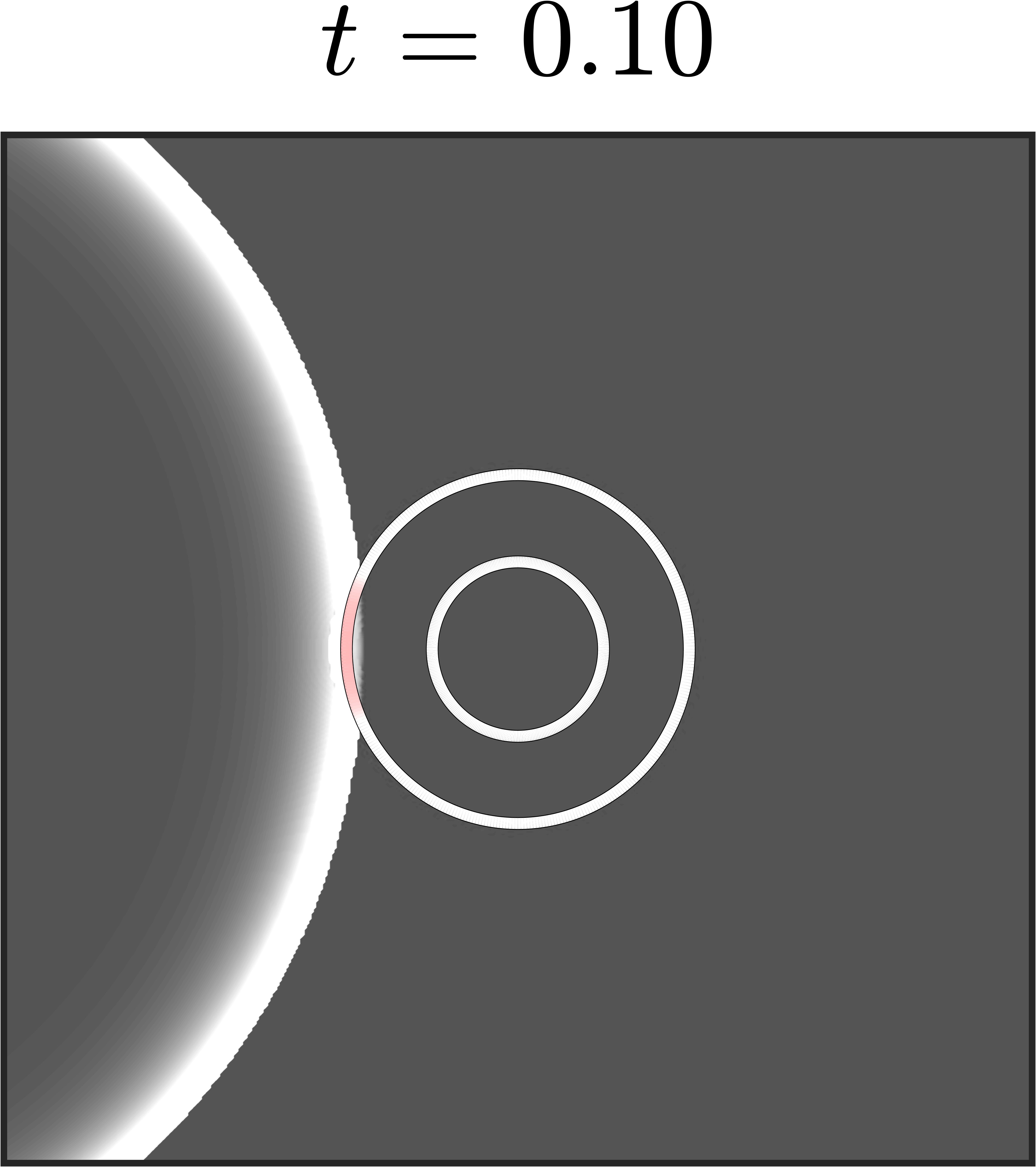}
    \end{minipage}%
    \begin{minipage}[t]{0.25\textwidth}
        \centering
        \includegraphics[height=0.18\textheight]{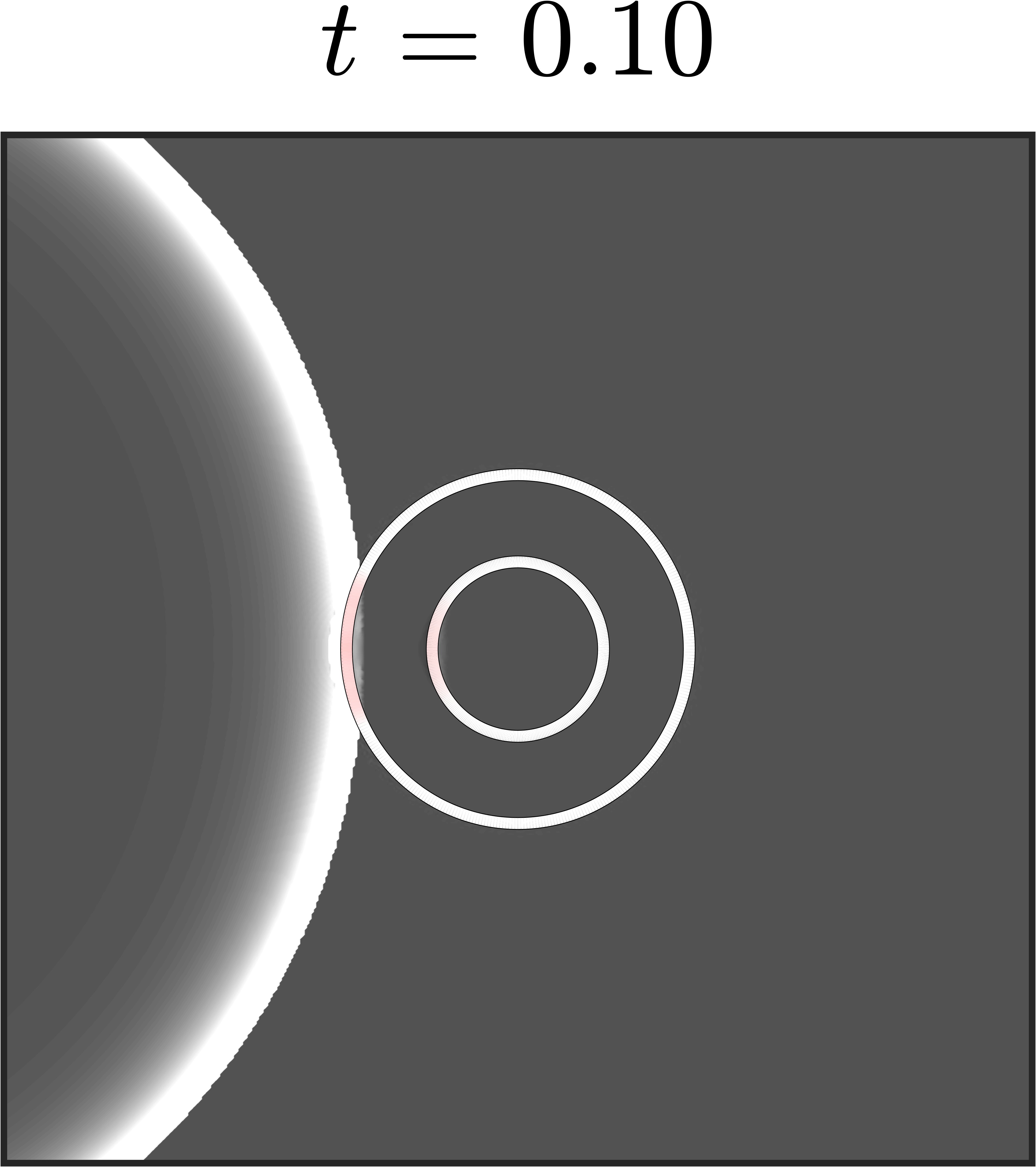}
    \end{minipage}%
    \vspace{0.12cm}
    \begin{minipage}[t]{0.25\textwidth}
        \centering
        \includegraphics[height=0.18\textheight]{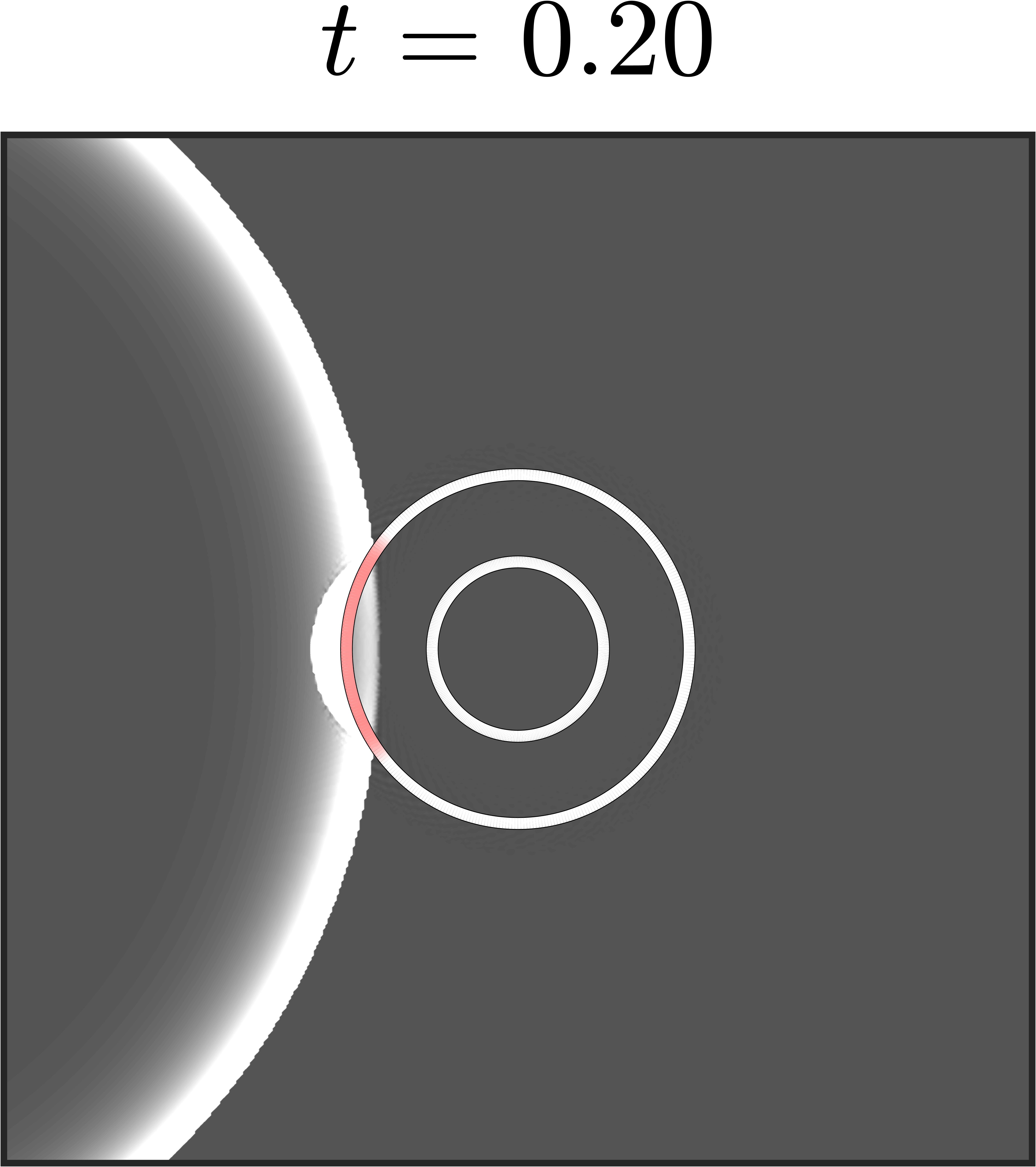}
    \end{minipage}%
    \begin{minipage}[t]{0.25\textwidth}
        \centering
        \includegraphics[height=0.18\textheight]{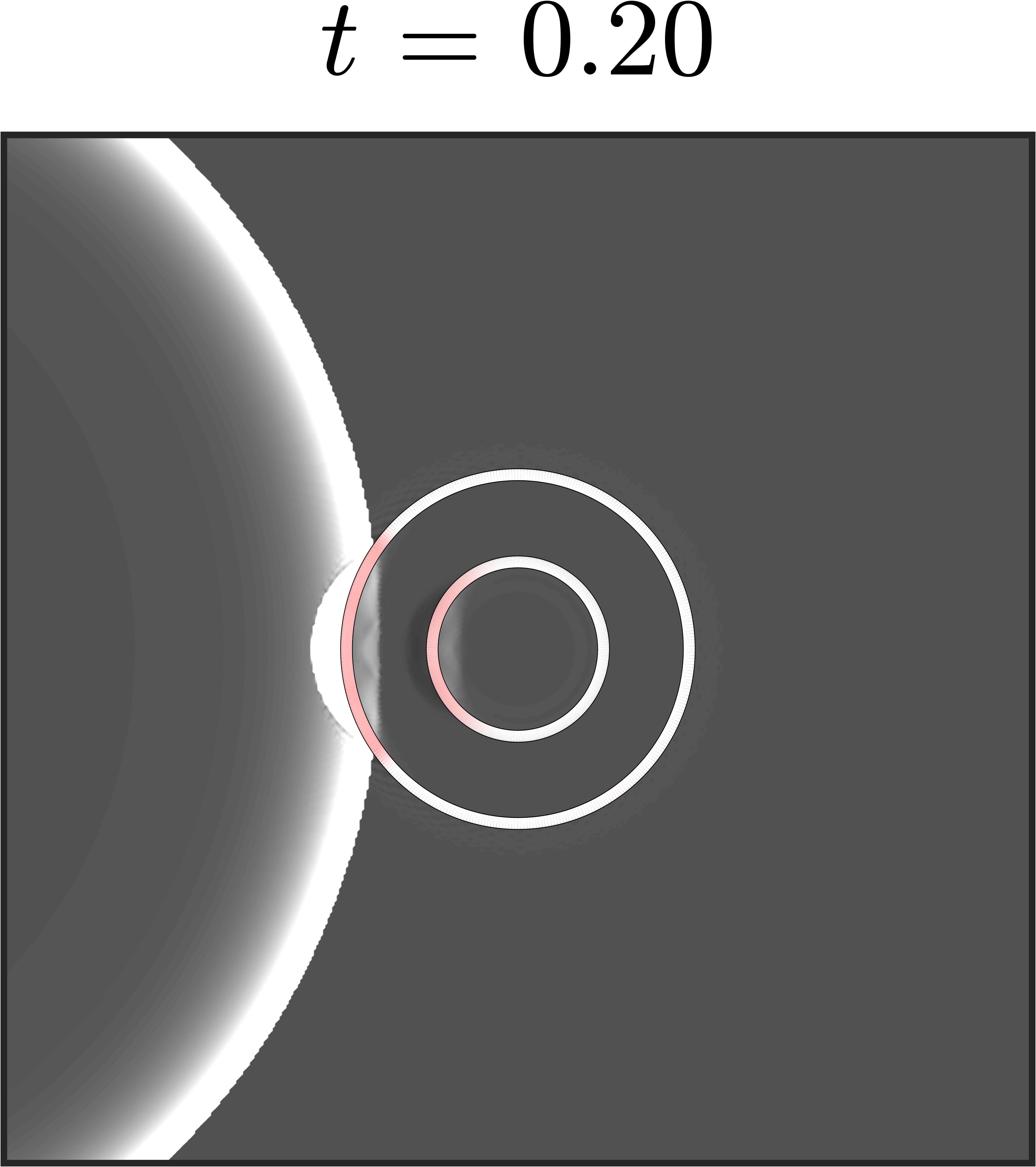}
    \end{minipage}%
    \begin{minipage}[t]{0.25\textwidth}
        \centering
        \includegraphics[height=0.18\textheight]{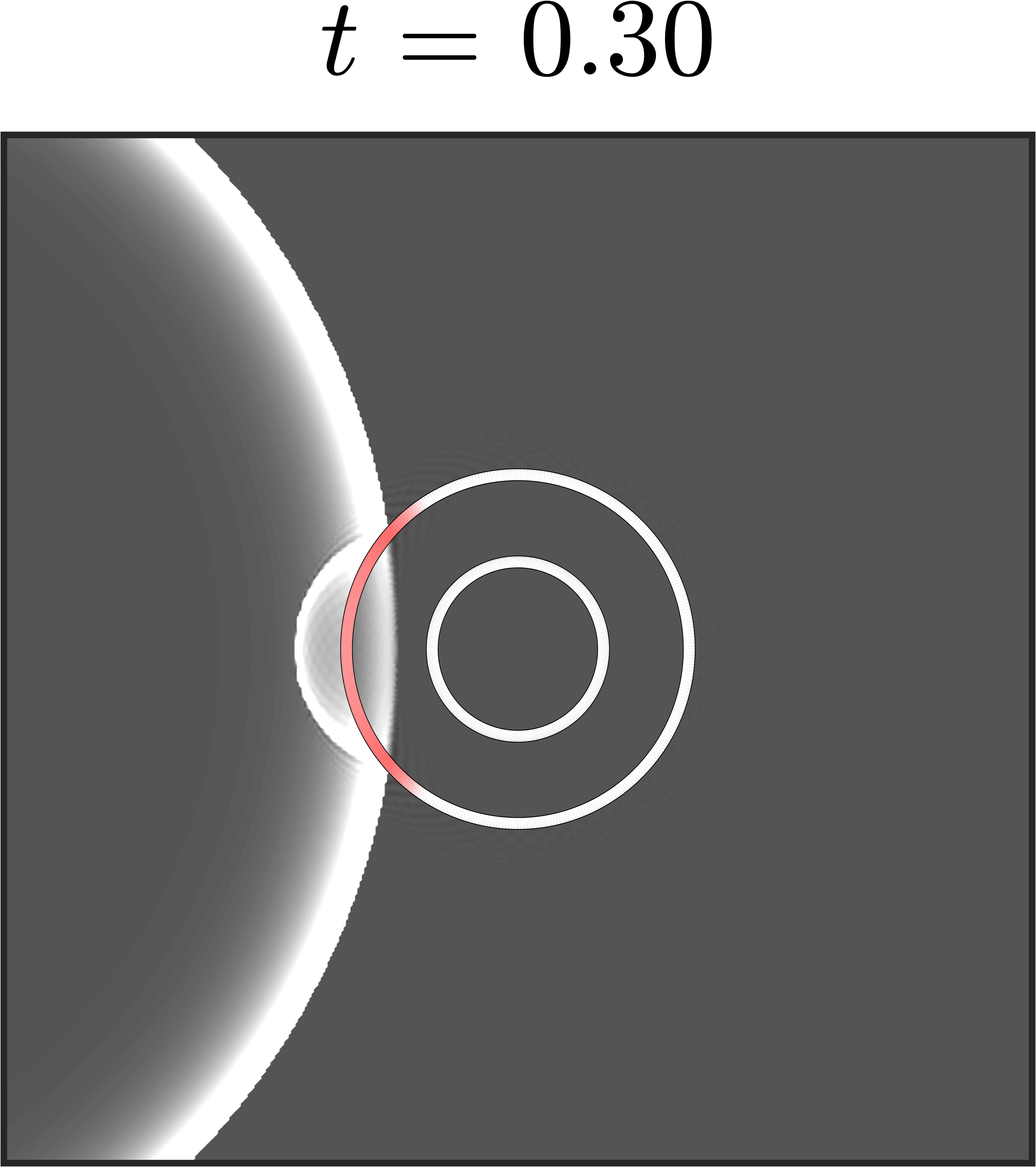}
    \end{minipage}%
    \begin{minipage}[t]{0.25\textwidth}
        \centering
        \includegraphics[height=0.18\textheight]{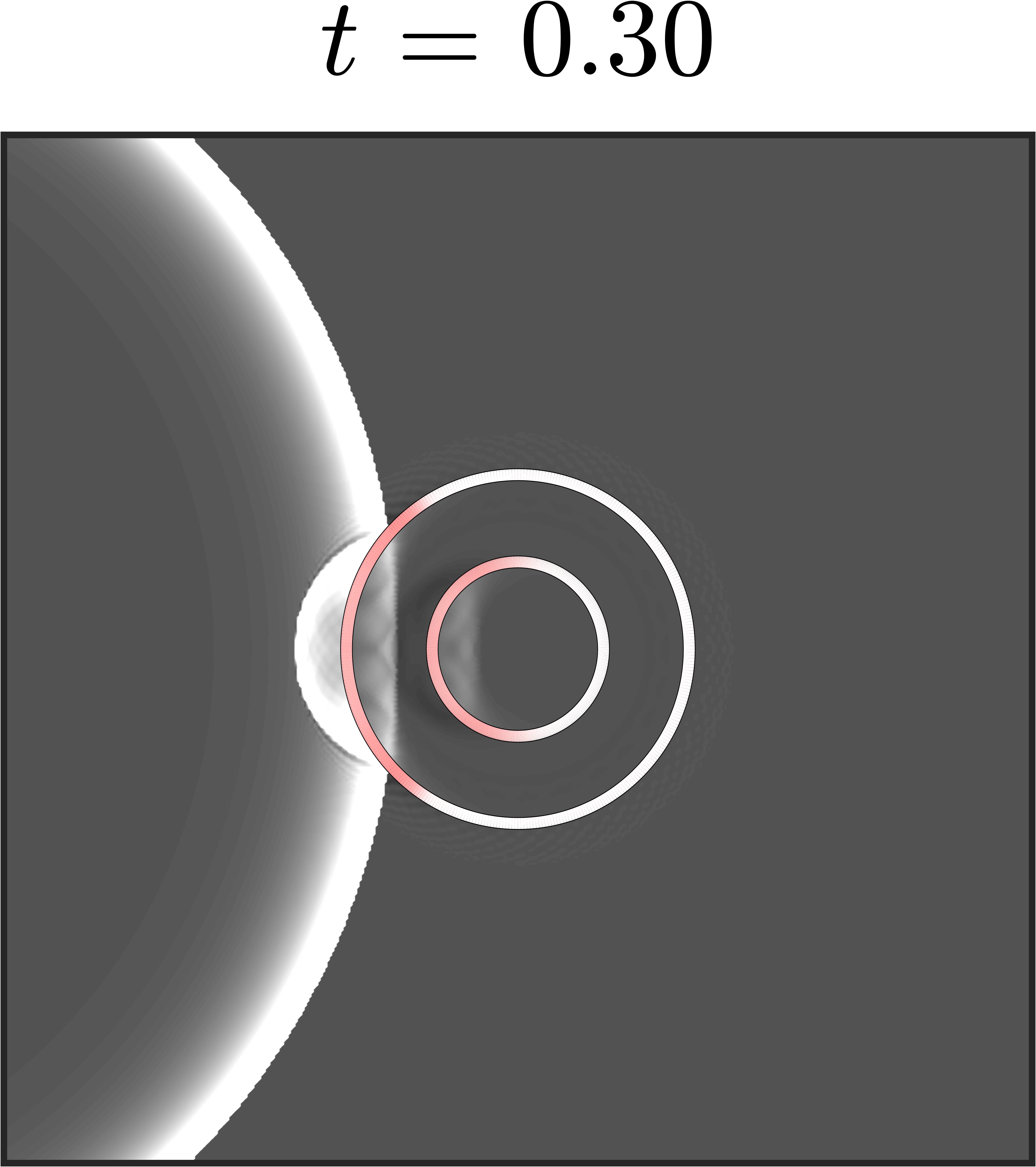}
    \end{minipage}%
    \vspace{0.12cm}
    \begin{minipage}[t]{0.25\textwidth}
        \centering
        \includegraphics[height=0.18\textheight]{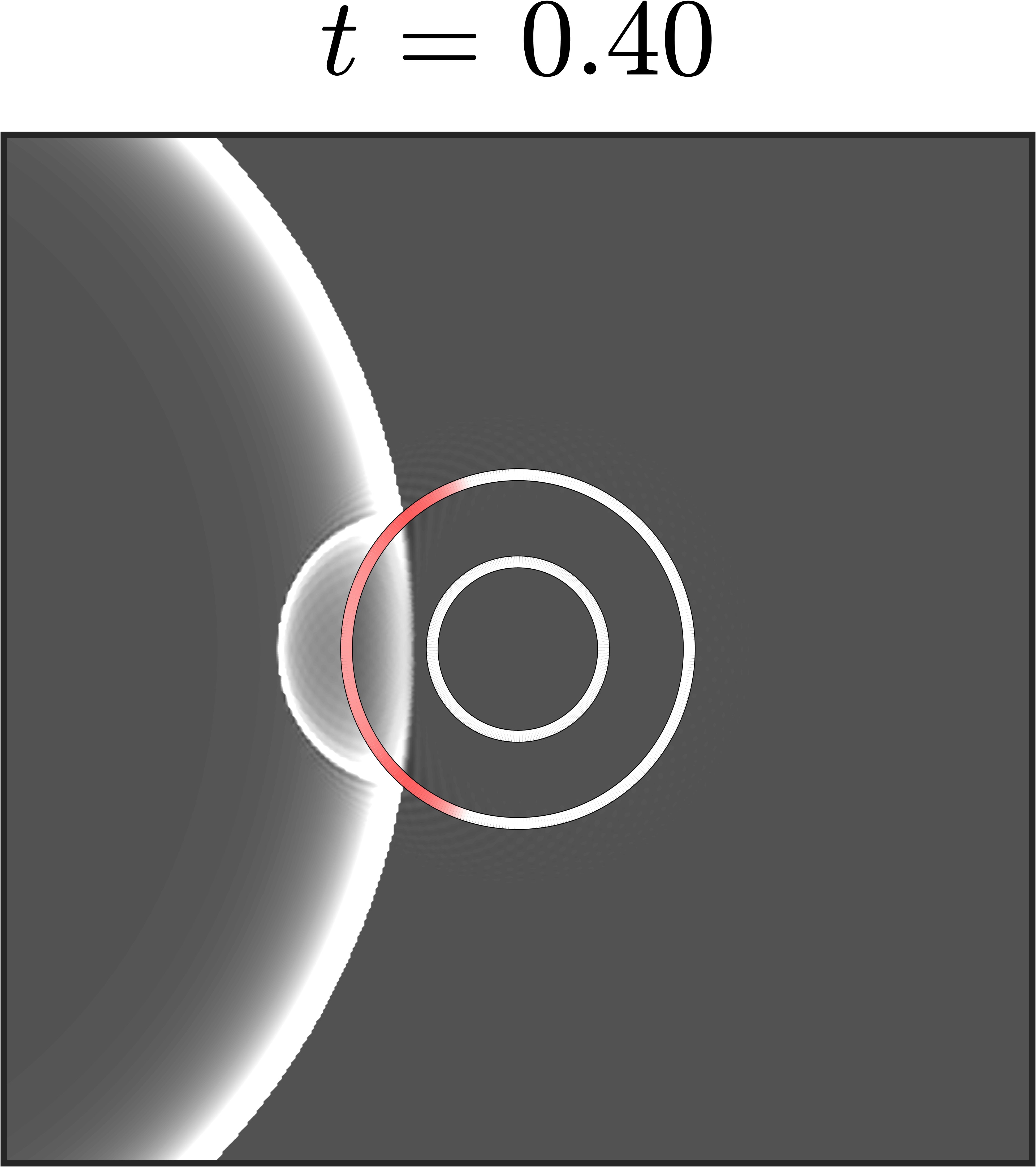}
    \end{minipage}%
    \begin{minipage}[t]{0.25\textwidth}
        \centering
        \includegraphics[height=0.18\textheight]{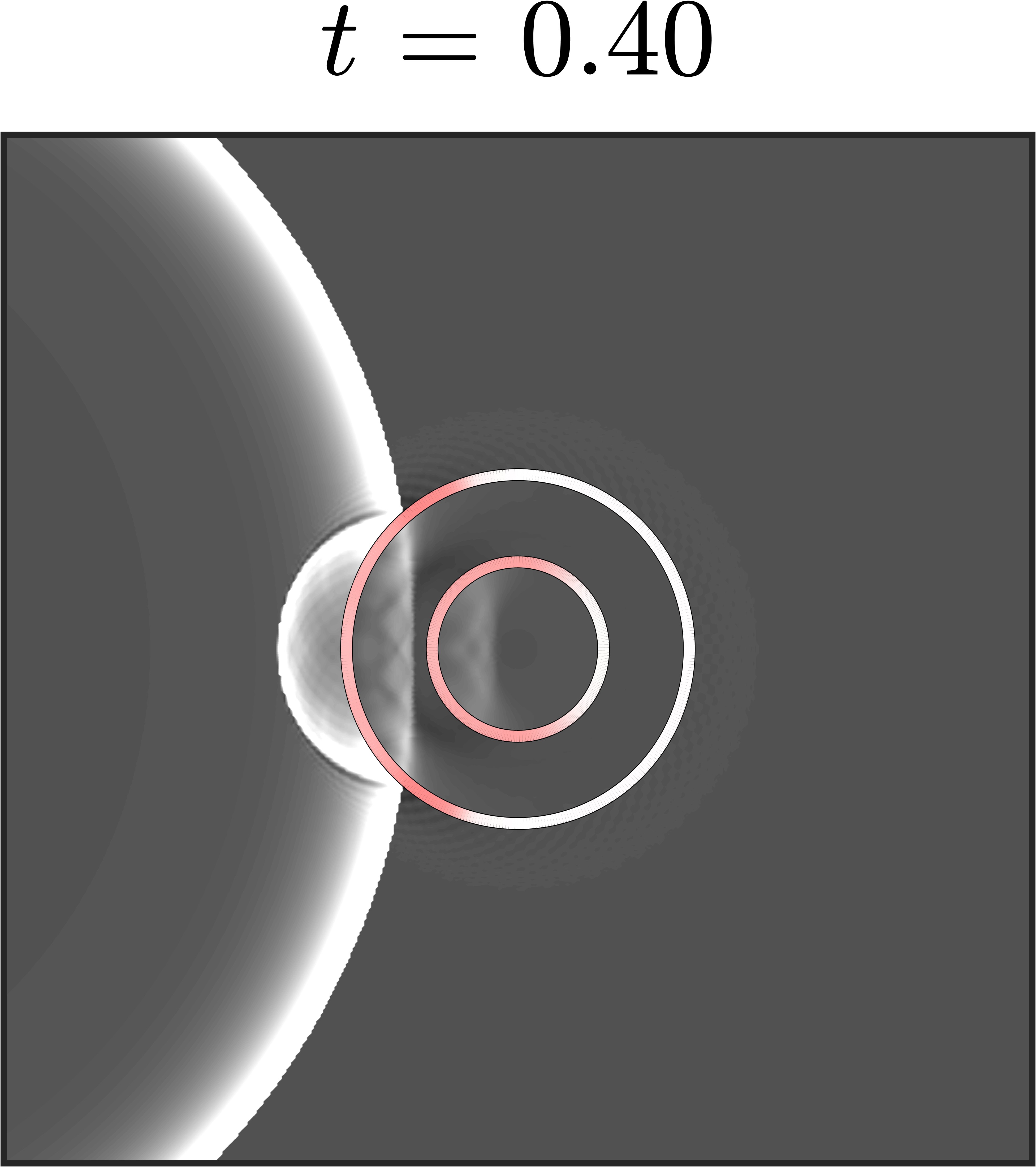}
    \end{minipage}%
    \begin{minipage}[t]{0.25\textwidth}
        \centering
        \includegraphics[height=0.18\textheight]{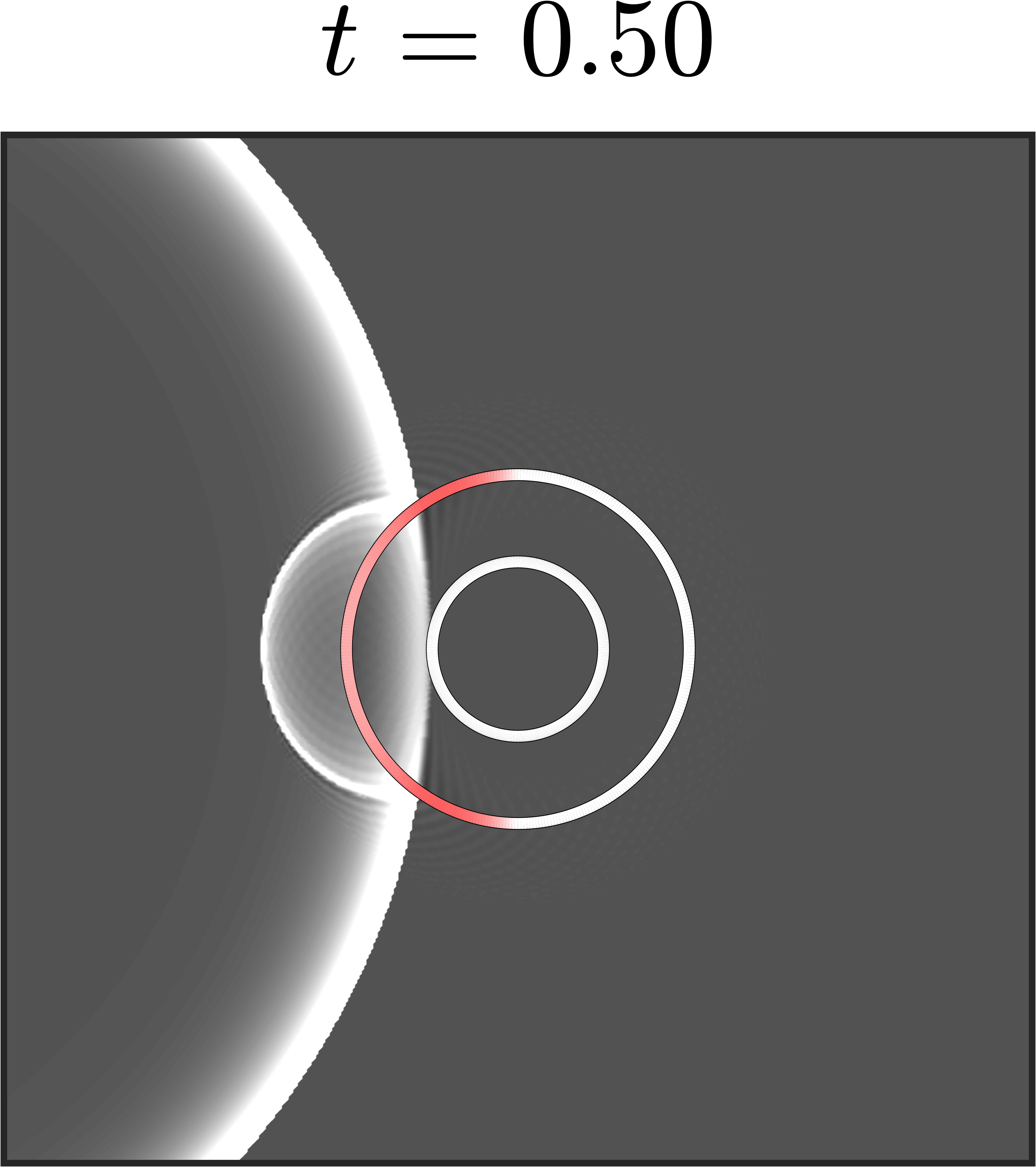}
    \end{minipage}%
    \begin{minipage}[t]{0.25\textwidth}
        \centering
        \includegraphics[height=0.18\textheight]{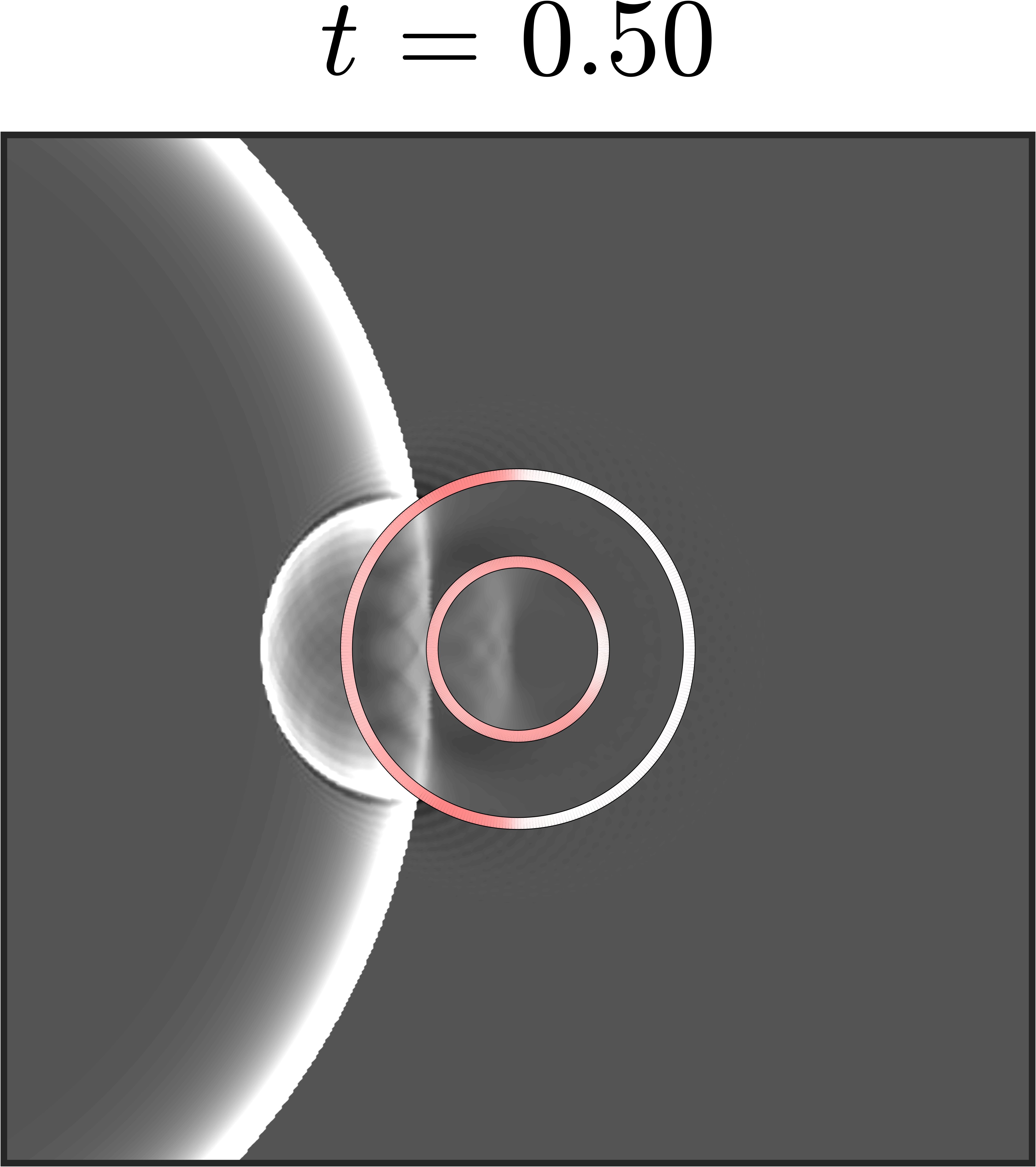}
    \end{minipage}%
    \vspace{0.12cm}
    \begin{minipage}[t]{0.25\textwidth}
        \centering
        \includegraphics[height=0.18\textheight]{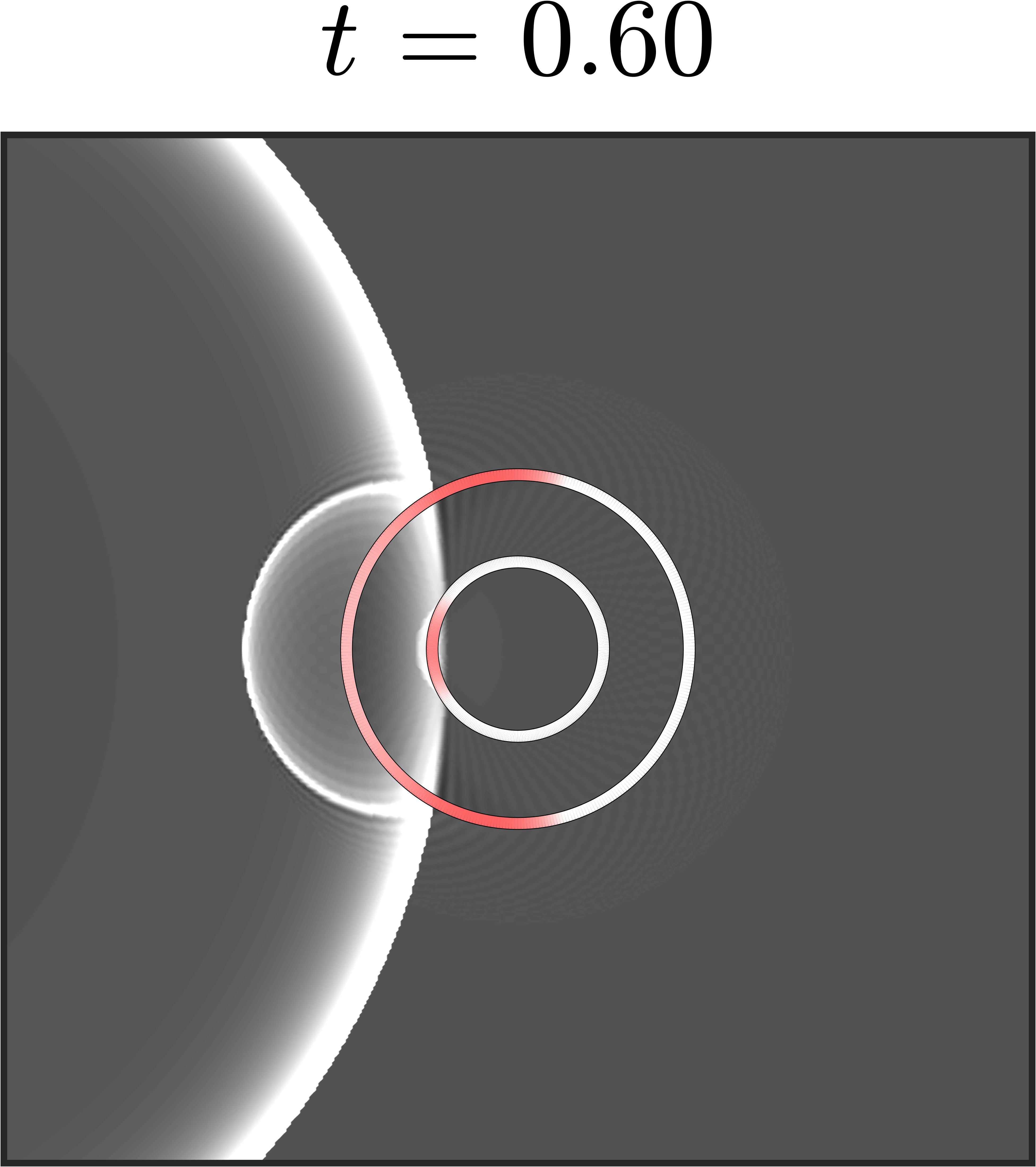}
    \end{minipage}%
    \begin{minipage}[t]{0.25\textwidth}
        \centering
        \includegraphics[height=0.18\textheight]{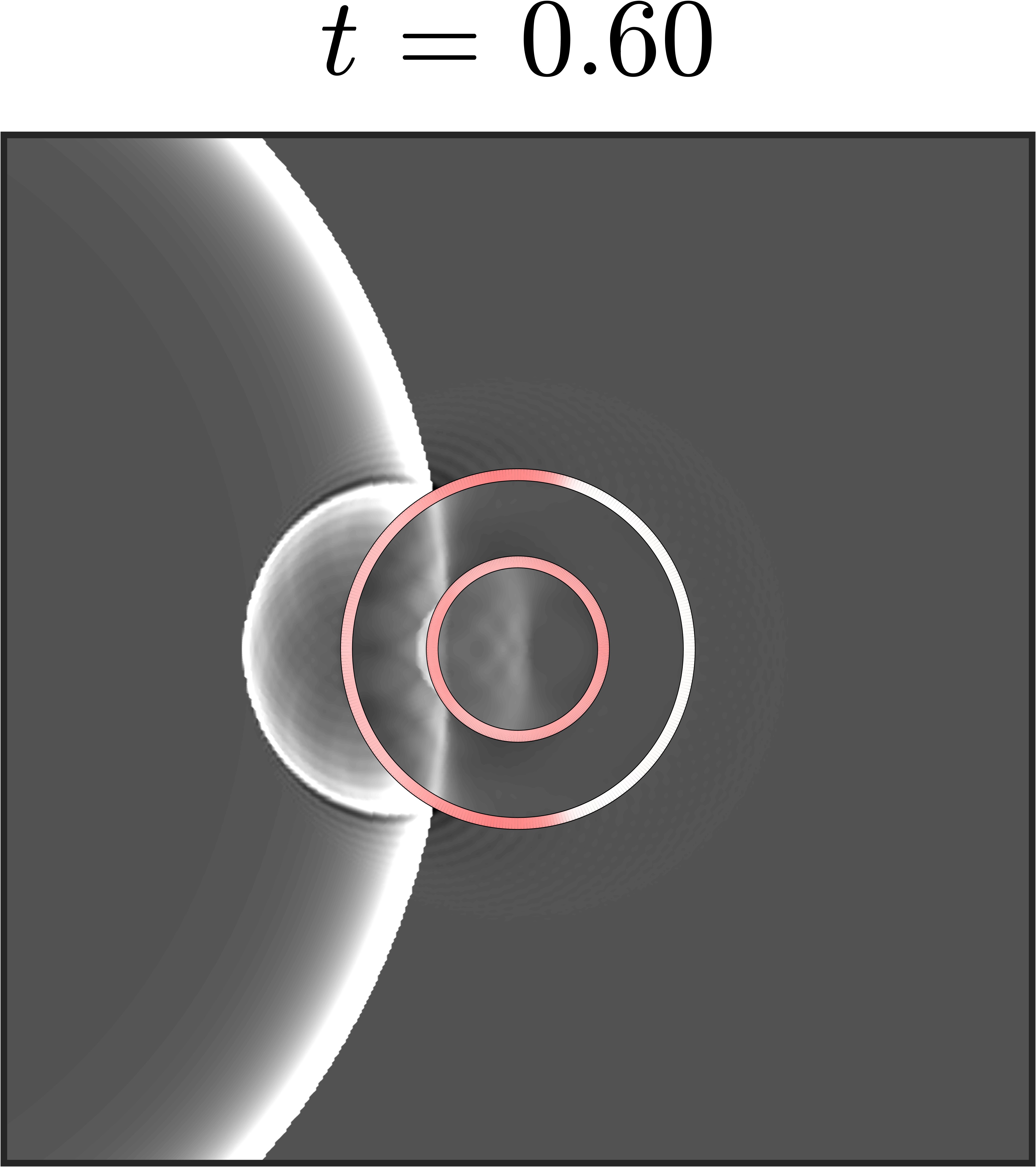}
    \end{minipage}%
    \begin{minipage}[t]{0.25\textwidth}
        \centering
        \includegraphics[height=0.18\textheight]{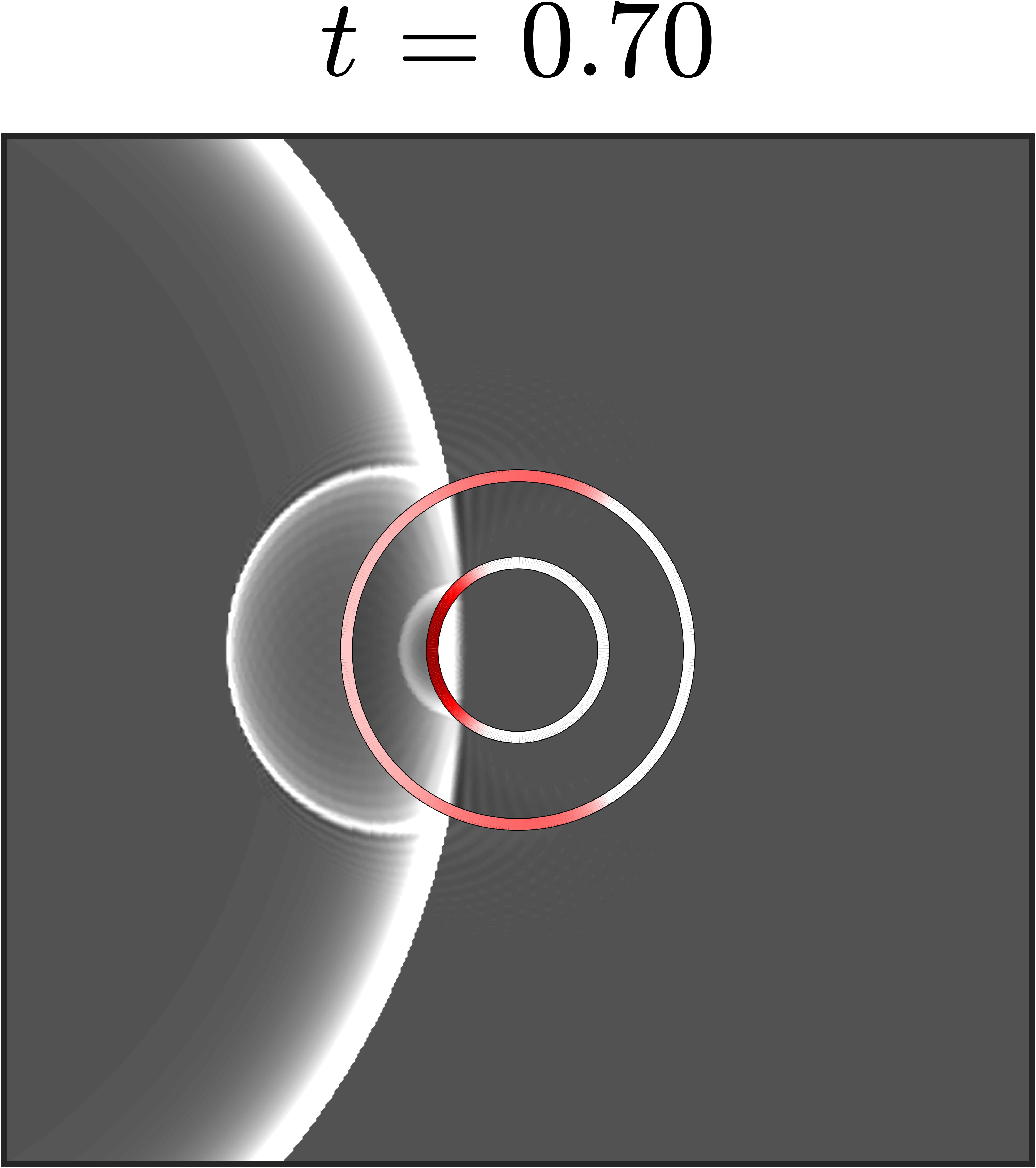}
    \end{minipage}%
    \begin{minipage}[t]{0.25\textwidth}
        \centering
        \includegraphics[height=0.18\textheight]{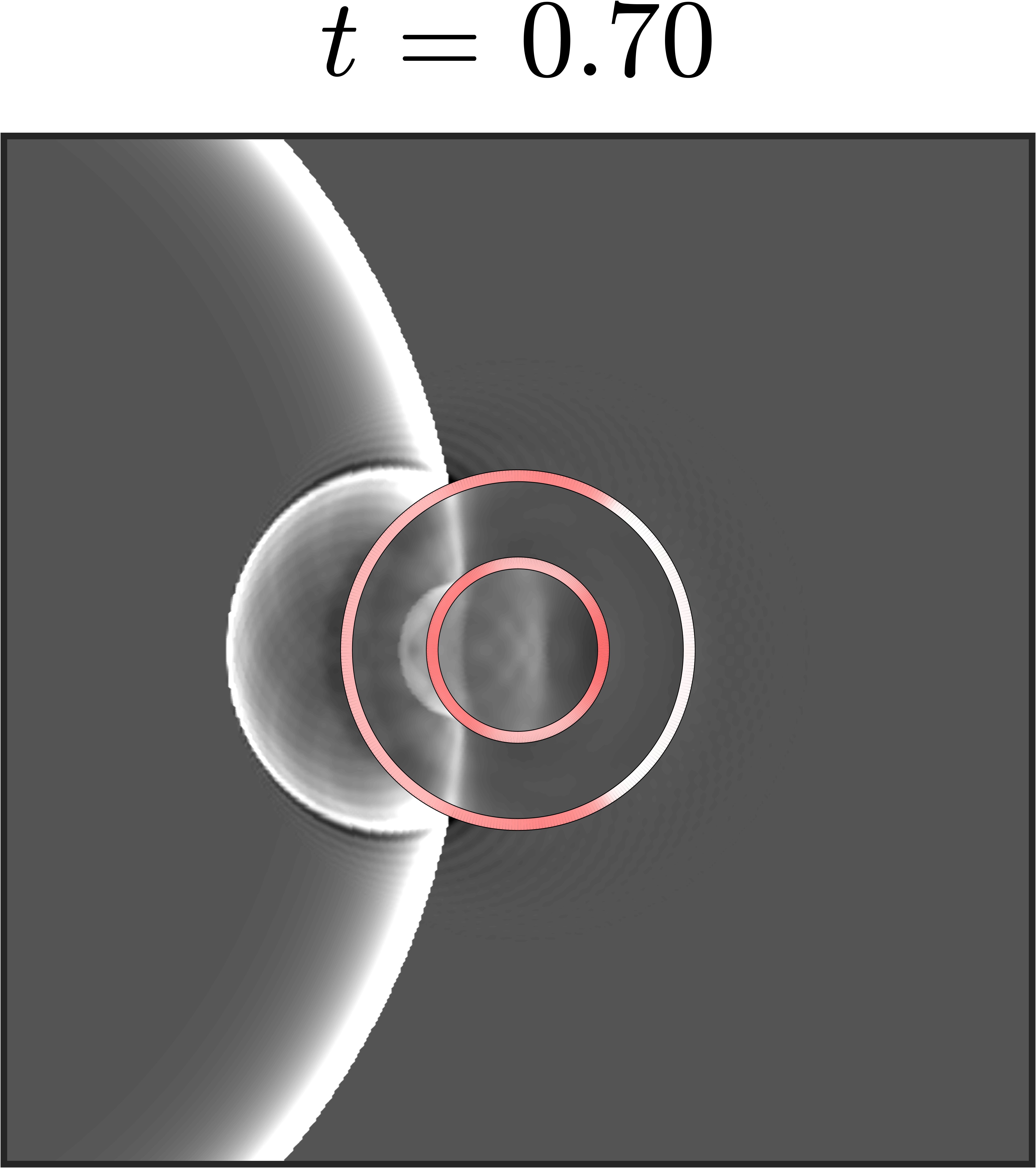}
    \end{minipage}%
    \vspace{0.12cm}
    \begin{minipage}[t]{0.25\textwidth}
        \centering
        \includegraphics[height=0.18\textheight]{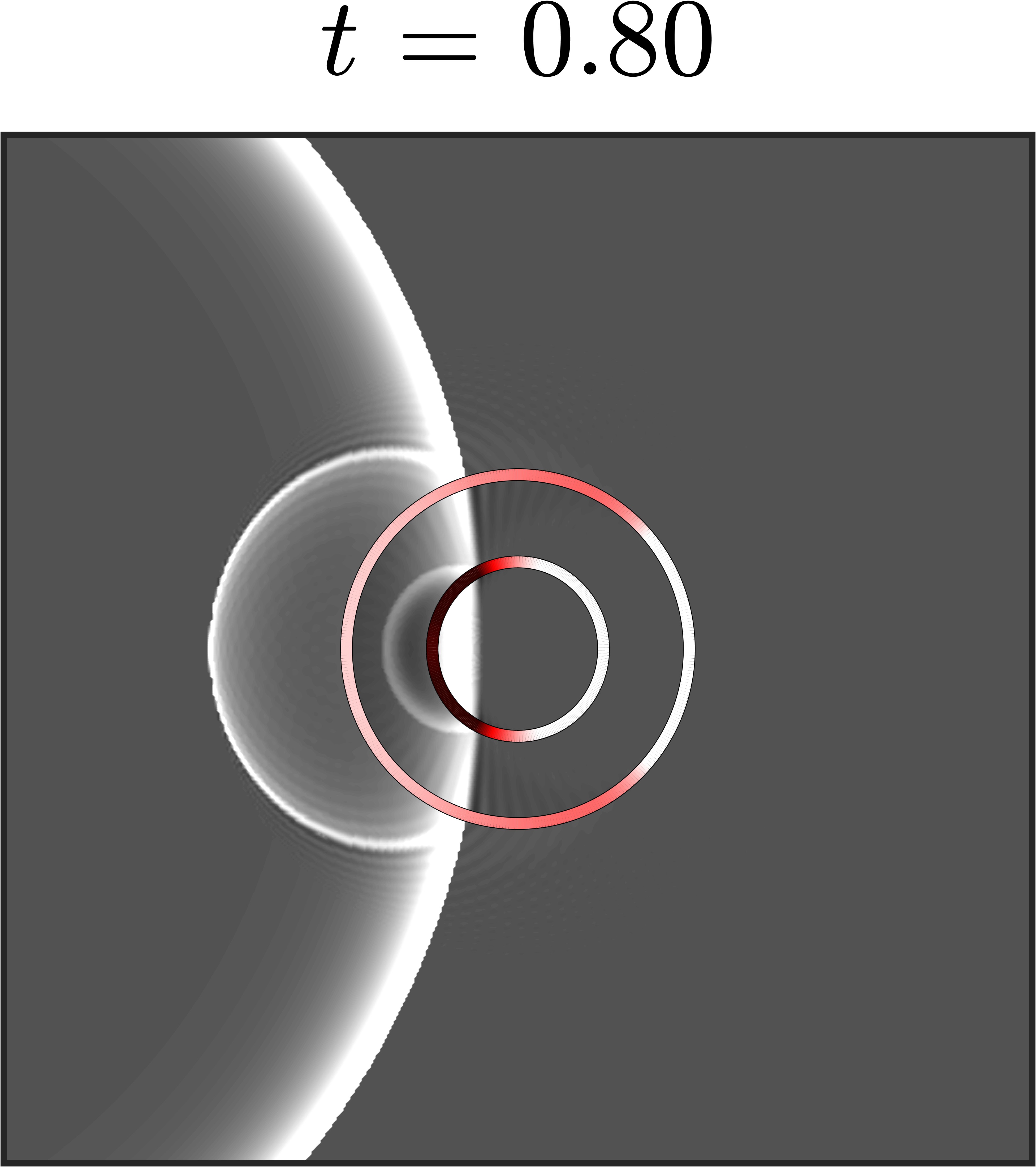}
    \end{minipage}%
    \begin{minipage}[t]{0.25\textwidth}
        \centering
        \includegraphics[height=0.18\textheight]{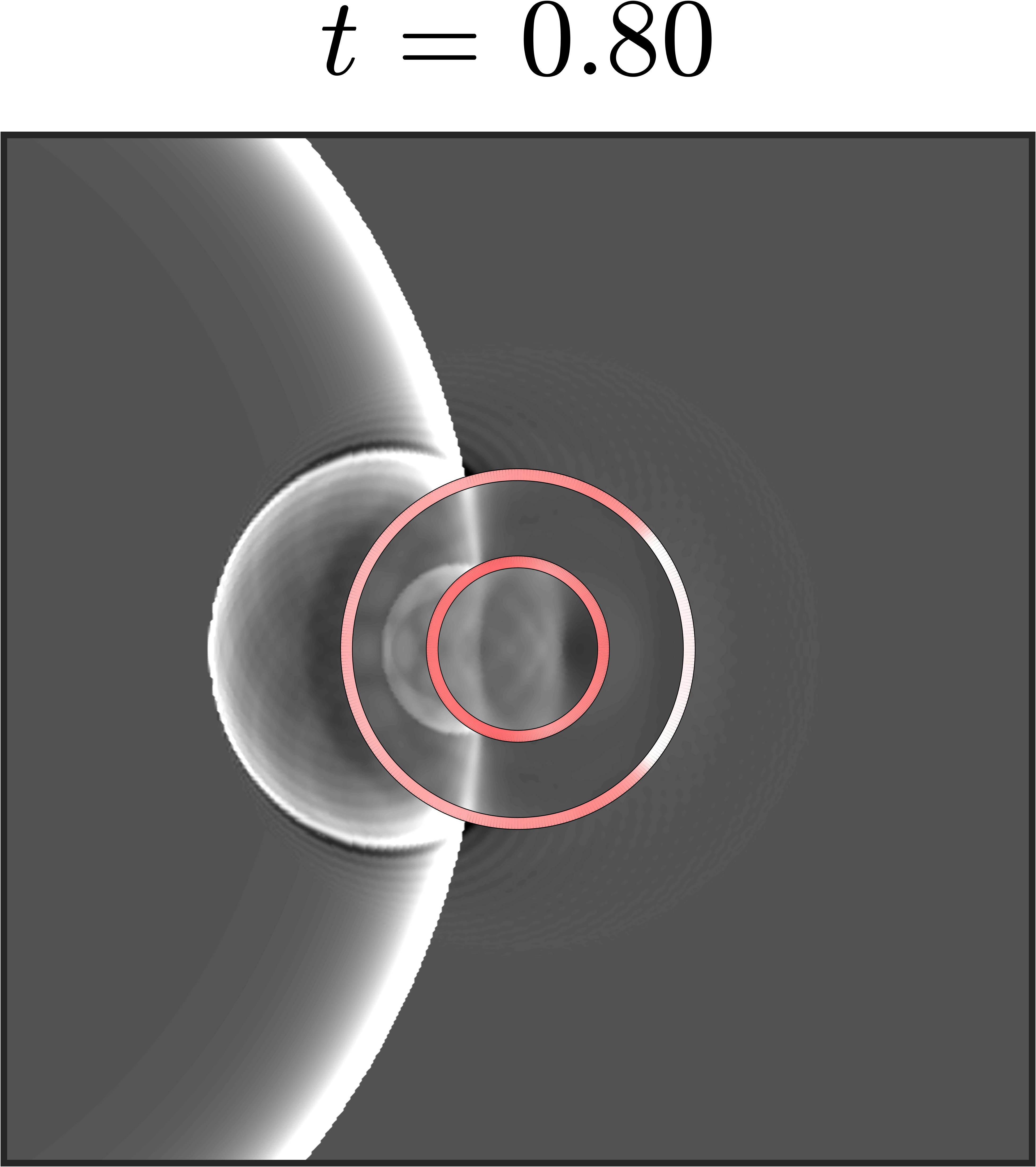}
    \end{minipage}%
    \begin{minipage}[t]{0.25\textwidth}
        \centering
        \includegraphics[height=0.18\textheight]{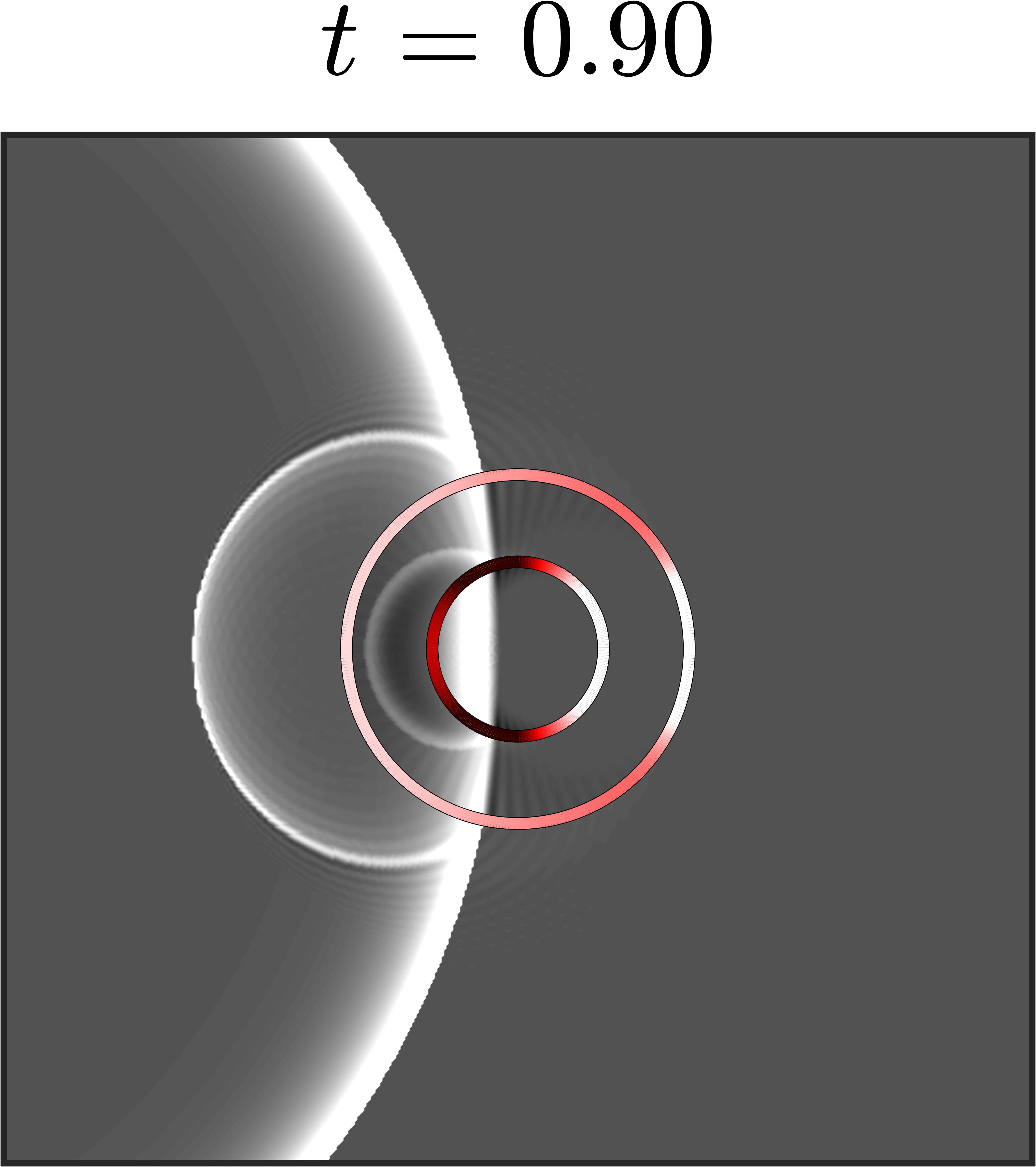}
    \end{minipage}%
    \begin{minipage}[t]{0.25\textwidth}
        \centering
        \includegraphics[height=0.18\textheight]{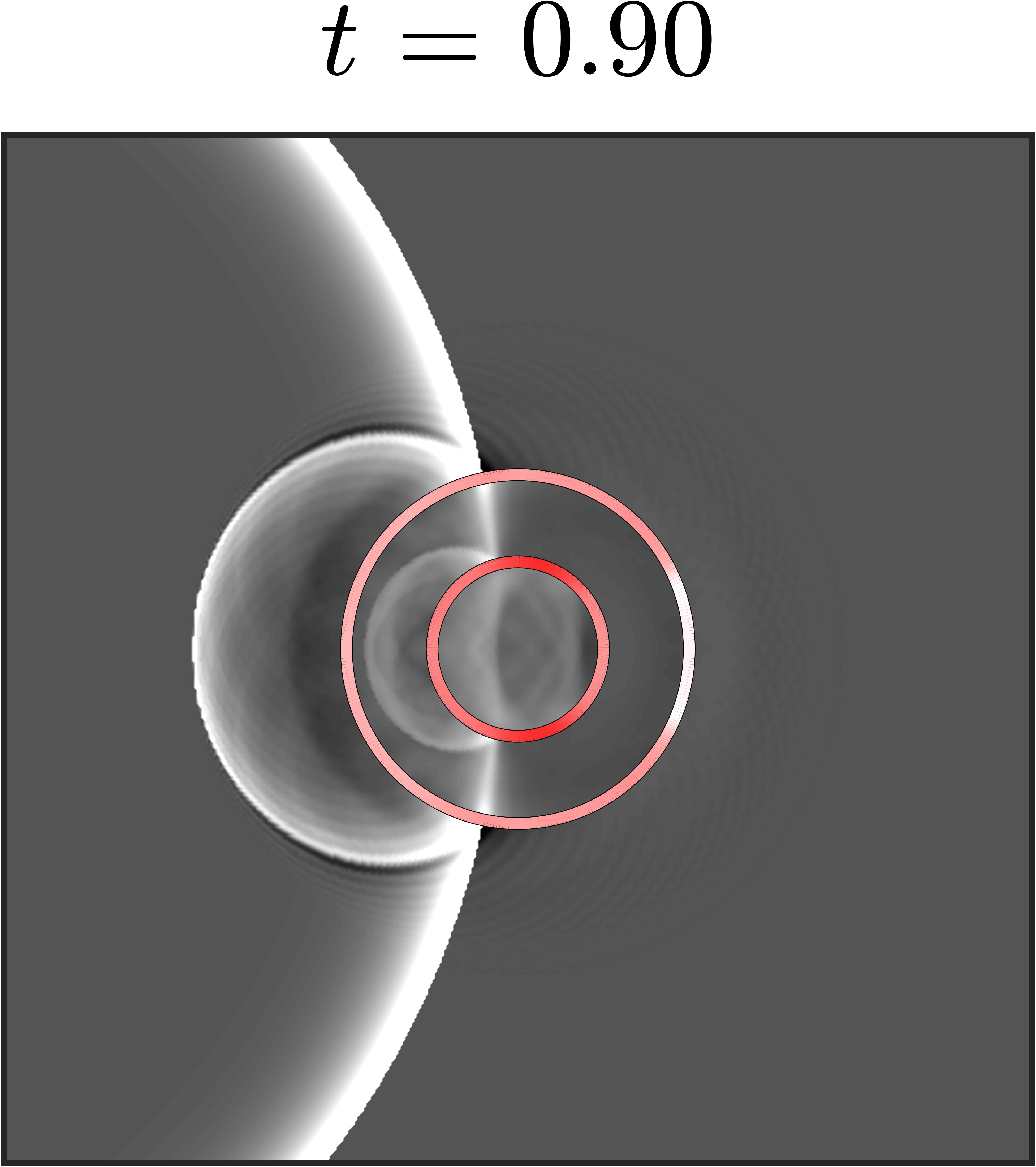}
    \end{minipage}%
    \caption{Comparison of transient acoustic field and induced mechanical stress between the bare system and hybrid control - early times.}
    \label{fig:13}
\end{figure}

\newpage
\begin{figure}[H]
\ContinuedFloat
\centering
\makebox[0.25\linewidth][c]{\text{Bare system}}%
\hfill
\makebox[0.25\linewidth][c]{\text{Hybrid control}}%
\hfill
\makebox[0.25\linewidth][c]{\text{Bare system}}%
\hfill
\makebox[0.25\linewidth][c]{\text{Hybrid control}}

\vspace{0.5cm} %
    \begin{minipage}[t]{0.25\textwidth}
        \centering
        \includegraphics[height=0.18\textheight]{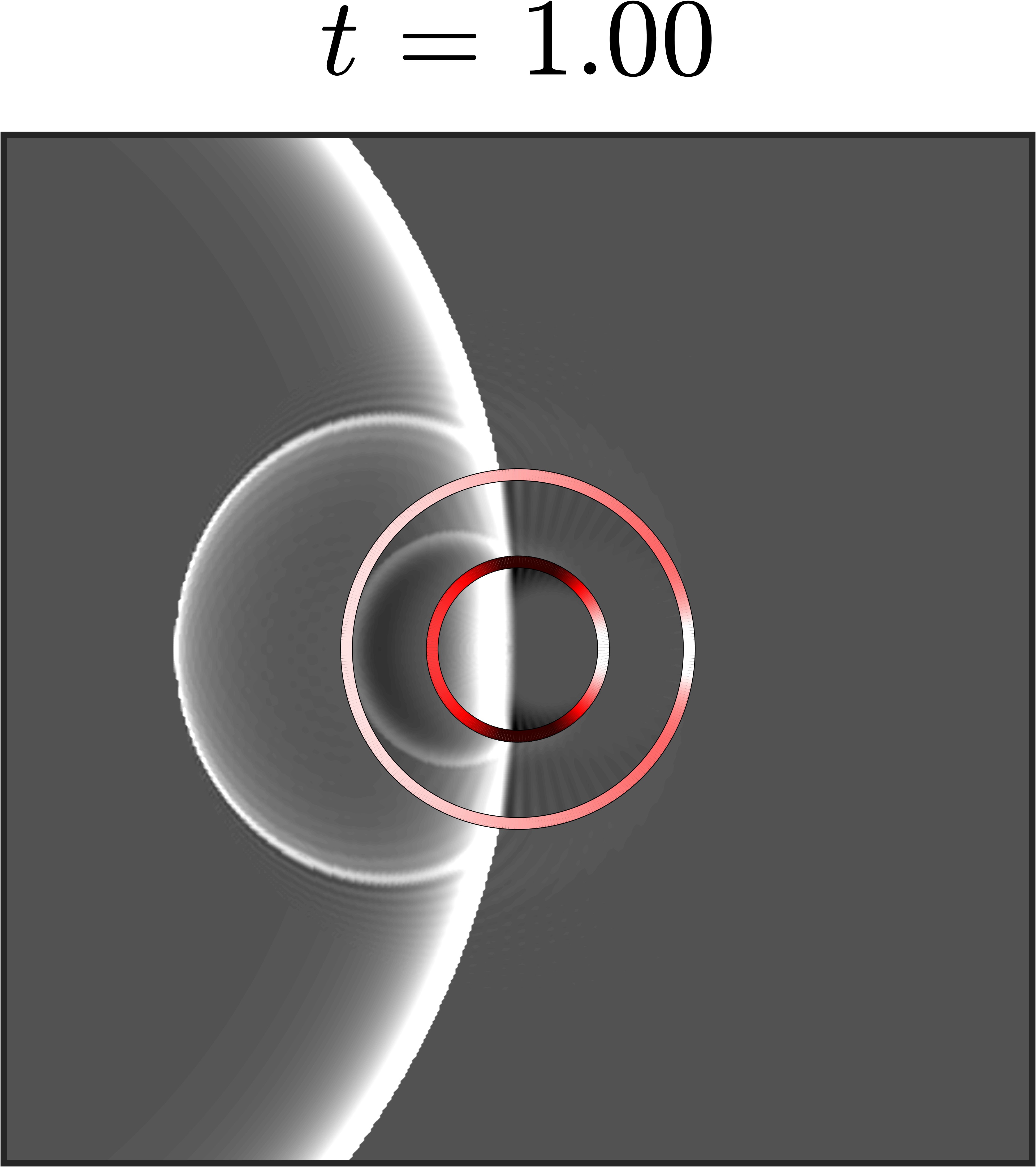}
    \end{minipage}%
    \begin{minipage}[t]{0.25\textwidth}
        \centering
        \includegraphics[height=0.18\textheight]{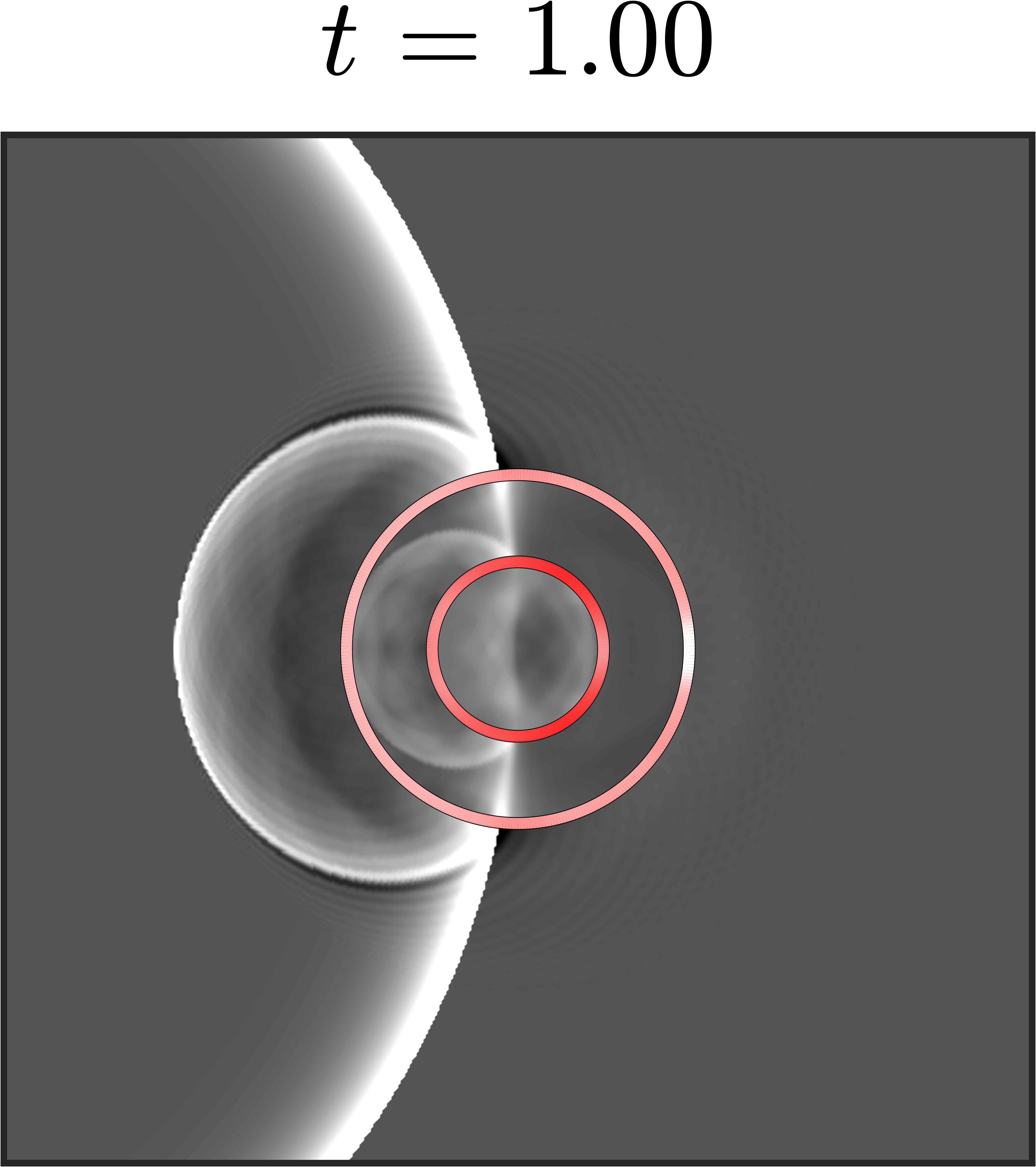}
    \end{minipage}%
    \begin{minipage}[t]{0.25\textwidth}
        \centering
        \includegraphics[height=0.18\textheight]{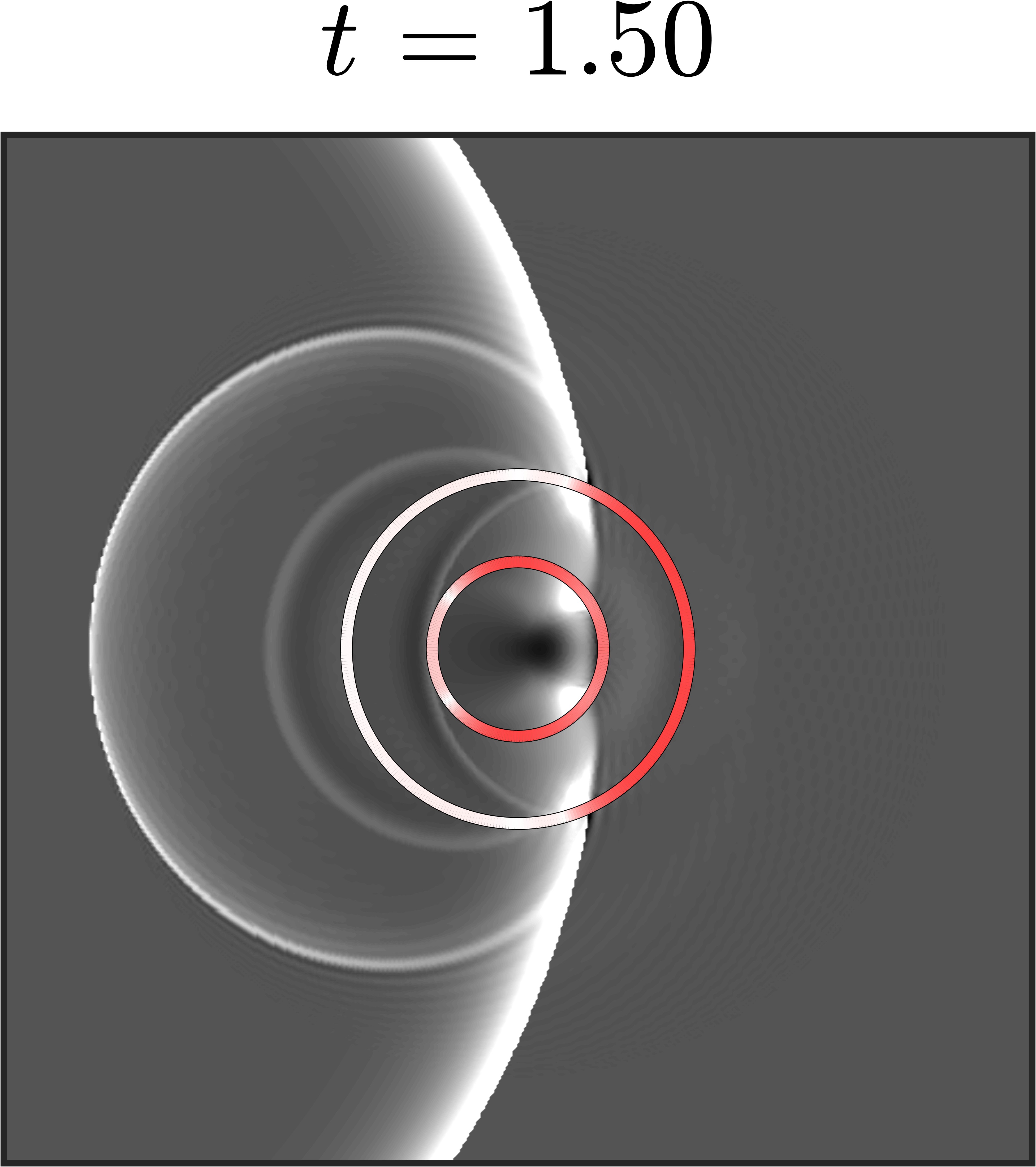}
    \end{minipage}%
    \begin{minipage}[t]{0.25\textwidth}
        \centering
        \includegraphics[height=0.18\textheight]{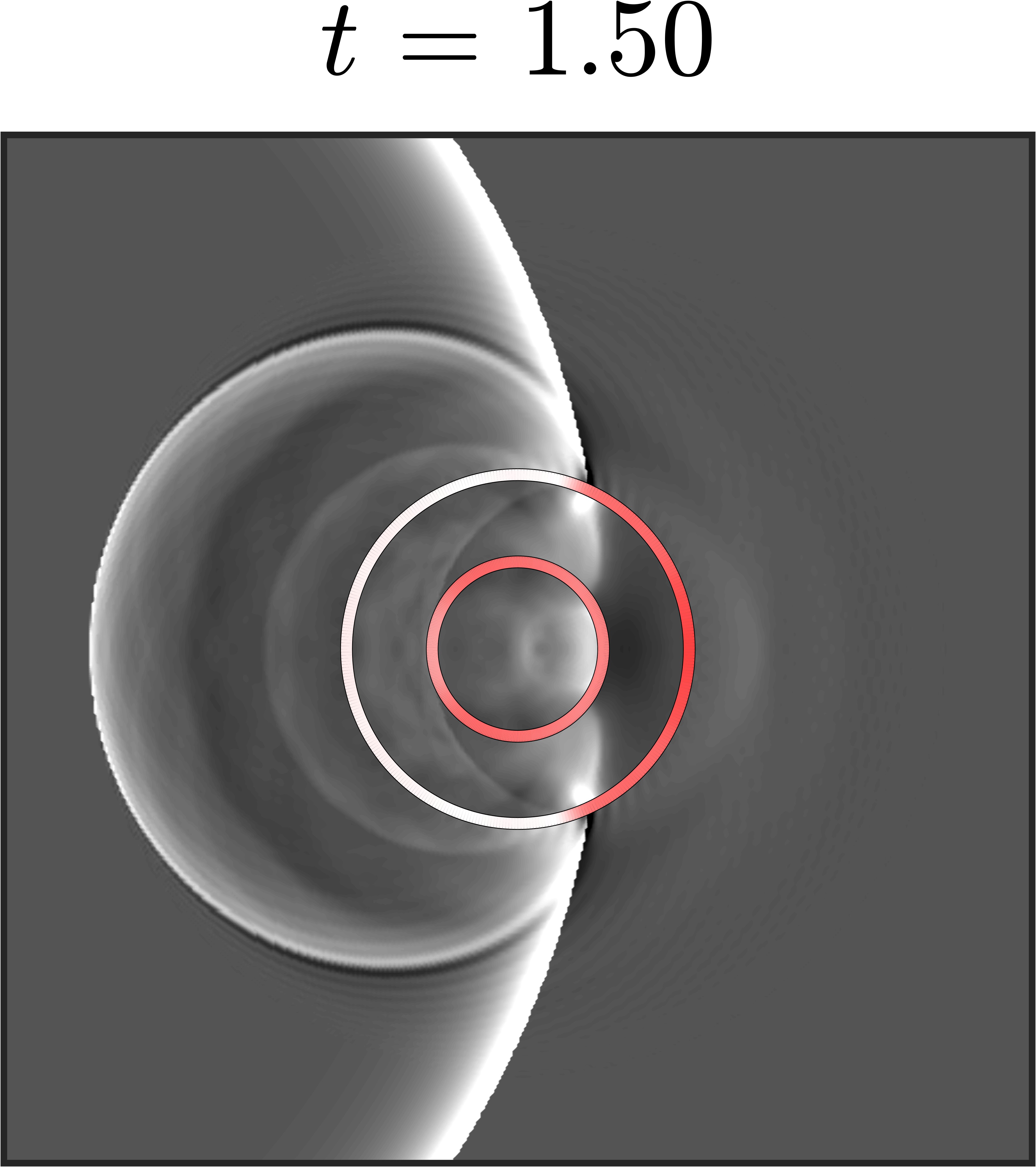}
    \end{minipage}%
    \vspace{0.12cm}
    \begin{minipage}[t]{0.25\textwidth}
        \centering
        \includegraphics[height=0.18\textheight]{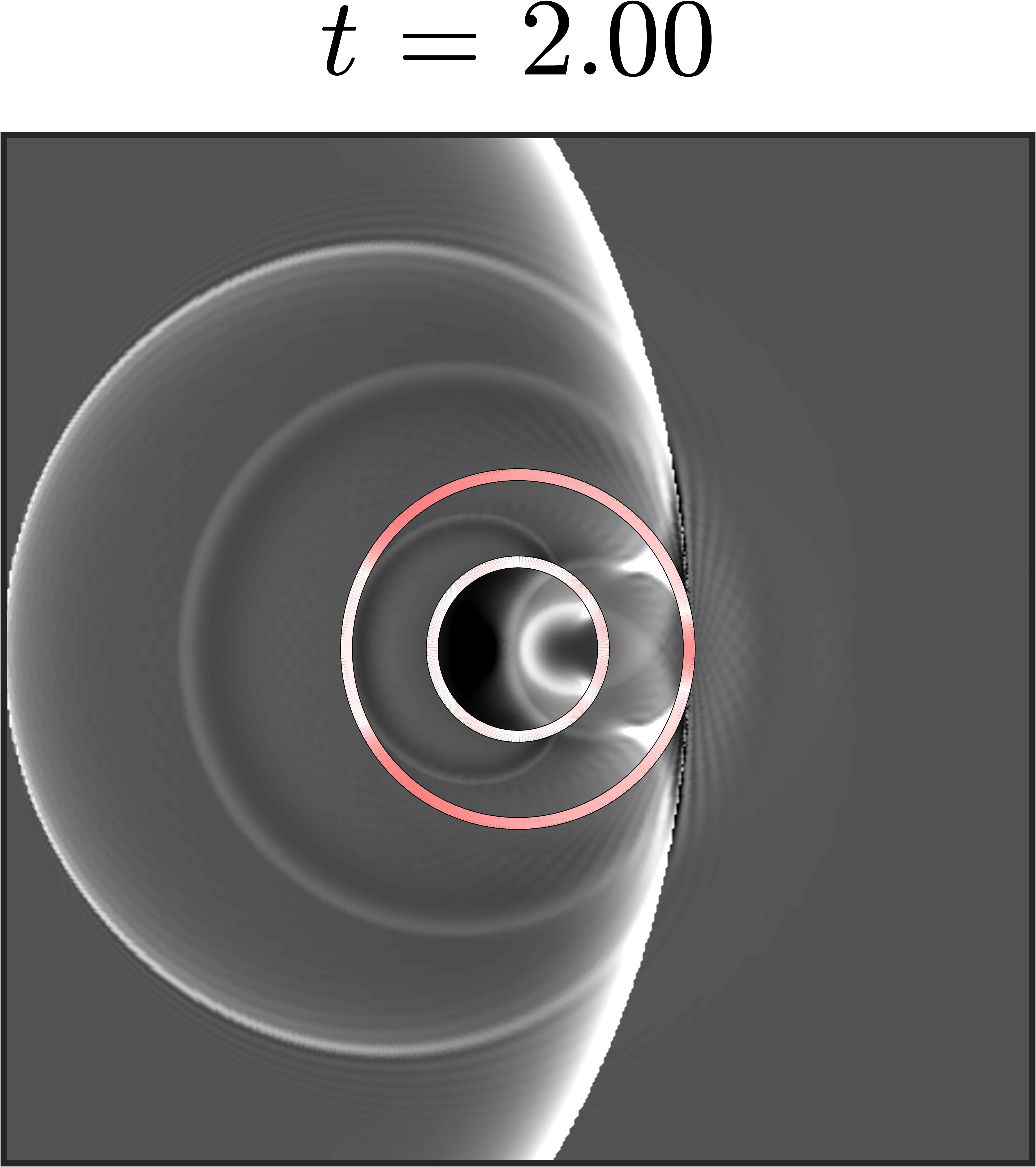}
    \end{minipage}%
    \begin{minipage}[t]{0.25\textwidth}
        \centering
        \includegraphics[height=0.18\textheight]{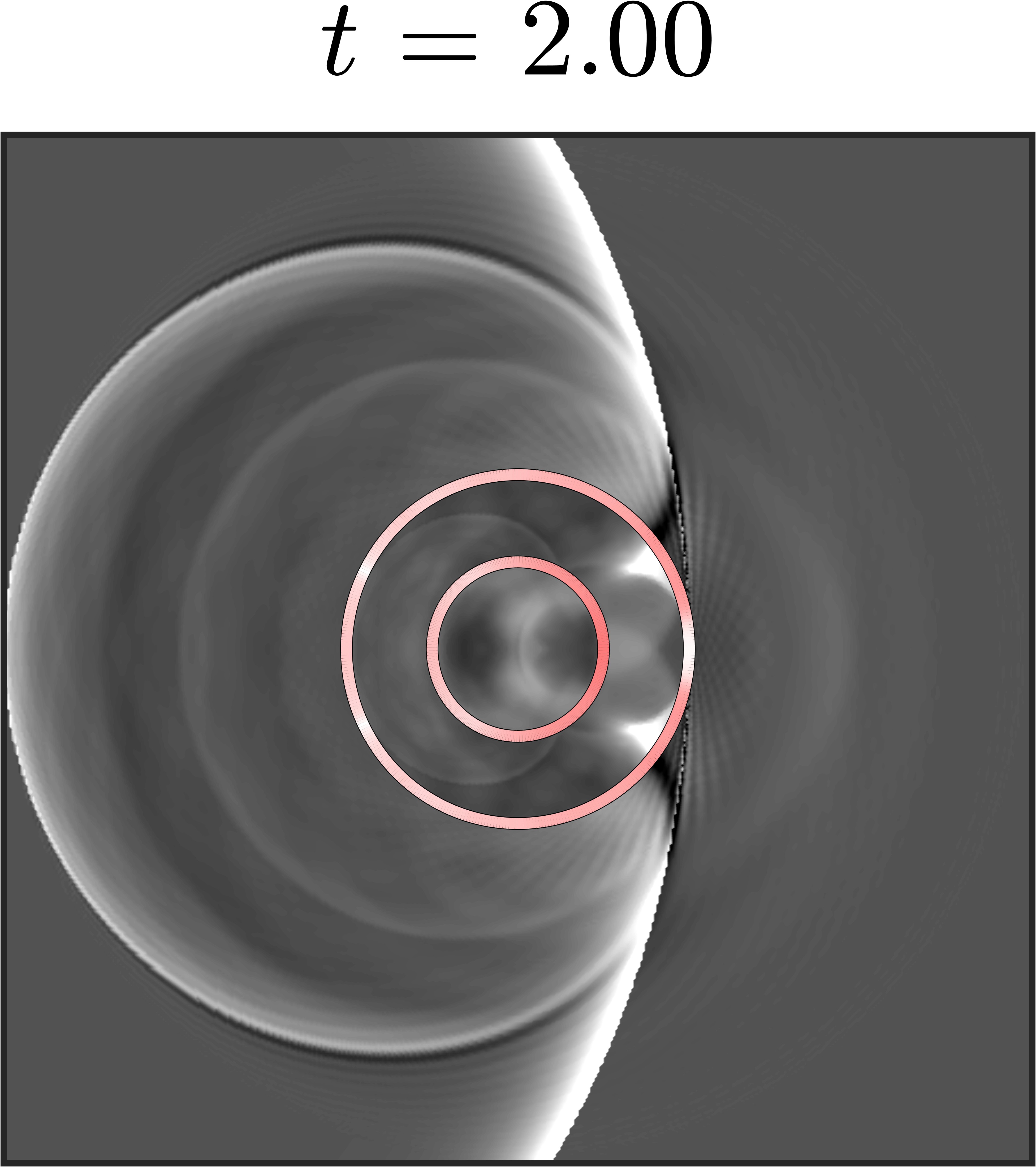}
    \end{minipage}%
    \begin{minipage}[t]{0.25\textwidth}
        \centering
        \includegraphics[height=0.18\textheight]{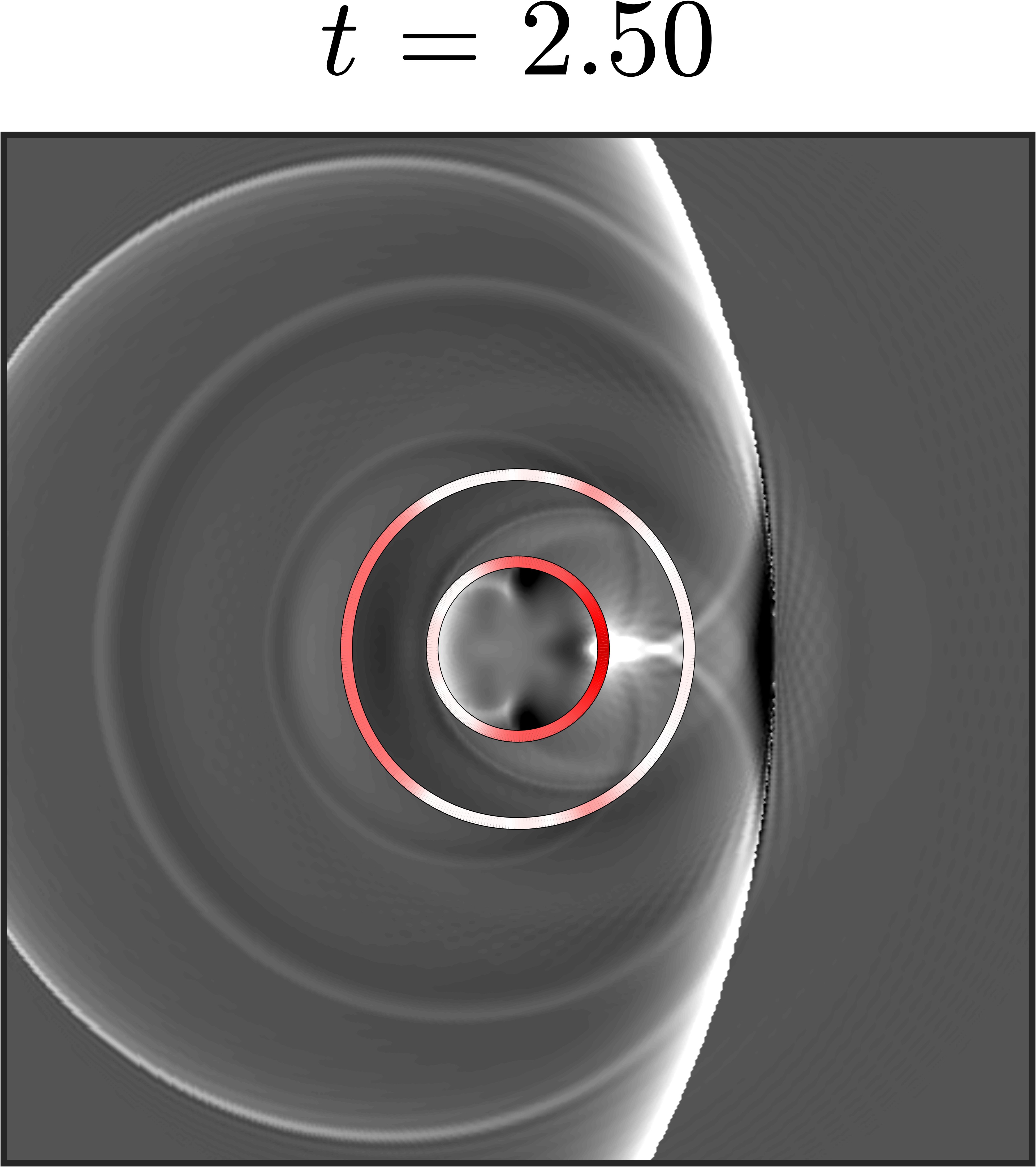}
    \end{minipage}%
    \begin{minipage}[t]{0.25\textwidth}
        \centering
        \includegraphics[height=0.18\textheight]{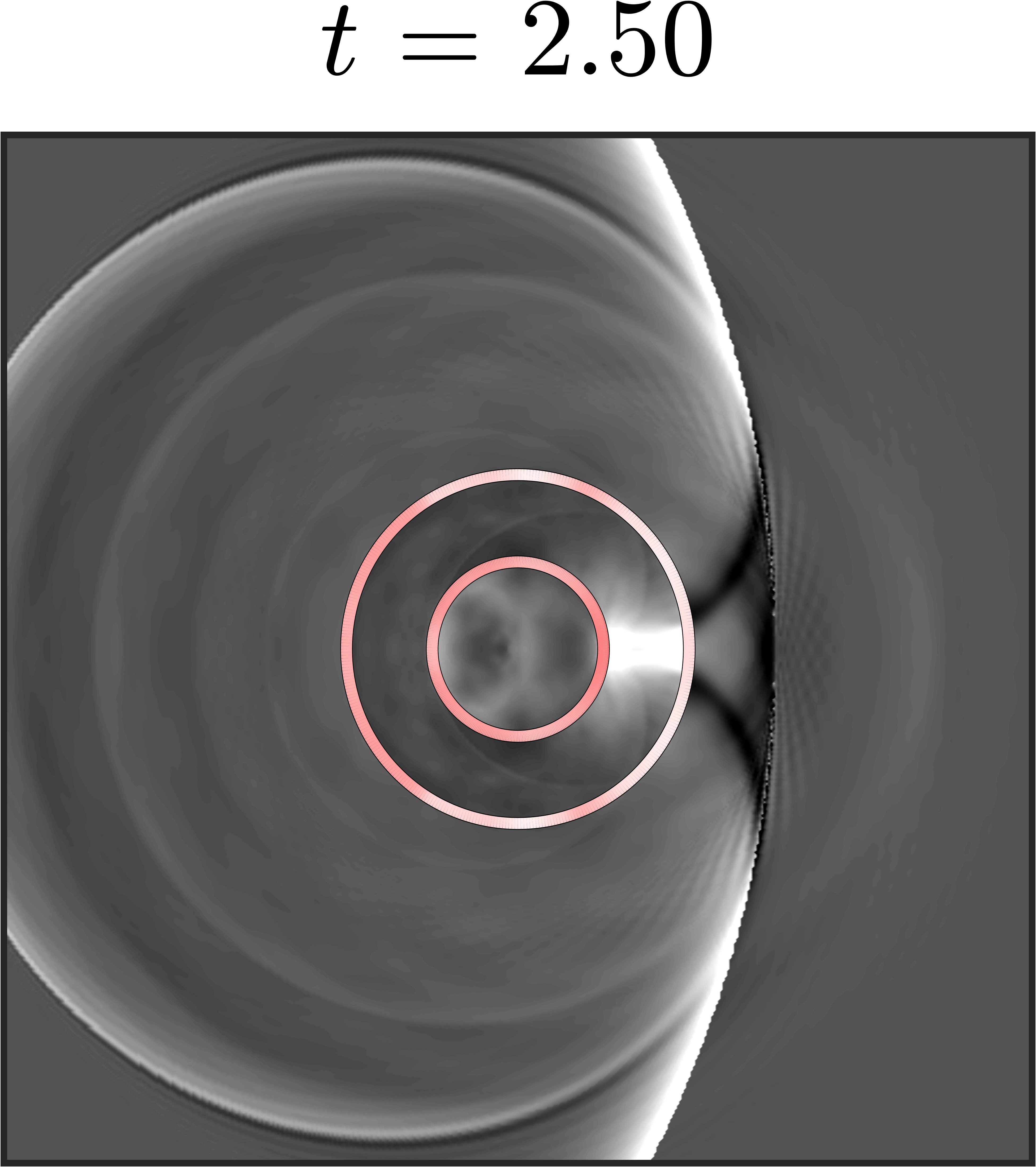}
    \end{minipage}%
    \vspace{0.12cm}
    \begin{minipage}[t]{0.25\textwidth}
        \centering
        \includegraphics[height=0.18\textheight]{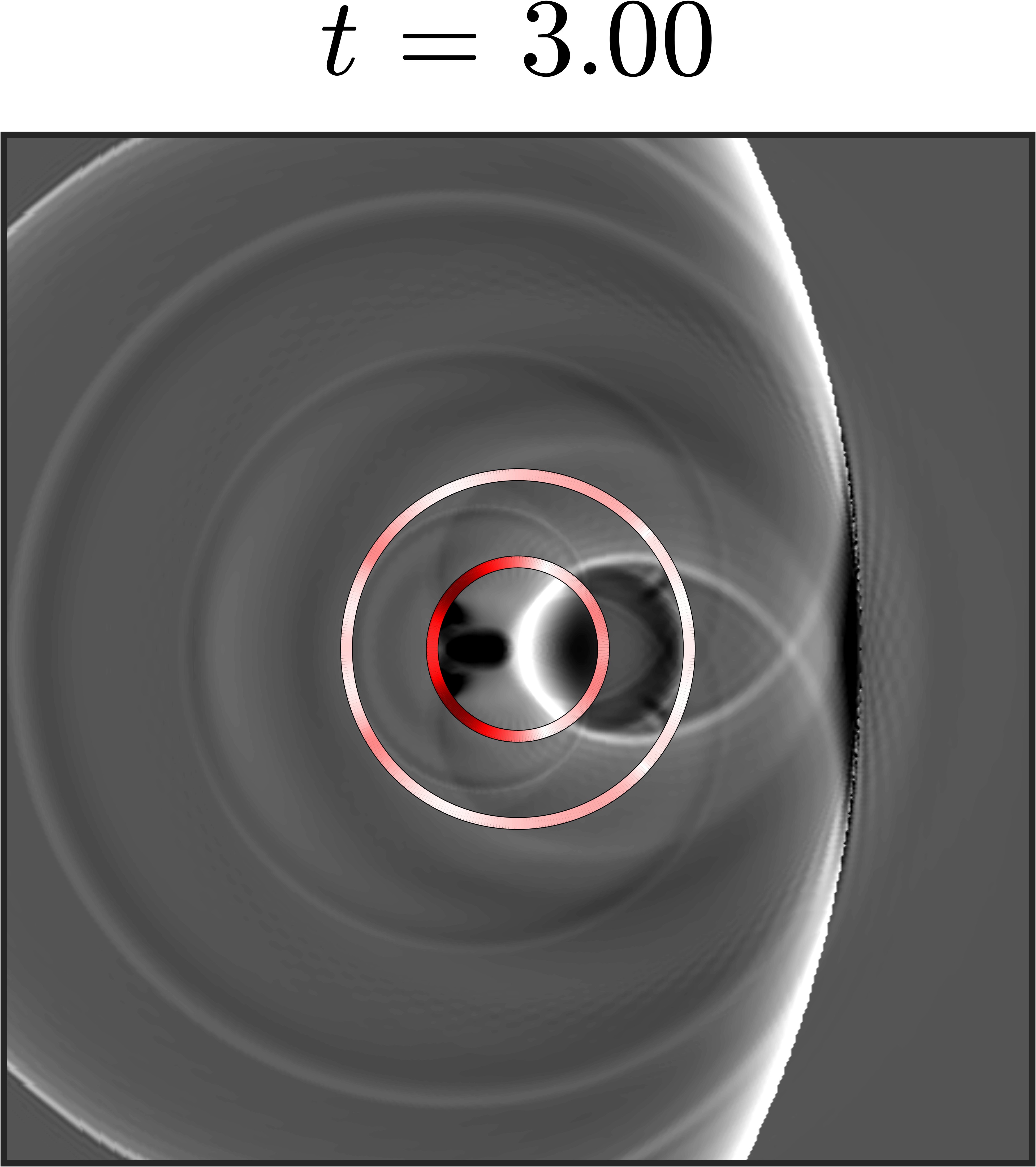}
    \end{minipage}%
    \begin{minipage}[t]{0.25\textwidth}
        \centering
        \includegraphics[height=0.18\textheight]{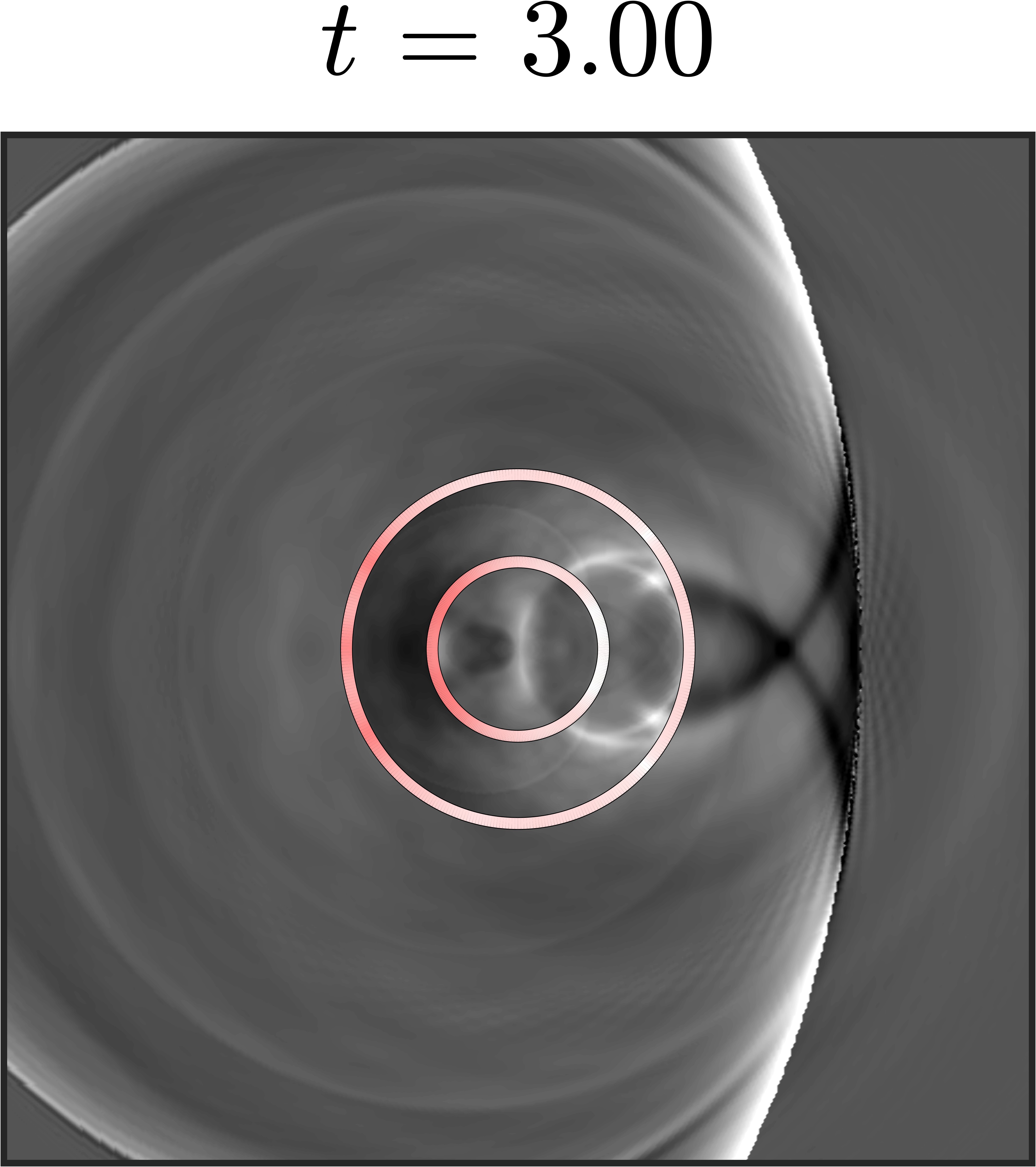}
    \end{minipage}%
    \begin{minipage}[t]{0.25\textwidth}
        \centering
        \includegraphics[height=0.18\textheight]{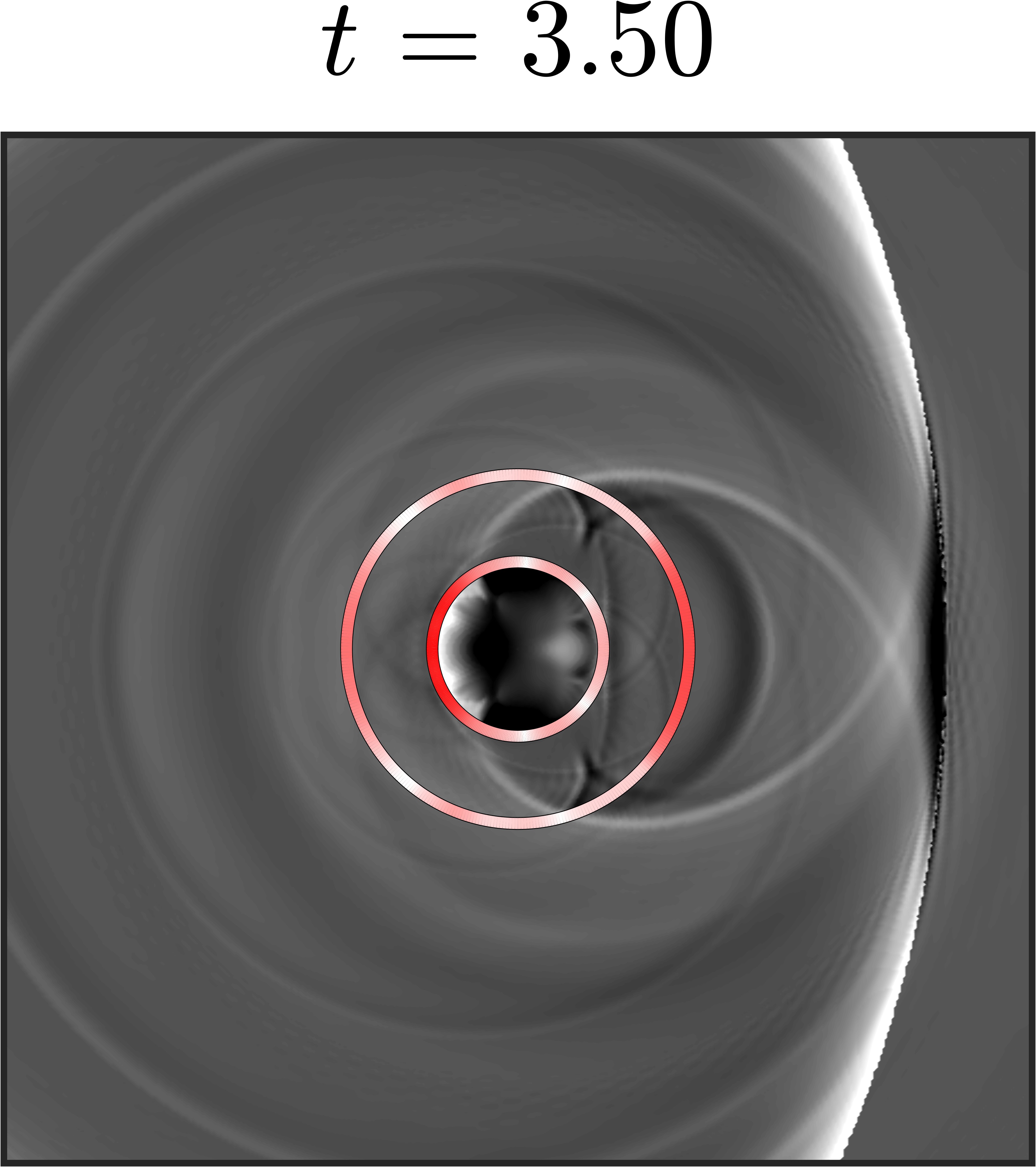}
    \end{minipage}%
    \begin{minipage}[t]{0.25\textwidth}
        \centering
        \includegraphics[height=0.18\textheight]{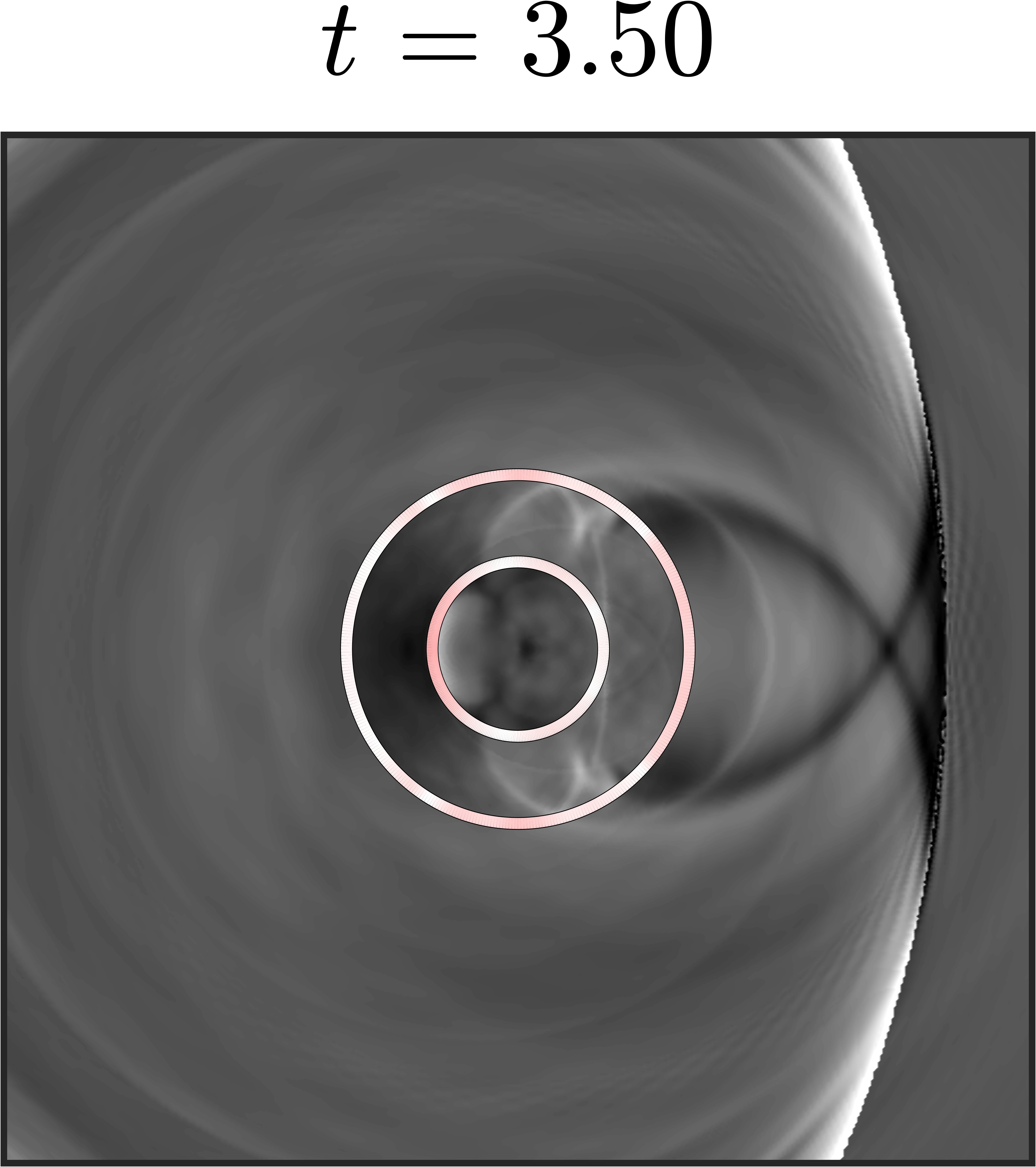}
    \end{minipage}%
    \vspace{0.12cm}
    \begin{minipage}[t]{0.25\textwidth}
        \centering
        \includegraphics[height=0.18\textheight]{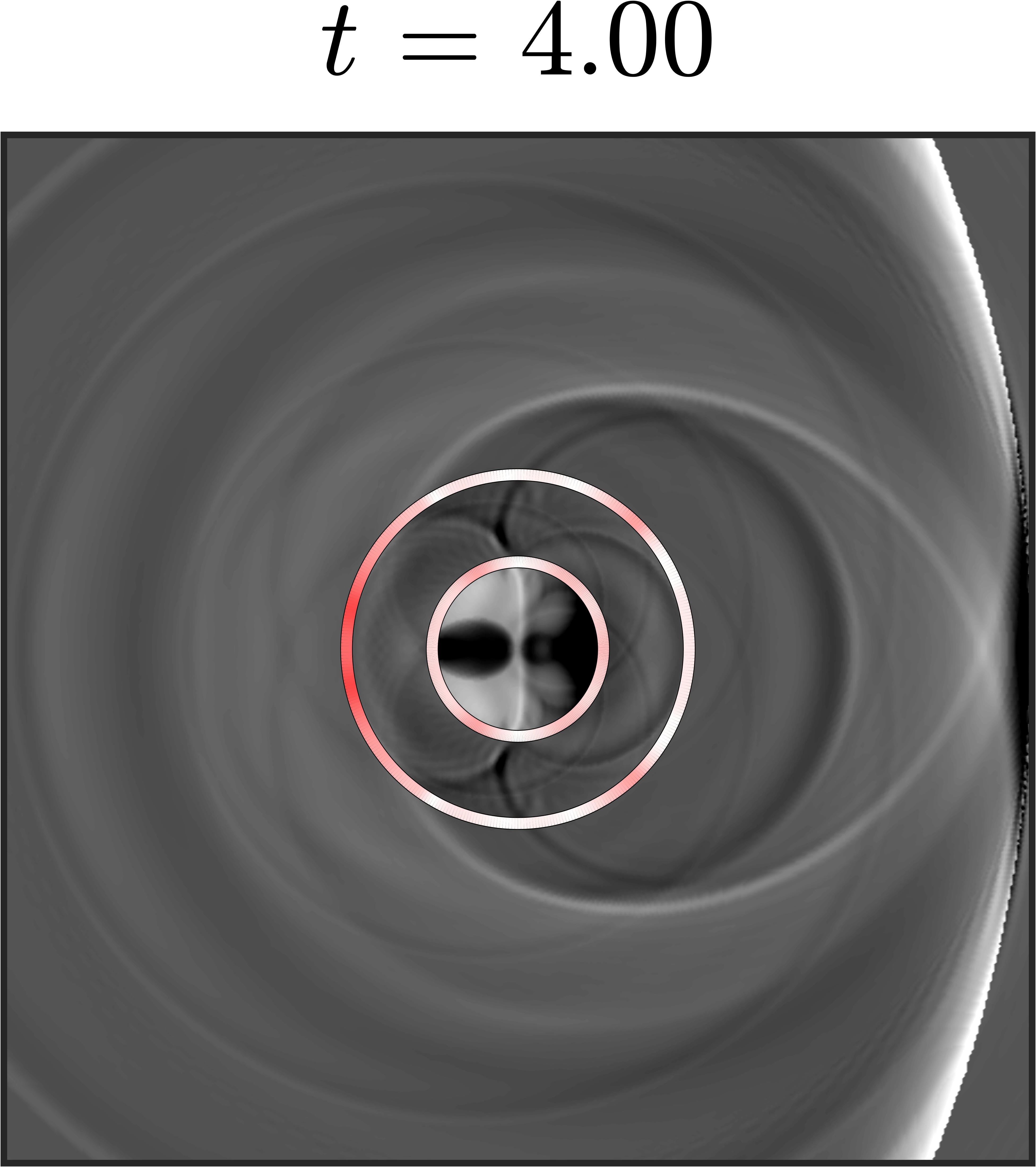}
    \end{minipage}%
    \begin{minipage}[t]{0.25\textwidth}
        \centering
        \includegraphics[height=0.18\textheight]{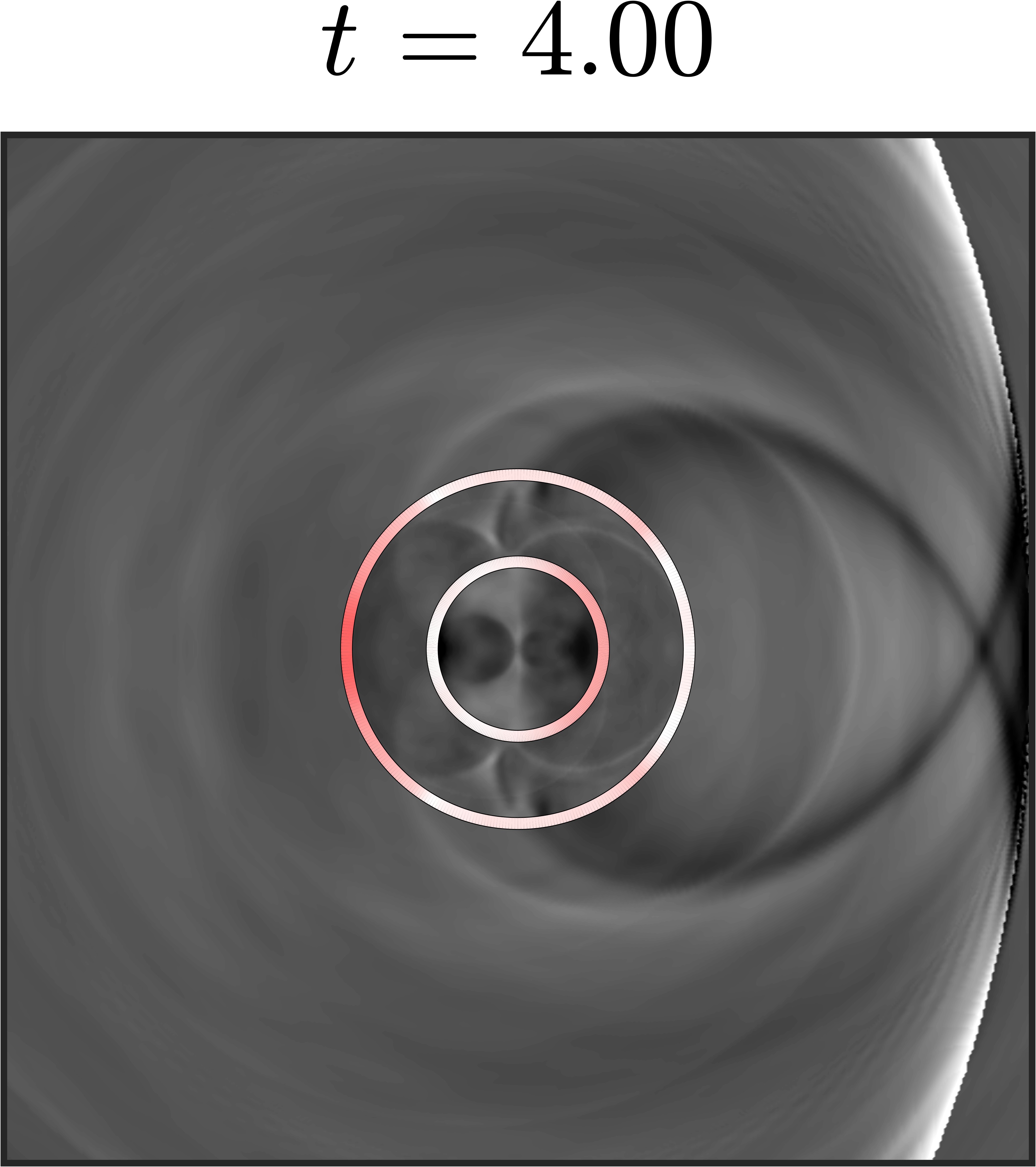}
    \end{minipage}%
    \begin{minipage}[t]{0.25\textwidth}
        \centering
        \includegraphics[height=0.18\textheight]{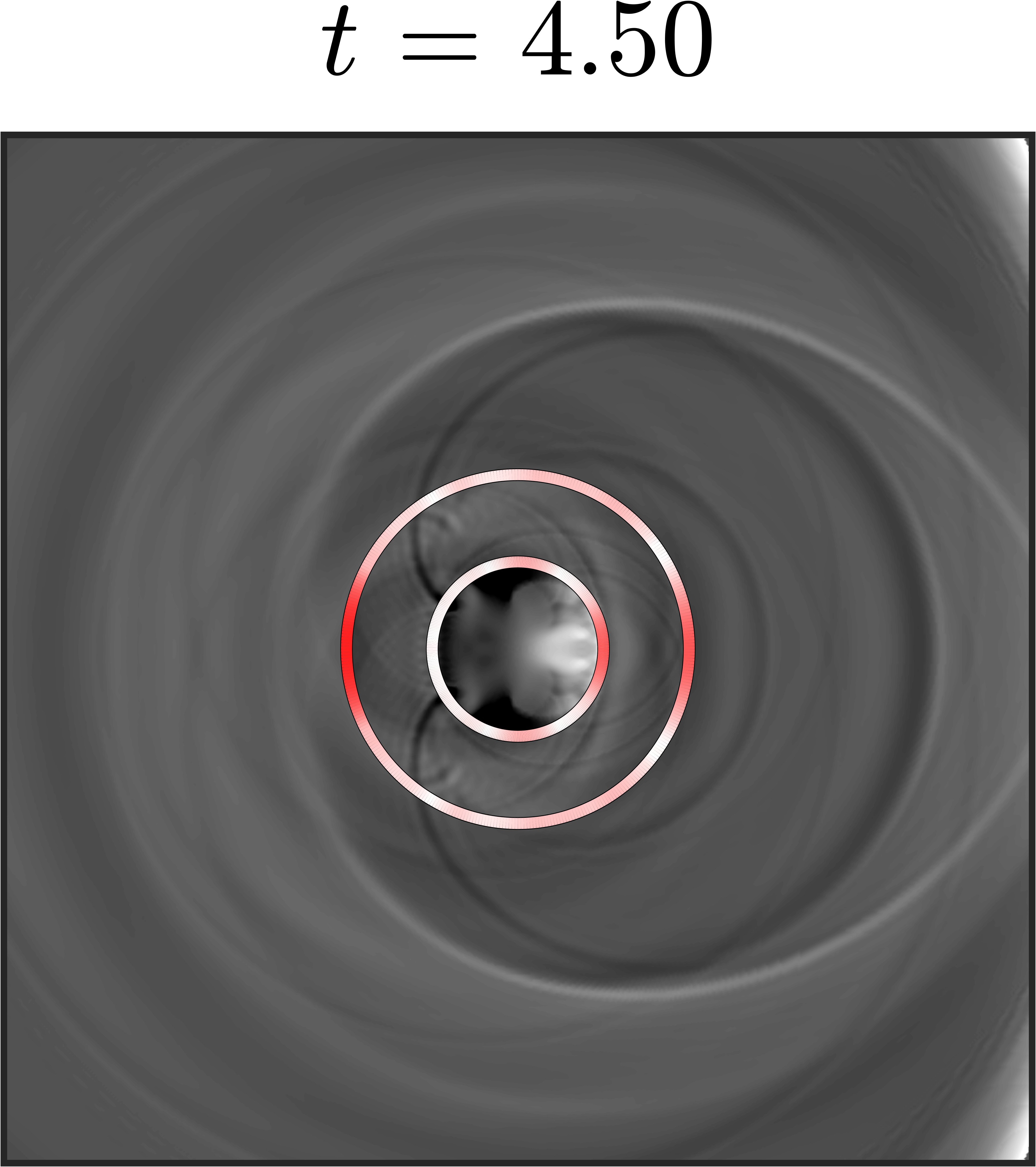}
    \end{minipage}%
    \begin{minipage}[t]{0.25\textwidth}
        \centering
        \includegraphics[height=0.18\textheight]{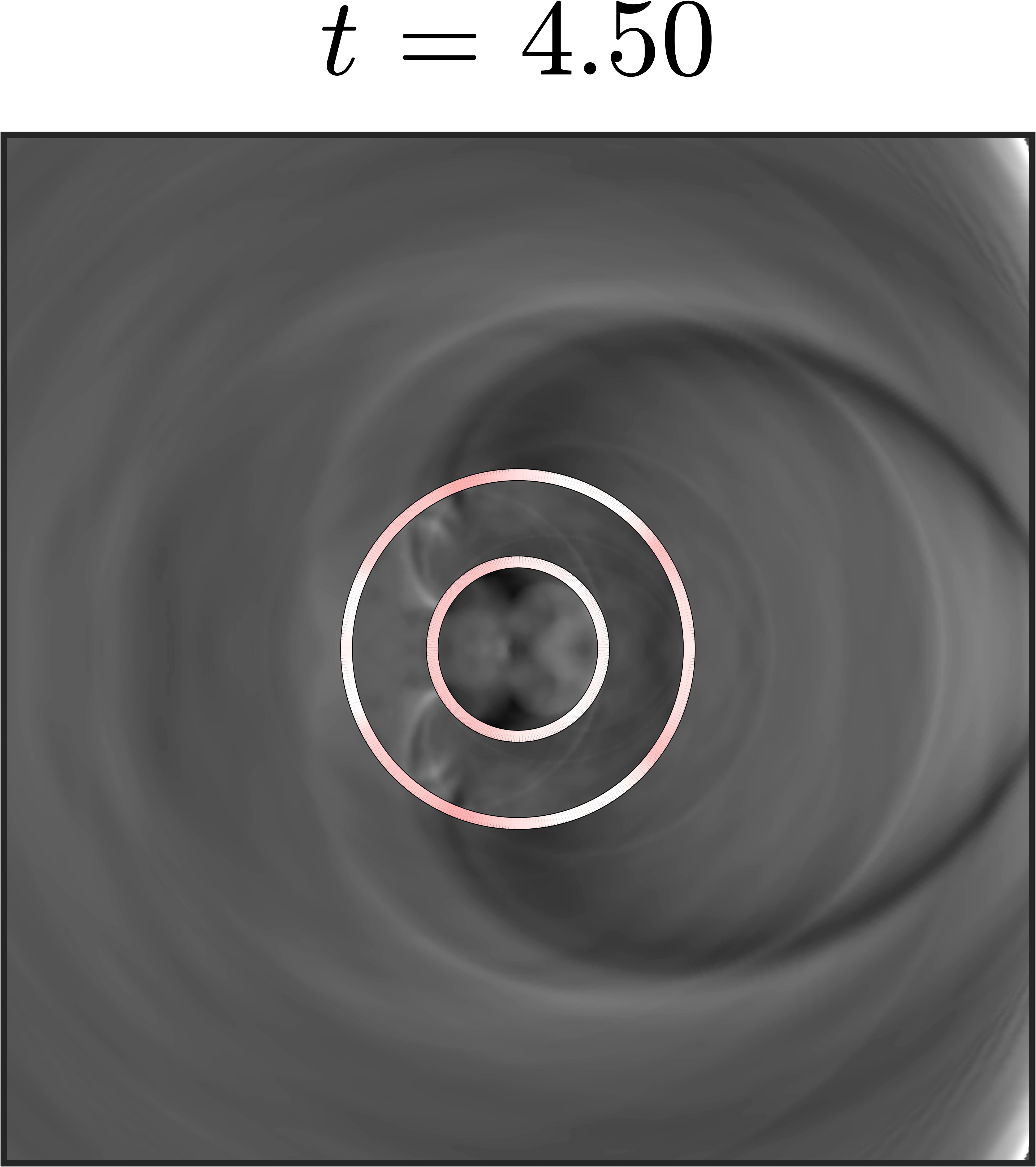}
    \end{minipage}%
    \vspace{0.12cm}
    \begin{minipage}[t]{0.25\textwidth}
        \centering
        \includegraphics[height=0.18\textheight]{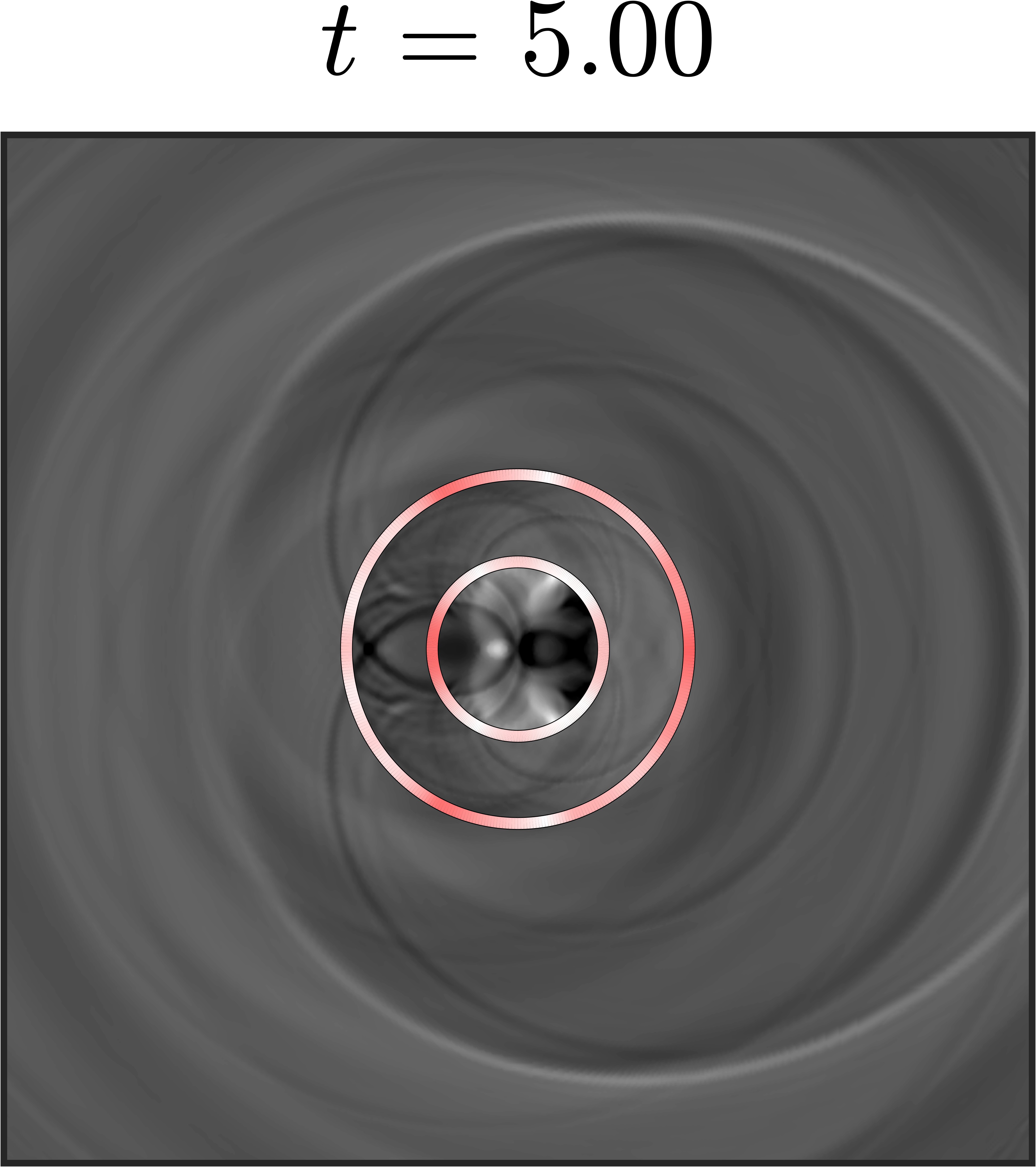}
    \end{minipage}%
    \begin{minipage}[t]{0.25\textwidth}
        \centering
        \includegraphics[height=0.18\textheight]{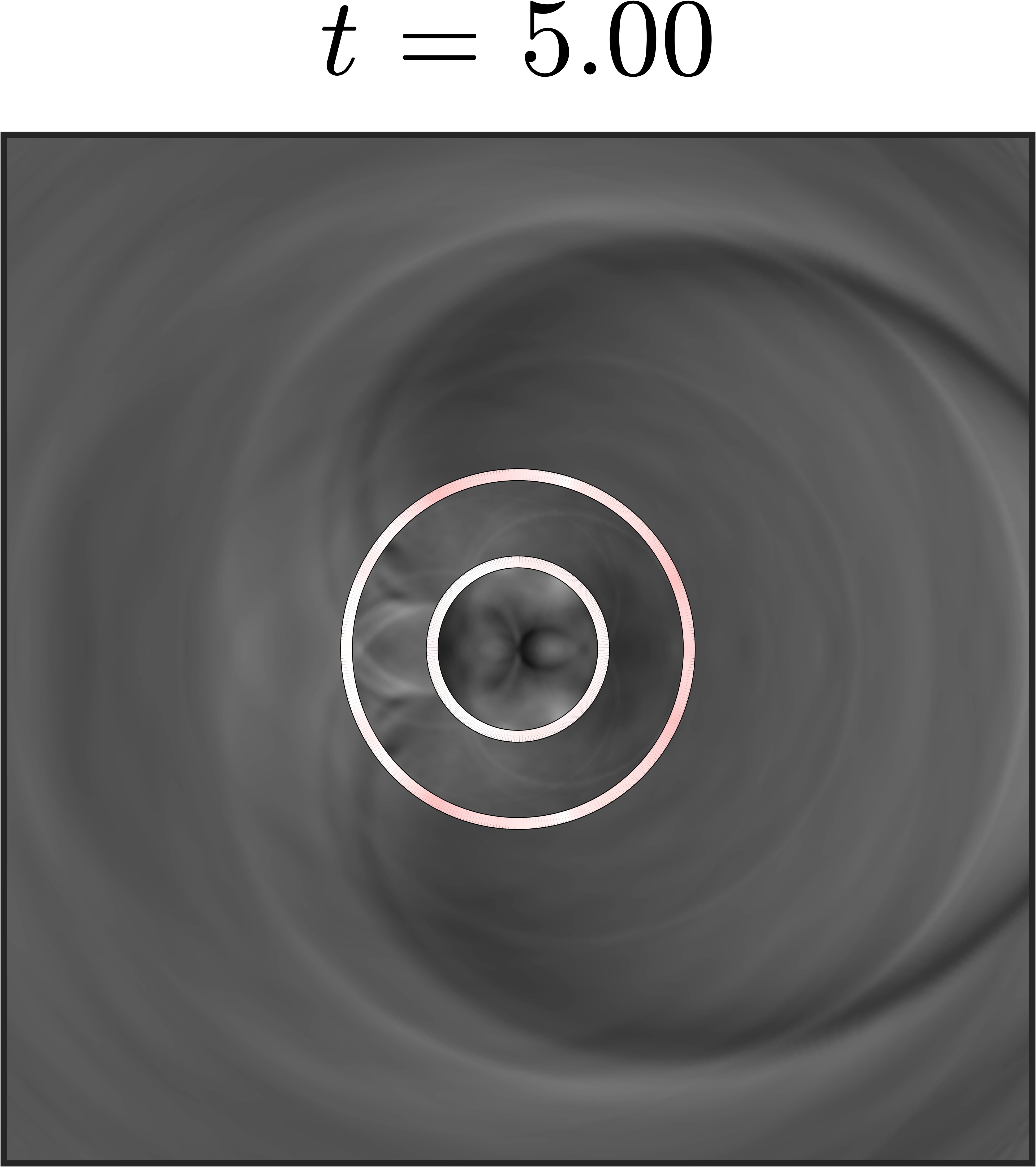}
    \end{minipage}%
    \begin{minipage}[t]{0.5\textwidth}
        \centering
        \raisebox{2.5ex}{\includegraphics[width=0.25\textheight]{Figures/Fig-4-colorbar.jpg}}
    \end{minipage}%
    \caption{continued - late times.}
    \label{fig:13-continue}
\end{figure}

\begin{figure}[H]
\centering
    \begin{minipage}[t]{0.5\textwidth}
        \centering
        \begin{overpic}[height=0.2\textheight,trim= 0cm 0cm 0cm 0.0cm,clip]{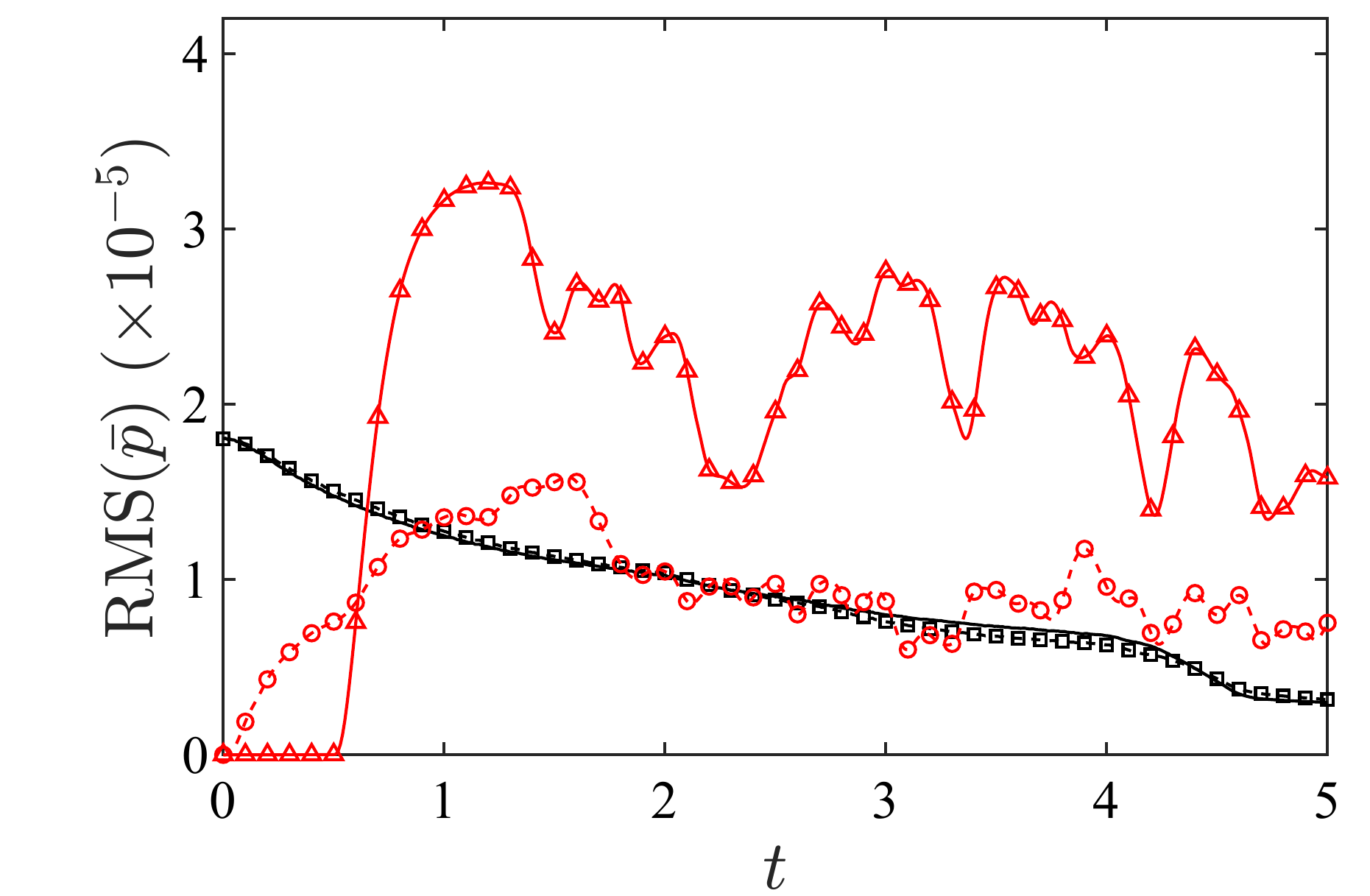}
        \put(88,42){\includegraphics[width=0.6\textwidth]{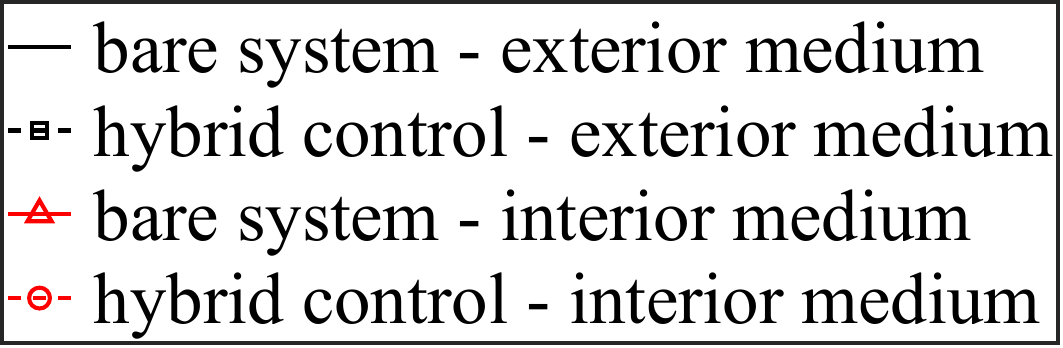}}
        \end{overpic}
    \end{minipage}%
    \caption{Influence of hybrid control on the RMS value of acoustic pressure at exterior and interior media.}
    \label{fig:14}
\end{figure}

Fig. \ref{fig:14} extends the visual insights of Fig. \ref{fig:13} with a complementary quantitative evaluation. In order to assess the hybrid system's effectiveness in attenuating the acoustic field intensity, the RMS value of the dimensionless pressure is tracked over time and the results are compared with those of the bare system. According to Fig. \ref{fig:14}, while the acoustic field intensity remains almost unaffected in the exterior medium, a marked reduction is observed within the interior fluid. In agreement with Fig. \ref{fig:13}, these findings suggest that the implemented hybrid mechanism is capable of mitigating the acoustic energy within the interior medium without redirecting it outward. Hence, the observed decline in stress amplitudes is best explained by internal energy losses, primarily through dissipation and absorption by the integrated vibration absorbers and piezoelectric layers.

\section*{Conclusions}
Despite the topic’s scientific significance, the mathematical complexity of circular double-shell structures has posed a major barrier to conducting comprehensive vibroacoustic analyses. The current work contributes to the field by developing an analytical model that captures the transient behavior of a piezoelectric double-shell subjected to an incident acoustic shockwave. It advances the state of research through incorporating the effect of the interior fluid within the inner shell on the model dynamics. The model consists of two shells, which are interconnected via a series of nonlinear vibration absorbers and partition acoustic space into three subdomains with homogeneous, inviscid, isotropic, and compressible fluids. The incident wave induces structural vibrations upon encountering the outer shell, which leads to acoustic radiation from its surface. The radiated waves propagate through the gap medium between the shells and interact with the inner shell. The governing equations of fluid–structure interaction are derived using Hamilton’s principle, Kirchhoff–Love thin shell theory, Maxwell’s electrodynamics, and the linear wave equation. Following the application of boundary conditions and non-dimensionalization, the equations are transformed into the Laplace domain. Time-domain results are subsequently retrieved via inverse Laplace transformation using Cauchy’s residue theorem and Jordan’s lemma implemented through a response function-based solution scheme.

Prior to presenting the main results, the solution methodology is validated against benchmark studies under limiting conditions. Subsequently, the passive response of the bare system is thoroughly examined in absence of vibration absorbers. The obtained results evidence the higher vulnerability of the inner shell to the incident wave, particularly at its tail point. Additionally, the analysis reveals the acoustic transparency of the shells to the incident wave, characterized by minimal variation in wave intensity across the shell interfaces and consistent pressure field geometry. Given the fast-paced nature of acoustic shockwave interactions with the structure beside the intricate dynamics involved, designing effective stress-mitigating control mechanisms remains a challenging task for double-shell systems. To tackle the challenge, the present work employs a combination of mass-spring-damper vibration absorbers as well as active control scheme based on piezoelectric actuators in order to control the vibroacoustic response of the system. The influence of each parameter affecting the performance of the control system, including the absorber’s mass, linear and nonlinear spring stiffness, attachment position, damping, and number of absorbers, as well as proportional and derivative feedback gains for the piezoelectric actuators, is individually evaluated. Finally, the passive vibration absorbers and active mechanism of piezoelectric actuators are integrated to form a hybrid control system that effectively reduces shell stress and acoustic field intensity. We hope that our findings on inter-shell connectors and active control mechanism will serve as a foundation for designing more efficient and resilient double-shell systems.

\appendix
\newcommand{\oldthesection}{\thesection}
\counterwithin{equation}{section}
\renewcommand{\theequation}{A.\arabic{equation}}
\makeatletter
\renewcommand{\@seccntformat}[1]{}
\makeatother

\section{Appendix A}
\label{sec:appendixA}
\setcounter{equation}{0}
The electromechanical equations governing the dynamics of the double-shell structure are detailed in this section. According to the Kirchhoff-Love theory for thin shells \cite{ke2014thermo}, the  tangential and normal displacements components of the displacement field, can be described by Eqs. (\ref{Eq:A.1}) and (\ref{Eq:A.2}), respectively:
\begin{equation}
\hat{v}_i(z_i,\theta,\tau) = \bar{v}_i(\theta,\tau) + z_i \frac{\partial}{\partial\theta} \bar{w}_i(\theta,\tau),
\label{Eq:A.1}
\end{equation}
\begin{equation}
\hat{w}_i(z_i,\theta,\tau) = \bar{w}_i(\theta,\tau), \quad i \in \{1,2\},
\label{Eq:A.2}
\end{equation}
where $ \bar{v}_i $ and $ \bar{w}_i $ denote the tangential and normal displacements of the neutral axis for each shell, respectively, and $ z_i $ is the radial distance from it. This leads to the following expression for the circumferential strain in each shell:
\begin{equation}
\varepsilon_{\theta\theta}^{(i)} = \frac{1}{R_i} \left[ \left(1 - \frac{z_i}{R_i} \right) \frac{\partial \bar{v}_i}{\partial\theta} - \bar{w}_i \right], \quad i \in \{1,2\}.
\label{Eq:A.3}
\end{equation}
The linear stress components in the elastic and piezoelectric layers, along with the electric displacement in each shell, are respectively expressed by Eqs. (\ref{Eq:A.4}) to (\ref{Eq:A.6}) as:
\begin{equation}
\sigma_\text{h}^{(i)} = Q_\text{h} \varepsilon_{\theta\theta}^{(i)},
\label{Eq:A.4}
\end{equation}
\begin{equation}
\sigma_{\text{a},\text{s}}^{(i)} = \tilde{c}_{22}^{(\text{a},\text{s})} \varepsilon_{\theta\theta}^{(i)} - \tilde{e}_{32}^{(\text{a},\text{s})} E_{\text{a},\text{s}}^{(i)},
\label{Eq:A.5}
\end{equation}
\begin{equation}
D_{\text{a},\text{s}}^{(i)} = \tilde{e}_{32}^{(\text{a},\text{s})} \varepsilon_{\theta\theta}^{(i)} - \tilde{\mu}_{33}^{(\text{a},\text{s})} E_{\text{a},\text{s}}^{(i)}, \quad i \in \{1,2\},
\label{Eq:A.6}
\end{equation}
where, the subscripts $\text{h}$, $\text{s}$, and $\text{a}$ refer to the elastic, piezoelectric sensor and actuator layers, respectively. $D$ represents the electric displacement, and $E = -\frac{\partial \bar{\varphi}}{\partial z}$ indicates the electric field intensity along the radial direction in piezoelectric layers. As detailed in Eq. (\ref{Eq:A.11}) later in this section, $Q_\text{h}$ and $\tilde{c}_{22}$ denote the elasticity matrix components in elastic and piezoelectric layers, while $\tilde{e}_{32}$ and $\tilde{\mu}_{33}$ are the reduced electromechanical coupling and dielectric permittivity parameters, respectively. Adopting a cosine distribution, the electric potential variation in each layer emerges as shown in Eq. (\ref{Eq:A.7}) \cite{hasheminejad2022control}:
\begin{equation}
\begin{gathered}
\bar{\Phi}_{\text{a}}^{(i)} = -\cos\left( \frac{\pi}{t_\text{a}} \left(z_i - \frac{t_\text{h} + t_\text{a}}{2} \right) \right) \phi_i^{(\text{a})}(\theta,t), \\
\bar{\Phi}_{\text{s}}^{(i)} = -\cos\left( \frac{\pi}{t_\text{s}} \left(z_i + \frac{t_\text{h} + t_\text{s}}{2} \right) \right) \phi_i^{(\text{s})}(\theta,t), \quad i \in \{1,2\}.
\end{gathered}
\label{Eq:A.7}
\end{equation}
Now, the system's final equations of motion can be obtained by applying Hamilton’s principle in the form of Eq.(\ref{Eq:A.8}):
\begin{equation}
\int_{\tau_1}^{\tau_2} \delta \bigl( T^{(i)} + W^{(i)} - U^{(i)} \bigr)\, d\tau = 0, 
\quad i \in \{1,2\},
\label{Eq:A.8}
\end{equation}
and the variations of potential energy ($U$), kinetic energy ($T$), and virtual work done by external forces ($W$) are described by Eq. (\ref{Eq:A.9}):
\begin{equation}
\begin{gathered}
\delta U^{(i)} = \iint_{A_\text{s}^{(i)}} \left( \sigma_{\text{s}}^{(i)} \, \delta \varepsilon_{\theta\theta}^{(i)} - D_{\text{s}}^{(i)} \, \delta E_{\text{s}}^{(i)} \right) \, \text{d}z_i \, \text{d}\theta + \iint_{A_\text{h}^{(i)}} \left( \sigma_{\text{h}}^{(i)} \, \delta \varepsilon_{\theta\theta}^{(i)} \right) \, \text{d}z_i \, \text{d}\theta \\
+ \iint_{A_\text{a}^{(i)}} \left( \sigma_{\text{a}}^{(i)} \, \delta \varepsilon_{\theta\theta}^{(i)} - D_{\text{a}}^{(i)} \, \delta E_{\text{a}}^{(i)} \right) \, \text{d}z_i \, \text{d}\theta, \\
\\
\delta T^{(i)} = \iint_{A_\text{s}^{(i)}} \rho_\text{s} \left( \dot{\bar{v}}_i \, \delta \dot{\bar{v}}_i + \dot{\bar{w}}_i \, \delta \dot{\bar{w}}_i \right) \, \text{d}z_i \, \text{d}\theta + \iint_{A_\text{h}^{(i)}} \rho_\text{h} \left( \dot{\bar{v}}_i \, \delta \dot{\bar{v}}_i + \dot{\bar{w}}_i \, \delta \dot{\bar{w}}_i \right) \, \text{d}z_i \, \text{d}\theta \\
+ \iint_{A_\text{a}^{(i)}} \rho_\text{a} \left( \dot{\bar{v}}_i \, \delta \dot{\bar{v}}_i + \dot{\bar{w}}_i \, \delta \dot{\bar{w}}_i \right) \, \text{d}z_i \, \text{d}\theta, \\
\\
\delta W^{(i)} = \int_0^{2\pi} \bar{F}_i^{\text{net}} \, \delta \bar{w}_i \, \text{d}\theta, \quad i \in \{1,2\},
\end{gathered}
\label{Eq:A.9}
\end{equation}
where, $A_\text{h}^{(i)}$, $A_\text{s}^{(i)}$, and $A_\text{a}^{(i)}$ denote the areas of the elastic, piezoelectric sensor and actuator layers, respectively. The term $\bar{F}_i^{\text{net}}$ captures the total external loading on each shell that comprises the net force induced by surrounding acoustic pressure, the actuation force generated by the piezoelectric layers, and the concentrated forces exerted by the absorbers, as shown in Eq. (\ref{Eq:A.10}).
\begin{equation}
\bar{F}_i^{\text{net}} = \bar{p}_i^{\text{net}} + \frac{1}{R_i} \sum_{n=1}^N \delta(\theta - \theta_n) \bar{f}_n^{(i)} - \frac{e_{32}^{(\text{a})}}{R_i} \bar{u}_i^{(\text{a})}.
\label{Eq:A.10}
\end{equation}
By substituting Eqs. (\ref{Eq:A.3})–(\ref{Eq:A.6}) and (\ref{Eq:A.9}) into Eq. (\ref{Eq:A.8}), applying integration by parts, and separating the coefficients of $\delta\bar{w}_i$, $\delta\bar{v}_i$, and $\delta\bar{\Phi}_i^{(\text{a},\text{s})}$, the final governing equations of motion for each of the piezoelectric shells could be achieved as the set of Eqs. (\ref{Eq:3})–(\ref{Eq:5}). Finally, the quantitative definitions of the parameters $I$, $\zeta$, $a$, $d$, and $E$, which encapsulate the geometric, material, and electromechanical characteristics of the system, are determined through the integral expressions detailed in Eq. (\ref{Eq:A.11}).
\begin{equation}
\begin{gathered}
I_1 = I_2 = \int_{-\frac{t_\text{h} + 2 t_\text{s}}{2}}^{-\frac{t_\text{h}}{2}} \rho_\text{s} \, \text{dz}
+ \int_{-\frac{t_\text{h}}{2}}^{\frac{t_\text{h}}{2}} \rho_\text{h} \, \text{d}z
+ \int_{\frac{t_\text{h}}{2}}^{\frac{t_\text{h} + 2 t_\text{a}}{2}} \rho_\text{a} \, \text{d}z, \\
(a_{22}, d_{22}) = \int_{-\frac{t_\text{h} + 2 t_\text{s}}{2}}^{-\frac{t_\text{h}}{2}} (1, z^2) \tilde{c}_{22} \, \text{d}z
+ \int_{-\frac{t_\text{h}}{2}}^{\frac{t_\text{h}}{2}} (1, z^2) Q_\text{h} \, \text{dz}
+ \int_{\frac{t_\text{h}}{2}}^{\frac{t_\text{h} + 2 t_\text{a}}{2}} (1, z^2) \tilde{c}_{22} \, \text{d}z, \\
E_{32}^{(\text{s})} = \int_{-\frac{t_\text{h} + 2 t_\text{s}}{2}}^{-\frac{t_\text{h}}{2}} \frac{\tilde{e}_{32} \pi}{t_\text{s}} z \sin\left(\frac{\pi}{t_\text{s}} \left(z + \frac{t_\text{h} + t_\text{s}}{2} \right)\right) \text{d}z, \\
E_{32}^{(\text{a})} = \int_{\frac{t_\text{h}}{2}}^{\frac{t_\text{h} + 2 t_\text{a}}{2}} \frac{\tilde{e}_{32} \pi}{t_\text{a}} z \sin\left(\frac{\pi}{t_\text{a}} \left(z - \frac{t_\text{h} + t_\text{a}}{2} \right)\right) \text{d}z, \\
\zeta_{i,22}^{(\text{s})} = \int_{-\frac{t_\text{h} + 2 t_\text{s}}{2}}^{-\frac{t_\text{h}}{2}} \tilde{\mu}_{22} \left[ \frac{1}{R_i + z} \cos\left(\frac{\pi}{t_\text{s}} \left(z + \frac{t_\text{h} + t_\text{s}}{2} \right) \right) \right]^2 \text{d}z, \\
\zeta_{i,22}^{(\text{a})} = \int_{\frac{t_\text{h}}{2}}^{\frac{t_\text{h} + 2 t_\text{a}}{2}} \tilde{\mu}_{22} \left[ \frac{1}{R_i + z} \cos\left(\frac{\pi}{t_\text{a}} \left(z - \frac{t_\text{h} + t_\text{a}}{2} \right) \right) \right]^2 \text{d}z, \\
\zeta_{i,33}^{(\text{s})} = \int_{-\frac{t_\text{h} + 2 t_\text{s}}{2}}^{-\frac{t_\text{h}}{2}} \tilde{\mu}_{33} \left[ \frac{\pi}{t_\text{s}} \sin\left(\frac{\pi}{t_\text{s}} \left(z + \frac{t_\text{h} + t_\text{s}}{2} \right) \right) \right]^2 \text{d}z, \\
\zeta_{i,33}^{(\text{a})} = \int_{\frac{t_\text{h}}{2}}^{\frac{t_\text{h} + 2 t_\text{a}}{2}} \tilde{\mu}_{33} \left[ \frac{\pi}{t_\text{a}} \sin\left(\frac{\pi}{t_\text{a}} \left(z - \frac{t_\text{h} + t_\text{a}}{2} \right) \right) \right]^2 \text{d}z, \quad i \in \{1,2\}, \\
Q_\text{h} = \frac{E_\text{h}}{1 - \nu_\text{h}^2}, \quad
\tilde{c}_{22} = c_{22} - \frac{c_{23}^2}{c_{33}}, \quad
\tilde{e}_{32} = e_{32} - \frac{c_{23} e_{33}}{c_{33}}, \\
\tilde{\mu}_{22} = \mu_{22}, \quad
\tilde{\mu}_{33} = \mu_{33} - \frac{e_{33}^2}{c_{33}}.
\end{gathered}
\label{Eq:A.11}
\end{equation}

\renewcommand{\theequation}{B.\arabic{equation}}
\section{Appendix B}
\label{sec:appendixB}
\setcounter{equation}{0}
This section outlines the key steps of the response function-based methodology used to solve the dimensionless wave equation (\ref{Eq:11}). Accordingly, the linear wave equation is first transformed into the Laplace domain ($s$), yielding the expression in Eq. (\ref{Eq:B.1}).
\begin{equation}
\begin{gathered}
\left( \frac{\partial^2}{\partial r^2} + \frac{1}{r} \frac{\partial}{\partial r} + \frac{1}{r^2} \frac{\partial^2}{\partial \theta^2} \right) \hat{p}_j - s^2 \alpha_j^2 \hat{p}_j = 0, \\
\left( \frac{\partial^2}{\partial r^2} + \frac{1}{r} \frac{\partial}{\partial r} + \frac{1}{r^2} \frac{\partial^2}{\partial \theta^2} \right) \hat{\psi}_j - s^2 \alpha_j^2 \hat{\psi}_j = 0, \quad j \in \{\text{i}, \text{g}, \text{e}\},
\label{Eq:B.1}
\end{gathered}
\end{equation}
where $\hat{p}$ and $\hat{\psi}$ represent the dimensionless acoustic pressure and velocity potential in the Laplace domain, respectively. Employing the classical method of separation of variables, the general solution for the Laplace-domain velocity potential at each acoustic medium corresponding to the wave equations in Eq. (\ref{Eq:B.1}), can be derived as shown in Eq. (\ref{Eq:B.2}) \cite{iakovlev2009interaction}:
\begin{equation}
\hat{\psi}_j(r,\theta,s) = \sum_{m=0}^{M} \left[ F_m^{(j)}(\text{s}) \text{K}_m(\alpha_j r s) + G_m^{(j)}(s) \text{I}_m(\alpha_j r s) \right] \cos(m\theta),
\label{Eq:B.2}
\end{equation}
where, $\text{I}_m$ and $\text{K}_m$ are the modified Bessel functions of the first and second kind of order $m$, while $F_m$ and $G_m$ are unknown functions. Taking into account the boundary conditions specified in Eq.(\ref{Eq:9}), along with the physical requirements of acoustic pressure attenuation with distance, its finite amplitude at the center, and angular periodicity, the Laplace-domain representations of the velocity potential components in each acoustic subdomain can be described by Eqs. (\ref{Eq:B.3}) to (\ref{Eq:B.6}).
\begin{equation}
\hat{\psi}_m^{(\text{d})}(r, \theta, s) = \hat{B}_m(s)\varXi_m^{(\text{e})}(r, s) \cos m\theta,
\label{Eq:B.3} \
\end{equation}
\begin{equation}
\hat{\psi}_m^{(r,\text{e})}(r, \theta, s) = s\hat{W}_m^{(2)}(s)\varXi_m^{(\text{e})}(r, s) \cos m\theta,
\label{Eq:B.4}
\end{equation}
\begin{equation}
\hat{\psi}_m^{(r,\text{g})}(r, \theta, s) = s\left[ \hat{W}_m^{(1)}(s) , \varXi_m^{(1)}(r, \alpha_\text{g} s) - \hat{W}_m^{(2)}(s) , \varXi_m^{(2)}(r, \alpha_\text{g} s) \right] \cos m\theta,
\label{Eq:B.5}
\end{equation}
\begin{equation}
\hat{\psi}_m^{(r,\text{i})}(r, \theta, s) = -s\hat{W}_m^{(1)}(s)\varXi_m^{(\text{i})}(r, \alpha_\text{i} s) \cos m\theta,
\label{Eq:B.6}
\end{equation}
where, $\varXi^{(\text{e})}$, $\varXi^{(1)}$, $\varXi^{(2)}$, and $\varXi^{(\text{i})}$ denote the counterparts of the corresponding response functions $\xi^{(\text{e})}$, $\xi^{(1)}$, $\xi^{(2)}$, and $\xi^{(\text{i})}$ in Laplace domain as shown in Eqs. (\ref{Eq:B.7}) through (\ref{Eq:B.10}) \cite{iakovlev2011modeling, iakovlev2009interaction, iakovlev2010hydrodynamic}:
\begin{equation}
\varXi_m^{(\text{e})}(r, s) = \frac{\text{K}_m(rs)}{s \text{K}_m^{'}(s)}, \label{Eq:B.7}
\end{equation}
\begin{equation}
\varXi_m^{(1)}(r, s) = \left(\frac{R_1}{R_2}\right)^2 \frac{\text{I}_m^{'}(s) \text{K}_m(rs) - \text{K}_m^{'}(s) \text{I}_m(rs)}{s \left[ \text{I}_m^{'}(r_1 s) \text{K}_m^{'}(s) - \text{I}_m^{'}(s) \text{K}_m^{'}(r_1 s) \right]}, \label{Eq:B.8}
\end{equation}
\begin{equation}
\varXi_m^{(2)}(r, s) = \frac{\text{I}_m^{'}(r_1 s) \text{K}_m(rs) - \text{K}_m^{'}(r_1 s) \text{I}_m(rs)}{s \left[ \text{I}_m^{'}(r_1 s) \text{K}_m^{'}(s) - \text{I}_m^{'}(s) \text{K}_m^{'}(r_1 s) \right]},
\label{Eq:B.9}
\end{equation}
\begin{equation}
\varXi_m^{(\text{i})}(r, s) = \left( \frac{R_1}{R_2} \right)^2 \frac{\text{I}_m(rs)}{s \text{I}_m^{'}(s)}.
\label{Eq:B.10}
\end{equation}
The response functions $\varXi(r,s) = \mathcal{L}\{\xi(r,t)\}$ characterize the medium's acoustic response to wave scattering or displacements at the shell surfaces. These functions are solely determined by the geometrical configuration of the model, and independent from the physical properties of the fluid and shells. The key advantage of the response function-based approach lies in its high computational efficiency. Without the need to solve the full system of wave differential equations, this method allows for direct evaluation of the acoustic pressure in all three subdomains once the shells' displacements are determined. This feature conceptually aligns these functions with Green’s functions. First introduced by Geers \cite{geers1969excitation} to obtain the transient acoustic response of circular shells, each of these functions possesses unique characteristics that have been thoroughly investigated in Yakovlev’s works, i.e. $\varXi^{(\text{e})}$ in \cite{iakovlev2008interaction_2}, $\varXi^{(1)}$ in \cite{iakovlev2015shock}, $\varXi^{(2)}$ in \cite{iakovlev2010hydrodynamic}, and $\varXi^{(\text{i})}$ in \cite{iakovlev2009interaction}.
In addition, $\hat{B}_m$, $\hat{W}_m^{(1)}$, and $\hat{W}_m^{(2)}$ are the Laplace transforms of $b_m$, $W_m^{(1)}$, and $W_m^{(2)}$, which represent the modal components of the incident wave velocity potential and the shells' normal displacements, as shown in Eqs.(\ref{Eq:B.11}) and (\ref{Eq:B.12}).
\begin{equation}
\left. \frac{\partial}{\partial r} \psi_0 (r, \theta, t) \right|_{r = r_2} = \sum_{m=0}^{M} b_m(t) \cos m\theta,
\label{Eq:B.11}
\end{equation}
\begin{equation}
w_i (\theta, t) = \sum_{m=0}^{M} W_m^{(i)}(t) \cos m\theta, \quad j \in \{1, 2\}.
\label{Eq:B.12}
\end{equation}
According to the Laplace-domain Eqs. (\ref{Eq:B.3}) through (\ref{Eq:B.6}), once $b_m$, $W_m^{(1)}$, and $W_m^{(2)}$ are determined considering the modal orthogonality,  acoustic pressure modal components in Eq. (\ref{Eq:16}) can be expressed as time domain convolutions as shown in Eq. (\ref{Eq:B.13}):
\begin{equation}
\begin{gathered}
P_m^{(\text{d})}(r,t) = -\frac{1}{\sqrt{r}} b_m(t) - \int_0^t b_m(\eta) \frac{\text{d}}{\text{d}\eta}  \xi_m^{(\text{e})}(r, t - \eta) \text{d}\eta, \\
P_m^{(\text{r},\text{e})}(r,t) = -\int_0^t \frac{\text{d}^2 W_m^{(2)}(\eta)}{\text{d}\eta^2} \xi_m^{(\text{e})}(r, t - \eta) \text{d}\eta, \\
P_m^{(\text{r},\text{g})}(r,t) = \frac{\rho_\text{g} c_\text{g}}{\rho_\text{e} c_\text{e}} \left[ \int_0^t \frac{\text{d}^2 W_m^{(2)}(\eta)}{\text{d}\eta^2} \xi_m^{(2)}\left(r, \frac{c_\text{g}}{c_\text{e}}(t - \eta)\right) \text{d}\eta \right. \\
\left. - \int_0^t \frac{\text{d}^2 W_m^{(1)}(\eta)}{\text{d}\eta^2} \xi_m^{(1)}\left(r, \frac{c_\text{g}}{c_\text{e}}(t - \eta)\right) \text{d}\eta \right], \\
P_m^{(\text{r},\text{i})}(r,t) = \frac{\rho_\text{i} c_\text{i}}{\rho_\text{e} c_\text{e}} \int_0^t \frac{\text{d}^2 W_m^{(1)}(\eta)}{\text{d}\eta^2} \xi_m^{(1)}\left(r, \frac{c_\text{i}}{c_\text{e}}(t - \eta)\right) \text{d}\eta,
\label{Eq:B.13}
\end{gathered}
\end{equation}
Moreover, based on Eq. (\ref{Eq:16}), the modal components of the dimensionless incident wave pressure, \( P_m^{(0)} \), can be achieved as shown in Eq. (\ref{Eq:B.14}):
\begin{equation}
P_m^{(0)}(r,t) = \frac{1}{\beta \pi} \int_0^{2\pi} p_0(r,\theta,t) \cos(m \theta) \, d\theta, \quad
\beta = \begin{cases} 2 \quad m=0, \\
1 \quad m \neq 0,
\end{cases}
\label{Eq:B.14}
\end{equation}
where \( p_0 = \rho_\text{e}^{-1} c_\text{e}^{-2} \bar{p}_0 \) is the dimensionless incident wave acoustic pressure. As demonstrated, adopting the response function–based approach considerably simplifies solving the governing vibroacoustic equations. Nevertheless, applying the inverse Laplace transform to Eqs. (\ref{Eq:B.7}) through (\ref{Eq:B.10}) to obtain these functions in the time domain is mathematically intricate and requires techniques such as the Cauchy Residue Theorem and Jordan’s Lemma. Since detailed derivations are available in the literature \cite{iakovlev2008interaction_2, iakovlev2006external, iakovlev2015shock}, they are omitted here for brevity.

\section*{References}
\printbibliography
\justifying

\end{document}